\documentclass[11pt, a4paper, oneside]{book}

\usepackage[english]{babel}
\usepackage[utf8]{inputenc}
\usepackage{graphicx} %include figure files
\usepackage{amsmath}
\usepackage{amssymb}
\usepackage{units}
\usepackage{tikz}
\usepackage{natbib}
\usepackage{color}
\usepackage{rotating}

\usepackage{feynmp}
\usepackage{ifpdf}
\ifpdf
\fi
\makeatletter
\def\endfmffile{%
  \fmfcmd{\p@rcent\space the end.^^J%
          end.^^J%
          endinput;}%
  \if@fmfio
    \immediate\closeout\@outfmf
  \fi
  \IfFileExists{\thefmffile.mp}{\immediate\write18{mpost \thefmffile}}{}
  \let\thefmffile\relax
}
\makeatother

\newcommand{\nB}{n_\mathrm{B}}

\title{Early evolution of newly born proto-neutron stars}
\author{Giovanni Camelio}

\begin{document}

\begin{titlepage}
\centering
\vspace*{.5cm} \Huge \textbf{Early evolution of newly born proto-neutron
stars}\\
\vspace*{.5cm}\LARGE\textbf{Giovanni Camelio}\\
\vspace*{2cm}
(minor modifications to the)\\
Thesis submitted for the degree of\\
Doctor of Philosophy in Physics\\
\vspace*{.5cm}
Defended on 25th January 2017\\
University of Rome ``Sapienza''\\
\vspace*{.5cm}
Advisors:\\
prof.~Valeria~Ferrari\\
prof.~Leonardo~Gualtieri\\
\end{titlepage}

\begin{titlepage}
\null
\vspace{\stretch{1}}
\begin{flushright}
a mia madre\\
stella fra le stelle
\end{flushright}
\vspace{\stretch{4}}
\null
\end{titlepage}

\chapter*{Acknowledgments}

I am grateful to Valeria~Ferrari, Leonardo~Gualtieri, Omar~Benhar,
Jos\'e~A.~Pons, Alessandro~Lovato, Tiziano~Abdelsalhin, and Noemi~Rocco
for their time, advises, teachings, and the scientific discussions we had.

I also whish to thank the external referees of this thesis, Juan~Antonio~Miralles
and Ignazio~Bombaci, and the unknown referees of my papers, for their
suggestions and comments.\\

During my PhD I have been partially supported by “NewCompStar”, COST Action MP1304.

\tableofcontents

\chapter{Introduction}

When a star with a mass greater than about $\unit[8]{M_\odot}$ reaches the end
of its evolution, it has a massive stratified core which keeps on accreting
material as the star consumes its fuel.  This core is supported by the electron
degeneracy pressure. When, because of the ongoing accretion, the core mass
overtakes the so-called Chandrasekhar limit
$M_\mathrm{Ch}\simeq\unit[1.4]{M_\odot}$, the electron degeneracy pressure can
not support the core anymore and it sudden collapses. During the collapse, the
matter density increases until it reaches the nuclear density, which is about
$\unit[0.16]{fm^{-3}}$ ($\unit[2.68\times10^{14}]{g/cm^3}$). At this point, the
nucleon degeneracy pressure sets in, the collapse stops, and the external
layer of the core bounces off.  Then, the core continues to contract at a much
milder rate as it loses thermal support, and the bouncing material forms a
shock wave that moves upward. However at a certain point the shock wave stalls
because it has not enough energy to wipe out the stellar envelope.  For a long
time the problem to identify the mechanism that revitalizes the shock wave and
permits to complete the SN explosion has been unsolved.  \citet{Bethe+Wilson.1985} for the first
time identified the neutrinos as the agents responsible of that, in a process
called \emph{delayed explosion mechanism}. In this mechanism, it is the
late-time energy deposition due to the neutrino flux that revitalize the shock
wave.  After the initial enthusiasm, it is now believed that neutrinos delayed
energy deposition is not enough to trigger the explosion, at least in many kind
of SN progenitors, and new mechanisms (like the convective instability) have
been explored. Even today the problem of SN explosion is not completely solved.

The explosion of a nearby supernova (SN) in the Large Magellanic Clouds in 1987
(the renowned event SN1987a) has been a milestone for both astrophysics and
particle physics. In fact, concurrently with the electromagnetic signal, about
19 neutrinos have been observed in two Cherenkov detectors that were operating
at that time. These neutrinos, even though too few to significantly constrain the
supernova physics, had deepened our understanding of both high density physics
and the astrophysical processes that undergo in the stellar interior.

Since then, great efforts have been undertaken in trying to understand the SN
explosion mechanism, and numerical codes with increasing complexity have been
set up, passing from one dimension to two and then to three dimensions and
increasing in number and complexity the physical ingredients adopted, for
example including general relativity and a more realistic treatment of neutrino
cross sections with matter.  However, the race to understand the explosion
mechanism resulted in a relatively smaller attention to the simpler and longer
evolution of the supernova core. This core, if the progenitor star is in the
mass range $8$--$\unit[25]{M_\odot}$, slowly contracts to a neutron star, and
it is therefore called proto-neutron star (PNS).  In the first tenths of
seconds after core-bounce, the PNS is turbulent and characterized by large
instabilities, but during the next tens of seconds, it undergoes a more quiet,
``quasi-stationary'' evolution (the Kelvin-Helmholtz phase), which can be
described as a sequence of equilibrium configurations. This phase is
characterized by an initial increase of the PNS temperature as the neutrino
degeneracy energy is transferred to the matter and the PNS envelope rapidly
contracts, and then by a general deleptonization and cooling. After tens of
seconds, the temperature becomes lower than about $\unit[5]{MeV}$
($\unit[5.8\times10^{10}]{K}$) and the neutrinos mean free path is greater than
the stellar radius. The PNS becomes transparent to neutrinos, and a ``mature''
neutron star is born.\\

For a long time, the longer timescales of the PNS evolution were prohibitive
for the complex core-collapse codes, which rarely explored more than few
fractions of second after the core-bounce. Just recently, some supernova codes
have been employed to explore the PNS evolution \citep[e.g.,][]{Fischer+2010,
Hudepohl+2010, Hudepohl+erratum}, in particular to explore the neutrino wind
and the subsequent nucleosynthesis.  These codes are not adapted to describe
the PNS phase, and have some limitations.  For example, \citet{Hudepohl+2010}
used a Newtonian core-collapse 1D code, including the general relativistic
corrections only in an effective way, and \citet{Fischer+2010} are more interested
in the nucleosynthesis in the supernova external layers whose dimension is far
bigger than the PNS one.

However, the quasi-stationary Kelvin-Helmholtz phase may be more easily studied
with ad-hoc, simpler and faster codes.  In 1986 Burrows and Lattimer
\citep{Burrows+Lattimer.1986} for the first time performed a numerical
simulation of the Kelvin-Helmholtz phase of the PNS evolution. They wrote a
general relativistic, one-dimensional and energy averaged (i.e., they assumed a
neutrino Fermi-Dirac distribution function) code.  They studied the qualitative
evolution of a PNS with a simplified EoS and a simplified treatment of the
neutrino cross-sections, that does not fully account for the particles finite
degeneracy. In a subsequent paper, \citet{Burrows.1988} studied how the
neutrino signal on a Cherenkov detector depends on the PNS physics, i.e., the
stiffness of the stellar equation of state (EoS), the accretion process,
and the possible formation of a black hole.

\citet{Keil+Janka.1995} wrote a PNS evolutionary code (general relativistic,
one-dimensional, and energy averaged) to study how the EoS, and in particular
the presence of hyperons, influences the evolution.  They included thermal
effects in the EoS in a simplified way (only at the free gas level), used
neutrino cross sections similar to those of \citet{Burrows+Lattimer.1986}, and
assumed beta-equilibrium. They also studied the black hole formation process,
finding that it would not produce a delayed neutrino outburst.

In 1999 Pons and collaborators \citep{Pons+.1999} wrote a  general
relativistic, one-dimensional, energy averaged code to study the dependence of
the PNS evolution on the EoS and the total baryonic PNS mass, and the black
hole formation process. They used nucleonic and hyperonic mean-field EoSs
\citep{Glendenning.1985}, including the thermal contribution consistently. The
effects of the temperature, finite degeneracy, and baryon interaction (at the
mean-field level) have been fully included in the treatment of the neutrino
cross-sections \citep{Reddy+Prakash+Lattimer.1998}.  In subsequent papers, they
also allowed for hadron-quark transition in the PNS \citep{Pons+2001} and
consistently included the convection in their evolution
\citep{Miralles+Pons+Urpin.2000, Roberts+2012}.

\citet{Roberts.2012} wrote a new PNS evolutionary code to study the effects of
the neutrino wind on the nucleosynthesis process. He included the GM3
mean-field nucleonic EoS \citep{Glendenning+Moszkowski.1991} with table
interpolation, and has determined the neutrino opacities accounting for the
effects of finite temperature, finite degeneracy, baryon interaction and weak
magnetism. His code is general relativistic, one-dimensional, multi-group
(i.e., different neutrino energy bins are evolved separately) and multi-flavour
(i.e., each neutrino flavour is evolved separately).\\

The accurate study of the PNS phase is important because most neutrinos are
emitted in this phase, and moreover the turbulent and rapid core-collapse and
core-bounce processes perturb the PNS, exciting its vibration modes, the
so-called quasi-normal modes (QNMs).  A quasi-normal mode in general relativity
is a solution of the stellar perturbation equations, which is regular at the
center of the star, continuous at the stellar border, and behaves as a pure
outgoing wave at infinity.  A relativistic star loses energy through the
gravitational wave (GW) emission associated with the QNMs. There are several
classes of QNMs that are classified on the basis of the restoring force which
prevails in restoring the equilibrium position of the perturbed fluid element.
The fundamental mode is the principal stellar oscillation mode and does not
exhibit radial nodes in the corresponding eigenfunction, generally excited in
most astrophysical processes.

\citet{Andersson+Kokkotas.1998a} and \citet{Andersson+Kokkotas.1998b} studied
the frequency and damping time dependence of some stellar pulsation modes on
global neutron star properties, like the radius and the gravitational mass, for
many EoSs. Later, \citet{Benhar+Ferrari+Gualtieri.2004} updated the analysis for
more modern EoSs. These analysis found general trends that are not EoS
dependent, for example, the fundamental mode frequency has a linear dependence
on the square root of the mean neutron star density, $\sqrt{M/R^3}$, where $M$ is
the neutron star gravitational mass and $R$ its radius. These results generalizes the results
of the Newtonian theory of stellar oscillations.

\citet{Ferrari+Miniutti+Pons.2003} studied the time evolution of the
quasi-normal mode frequencies during the first minute of the PNS life, for the
mean-field GM3 EoS of \citet{Pons+.1999} and for the hadron-quark EoS of
\citet{Pons+2001}. The PNS was evolved consistently using the code described in
\citet{Pons+.1999}.  They found that in the first second, the QNMs do not show
the scaling with the mass and radius typical of cold neutron stars; for example
the fundamental mode frequency is not proportional to the square root mean
density. Later, \citet{Burgio+2011} studied the QNMs of a PNS adopting the
many-body EoS of \citet{Burgio+Schulze.2010}. They simulated the PNS time
evolution adopting some reasonable time dependent thermal and composition
profiles, that are qualitatively similar to those obtained by a consistent
evolution, finding results similar to those of
\citet{Ferrari+Miniutti+Pons.2003}.
The entropy and lepton fraction profiles were included in a similar way
in~\cite{Sotani:2016uwn}, in order to mimic the profiles obtained, in the
first second after bounce, by numerical core-collapse simulations. However,
the EoSs they employed (such as that of Lattimer and
Swesty~\cite{Lattimer+Swesty.1991}) are more appropriate to describe the
core-collapse phase than the PNS evolution. 
We remark that, up to now, finite temperature, many-body nuclear dynamics have
not been included in a consistent way (i.e., accounting for the modifications in
the neutrino cross sections) in PNS evolution.\\

When a supernova explodes, the contracting core is thought to be rapidly
rotating. In the PNS phase, a huge amount of angular momentum is released
through neutrino emission.  An accurate modeling  of this phase is needed, for
instance, to compute the frequencies of the PNS quasi-normal modes,
and the rotational contribution to gravitational waves.  Moreover, it provides
a link between supernova explosions, a phenomenon which is still not fully
understood, and the properties of the observed population of young pulsars.
Current models of the evolution of progenitor stars
\citep{Heger+Woosley+Spruit.2005}, combined with numerical simulations of core
collapse and explosion (see e.g.~\citealp{Thompson+Quataert+Burrows.2005,
Ott+2006, Hanke+2013, Couch+Ott.2015, Nakamura+2014}), do not allow
sufficiently accurate estimates of the expected rotation rate of newly born
PNSs; they only show that the minimum rotation period at the onset of the
Kelvin-Helmoltz phase can be as small as few ms, if the spin rate of the
progenitor is sufficiently high.  On the other hand, astrophysical observations
of young pulsar populations (see \citealp{Miller+Miller.2014}, and references
therein) show typical periods $\gtrsim~100$ ms.

The evolution of rotating PNSs has been studied in \cite{Villain+.2004}, where
the  thermodynamic profiles obtained in \cite{Pons+.1999} for a non-rotating
PNS evolved with the GM3 EoS were employed as effective one-parameter EoSs; the
rotating configurations were obtained using the non-linear BGSM code
\citep{Gourgoulhon+1999} to solve Einstein's equations. A similar approach has
been followed in \cite{Martinon+2014}, which used  the profiles of
\cite{Pons+.1999} and \cite{Burgio+2011}. The main limitations of these works
is that the evolution of the PNS rotation rate is due not only to the change of
the moment of inertia (i.e., to the contraction), but also to the angular
momentum change due to neutrino emission \citep{Epstein.1978}.  This was
neglected in \citet{Villain+.2004}, and described with a heuristic formula in
\citet{Martinon+2014}.  Moreover, when the PNS profiles describing a
non-rotating star are treated as effective EoSs, one can obtain
configurations which are unstable to radial perturbations, unless particular
care is taken in modelling the effective EoS. In fact, this instability is not physical and
depends on the procedure adopted to obtain the effective EoS.\\

The main goal of this thesis is to study the frequencies of the QNMs associated
to the gravitational wave emission in the PNS phase.  To accomplish that, we
have written a new one-dimensional, energy-averaged and flux-limited PNS
evolutionary code. We have studied the PNS evolution and the QNMs for three
nucleonic EoSs and for three baryon masses \citep{Camelio+2017}. In particular, for the first time we have
consistently evolved a PNS with a many-body EoS, found by \citet{Benhar+Lovato2017};
\citeauthor[][forthcoming]{Lovato+prep}. We have also used the evolutionary profiles
obtained with our code to study the evolution of a rotating star, with rotation
included in an effective way. In so doing, we have adopted a procedure to
include rotation that does not give rise to nonphysical instabilities
and we have consistently accounted for the angular momentum loss due to neutrinos \citep{Camelio+2016}.
Some of the results discussed in this thesis are published in \citet{Camelio+2016, Camelio+2017},
and others will be reported in 
\citeauthor[][forthcoming]{Lovato+prep}.

In Chapter~\ref{cha:EoS} we introduce the nucleonic EoSs used in this thesis,
and we describe a new fitting formula to model the interacting part of the
baryon free energy. In Chapter~\ref{cha:evolution} we describe our PNS
evolutionary code and we study the evolution and neutrino signal in terrestrial
detectors for the three nucleonic EoSs described in Chapter~\ref{cha:EoS} and
for three stellar baryon masses.  In Chapter~\ref{cha:GW} we illustrate the
theory of quasi-normal modes from stellar perturbations in general relativity,
and show our results for the EoSs analyzed in Chapter~\ref{cha:EoS}.  In
Chapter~\ref{cha:rotation} we include in an effective way slow rigid rotation
in the PNS and study the evolution of the rotation rate and of the angular momentum. In
Chapter~\ref{cha:conclusions} we draw our conclusions and the outlook of this
work. In Appendix~\ref{app:eos}, we derive some analytic Fermion
non-interacting EoSs. In Appendix~\ref{app:code_checks}, we make some code
checks and demonstrate the validity of our approximations. In
Appendix~\ref{app:inverse}, we elucidate the formulae needed to compute the
neutrino inverse processes.\\

The recent detection of the gravitational wave emission from two merging black
holes \citep{Abbot+2016} has opened a new observational window on our universe,
as the neutrino detection from a supernova in 1987 did.  Our hope is that this
thesis would contribute to exploit the great opportunity that gravitational
waves give us to understand some still unsolved problems in fundamental physics and
astrophysics.

\section{Units and constants}
\label{sec:units}

Unfortunately, astrophysics, nuclear physics and GR communities do not
``speak'' the same language, in the sense that astrophysics use $\unit{cgs}$
units (but energies are often reported in $\unit{MeV}$ and masses in
$\unit{MeV\,c^{-2}}$), nuclear physics use natural units with $c=\hbar=1$ and
lengths measured in $\unit{fm}$ (but sometimes $h=1$ instead of $\hbar=1$), GR physics use natural units with $c=G=1$ and
lengths measured in $\unit{km}$ (but sometimes also the sun mass is set equal to one $M_\odot=1$ and therefore
lengths are measured in units of $c^{-2}GM_\odot$).  Since this thesis is at the interface between
micro- and macro-physics, it is necessary to relate results reported in
different units.  In Tab.~\ref{tab:units} we report the dimensions of some physical quantities
in the different unit systems. In Tab.~\ref{tab:constants} we report the value of some physical constants,
expressed in various units; this table can be used to convert the physical quantities between
the different units\footnote{An interesting discussion on the role of physical
units and dimensions may be found in \cite{Duff+Okun+Veneziano.2002} and
\cite{Duff.2015}.}.

Unless stated differently, in this thesis we set to unity the speed of light $c=1$, the gravitational
constant $G=1$, and the Boltzmann constant $k_B=1$. However, microphysical masses
and energies (like those of particles) are expressed in $\unit{MeV\,c^{-2}}$
and $\unit{MeV}$, respectively, whereas macrophysical masses and energies will be
expressed in units of Sun masses $M_\odot$ and in $\unit{erg}$, respectively.

\begin{table}
\centering
\caption{Dimensions of some physical quantities in cgs (fundamental units: cm, g, s), micro (c,
$\hbar$, fm), micro$^\ast$ (c, MeV, fm) and macro (c, G, km) unit systems.
Note that placing $k_\mathrm{B}=1$ is equivalent to express the temperature with
energy dimension ($k_\mathrm{B}$ play the role of a definition for the
temperature).}
\begin{tabular}{ccccc}
\hline
quantity & $\unit{cgs}$ & micro & micro$^\ast$& macro \\
\hline
length & $\unit{cm}$ & $\unit{fm}$ & $\unit{fm}$ & $\unit{km}$ \\
time & $\unit{s}$  & $\unit{c^{-1}\,fm}$ & $\unit{c^{-1}\,fm}$ &
$\unit{c^{-1}\,km}$\\
mass & $\unit{g}$ & $\unit{c^{-1}\,\hbar\,fm^{-1}}$ & $\unit{c^{-2}\,MeV}$ &
$\unit{c^2\,G^{-1}\,km}$ \\
energy & $\unit{erg}\equiv\unit{cm^2\,g\,s^{-2}}$ & $\unit{c\,\hbar\,fm^{-1}}$ & $\unit{MeV}$ &
$\unit{c^4\,G^{-1}\,km}$ \\
action & $\unit{cm^2\,g\,s^{-1}}$ & $\unit{\hbar}$ & $\unit{MeV\,c^{-1}\,fm}$ &
$\unit{c^3\,G^{-1}\,km^2}$ \\
temperature & $\unit{k_B^{-1}\,cm^2\,g\,s^{-2}}$ &
$\unit{k_B^{-1}\,c\,\hbar\,fm^{-1}}$ & $\unit{k_B^{-1}\,MeV}$ &
$\unit{k_B^{-1}\,c^4\,G^{-1}\,km}$ \\
pressure & $\unit{cm^{-1}\,g\,s^{-2}}$ & $\unit{c\,\hbar\,fm^{-4}}$ &
$\unit{MeV\,fm^{-3}}$ &
$\unit{c^4\,G^{-1}\,km^{-2}}$\\
entropy & $\unit{k_B}$ & $\unit{k_B}$ & $\unit{k_B}$ & $\unit{k_B}$ \\
\hline
\end{tabular}
\label{tab:units}
\end{table}
\begin{table}
\centering
\caption{Value of some physical constants in different units
($M_\odot$ is the sun mass and $g_\nu$ is the neutrino degeneracy).}
\begin{tabular}{crl}
\hline
quantity & value & units\\
\hline
$c$ & $2.99792458\times10^{10}$ & $\unit{cm\,s^{-1}}$ \\
$\hbar$ & $1.054571726(47)\times10^{-27}$ & $\unit{cm^2\,g\,s^{-1}}$\\
$G$ & $6.67384(80)\times10^{-8}$ & $\unit{cm^3\,g^{-1}\,s^{-2}}$\\
$\hbar c$ & $197.3269631(49)$ & $\unit{MeV\,fm}$\\
$\unit{MeV}$ & $1.602176565(35)\times10^{-6}$ & $\unit{erg}$ \\
$\unit{MeV\,c^{-2}}$ & $1.782661758(44)\times10^{-27}$ & $\unit{g}$ \\
$\unit{M_\odot}$ & $1.9884(2)\times10^{33}$ & $\unit{g}$\\
& $1.47664$ &$\unit{km\,c^2\,G^{-1}}$\\
$m_{n}$& $939.565(36)$ &$\unit{MeV\,c^{-2}}$\\
$m_{e}$& $0.510998910(13)$ &$\unit{MeV\,c^{-2}}$\\
$k_\mathrm{B}$& $1.3806504(24)\times10^{-16}$ & $\unit{erg\,K^{-1}}$\\
$g_\nu$&1&\#\\
\hline
\end{tabular}
\label{tab:constants}
\end{table}

\section{Abbreviations}

\begin{itemize}
\item GR = general relativity;
\item NS = neutron star;
\item PNS = proto-neutron star;
\item SN(e) = supernova(e);
\item EoS = equation of state (see Chapter~\ref{cha:EoS});
\item LS-bulk = bulk equation of state of \cite{Lattimer+Swesty.1991} (see
Sec.~\ref{sec:LSbulk} of this thesis);
\item GM3 = third choice of parameter of the \cite{Glendenning+Moszkowski.1991}
equation of state (see Sec.~\ref{sec:GM3} of this thesis);
\item CBF-EI = correlated basis function -- effective interaction (equation of
state, \citet{Benhar+Lovato2017}; \citeauthor[][forthcoming]{Lovato+prep}; see Sec.~\ref{sec:CBFEI}
of this thesis);
\item SNM = symmetric neutron matter, see Sec.~\ref{sec:fBfit};
\item PNM = pure neutron matter, see Sec.~\ref{sec:fBfit};
\item TOV = Tolman--Oppenheimer--Volkoff equation(s) (see Sec.~\ref{sec:tov});
\item BLE = Boltzmann--Lindquist equation (see Sec.~\ref{sec:BLE}).
\item GW = gravitational wave;
\item QNM = quasi-normal mode (see Chapter~\ref{cha:GW}).
\end{itemize}

\chapter{The equation of state}
\label{cha:EoS}

An ``old'' neutron star (i.e., after minutes from its birth) has a temperature
of $T\simeq\unit[10^9]{K}$. Even if this temperature may seem very high, the
corresponding thermal energy is only a tiny fraction of the nucleons internal
Fermi energy, $T\simeq\unit[10^9]{K} \simeq \unit[0.1]{MeV} \ll \mathcal
E_\mathrm{F} - m_n \simeq \unit[30]{MeV}$, where $m_n$ is the neutron mass.
Then, one can use a zero temperature approximation to describe the equation of
state (EoS) of an old neutron star (NS). In this way, the EoS depends only on
one independent variable (e.g.{} the baryon number or the pressure) and it is
called \emph{barotropic}.

Conversely, after about $\unit[200]{ms}$ from the core bounce following a
supernova explosion, the contracting core (i.e., the proto-neutron star, PNS)
thermal energy is higher or comparable to the nucleons Fermi energy
($T\simeq\unit[40]{MeV} \geq E_\mathrm{F}\simeq\unit[30]{MeV}$, see Fig.~\ref{fig:thermo_n})
and hence one cannot use the zero temperature approximation to describe
the EoS. One of the consequences is that at such high temperatures the matter EoS
depends on more than one independent variable, and it is therefore
called \emph{non-barotropic}.  In addition, in the PNS phase neutrinos are the
only particles that diffuse in the star, moving energy and lepton number
through the stellar layers. However, they do not leave immediately the PNS as
they are produced, since at such high temperatures and densities their
mean free path is far smaller than the stellar radius.  The neutrino mean free
path depends on the microphysical theory adopted to describe the matter;
therefore, to understand the PNS evolution, one has to consistently account for
the underlying EoS both to determine the PNS structure and to asses the
neutrino diffusion magnitude.

In this chapter we describe and compare the three EoSs adopted in this thesis.
In Sec.~\ref{sec:thermodynamical}, we obtain several useful general and less
general thermodynamic relations.  In
Sec.~\ref{sec:diffusion}, we describe the nuclear reactions that we account for
in neutrino diffusion, and we introduce the effective description of the baryon
single particle spectra, which permit to include the microphysical effects of
interaction in the determination of the neutrino mean free path in a general
way (i.e., it may be applied to any EoS).  In the following three sections we
describe the three EoSs that we consider in this thesis. In
Sec.~\ref{sec:LSbulk} we describe the LS-bulk EoS (corresponding to the bulk of
\citealp{Lattimer+Swesty.1991}), which is obtained from the extrapolation of
the nuclear properties of terrestrial nuclei at high temperature and density.
In Sec.~\ref{sec:GM3} we describe the GM3 mean-field EoS
\citep{Glendenning+Moszkowski.1991}. In Sec.~\ref{sec:CBFEI} we present a
many-body EoS based on the correlated basis function and the effective
interaction theory (CBF-EI EoS, see \citealp{Benhar+Lovato2017}; and
\citeauthor[][forthcoming]{Lovato+prep}).
In Sec.~\ref{sec:fBfit} we develop a general fitting formula for the baryon
part of the EoS, that can be used to speed up the determination of the thermodynamical
quantities inside the star.  This fitting formula is the main original
contribution presented in this chapter, and it is reported also in
\citet{Camelio+2017}. In Sec.~\ref{sec:totaleos} we explain
how to obtain the total EoS from the fitting formula for the baryon part of the
EoS.  Finally, in Sec.~\ref{sec:comparison} we compare the thermodynamical
quantities, the mean free paths and the diffusion coefficients of the three
EoSs described in this chapter.

In this thesis, the particle energies and chemical potentials are defined
including the rest mass. The (total) EoSs we consider are composed by protons, neutrons,
electrons, positrons, and neutrinos and antineutrinos of all three flavours.

\section{Thermodynamical relations}
\label{sec:thermodynamical}
In this section we enunciate and discuss some useful thermodynamical relations, that
will be used later in this thesis.

The first principle of the thermodynamics states that the variation of the \emph{energy} $E$ of a system is given by
\begin{equation}
\label{eq:dE}
\mathrm dE=T\mathrm dS-P\mathrm dV+\sum_i\mu_i\mathrm dN_i,
\end{equation}
where $T$ is the temperature, $S$ the total entropy, $P$ the pressure, $V$ the volume, and
$\mu_i$ and $N_i$ the chemical potential and total number of the particle $i$.
Therefore, the most natural choice for the independent variables on which the energy depends is $E=E(S,V,\{N_i\})$
and
\begin{align}
\label{eq:dE_dS}
T={}&\left.\frac{\partial E}{\partial S}\right|_{V,\{N_i\}},\\
\label{eq:dE_dV}
P={}&\left.-\frac{\partial E}{\partial V}\right|_{S,\{N_i\}},\\
\label{eq:dE_dNi}
\mu_i={}&\left.\frac{\partial E}{\partial N_i}\right|_{V,S,\{N_{j\ne i}\}}.
\end{align}
One can define the \emph{free energy} by means of a Legendre
transformation:
\begin{equation}
\label{eq:F}
F=E-TS,
\end{equation}
and therefore,
\begin{align}
\label{eq:dF}
\mathrm dF={}&-S\mathrm dT-P\mathrm dV+\sum_i\mu_i\mathrm dN_i,\\
F={}&F(T,V,\{N_i\}),\\
\label{eq:dF_dT}
S={}&\left.-\frac{\partial F}{\partial T}\right|_{V,\{N_i\}},\\
\label{eq:dF_dV}
P={}&\left.-\frac{\partial F}{\partial V}\right|_{T,\{N_i\}},\\
\label{eq:dF_dNi}
\mu_i={}&\left.\frac{\partial F}{\partial N_i}\right|_{V,T,\{N_{j\ne i}\}}.
\end{align}

In a stellar system, the total number of baryons is very big
$A\simeq10^{57}$, and moreover the intensive quantities (pressure, temperature, and chemical potential) change
from point to point inside the star. Therefore, it is more natural to introduce
the \emph{average} of the extensive thermodynamical quantities at a certain point
of the star. To do so, one should average over an amount of particles which is big enough to permit a statistical description
of their properties, but whose extent is far smaller than the stellar scale length (e.g., the pressure scale height).
In this context, it is useful to consider the average thermodynamical quantities \emph{per baryon},
since the number of baryons is conserved by all type of microphysical interactions. This approach is equivalent to consider a system
with a fixed number of baryons $N_\mathrm{B}$.  The first law of the thermodynamics becomes
\begin{align}
\label{eq:de}
\mathrm de={}&T\mathrm ds+\frac{P}{n_\mathrm{B}^2}\mathrm
dn_\mathrm{B}+\sum_i\mu_i\mathrm dY_i\qquad&\mbox{$N_\mathrm{B}$ constant},\\
\label{eq:flegendre}
f={}&e-Ts\qquad&\mbox{$N_\mathrm{B}$ constant},\\
\label{eq:df}
\mathrm df={}&-s\mathrm dT+\frac{P}{n_\mathrm{B}^2}\mathrm
dn_\mathrm{B}+\sum_i\mu_i\mathrm dY_i\qquad&\mbox{$N_\mathrm{B}$ constant},\\
\label{eq:de_ds}
T={}&\left.\frac{\partial e}{\partial s}\right|_{n_\mathrm{B},\{Y_i\}}\qquad&\mbox{$N_\mathrm{B}$ constant},\\
\label{eq:df_dT}
s={}&\left.-\frac{\partial f}{\partial T}\right|_{n_\mathrm{B},\{Y_i\}}\qquad&\mbox{$N_\mathrm{B}$ constant},\\
\label{eq:P1}
P={}&\left.n_\mathrm{B}^2\frac{\partial e}{\partial
n_\mathrm{B}}\right|_{s,\{Y_i\}} =\left.n_\mathrm{B}^2\frac{\partial
f}{\partial n_\mathrm{B}}\right|_{T,\{Y_i\}}\qquad&\mbox{$N_\mathrm{B}$ constant},\\
\label{eq:mu1}
\mu_i={}&\left.\frac{\partial f}{\partial
Y_i}\right|_{T,n_\mathrm{B},\{Y_{j\neq i}\}}= \left.\frac{\partial e}{\partial
Y_i}\right|_{s,n_\mathrm{B},\{Y_{j\neq i}\}}\qquad&\mbox{$N_\mathrm{B}$ constant},
\end{align}
where $e=E/N_\mathrm{B}$, $f=F/N_\mathrm{B}$, $s=S/N_\mathrm{B}$, and $Y_i=N_i/N_\mathrm{B}$
are the energy
per baryon, the free energy per baryon,
the entropy per baryon, and the number of particle $i$ per
baryon (that
is also called particle number fraction, particle fraction, or particle
abundance), respectively. The $i$-th particle number density is $n_i=N_i/V$, and 
the baryon number density is $\nB=N_\mathrm{B}/V\equiv n_p+n_n$ (if the only baryons
are protons and neutrons)\footnote{Nuclear physicists adopt the notation $A\equiv
N_\mathrm{B}$, $\rho\equiv n_\mathrm{B}$ and $x_i\equiv Y_i$, and often
$x\equiv Y_p$.}.

Eqs.~\eqref{eq:de} and \eqref{eq:df} suggest that the energy and the free
energy per baryon may be written as $e=e(s,n_\mathrm{B},\{Y_i\})$ and
$f=f(T,n_\mathrm{B},\{Y_i\})$. However, the number fractions are
not totally independent from each other. In fact, having fixed the baryon number implies that $\mathrm
dY_p=-\mathrm d Y_n$ (in an
EoS where the only baryons are neutrons and protons). Moreover, in a realistic EoS, there are additional
relations between the number fractions $Y_i$ (see below). Therefore, it is impossible to differentiate with respect
to $Y_p$ (or $Y_n$), fixing at the same time $n_\mathrm{B}=n_n+n_p$ and $Y_n$
(or $Y_p$).  In this case, one can consider the average thermodynamical quantities
in a given volume $V$ (that is, one can fix the volume), and rewrite Eqs.~\eqref{eq:dE} and
\eqref{eq:dF} as
\begin{align}
\label{eq:depsilon}
\mathrm d \epsilon={}& T \mathrm d\sigma +\sum_i \mu_i \mathrm d
n_i\qquad&\mbox{$V$ constant},\\
\label{eq:dvarphi}
\mathrm d \varphi={}& -\sigma \mathrm dT+\sum_i \mu_i \mathrm d n_i&\mbox{$V$
constant},
\end{align}
where $\epsilon=E/V$, $\varphi=F/V$, and $\sigma=S/V$ are the energy density, the
free energy density, and the entropy density, respectively.  Since
$n_\mathrm{B}=N_\mathrm{B}/V$, we can also write the densities as
$\epsilon=n_\mathrm{B}e$, $\varphi=n_\mathrm{B}f$, and $\sigma=n_\mathrm{B}s$
(beware that fixing the volume $V$ is not equivalent of fixing the
baryon number in that volume, and therefore is not equivalent of fixing $\nB$).
From Eqs.~\eqref{eq:depsilon} and \eqref{eq:dvarphi}, one obtains the chemical potential of the
$i$-th species
\begin{equation}
\label{eq:mu2}
\mu_i=\left. \frac{\partial ( n_\mathrm{B} f )}{\partial n_i}
\right|_{T,\{n_{j\neq i}\}} =\left. \frac{\partial (n_\mathrm{B} e ) }{\partial
n_i} \right|_{s,\{n_{j\neq i}\}}.
\end{equation}

There is another important relation that is worth introducing before specialize
the discussion to the stellar case.
One can write Eq.~\eqref{eq:dE} in terms of $E=V\epsilon$, $S=V\sigma$, $N_i=Vn_i$,
\emph{without} fixing the volume or the baryon number,
\begin{align}
 \epsilon\mathrm dV+V\mathrm d\epsilon={}&
T\sigma\mathrm dV + TV\mathrm d\sigma - P\mathrm dV + \sum_i\big(
\mu_in_i\mathrm dV+ \mu_iV\mathrm dn_i\big),\\
\mathrm d\epsilon= {}& T\mathrm d\sigma +\sum_i\mu_i\mathrm dn_i 
+ \left(-\epsilon-P+T\sigma+\sum_i\mu_in_i\right) \frac{\mathrm dV}{V},
\end{align}
and if the system is \emph{scale invariant}\footnote{In a scale invariant system the quantity densities do
not change considering a larger amount of matter. In a star the scale invariance is respected
as far as one considers a stellar region small with respect to the stellar scale-lengths, see discussion above.},
that is, if $\partial \epsilon/\partial V=0$, then
\begin{equation}
\label{eq:fundamental}
\epsilon + P = T\sigma+\sum_i\mu_in_i.
\end{equation}
Eq.~\eqref{eq:fundamental} may be applied to the whole particle system, or to
the subsystem made up only by particles $i$; moreover, we may redo the
discussion using the quantities per baryon instead of the quantities per unit
volume, obtaining an analogous relation.

In real
matter the abundances $\{Y_i\}$ and their variations $\{\mathrm dY_i\}$ are
related to each other. For example, we have already discussed that one cannot
differentiate $f$ with respect to $Y_p$ keeping at the same time $n_\mathrm{B}$
and $Y_n$ fixed. This is due to the definition of $n_\mathrm{B}$, that implies
\begin{align}
\label{eq:defnB}
Y_p={}&1-Y_n\qquad\mbox{definition of $n_\mathrm{B}$}.\\
\label{eq:diffdefnB}
\mathrm dY_p={}&-\mathrm dY_n\qquad\mbox{(differential of the) definition of
$n_\mathrm{B}$}.
\end{align}
Similarly, the charge neutrality of matter is equivalent (if the only charged
particles are protons and electrons) to
\begin{equation}
\label{eq:charge-neutrality}
Y_{e^-}=Y_p\qquad\mbox{charge neutrality},
\end{equation}
which imply (but is not implied) by the request of charge conservation
\begin{equation}
\label{eq:charge-conservation}
\mathrm d Y_{e^-}=\mathrm dY_p\qquad\mbox{charge conservation}.
\end{equation}
Using Eqs.~\eqref{eq:diffdefnB} and \eqref{eq:charge-conservation} one obtains
(in an EoS with protons, neutrons, and electrons)
\begin{align}
\label{eq:df2}
\mathrm df={}&-s\mathrm dT+\frac{P}{n_\mathrm{B}^2}\mathrm dn_\mathrm{B}
+(\mu_p+\mu_{e^-}-\mu_n)\mathrm dY_p,\\
\label{eq:dfdYp}
\left.\frac{\partial f}{\partial
Y_p}\right|_{T,n_\mathrm{B}}={}&\mu_p+\mu_{e^-}-\mu_n.
\end{align}
We remark the difference of Eq.~\eqref{eq:dfdYp} with respect to
Eqs.~\eqref{eq:mu1} and \eqref{eq:mu2}: in Eq.~\eqref{eq:dfdYp} $n_\mathrm{B}$
is kept constant, while the number fractions $\mathrm dY_{i\neq p}$ cannot be
independently
fixed because they are related to each other.

The nuclear theory adds further constraints to the thermodynamical
relations. We have already stated charge conservation,
Eq.~\eqref{eq:charge-conservation}.  If the matter is in equilibrium with
respect to one nuclear reaction, for example neutrino emission/absorption (i.e., \emph{beta-equilibrium}),
\begin{equation}
\label{eq:beta}
p+e^-\rightleftharpoons n+\nu_e,
\end{equation}
the chemical potentials of the corresponding particles are simply related to
each other, for example (for the case of beta-equilibrium)
\begin{equation}
\label{eq:beta-mu}
\mu_p+\mu_{e^-}=\mu_n+\mu_{\nu_e}.
\end{equation}
To obtain Eq.~\eqref{eq:beta-mu}, one first notices that the structure of
Reaction~\eqref{eq:beta} implies that \emph{exactly} one neutron and one
neutrino are produced from \emph{exactly} one proton and one electron,
\begin{equation}
\label{eq:variations}
\mathrm dY_p=\mathrm dY_{e^-}=-\mathrm d Y_n = -\mathrm dY_{\nu_e}.
\end{equation}
If we assume that the
reaction occurs at constant temperature and volume (i.e., the matter
temperature and baryon density do not change during the reaction timescale),
the free energy changes by
the amount
\begin{equation}
\label{eq:df_beta}
\mathrm df=(\mu_p+\mu_{e^-}-\mu_n-\mu_{\mu_e})\mathrm dY_p,
\end{equation}
and since at equilibrium the free energy is at a minimum,
\begin{equation}
\left.\frac{\partial f}{\partial
Y_p}\right|_{T,n_\mathrm{B}}\equiv0\qquad\mbox{at equilibrium},
\end{equation}
one finally obtains Eq.~\eqref{eq:beta-mu}. In general,
\begin{equation}
\label{eq:equilibrium}
\sum_{i\in\mbox{reaction}}\mu_i\mathrm dY_i=0.
\end{equation}

Similarly, we can consider the following reactions (electron-positron annihilation)
\begin{align}
\label{eq:reac1}
e^++e^-\rightleftharpoons{}& \gamma+\gamma,\\
\label{eq:reac2}
e^++e^-\rightleftharpoons{}& \gamma+\gamma+\gamma.
\end{align}
First of all, we notice that if both reactions are in chemical equilibrium (as
happens in the conditions present in a PNS),
\begin{equation}
\label{eq:mugamma}
\mu_\gamma=0,
\end{equation}
from which follows
\begin{equation}
\label{eq:mupositron}
\mu_{e^+}=-\mu_{e^-}.
\end{equation}
In general, the chemical potential of the antiparticle $\bar\imath$ is related
to that of the particle $i$ by
\begin{equation}
\label{eq:mubar}
\mu_{\bar\imath}=-\mu_i,
\end{equation}
if there is equilibrium with respect to the reaction of pair
annihilation/creation of particle $i$.  Eq.~\eqref{eq:mubar} suggests to
redefine the number fractions as
\begin{equation}
\label{eq:Yi}
\{Y_i,Y_{\bar\imath}\}\rightarrow Y_\imath=Y_i-Y_{\bar\imath},
\end{equation}
that is, to consider the net abundances of the particles. For example, for
electrons and positrons, one has
\begin{equation}
\mu_e\mathrm dY_e\equiv\mu_{e^-}\mathrm dY_e=\mu_{e^-}(\mathrm dY_{e^-}-\mathrm dY_{e^+})=
\mu_{e^-}\mathrm dY_{e^-}+\mu_{e^+}\mathrm dY_{e^+}.
\end{equation}
In the following, unless explicitly stated and apart from neutrons and protons\footnote{Since
their mass is far greater than thermal energy, their antiparticle
densities are negligible and therefore we do not consider them}, we will use
the notation expressed in Eq.~\eqref{eq:Yi}, that is, electrons and neutrinos
abundances are defined subtracting their antiparticle abundances.

We now derive a relation that will be useful in Chapter~\ref{cha:evolution}. We
consider an EoS with protons, neutrons, electrons, positrons, and neutrinos. We
include all neutrino flavours, but we will assume that muon and tauon neutrinos
have vanishing chemical potential. Finally, we assume beta-equilibrium. We
obtain
\begin{multline}
\label{eq:mu_nudYL}
\sum_i\mu_i\mathrm dY_i=\mu_p\mathrm dY_p+\mu_n\mathrm dY_n +\mu_e \mathrm dY_e
+\mu_{\nu_e} \mathrm dY_{\nu_e}=\\
=(\mu_p-\mu_n+\mu_e)\mathrm dY_e+\mu_{\nu_e} \mathrm
dY_{\nu_e}=\mu_{\nu_e}\mathrm d(Y_e+Y_\nu) = \mu_{\nu_e}\mathrm
dY_{\mathrm{L}_e},
\end{multline}
where we have used Eqs.~\eqref{eq:diffdefnB}, \eqref{eq:charge-conservation}, and
\eqref{eq:beta-mu}\footnote{Eq.~\eqref{eq:mu_nudYL} is true also if there are
also muons and $\mu_{\nu_\mu}=\mu_{\nu_\tau}=0$, since
\begin{align}
\mathrm dY_p={}&\mathrm dY_e+\mathrm dY_\mu\qquad&\mbox{charge conservation
with muons},\\
\mu_n-\mu_p={}&\mu_\mu\qquad&\mbox{beta-equilibrium for muons},
\end{align}
and therefore
\begin{multline}
\sum_i\mu_i\mathrm dY_i=\mu_p\mathrm dY_p+\mu_n\mathrm dY_n +\mu_e \mathrm dY_e
+ \mu_\mu \mathrm dY_\mu + \mu_{\nu_e} \mathrm dY_{\nu_e}=\\
=(\mu_p-\mu_n+\mu_e)\mathrm dY_e + (\mu_p-\mu_n+\mu_\mu)\mathrm dY_\mu
+\mu_{\nu_e} \mathrm dY_{\nu_e}= \mu_{\nu_e}\mathrm dY_{\mathrm{L}_e}.
\end{multline}
}. We observe that Eq.~\eqref{eq:mu_nudYL} may seem in contrast with
Eq.~\eqref{eq:equilibrium} and the request of beta-equilibrium. This apparent
paradox arises because we have not used Eq.~\eqref{eq:variations}, that is, the
number fraction variations do not respect the stoichiometry of the
beta-equilibrium reaction even though the chemical potentials are derived from
beta-equilibrium. This is due to the fact that in deriving
Eq.~\eqref{eq:mu_nudYL} we are implicitly interested in the process of neutrino
diffusion in the star, which has a timescale longer than that of
beta-equilibrium (see Appendix~\ref{app:code_checks}).  Therefore, since
beta-equilibrium is respected on these timescales, the relation
between chemical potentials due to beta-equilibrium still holds
[Eq.~\eqref{eq:beta-mu}], but Eq.~\eqref{eq:variations} is not valid on these
timescales since the neutrino number change is due to a process different from
beta-equilibrium (i.e., neutrino diffusion).  Charge
conservation [Eq.~\eqref{eq:charge-conservation}] is still valid since baryons
and electrons are locked on the timescales of PNS evolution: only neutrinos
diffuse through the star.

\section{Baryons effective spectra and neutrino diffusion}
\label{sec:diffusion}

In a PNS the massive particles are locked, that is, they cannot diffuse. Therefore,
energy and composition (i.e., the number fractions $Y_i$) changes are driven only by
neutrino diffusion, since we neglect the contribution of photons.  To consistently determine how the PNS evolves, it is
therefore fundamental to determine the neutrino cross sections in high density,
finite temperature matter.  In this section we describe how we have treated
neutrino diffusion consistently with the underlying EoS.

As we explain in Chapter~\ref{cha:evolution}, the diffusion coefficients $D_2$,
$D_3$, and $D_4$ employed in the PNS evolution are (\citealp{Pons+.1999}; all
non-electronic neutrinos are treated as muon neutrinos)
\begin{align}
\label{eq:D2}
D_2={}&D_2^{\nu_e}+D_2^{\bar\nu_e},\\
\label{eq:D3}
D_3={}&D_3^{\nu_e}-D_3^{\bar\nu_e},\\
\label{eq:D4}
D_4={}&D_4^{\nu_e}+D_4^{\bar\nu_e}+4D_4^{\nu_\mu},\\
\label{eq:Dn}
D^{\nu_i}_n={}&\int_0^\infty\mathrm dx
x^n\lambda_\mathrm{tot}^{\nu_i}(\omega)f^{\nu_i}(\omega)\big(1-f^{\nu_i}(\omega)\big),\\
\label{eq:mean_free_path}
\frac 1{\lambda^{\nu_i}_\mathrm{tot}(\omega)}={}&\sum_{j\in\{\mathrm{reactions\}}}\frac {\sigma_j^{\nu_i}(\omega)}{V},
\end{align}
where $f^{\nu_i}(\omega)$, $\lambda_\mathrm{tot}^{\nu_i}(\omega)$, and $\sigma_j^{\nu_i}$ are the
distribution function, the total mean free path, and the cross-section of a $\nu_i$ neutrino of energy
$\omega=xT$ and each quantity depends on the temperature and the
particle chemical potentials, which are determined by the underlying EoS.

The nuclear processes that we consider in
Eq.~\eqref{eq:mean_free_path} are the scattering of all neutrino types on
electrons, protons, and neutrons and the absorption of electron neutrino and
electron anti-neutrino on neutron and proton, respectively, with their inverse
processes (for details regarding the inverse processes, see
Appendix~\ref{app:inverse})
\begin{align}
\label{eq:reaction_start}
\nu_i + n \rightleftharpoons{}& \nu_i + n,\\
\nu_i + p \rightleftharpoons{}& \nu_i + p,\\
\nu_i + e^- \rightleftharpoons{}& \nu_i + e^-,\\
\nu_e + n \rightleftharpoons{}& e^- + p,\\
\bar \nu_e + p \rightleftharpoons{}& e^+ + n.
\label{eq:reaction_end}
\end{align}

To determine the cross-sections $\sigma(E)$ we use Eq.~(82) of \citet{Reddy+Prakash+Lattimer.1998}; the coupling constants for
the different reactions are reported in \citet[][Tabs.~I and II]{Reddy+Prakash+Lattimer.1998}.  In
order to compute the neutrino scattering and absorption cross-sections on
interacting baryons, \citet{Reddy+Prakash+Lattimer.1998} make use of the baryon effective
parameters (effective masses and single particle potentials, see below), that allow to
approximate the relativistic single particle spectra of the interacting baryons.
The use of the single particle spectrum approximation is applicable only when it is reasonable to describe matter
in terms of quasi-particles, that is, near the Fermi surface. This approximation is justified since, due to the Pauli blocking
effect, neutrinos may interact only with baryons near their Fermi surface.
To
compute the diffusion
coefficients in a point of the star, we need therefore to determine the approximated
baryon effective spectra corresponding to the thermodynamical conditions in that point of the star.

The spectrum of a free relativistic particle $\mathcal E^K(k)$ is given by its kinetic
energy
\begin{equation}
\label{eq:E_k}
\mathcal E^K(k) = \sqrt{k^2+m^2},
\end{equation}
where $m$ is the particle mass. The interaction between particles changes the
single-particle spectrum, $\mathcal E(k)$. One can effectively describe the
single particle spectrum introducing the particle effective mass $m^\ast$ and
the single particle potential $U$,
\begin{equation}
\label{eq:Estar}
\mathcal E(k) \simeq \sqrt{k^2+m^\ast{}^2} + U.
\end{equation}
Eq.~\eqref{eq:Estar} is exact in the case of a mean-field EoSs [like GM3, see
Sec.~\ref{sec:GM3}, in particular Eq.~\eqref{eq:effective_E}], and it is only
approximate for a more realistic EoS, like a many-body EoS (see discussion in
Sec.~\ref{sec:CBFEI}).
This is due to the fact that, in the
many-body formalism, the concept of single particle is not well-posed;
moreover, assuming that it is possible to define a single-particle spectrum, it
is only approximately given by Eq.~\eqref{eq:Estar}.  As a
consequence, the thermodynamical quantities ($f$, $s$, $P$, and so on) are only
approximately recovered by integrating the effective spectra of protons
and neutrons (that is, by inserting the effective masses and single particle
potentials in the Fermi gas expressions, as done in Eqs.~\eqref{eq:eps_GM3} and
\eqref{eq:P_GM3} for the GM3 EoS).  This is expected and is not a relevant
issue; in fact, in our work the effective masses and single particle potentials are used
only to compute the diffusion coefficients and not the other EoS quantities,
for which we have performed a different fit (see Secs.~\ref{sec:fBfit} and
\ref{sec:totaleos}), and therefore there are no consistency problems. On the
contrary, it is important to recover the baryon
densities $n_n$ and $n_p$ from the effective spectrum description (as done in
Eqs.~\eqref{eq:n_baryon} for the GM3 EoS), since the mean free paths
are ``intensive'' quantities, Eq.~\eqref{eq:mean_free_path}.  If the baryon densities are poorly recovered from the effective
spectrum description, one is erroneously using diffusion coefficients at baryon
density $\nB$ and proton fraction $Y_p$, while they would correspond to those at
baryon density $\nB'$ and proton fraction $Y_p'$, compromising the consistency of the PNS evolution.
To be more explicit,
\begin{equation}
\label{eq:density_condition}
n_i=\frac{4\pi}{h^3}\int\frac{p^2\mathrm dp}{1+\mathrm e^\frac{\sqrt{p^2+m^2}-\mu_i^K}{T}}\simeq
\frac{4\pi}{h^3}\int\frac{p^2\mathrm dp}{1+\mathrm e^\frac{\sqrt{p^2+m^{\ast 2}}+U_i-\mu_i}{T}},
\end{equation}
where $\mu_i^K$ is the chemical potential of a free Fermi gas with the same baryon density $\nB$,
temperature $T$, and proton fraction $Y_p$ of the interacting Fermi gas. The \emph{raison d'\^etre} of Eq.~\eqref{eq:density_condition}
will become clear in Secs.~\ref{sec:fBfit} and \ref{sec:totaleos}, here we anticipate that to determine the total EoS
we consider first the non-interacting baryon EoS, whose density, temperature and proton fraction are the same of the total interacting baryon EoS.
Eq.~\eqref{eq:density_condition} is a consequence of this procedure and of the
request that the baryon densities recovered by the effective spectra are
as those given by the EoS.

In the PNS evolution code the neutrino diffusion coefficients are evaluated by
linear interpolation of a three-dimensional table, evenly spaced in $Y_\nu$ (the
neutrino number fraction), $T$, and $\nB$. The table has been produced
consistently with the underlying EoS.
To generate the table, we have first solved the EoS using the method described
in Sec.~\ref{sec:totaleos}, obtaining the particles chemical potentials. The
proton and neutron effective masses and single particle potentials
have been obtained
by linear interpolation of a table evenly spaced in $Y_p$, $T$, and $\nB$. The neutrino cross sections (Eq.~(82) of
\citealp{Reddy+Prakash+Lattimer.1998}) and
the neutrino diffusion coefficients
[Eq.~\eqref{eq:Dn}] have been integrated by Gaussian quadrature.

\section{An extrapolated baryon EoS: LS-bulk}
\label{sec:LSbulk}

In this section we describe a high density EoS obtained from the extrapolation of the
properties of nuclear matter obtained by experiments on terrestrial nuclei. In particular, this EoS
has been used by \cite{Lattimer+Swesty.1991} for the bulk nuclear matter, that is,
baryons are treated as an interacting gas of protons and neutrons (there are neither alpha
particles, pasta phases, nor lattice). We call this EoS LS-bulk.

The well-known semi-empirical formula of the nuclear binding energy permits to fit with
a few parameters the binding energy of terrestrial nuclei with an astonishing
precision \citep{Krane.book}
\begin{multline}
\label{eq:krane}
E(N_p\equiv Z,N_\mathrm{B}\equiv A) = N_p m_p + N_n m_n - a_v N_\mathrm{B} +a_s
N_\mathrm{B}^{2/3}\\
+ a_c N_p(N_p-1)N_\mathrm{B}^{-1/3}+a_\mathrm{sym}
\frac{(N_\mathrm{B}-2N_p)^2}{N_\mathrm{B}} - \delta a_pN_\mathrm{B}^{-3/4},
\end{multline}
where $N_p\equiv Z$ and $N_n$ are the number of protons and neutrons in the nucleus,
respectively, $N_\mathrm{B}=N_p+N_n\equiv A$ is the number of baryons, and
$a_v$, $a_s$, $a_c$, $a_\mathrm{sym}$, and $\delta a_p$ (with $\delta=\{-1;0;1\}$) are parameters that assess the
importance of the volume, surface, Coulomb, symmetry and pairing force terms,
respectively, and whose particular value may be determined by fitting the
measured masses of terrestrial nuclei \citep{Krane.book}.  To extrapolate a
high density stellar EoS from Eq.~\eqref{eq:krane}, one considers the limit of
infinite nuclear matter, retaining only the terms proportional to the volume,
and adds some terms that account for other nuclear properties, like the
nuclear imcompressibility.
The
semi-empirical formula for the free energy of infinite nuclear matter may therefore be written as \citep[Eq.~(2.3)
of][]{Lattimer+Swesty.1991}
\begin{multline}
\label{eq:semi-empirical}
f_{B}(T,\nB,Y_p) = m_n -\mathrm{BE} -\Delta_m Y_p \\ +
\frac{K_s}{18}\left(1-\frac{\nB}{n_s}\right)^2 + S_v(1-2 Y_p)^2 - a_T T^2,
\end{multline}
where
\begin{itemize}
\item $\Delta_m\equiv m_n-m_p$ is the neutron-proton mass difference; at
variance with \cite{Lattimer+Swesty.1991} we set this term to zero;
\item $n_s$ is the saturation density of symmetric nuclear matter, that is, the
density for which
\begin{equation}
\label{eq:saturation}
\left.\frac{\partial P_{B}}{\partial \nB}\right|_{T=0,Y_p=1/2}=0.
\end{equation}
We adopt the same value of \citet{Lattimer+Swesty.1991},
$n_s=\unit[0.155]{fm^{-3}}$.
\item $\mathrm{BE}=-f_{B}(T=0,n_s,Y_p=1/2) - \Delta_m/2 \equiv a_v$ is the
binding energy of saturated, symmetric nuclear matter. Its value can be derived
from nuclear mass fits; we set it to the same value of
\citet{Lattimer+Swesty.1991}, $\mathrm{BE}=\unit[16]{MeV}$.
\item $K_s$ is the imcompressibility of bulk nuclear matter\footnote{Beware
that there is a typo in the definition of $K_s$ that appears under Eq.~(2.3) of
\citet{Lattimer+Swesty.1991}.},
\begin{equation}
\label{eq:Ks}
K_s=\left.9\nB^2\frac{\partial^2 f_{B}}{\partial
\nB^2}\right|_{T=0,\nB=n_s,Y_p=1/2}.
\end{equation}
$K_s$ can be determined by isoscalar breathing modes and isotopic differences
in charge densities of large nuclei.
At variance with \citet{Lattimer+Swesty.1991}, we take $K_s=\unit[220]{MeV}$, a
value which is more similar to recent measurements and to those of the
other EoSs we are considering.
\item $S_v$ is the symmetry energy parameter of bulk nuclear matter,
\begin{equation}
\label{eq:Sv}
S_v=\left.\frac18\frac{\partial^2f_\mathrm{B}}{\partial
Y_p^2}\right|_{T=0,\nB,Y_p=1/2}\equiv a_\mathrm{sym},
\end{equation}
it can be derived from the fit of the mass formula and from giant dipole
resonances; as \citet{Lattimer+Swesty.1991} we set its value to
$S_v=\unit[29.3]{MeV}$.
\item $a_T$ is the bulk level density parameter,
\begin{equation}
a_T=-\left.\frac12\frac{\partial^2f_\mathrm{B}}{\partial^2T}\right|_{T=0,\nB,Y_p=1/2}\simeq\frac1{15}
\frac{m^\ast}{m_n}\unit{MeV^{-1}},
\end{equation}
and it is related to the nucleon effective mass $m^\ast$. However, following
\citet{Lattimer+Swesty.1991}, we set the effective masses of protons and
neutrons for the LS-bulk EoS equal to their rest masses; and in addition the thermal
contribution to the LS-bulk EoS is given only by the kinetic term (see
below).  Therefore the parameter $a_T$ is not relevant to the following
discussion.
\end{itemize}
With this choice of parameters, we obtain a maximum mass for a non-rotating
cold star of $M_\mathrm{LSbulk}\simeq \unit[2.02]{M_\odot}$ \citep{Camelio+2017}.

\citet{Lattimer+Swesty.1991} give the following parametrization for the baryon
free energy at finite temperature
\begin{align}
\label{eq:fLS}
f_\mathrm{B}(T,\nB,Y_p)={}&f_\mathrm{B}^K(T,\nB,Y_p)+f_\mathrm{B}^I(T,\nB,Y_p),\\
\label{eq:fILS}
f_\mathrm{B}^I(T,\nB,Y_p)={}& \big(a+
4bY_p(1-Y_p)\big)\nB+c\nB^\delta-Y_p\Delta_m,
\end{align}
where $f_\mathrm{B}^K$ is the free energy per baryon of a free gas of protons
and neutrons, but with the same baryon density, temperature, and proton
fraction of the total interacting baryon free energy, and $f_\mathrm{B}^I$ is the
interacting contribution to the free energy. The proton-neutron mass difference can appears only in the interaction term,
while in the kinetic term proton and neutrons have the same mass. To obtain the values of the
parameters in Eq.~\eqref{eq:fILS}, one should take the expression for the
interacting free energy [Eq.~\eqref{eq:semi-empirical}] at zero temperature and
subtract the (non-relativistic) zero temperature non-interacting contribution,
$f_\mathrm{B}^K$, see Appendix~\ref{sec:UDNR}. One obtains (Eq.~(2.21) of
\citealp{Lattimer+Swesty.1991}; beware that Eq.~(2.19c) of
\citealp{Lattimer+Swesty.1991} has a typo)
\begin{align}
\alpha={}&\frac{3\hbar^2}{10m_n}(3\pi^2n_s/2)^{2/3},\\
\label{eq:pardelta}
\delta={}&\frac{K_s+2\alpha}{3\alpha+9\mathrm{BE}}=1.260,\\
b={}&\frac{\alpha(2^{2/3}-1)-S_v}{n_s}=\unit[-107.1]{MeV\,fm^3},\\
a={}&\frac{\delta(\alpha+\mathrm{BE})-2\alpha/3}{n_s(1-\delta)}-b=\unit[-711.0]{MeV\,fm^3},\\
c={}&\frac{K_s+2\alpha}{9\delta(\delta-1)n_s^\delta}=\unit[934.6]{MeV\,fm^{3\delta}},\\
\label{eq:parDelta}
\Delta_m={}&0,
\end{align}
where we set the effective masses of protons and neutrons equal to the rest
mass of neutrons, $m^\ast=m_n$, and $\alpha$ is the internal energy (i.e., the energy without the
contribution of the bare rest mass $m_n$) of a non-relativistic free gas of protons
and neutrons at $T=0$, $\nB=n_s$, and $Y_p=1/2$.  Since we have not considered
thermal effects in the determination of the interacting contribution to the
free energy, those are included in the LS-bulk EoS only by the kinetic term,
$f_\mathrm{B}^K$.

As we have discussed in Sec.~\ref{sec:diffusion}, it is important that the
baryon densities recovered from the effective single particle spectra are equal
to the densities obtained solving the EoS.  Since for the LS-bulk we have set
$m^\ast_i=m_n$, this means that [see Eq.~\eqref{eq:density_condition}]
\begin{equation}
\label{eq:U_LS}
U_i = \mu^I_i=\mu_i-\mu_i^K,
\end{equation}
where $\mu_i^I$ and $\mu^K_i$ are the kinetic and interacting parts of the
chemical potential, obtained by differentiating the kinetic $f^K_\mathrm{B}$
and interacting $f^I_\mathrm{B}$ parts of the free energy, see
Eqs.~\eqref{eq:fLS} and \eqref{eq:mu2}, and the index refers to protons and
neutrons, $i\in\{p,n\}$.

\section{A mean-field baryon EoS: GM3}
\label{sec:GM3}

At the center of a neutron star, the baryon density may easily reach 4 or 5
times the nuclear saturation density, $n_s$. At such high densities, the Fermi
momentum is expected to be comparable to the baryon masses ($k_{\mathrm F
n}\simeq \unit[450]{MeV}$ at $n_n\simeq \unit[0.4]{fm^{-3}}$, and
$m_n\simeq\unit[939]{MeV}$), and therefore it would be preferable to adopt a
relativistic description of baryons \citep{Prakash+1997}.  In the following, we
consider a model \citep{Walecka.1974, Glendenning.1985} where the nuclear
forces between baryons are mediated by the exchange of the $\sigma$, $\rho$,
and $\omega$ mesons. This model is easily extended with the presence of hyperons; however we do not include them because the other
EoSs considered in this thesis (LS-bulk and CBF-EI) are composed only by protons and neutrons.
To simplify the notation, in this section we set $\hbar=c=1$.

The baryon Lagrangian is \citep{Glendenning.1985, Prakash+1997}
\begin{multline}
\label{eq:lagrangian}
\mathcal L_\mathrm{B}=\sum_{i\in\{p,n\}} \bar\psi_i (\imath
\gamma^\mu\partial_\mu -g_{\omega i}\gamma^\mu\omega_\mu -g_{\rho
i}\gamma^\mu\vec\rho_\mu \cdot \vec t -m_i +g_{\sigma i}\sigma)\psi_i\\
+\frac12 \partial_\mu\sigma\partial^\mu\sigma -\frac12m_\sigma^2\sigma^2
-U(\sigma) -\frac14 W_{\mu\nu}W^{\mu\nu}
+\frac12m_\omega^2\omega_\mu\omega^\mu\\
-\frac 14 \vec R_{\mu\nu}\cdot\vec R^{\mu\nu} + \frac12 m_\rho^2\vec
\rho_\mu\cdot\vec\rho^\mu,
\end{multline}
where $\psi_i=\psi_i(x)$ is the wave function of the proton or the neutron,
$\sigma=\sigma(x)$, $\omega_\mu=\omega_\mu(x)$ and $\vec \rho_i=\vec \rho_i(x)$
are the meson wave functions,
$W_{\mu\nu}=\partial_\mu\omega_\nu-\partial_\nu\omega_\mu$, $\vec
R_{\mu\nu}=\partial_\mu\vec\rho_\nu - \partial_\nu\vec \rho_\nu$, $g_{ji}$ are
the coupling constants between the meson $j$ and the baryon $i$, $\vec t$ is
the baryon isospin operator and the potential
\begin{equation}
\label{eq:Usigma}
U(\sigma)=(b m_n + cg_\sigma\sigma)(g_\sigma\sigma)^3
\end{equation}
represents the self-interaction of the $\sigma$ field ($b$, $c$ and $g_\sigma$
are parameters).

With \cite{Glendenning.1985}, we define the \emph{normal} state of infinite
matter as ``uniform and isotropic, and [\ldots] the baryon eigenstates in
the medium carry the same quantum numbers as they do in vacuum''
\citep{Glendenning.1985}.  In addition, we apply the mean-field approximation, that
is, we replace the meson fields by their mean values,
$\sigma(x)\rightarrow\sigma$, $\omega_\mu(x)\rightarrow\omega_\mu$, and
$\vec\rho_\mu(x) \rightarrow\vec \rho_\mu$.

We remind that the Euler-Lagrange equation for the field $\psi$ is
\begin{equation}
\label{eq:euler-lagrange}
\partial_\mu \left(\frac{\partial \mathcal L}{\partial(\partial_\mu
\psi)}\right) = \frac{\partial \mathcal L}{\partial \psi}.
\end{equation}
Evaluating Eq.~\eqref{eq:euler-lagrange} for the meson fields we obtain (since the
field wave functions are constants in the mean-field approximation, the spatial
derivatives vanish)
\begin{align}
\label{eq:sigma_field}
m_\sigma^2\sigma={}&-\frac{\mathrm dU(\sigma)}{\mathrm d\sigma}
+\sum_{i\in\{p,n\}}g_{\sigma i}<\bar\psi_i\psi_i>,\\
m_\omega^2\omega^\mu={}&\sum_{i\in\{p,n\}}g_{\omega
i}<\bar\psi_i\gamma^\mu\psi_i>,\\
\label{eq:rho1}
m_\rho^2\rho_1^\mu={}&\frac12\big(g_{\rho
p}<\bar\psi_p\gamma^\mu\psi_n>+g_{\rho n}<\bar\psi_n\gamma^\mu\psi_p>\big)=0,\\
\label{eq:rho2}
m_\rho^2\rho_2^\mu={}&\frac12\big( -\imath g_{\rho
p}<\bar\psi_p\gamma^\mu\psi_n>+\imath g_{\rho
n}<\bar\psi_n\gamma^\mu\psi_p>\big) =0,\\
m_\rho^2\rho_3^\mu={}&\sum_{i\in\{p,n\}}g_{\rho
i}t_{3i}<\bar\psi_i\gamma^\mu\psi_i>,
\end{align}
where the third component of the isospin is $t_{3p}=+1/2$ for the proton and
$t_{3n}=-1/2$ for the neutron. The expectation value of the charged $\rho$
mesons vanishes since in nuclear normal matter baryons carry the same
quantum numbers as they do in vacuum, and the first two isospin operators change
the isospin of the wavefunction they are applied to [flip protons in neutrons,
see Eqs.~\eqref{eq:rho1} and \eqref{eq:rho2}].  Moreover, since nuclear normal
matter is isotropic, the expectation value of $<\bar\psi\gamma^i\psi>$, with
$i=\{1,2,3\}$, vanishes \citep{Walecka.1974}, and therefore
\begin{align}
\label{eq:omega}
m_\omega^2\omega^0={}&\sum_{i\in\{p,n\}}g_{\omega
i}<\bar\psi_i\gamma^0\psi_i>,\\
\label{eq:rho3}
m_\rho^2\rho_3^0={}&\sum_{i\in\{p,n\}}g_{\rho
i}t_3<\bar\psi_i\gamma^0\psi_i>,\\
\omega^j={}&\rho_3^j=0,
\end{align}
where $n_i$ is the density of the baryon $i$ and $j$ is a spatial index. It is possible to show that
\citep{Glendenning.1985, Prakash+1997}
\begin{align}
\label{eq:n_baryon}
<\bar\psi_i\gamma^0\psi_i>={}&<\psi_i^\dag\psi_i>\equiv n_i=4\pi\int_0^\infty
f_i(p) p^2\mathrm dp,\\
\label{eq:scalar_baryon}
<\bar\psi_i\psi_i>={}&4\pi\int_0^\infty \frac{m_i-g_{\sigma
i}\sigma}{\sqrt{p^2+(m_i-g_{\sigma i}\sigma)^2}} f_i(p) p^2 \mathrm dp,
\end{align}
where $f_i(p)$ is the distribution function of the \emph{interacting} baryon $i$,
Eq.~\eqref{eq:f_baryon}, that is different from the non-interacting distribution function, see below.

Applying the Euler-Lagrange equation to the baryon fields one obtain the Dirac
equations for baryons,
\begin{equation}
\left[\gamma^0(p_0 -g_{\omega i} \omega_0 - t_3 g_{\rho i} \rho_{03}) +
\gamma^j p_j -(m_i - g_{\sigma i} \sigma)\right] \psi_i = 0,
\end{equation}
and from them the baryon spectra
\begin{equation}
\label{eq:baryon_spectrum}
\mathcal E_i(\vec p)= g_{\omega i} \omega_0 + t_3 g_{\rho i} \rho_{03} +
\sqrt{p^2+(m_i - g_{\sigma i} \sigma)^2}.
\end{equation}
We notice that, defining the \emph{effective mass} $m^\ast$ and the
\emph{single particle potential} $U$,
\begin{align}
\label{eq:mstar}
m^\ast_i={}&m_i-g_{\sigma i} \sigma,\\
\label{eq:U}
U_i={}&g_{\omega i} \omega_0 + t_3 g_{\rho i} \rho_{03},
\end{align}
we can write the mean-field baryon spectra in a way that is formally identical
to Eq.~\eqref{eq:Estar},
\begin{equation}
\label{eq:effective_E}
\mathcal E_i(p) = \sqrt{p^2 +m_i^{\ast 2}} + U_i,
\end{equation}
where in general the effective masses and single particle potentials depend on
the density and the temperature.  The baryon interacting distribution function
is
\begin{equation}
\label{eq:f_baryon}
f_i(p)=\frac{g_i/h^3}{1+\exp\left(\frac{\mathcal E_i(k)-\mu_i}{T}\right)}
=\frac{g_i/h^3}{1+\exp\left(\frac{\sqrt{p^2 +m_i^{\ast 2}} + U_i
-\mu_i}{T}\right)},
\end{equation}
where $g_i$ is the baryon degeneracy (for protons and neutrons, $g_i=2$).

For given temperature $T$ and chemical potentials $\mu_p$ and $\mu_n$,
Eqs.~\eqref{eq:sigma_field}, \eqref{eq:omega}, \eqref{eq:rho3}, \eqref{eq:n_baryon},
\eqref{eq:scalar_baryon}, \eqref{eq:f_baryon} may be solved iteratively (e.g., with a Newton-Raphson algorithm) to give
the mean-field values of the meson fields and the baryon effective masses and
single particle potentials.  From the stress-energy tensor and the partition
function one can then obtain the other thermodynamical quantities
\citep{Glendenning.1985, Prakash+1997}.  However, we find instructive to adopt
here a heuristic argument to obtain the thermodynamical quantities.  First of
all, the contribution to the thermodynamical quantities given by the baryons
may be obtained from the Eqs.~\eqref{eq:eps} and \eqref{eq:P}, substituting the
free baryon energy spectra with the interacting ones,
Eq.~\eqref{eq:effective_E}.  However, one has to consider also the contribution
given by the meson fields.  From the
Lagrangian~\eqref{eq:lagrangian}, it is apparent that the contribution at the
mean-field level is
\begin{equation}
\epsilon_\mathrm{mesons}= +\frac12 m_\sigma^2\sigma^2 + U(\sigma) - \frac12
m_\omega^2 \omega_0^2
- \frac 12 m_\rho^2 \rho_{03}^2.
\end{equation}
Since the meson fields are treated at the mean-field level,
\begin{equation}
P_\mathrm{mesons}=-\epsilon_\mathrm{mesons}.
\end{equation}
Then, the total baryon energy is \citep{Prakash+1997}
\begin{align}
\label{eq:eps_GM3}
\epsilon_\mathrm{B} ={}& \epsilon_\mathrm{mesons}+\epsilon_p +\epsilon_n =
+\frac12 m_\sigma^2\sigma^2 + U(\sigma) - \frac12 m_\omega^2 \omega_0^2
- \frac 12 m_\rho^2 \rho_{03}^2\notag\\
{}&+ 4 \pi \sum_{i\in\{p,n\}} \int \left( \sqrt{p^2+m_i^{\ast 2}} + U_i \right) f_i(p) p^2 \mathrm d p\notag\\
={}&\frac12 m_\sigma^2\sigma^2 + U(\sigma) + \frac12 m_\omega^2 \omega_0^2+
\frac 12 m_\rho^2 \rho_{03}^2 \notag\\
{}&+ 4 \pi \sum_{i\in\{p,n\}} \int \sqrt{p^2+m_i^{\ast 2}} f_i(p) p^2 \mathrm
dp,
\end{align}
where in the last step we have used Eqs.~\eqref{eq:U}, \eqref{eq:omega} and
\eqref{eq:rho3}. The total baryon pressure is \citep{Prakash+1997}
\begin{equation}
\label{eq:P_GM3}
P_\mathrm{B} = P_\mathrm{field}+P_p +P_n = -\frac12 m_\sigma^2\sigma^2 -
U(\sigma) + \frac12 m_\omega^2 \omega_0^2 + \frac 12 m_\rho^2
\rho_{03}^2+P_p+P_n,
\end{equation}
and the total baryon entropy can be obtained from Eq.~\eqref{eq:fundamental}.
We remark that the condition in Eq.~\eqref{eq:density_condition} is automatically fulfilled, since by construction
the baryon thermodynamical quantities are obtained from their single particle effective spectra.

In this thesis, we have adopted the set of parameters denoted as GM3
\citep[][see Table~\ref{tab:GM3} of this thesis]{Glendenning+Moszkowski.1991, Pons+.1999}, which correspond to a
saturation density $n_s=\unit[0.153]{fm^{-3}}$, a binding energy
$\mathrm{BE}=\unit[16.3]{MeV}$, a bulk imcompressibility
parameter $K_s=\unit[240]{MeV}$, and a symmetry energy $S_v=\unit[32.5]{MeV}$.
The maximum mass of a cold NS with the GM3 EoS is $M_\mathrm{GM3}\simeq\unit[2.02]{M_\odot}$
\citep{Camelio+2017}.

\begin{table}
\caption{GM3 parameters \citep{Glendenning+Moszkowski.1991, Pons+.1999}. Since the
meson couplings are the same for protons and neutrons, we drop the baryon index (e.g., $g_{\sigma p}=g_{\sigma
n}\equiv g_\sigma$).}
\label{tab:GM3} \centering
\begin{tabular}{ccl}
\hline
$g_\sigma/m_\sigma$ & $3.151$ & fm\\
$g_\omega/m_\omega$ & $2.195$ & fm\\
$g_\rho/m_\rho$     & $2.189$ & fm\\
$b$                 & $0.008659$ & \# \\
$c$                 & $-0.002421$ & \# \\
\hline
\end{tabular}
\end{table}

As a final remark, we notice that since $g_{\sigma p}=g_{\sigma n}$ (see
Table~\ref{tab:GM3}), the effective masses
of proton and neutron are the same [Eq.~\eqref{eq:mstar}];
whereas their single particle potential is different because of the presence of
the third component of the isospin in Eq.~\eqref{eq:U}.  This behaviour is due
to the way baryons couple to the (neutral) rho meson $\rho_3$
[Eqs.~\eqref{eq:lagrangian} and \eqref{eq:rho3}], which therefore is
responsible of the symmetry energy that drives the neutron excess in nuclear
matter.

\section{A many-body baryon EoS: CBF-EI}
\label{sec:CBFEI}
The mean-field approximation we have described in Sec.~\ref{sec:GM3} consists
in the assumption that meson fields may be replaced by their mean value, that
is, meson wavefunctions oscillate many times on the scale length of the baryon
wavefunctions.  However, in a neutron star density may easily reach
$\nB\simeq\unit[0.4]{fm^{-3}}$, that corresponds to an average distance between
nucleons of the order of $d\simeq\unit[1.7]{fm}$, which is comparable to the
meson Compton wavelengths $\lambda_\mathrm{c}=h/m$
($\lambda_\sigma=\unit[0.33]{fm}$, $\lambda_\omega=\unit[0.25]{fm}$, and
$\lambda_\rho=\unit[0.26]{fm}$).  In addition, at the mean-field level the pion
meson expectation value is zero\footnote{Apart for the case of pion condensate,
that in any case we do not consider \citep{Glendenning.1985}.}, whereas pion
exchange is the main process that determines the baryon interaction.
Therefore, the mean-field approximation is poorly justified in this regime. In
this section we describe a non-relativistic many-body EoS that is based on the
semi-phenomenological nuclear potentials Argonne $v'_6$ and Urbana IX
\citep{Lovato.PhD, Benhar+Lovato2017}, that takes into
account aspects of the nuclear dynamics that are neglected by the mean-field
approximation.  This EoS is based on the correlated basis function theory and
makes use of the Hartree-Fock effective interaction, and therefore we call it
CBF-EI EoS.

There are strong numerical and experimental evidence that the Hamiltonian of a
many body nuclear system is given by
\begin{equation}
\label{eq:hamiltonian}
\hat H= -\sum_i \frac{\nabla^2_i}{2m\hbar} + \sum_{i>j}\hat v_{ij} +
\sum_{k>j>i}\hat V_{ijk},
\end{equation} where sums are performed over the nucleons, and $\hat v_{ij}$
and $\hat V_{ijk}$ are two- and three-body potentials.  The inclusion of the
additional three-nucleon term, $V_{ijk}$, is needed to explain the binding
energies of three-nucleon systems and the saturation properties of symmetric
neutron matter. The three-nucleon force is the consequence of having
neglected the quark degrees of freedom, that is implicit in the formulation of
the problem in terms of nucleons (each nucleon is composed at a more
fundamental level by three quarks).  Theoretical, numerical and
phenomenological constraints permit to
determine the form of the terms entering in the potentials $\hat v_{ij}$ and
$\hat V_{ijk}$.  The two-body form of the potential of the CBF-EI EoS
considered in this thesis is the so called \emph{Argonne} $v'_6$
potential\footnote{The potential $v'_6$ does not include spin-orbit terms, nor
charge asymmetry terms.  It is not a simple truncation of the Argonne $v_{18}$
potential (which has 18 terms that accounts for spin-orbit and charge
asymmetry, \citealp{Wiringa+Stoks+Schiavilla.1995}); in fact Argonne nuclear
data have been refitted to produce the $v'_6$ potential
\citep{Wiringa+Pieper.2002}.}, $\hat v_{ij}\equiv \hat v'_6(r_{ij})$
\begin{align}
\hat v'_6(r_{ij})={}&\sum_{p=1}^6 v^p(r_{ij})\hat O^p_{ij},\\
\hat O^{p=1-6}_{ij}={}&(1,\vec\sigma_{i}\cdot \vec\sigma_j,S_{ij})\otimes
(1,\vec\tau_i\cdot\vec\tau_{j}),\\
S_{ij}={}&\frac 3{r_{ij}^2} (\vec \sigma_i\cdot \vec r_{ij}) (\sigma_j\cdot
\vec r_{ij}) - (\vec \sigma_i\cdot\vec \sigma_j),
\end{align}
where $\sigma_i$ and $\tau_i$ are the Pauli matrices acting on the spin and
isospin of particle $i$, ans $r_{ij}$ is the distance between the two
particles. The CBF-EI EoS uses as three-body potential the Urbana IX potential
\citep{Fujita+Miyazawa.1957, Pudliner+1995}, whose expression may be found for
example in \cite{Lovato.PhD}.

Within the Hartree-Fock approximation, the many-body ground state is assumed to
be the Slater determinant of a system of $N_\mathrm{B}$ interacting baryons,
\begin{equation}
\label{eq:hamiltonian2}
\hat H\Psi_0(x_1,\ldots,x_{N_\mathrm{B}}) =
E_0\Psi_0(x_1,\ldots,x_{n_\mathrm{B}}),
\end{equation}
where $\Psi_0(x_1,\ldots,x_{N_\mathrm{B}})$ is the many-body wave eigenfunction
corresponding to the energy ground state $E_0$. Usually in the Hartree-Fock
procedure one adopts as many-body trial wavefunction the Slater determinant of
$N_\mathrm{B}$ one-nucleon wavefunctions,
\begin{equation}
\label{eq:slater}
\Phi_0=\mathcal A
\big(\phi_{n_1}(x_1),\ldots,\phi_{n_{N_\mathrm{B}}}(x_{N_\mathrm{B}})\big),
\end{equation}
where $\mathcal A$ is the antisymmetrization operator. The standard variational
methods applied in the Hartree-Fock procedure fail to converge with potentials
having a repulsive core and strong tensor interactions, like in the nuclear
case. One way to circumvent this problem, is the so called correlated basis
function (CBF) theory, that consists in considering correlated wave functions
constructed by means of a \emph{correlation operator} $\mathcal{\hat F}$,
\begin{align}
|\Psi_0> ={}& \frac{\hat{\mathcal F} |\Phi_0>} {<\Phi_0| \hat{\mathcal F}^\dag
\hat{\mathcal F}|\Phi_0>},\\
\label{eq:correlator}
\hat{\mathcal F}= {}& \mathcal S\left( \prod_{j>i}^{N_\mathrm{B}} \sum_{p=1}^6
f^p(r_{ij})\hat O_{ij}^p\right),
\end{align}
where $\mathcal S$ is the symmetrization operator, $r_{ij}$ is the distance
between the nucleons $i$ and $j$, and $f^p(r_{ij})$ are correlation functions
to be determined.  We have applied the symmetrization operator to keep the
wavefunction anti-symmetric, since in general the operators $O^p$ do not
commute. The point of using the correlated basis function approach is that the
correlation functions $f^p_{ij}$ make the wavefunctions $|\Psi_0>$ small where the potential is stronger,
that is, in the repulsive region of the nuclear potential; in this way the
variational procedure converges even in presence of non-perturbative potentials.  The
correlation functions $f^p_{ij}$ are first determined by exploiting the
variational principles.  An efficient way of computing $E_0$ of
Eq.~\eqref{eq:hamiltonian2} consists in expanding the expectation values in
clusters including an increasing number of correlated
particles \citep{CLARK197989}, which can be represented by diagrams and
classified according to their topological structures.  Selected classes of
diagrams can then be summed to all orders, solving a set of integral equations
referred to as Fermi Hyper-Netted Chain/ Single Operator Chain
equations \citep{Fantoni:1974jv,Pandharipande:1979bv} to obtain an accurate
estimate of $E_0$.  This latter step can be done only for symmetric and pure
neutron matter. One can define an effective two-body potential (or Hartree-Fock potential)
$\hat v^\mathrm{eff}$, that results from integrating the degrees of freedom of
$N_\mathrm{B}-2$ nucleons,
\begin{align}
\label{eq:effective}
e_\mathrm{B}={}&\frac {E_0}{N_\mathrm{B}}=\frac{<\Psi_0|\hat
H|\Psi_0>}{N_\mathrm{B}}
=\frac{K+<\Phi_0|v^\mathrm{eff}|\Phi_0>}{N_\mathrm{B}},\\
\label{eq:veff}
v^\mathrm{eff}={}&\sum_{i<j}v^\mathrm{eff}_{ij}=\sum_{i<j}\sum_{p=1}^6
v^{\mathrm{eff},p}(r_{ij})O^p_{ij},
\end{align}
where $K$ is the kinetic energy of the uncorrelated state $|\Phi_0>$ defined in
Eq.~\eqref{eq:slater}. We remark that the operators $O^p$ that appear in
Eq.~\eqref{eq:veff} are the same of the two-body Argonne $v'_6$ potential. At
variance with previous implementation, the effective potential $v^\mathrm{eff}$
we have employed simultaneously reproduces the EoS of both pure neutron matter (PNM) and symmetric neutron matter (SNM). The
effective potential of Eq.~\eqref{eq:effective} can be used for intermediate
proton fraction and at finite temperature (with the condition that for SNM
and PNM at zero temperature it gives the same results as variational
calculations with the full Hamiltonian). The CBF-EI EoS has a saturation density
$n_s=\unit[0.16]{fm^{-3}}$, an imcompressibility parameter
$K_s=\unit[180]{MeV}$, a binding energy at saturation
$\mathrm{BE}=\unit[10.95]{MeV}$ and a symmetry energy $S_v=\unit[30]{MeV}$.
The maximum gravitational mass of a cold NS with the CBF-EI EoS is
$M_\mathrm{CBF-EI}\simeq\unit[2.34]{M_\odot}$ \citep{Camelio+2017}.

From the effective two body potential $v^\mathrm{eff}$, the single particle energy can be written in terms of
\begin{multline}
\label{eq:single_CBFEI}
\mathcal E_{n_i} = \frac{p^2_i}{2m} + m + \sum_{n_j=1}^{N_\mathrm{B}} \int\mathrm d x_i\mathrm d x_j
\phi^\ast_{n_i}(x_i) \phi^\ast_{n_j}(n_j) v^\mathrm{eff}(r_{ij})\\
\cdot\big( \phi_{n_i}(x_i) \phi_{n_j}(x_j) - \phi_{n_j}(x_i) \phi_{n_i}(x_j) \big),
\end{multline}
where $\phi_{n_i}$ are the one-nucleon wavefunction entering in Eq.~\eqref{eq:slater}.
The total energy in Eq.~\eqref{eq:effective} is given by
\begin{multline}
\label{eq:totalE_CBFEI}
E_0 = \sum_{n_i}\mathcal E_{n_i} - \frac12 \sum_{n_i,n_j} \int\mathrm d x_i\mathrm d x_j
\phi^\ast_{n_i}(x_i) \phi^\ast_{n_j}(n_j) v^\mathrm{eff}(r_{ij})\\
\cdot\big( \phi_{n_i}(x_i) \phi_{n_j}(x_j) - \phi_{n_j}(x_i) \phi_{n_i}(x_j) \big),
\end{multline}
where the second term in the right hand side has the same role of the term $\epsilon_\mathrm{mesons}$ appearing in Eq.~\eqref{eq:eps_GM3}.

In the non-relativistic limit, the single-particle spectrum of Eq.~\eqref{eq:Estar} is given by
\begin{equation}
\label{eq:effSpectrum}
\mathcal E(k) = \sqrt{p^2 + m^{\ast 2}}+U\simeq \frac {k^2} {2m^\ast} + m^\ast + U\equiv\frac{k^2}{2m^\ast}+m+U',
\end{equation}
where we have written the expression in two ways, first using as rest mass the
effective mass and then using the bare mass. Off course,
\begin{equation}
U=U'-m^\ast+m.
\end{equation}
The most relevant contribution of baryons to the neutrino mean free path and diffusion
coefficients arises from particles whose energies are close to their
chemical potential, that is, whose momentum is close to the Fermi momentum.
Therefore, the effective masses and single particle potentials of the CBF-EI EoS have been
determined from the behaviour of the baryon spectrum of Eq.~\eqref{eq:single_CBFEI} near the Fermi momentum,
\begin{align}
\frac 1{m^\ast_i}={}&\frac1{k_F}\frac{\partial \mathcal
E_i}{\partial k}(k_F),\\
U'_i={}&\mathcal E_i(k_F) -\frac{k_F^2}{2m^\ast_i}-m,
\end{align}
where $i=(\{p;n\},Y_p,T,\nB)$.

\section{A fitting formula for the baryon EoS}
\label{sec:fBfit}

It is numerically feasible to directly compute the GM3 or the LS-bulk EoSs whenever they are
needed in the PNS evolution code. However, the great
computational cost to evaluate a many-body EoS (like CBF-EI) makes it
impossible to directly evaluate it during the simulation. Therefore, one should
use (i) an interpolation, or (ii) a fit.  Since we are studying the
evolution of a PNS, we need thermodynamical consistency and continuity of the
second order derivatives of the free energy \citep{Swesty.1996}. The EoSs that
we use to describe the PNS have three independent variables (see Sec.~\ref{sec:totaleos}); this
makes it difficult to interpolate a table in a thermodynamical consistent way
\citep{Swesty.1996}.  Therefore we have chosen to use a fitting formula to
describe the baryon
interaction. In this section we describe how we have constructed the fitting formula
and we describe the results of the fit for the GM3 and CBF-EI EoSs [for the
LS-bulk EoS, we have adopted the fitting formula of Eqs.~\eqref{eq:fLS} and
\eqref{eq:fILS}]. The content of this section is the main original contribution presented in this
chapter.

Since we are
interested in the evolution of a proto-neutron star, that is a ``hot'' neutron
star, we do not consider the formation of any kind of crust or envelope
(alpha particles, pasta phases and/or lattice). An \emph{a posteriori}
justification of this approximation will be given in Appendix~\ref{sec:gas}. In
addition, in this section we consider only the baryon part of the EoS (and
therefore our discussion is based on the baryon free energy), since we will add
the lepton part in the next section.  We consider only protons and neutrons, we
neglect any electromagnetic contribution to the baryon energy, and we
assume isospin invariance. An immediate consequence of this is that the proton bare mass
is equal to the neutron bare mass $m_p\equiv m_n\equiv m$.
Using Eq.~\eqref{eq:diffdefnB}, the baryon free energy variation may therefore be written as
\begin{multline}
\label{eq:dfB}
\mathrm df_\mathrm{B}=-s\mathrm dT + \frac P{n_\mathrm{B}^2} \mathrm dn_\mathrm{B} +\mu_p\mathrm dY_p + \mu_n\mathrm d Y_n\\
=-s\mathrm dT + \frac P{\nB^2}\mathrm d\nB + (\mu_p-\mu_n)\mathrm dY_p,
\end{multline}
and it is therefore a function of three variables,
\begin{equation}
\label{eq:fB}
f_\mathrm{B}=f_\mathrm{B}(T,n_\mathrm{B},Y_p),
\end{equation}
where we have taken the customary choice of $Y_p$ as third independent variable.

\begin{figure}
\centering
\includegraphics[width=\textwidth]{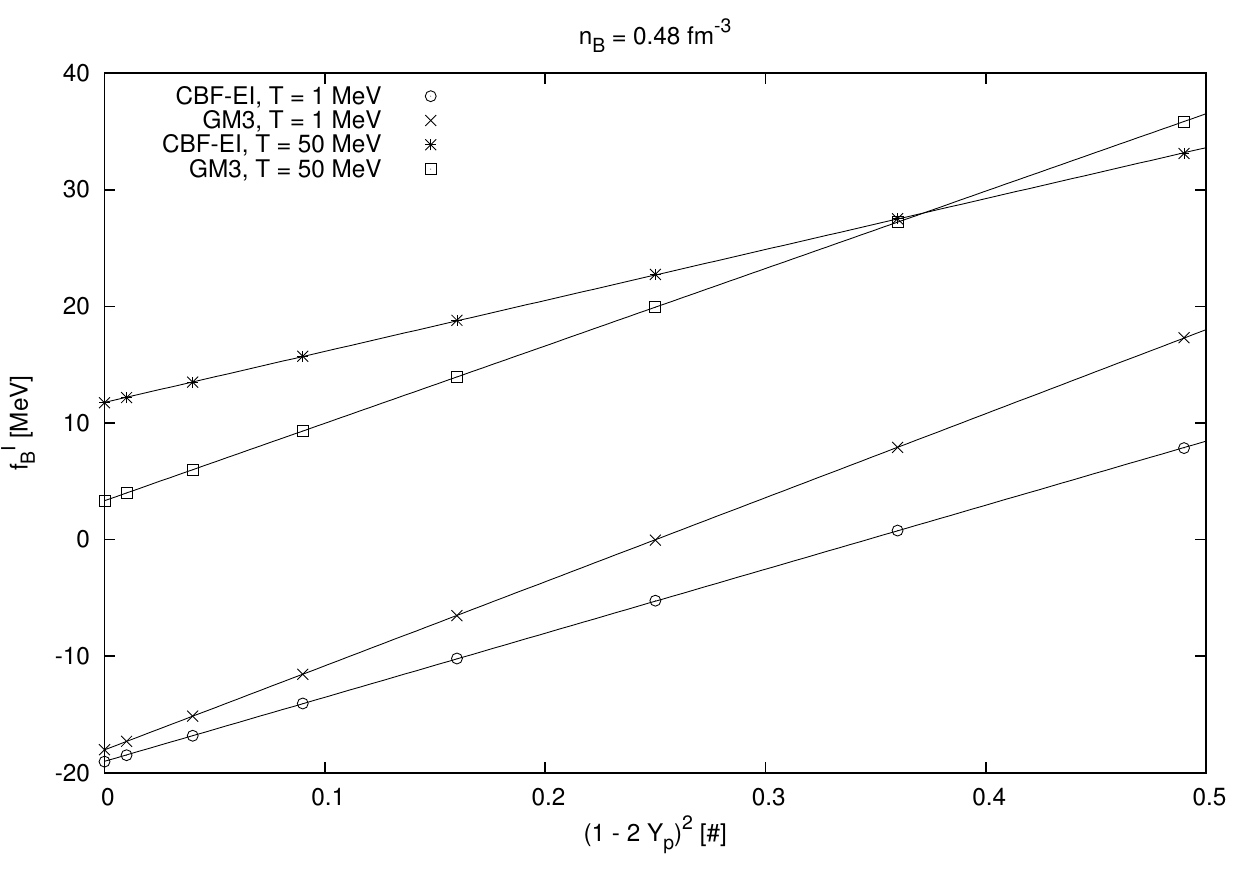}
\includegraphics[width=\textwidth]{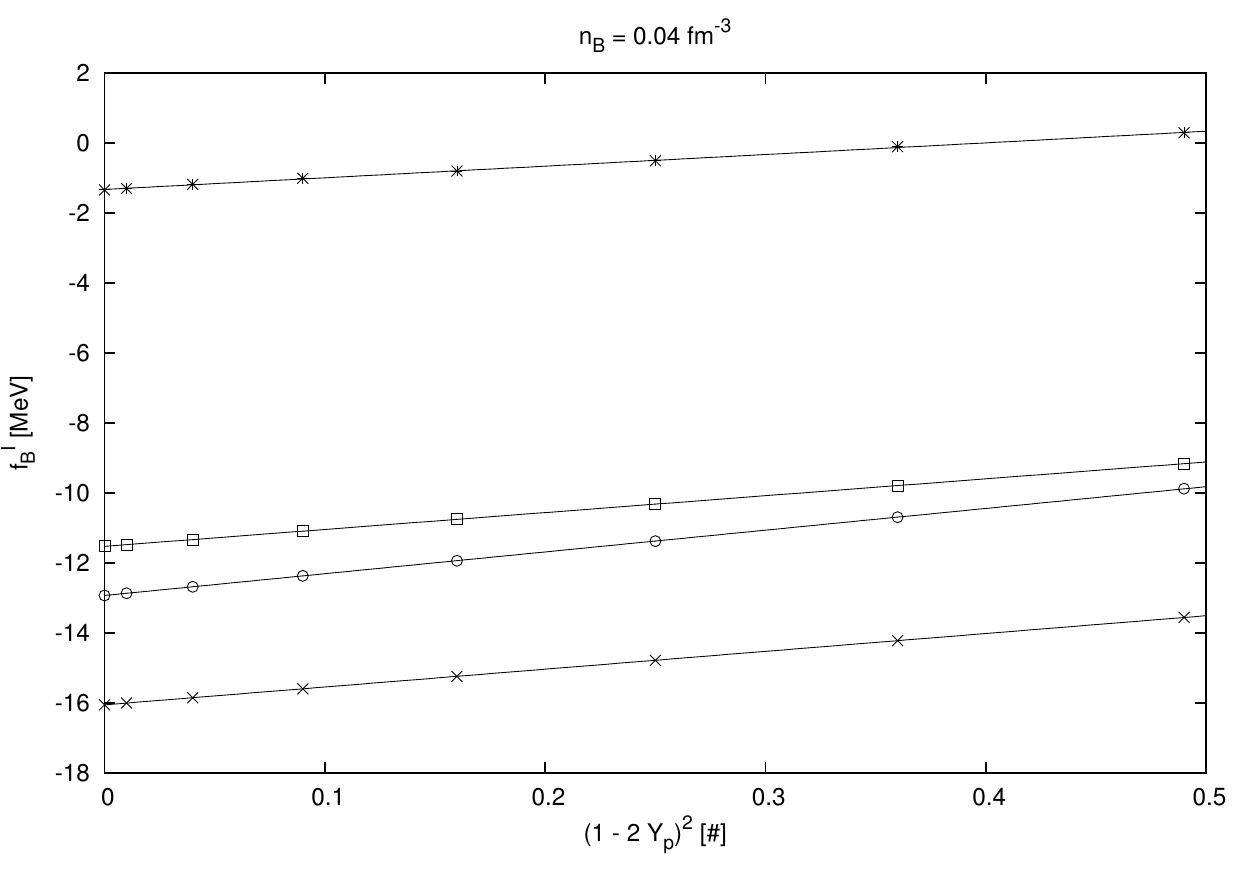}
\caption{Interacting part of the baryon free energy per baryon as a function of $(1-2Y_p)^2$ at different temperatures, for GM3 and CBF-EI EoSs,
fitted with the quadratic dependence on $Y_p$ of Eq.~\eqref{eq:ffit} (this fit has been performed only with the points plotted in this figure).
In the upper panel the baryon density is $n_\mathrm{B}=\unit[0.48]{fm^{-3}}$, in the lower panel it is
$n_\mathrm{B}=\unit[0.04]{fm^{-3}}$. The lines correspond to the fit.}
\label{fig:Yp}
\end{figure}

At zero temperature, the baryon energy per baryon $e_\mathrm{B}$ is usually written as a sum of a
kinetic contribution $e_\mathrm{B}^K$, that is, the
energy of a non interacting gas of free Fermions, and an interacting part $e^I_\mathrm{B}$,
\begin{equation}
\label{eq:Tzero_e}
e_\mathrm{B}(T=0,n_\mathrm{B},Y_p)=e^K_\mathrm{B}(T=0,n_\mathrm{B},Y_p) + e^I_\mathrm{B}(T=0,n_\mathrm{B},Y_p).
\end{equation}
The interacting energy per baryon $e^I_\mathrm{B}$
dependence on the proton fraction is well approximated
\citep{Bombaci+Lombardo.1991} by (we drop the dependence on $n_\mathrm{B}$ to simplify the
notation)
\begin{multline}
\label{eq:e_I}
e_\mathrm{B}^I(Y_p,T=0) = e^I_\mathrm{SNM}+(1-2Y_p)^2(
e^I_\mathrm{PNM}-e^I_\mathrm{SNM})\\
=4Y_p(1-Y_p)e^I_\mathrm{SNM} + (1-2Y_p)^2e^I_\mathrm{PNM},
\end{multline}
where $e^I_\mathrm{SNM}=e_\mathrm{B}(Y_p=1/2,T=0)$ and $e^I_\mathrm{PNM}=e_\mathrm{B}(Y_p=0,T=0)$
are the baryon interacting energies of the symmetric (SNM) and pure neutron
matter (PNM), respectively, at zero temperature.  To our knowledge, there are
no studies of the dependence of the baryon interacting energy on the proton
fraction at finite temperature. We assume the same dependence of the
zero temperature case, as done for example in \cite{Burgio+Schulze.2010},
\begin{align}
\label{eq:f}
f_\mathrm{B}(Y_p,T,\nB)={}&f^K_\mathrm{B}(Y_p,T,\nB)+f^I_\mathrm{B}(Y_p,T,\nB),\\
\label{eq:ffit}
f^I_\mathrm{B}(Y_p,T,\nB)={}&4Y_p(1-Y_p)f^I_\mathrm{SNM}(T,\nB)
+ (1-2Y_p)^2f^I_\mathrm{PNM}(T,\nB),
\end{align}
where $f_\mathrm{B}$, $f^K_\mathrm{B}$, and $f_\mathrm{B}^I$ are the baryon
total, kinetic (that of a free Fermi gas), and interacting free energy per
baryon, respectively, and $f^I_\mathrm{SNM}$ and $f^I_\mathrm{PNM}$ are the
baryon interacting free energies per baryon for symmetric and pure neutron
matter. We have checked for the GM3 and CBF-EI EoSs that the dependence on the proton
fraction is well described by this quadratic dependence, see Fig.~\ref{fig:Yp} (in the LS-bulk EoS this dependence
is fulfilled by construction, see Eq.~\eqref{eq:fILS} with $\Delta_m=0$).

Now that we have fixed the dependence of the free energy on the proton
fraction, we have to determine its value on the symmetric and pure neutron
matter planes, that is, we have to fit the baryon free energy on a plane of
constant proton fraction $Y_p$.  However, it does not exist a fitting formula
on which there is a general consensus in the literature. We have tried the
fitting formula used in \cite{Burgio+Schulze.2010}; however it
behaves badly as $\nB\rightarrow 0$, and therefore we have discarded it.  After several
attempts, we have chosen for our fitting formula a polynomial dependence on $T$
and $\nB$.

Since we want to use this fitting formula in an evolutionary code, we also want
to accurately evaluate, beyond the free energy, its first and second order
thermodynamical derivatives \citep{Swesty.1996}, and therefore to determine the
free energy fitting formula we take into account some considerations on these quantities.  The
second law of thermodynamics requires that
\begin{equation}
s\xrightarrow{T\to0}0,
\end{equation}
and therefore there could not be terms proportional to the temperature $T$ or
to negative powers of the temperature in the fitting formula.  We also want that
as the density tends to zero, the total thermodynamical quantities reduce to the
free gas ones, that is,
\begin{align}
f_\mathrm{B}^I\xrightarrow{n_\mathrm{B}\to0}{}&0,\\
s_\mathrm{B}^I\xrightarrow{n_\mathrm{B}\to0}{}&0,\\
P_\mathrm{B}^I\xrightarrow{n_\mathrm{B}\to0}{}&0,
\end{align}
and therefore there could not be terms proportional to powers of the baryon
density $n_\mathrm{B}$ equal or lower than one in the fitting formula.  With these
considerations, and visually inspecting the behaviour of the first and second
order thermodynamical derivatives of the GM3 and CBF-EI EoSs, we empirically
find that a good trade-off between number of parameters and precision of
the thermodynamical quantities determination is given by
\begin{multline}
\label{eq:fSNM_PNM}
f^I_{i}(\nB,T) = a_{1,i}\nB + a_{2,i}\nB^2 + a_{3,i}\nB^3+a_{4,i}\nB^4\\
+\nB T^2(a_{5,i}+a_{6,i}T+a_{7,i}\nB + a_{8,i}\nB T),
\end{multline}
with $i=\{\mathrm{SNM};\mathrm{PNM}\}$.  The fitting formula for the GM3 and
CBF-EI EoSs is therefore given by Eqs.~\eqref{eq:f}, \eqref{eq:ffit}, and
\eqref{eq:fSNM_PNM}.

\begin{figure}
\centerline{
\includegraphics[width=0.65\textwidth]{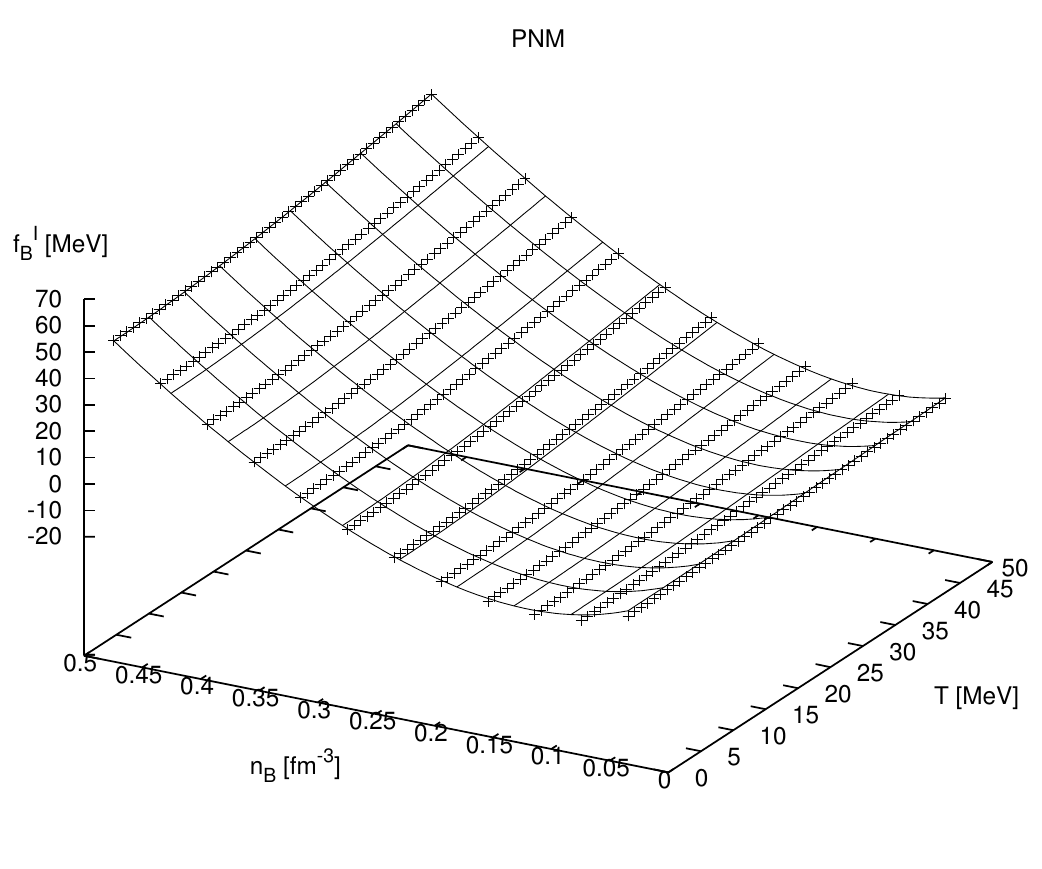}
\includegraphics[width=0.65\textwidth]{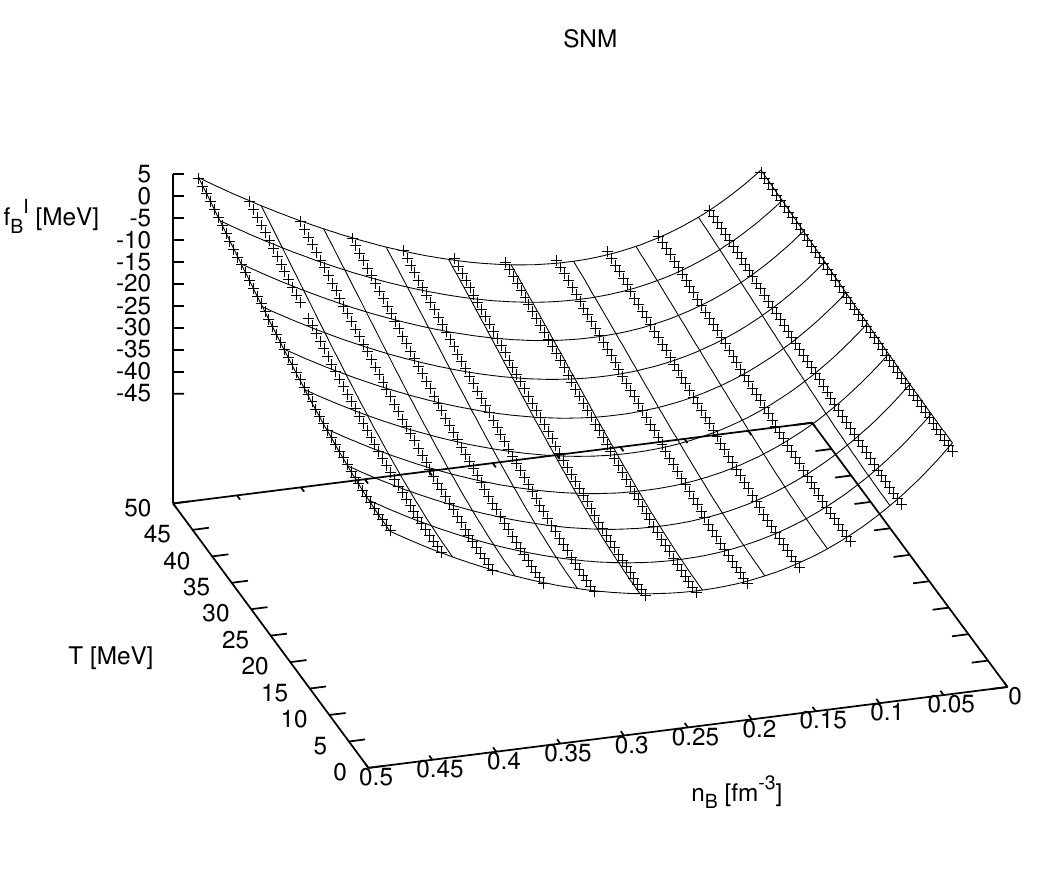}
}
\centerline{
\includegraphics[width=0.65\textwidth]{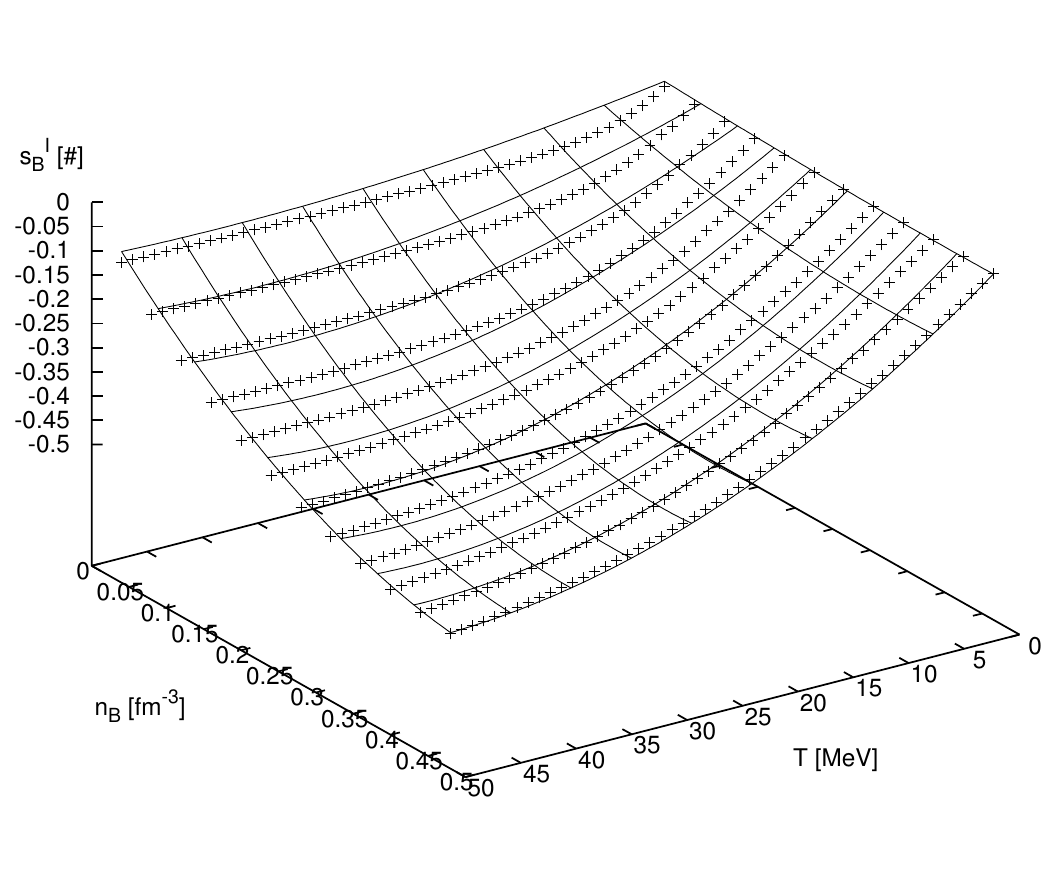}
\includegraphics[width=0.65\textwidth]{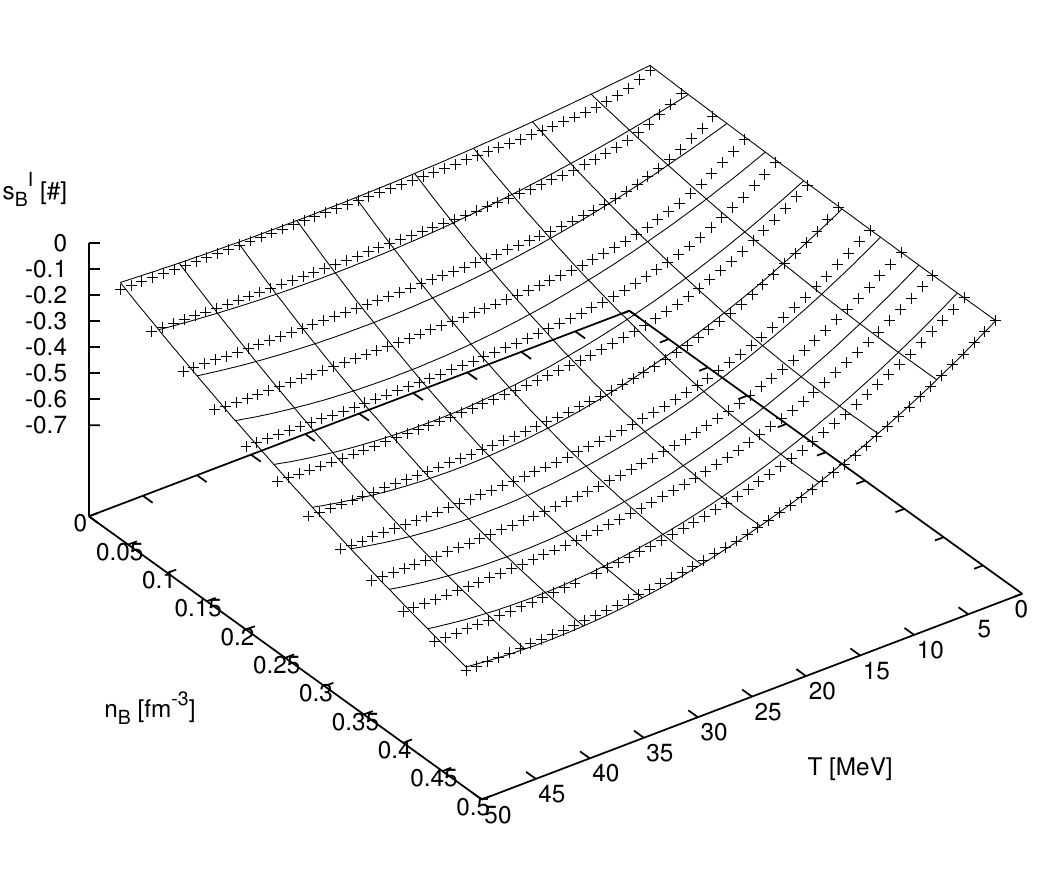}
}
\centerline{
\includegraphics[width=0.65\textwidth]{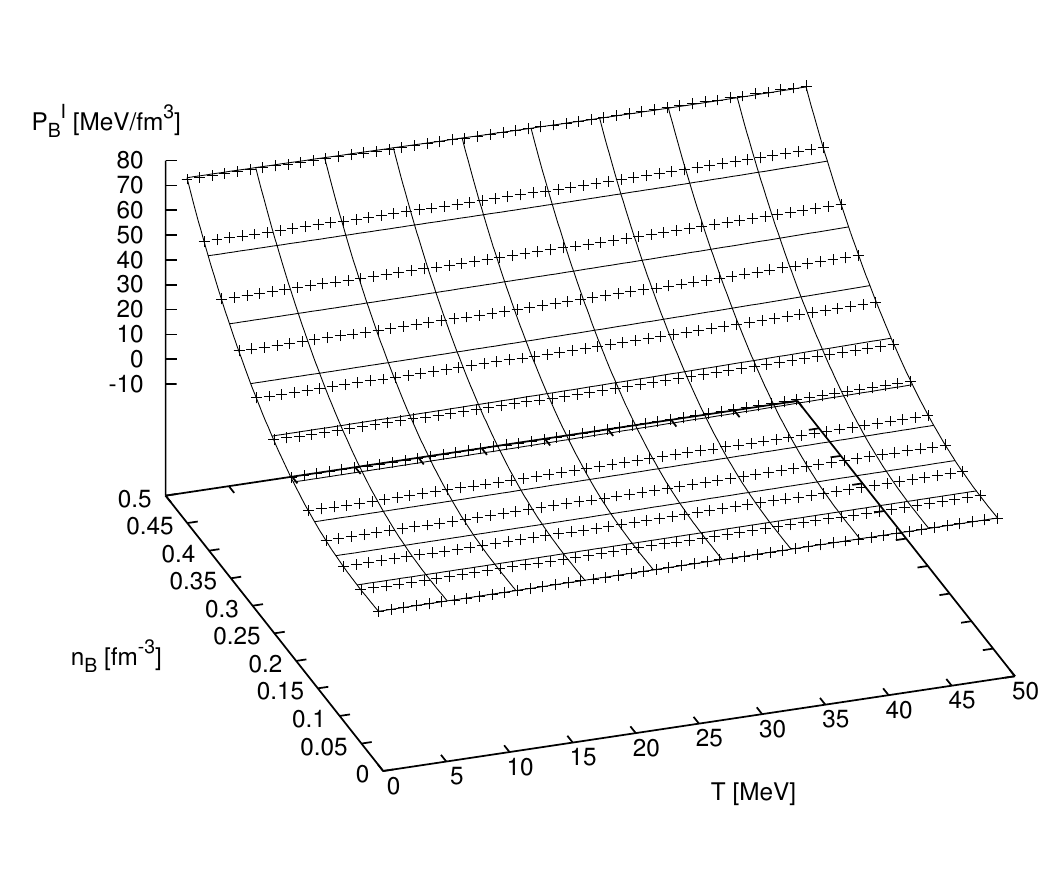}
\includegraphics[width=0.65\textwidth]{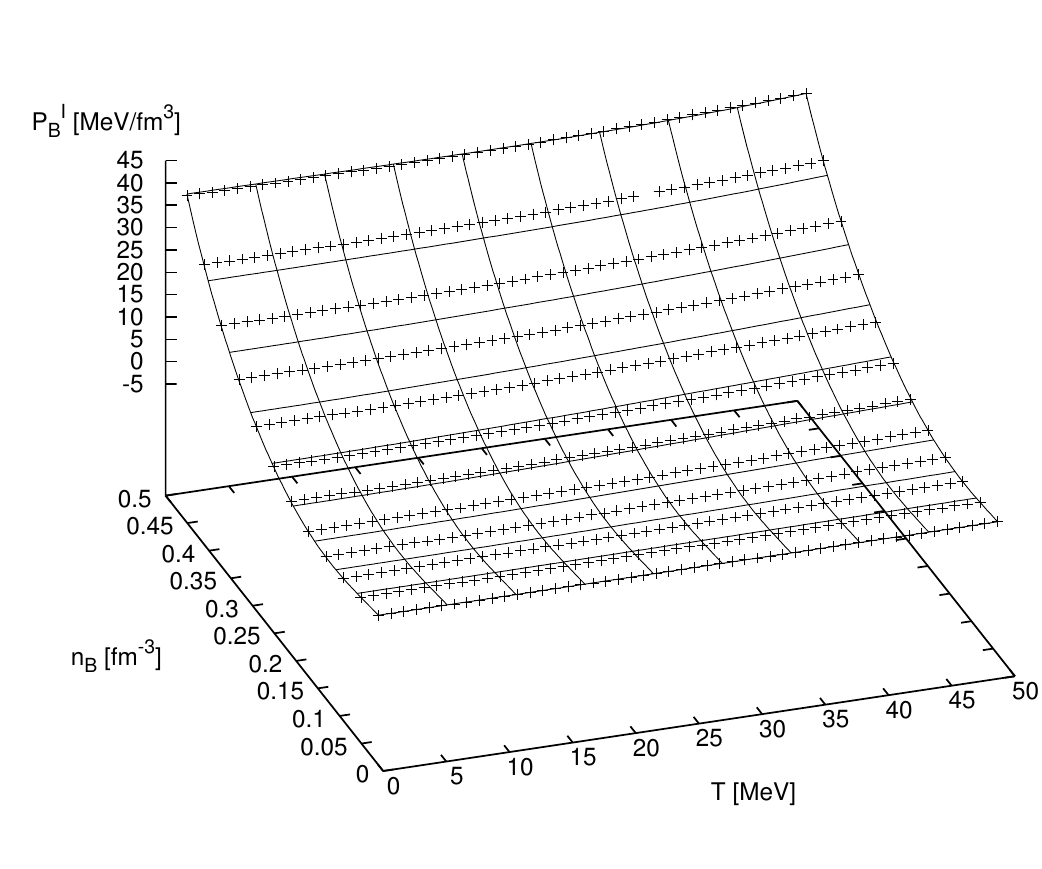}
}
\caption{Result of the fit for the GM3 EoS. We show for PNM (left column) and SNM (right column),
the interacting baryon free energy (upper row), entropy (middle row), and pressure (bottom row). The
fitting points are shown in the plots.}
\label{fig:fit_GM3}
\end{figure}

\begin{figure}
\centerline{
\includegraphics[width=0.65\textwidth]{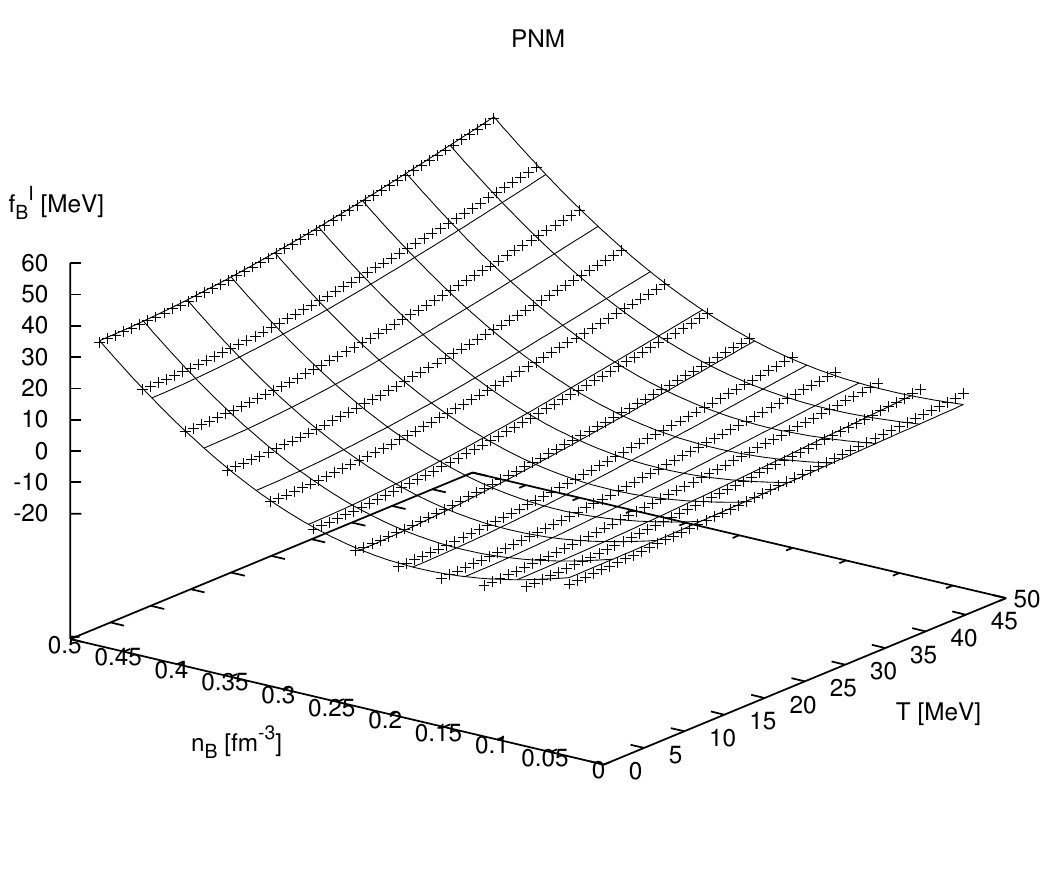}
\includegraphics[width=0.65\textwidth]{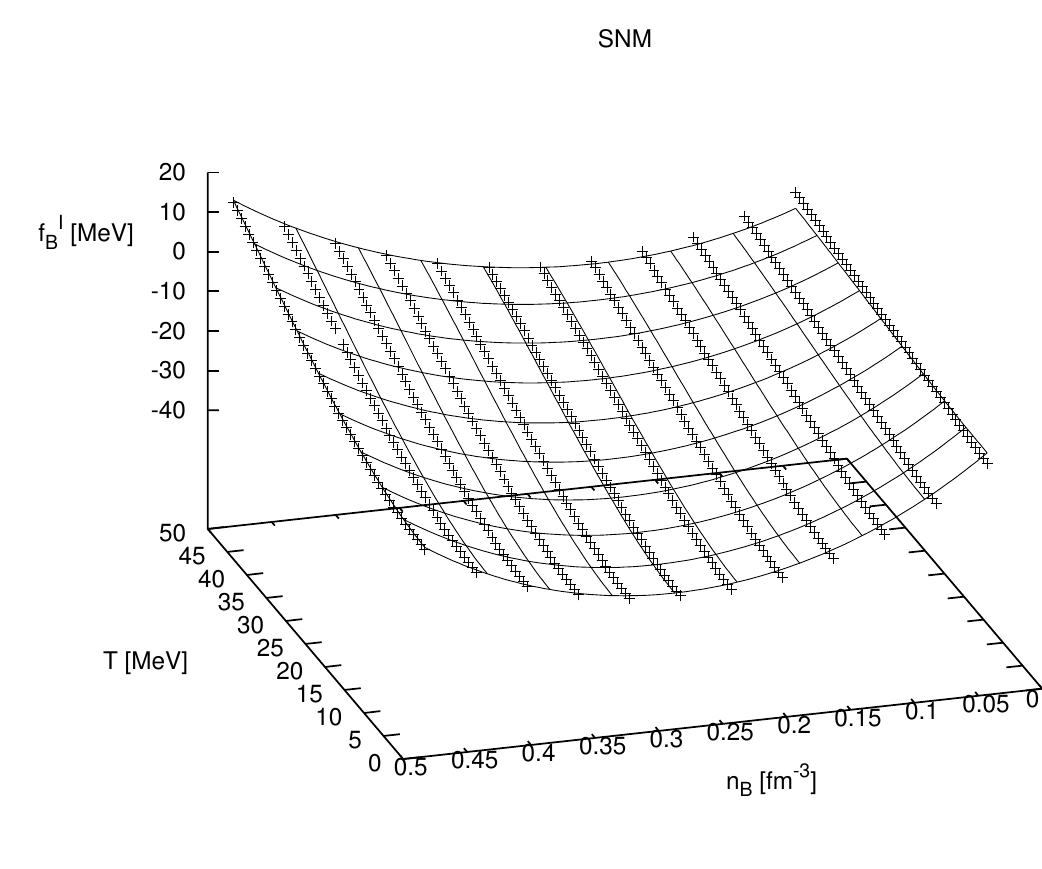}
}
\centerline{
\includegraphics[width=0.65\textwidth]{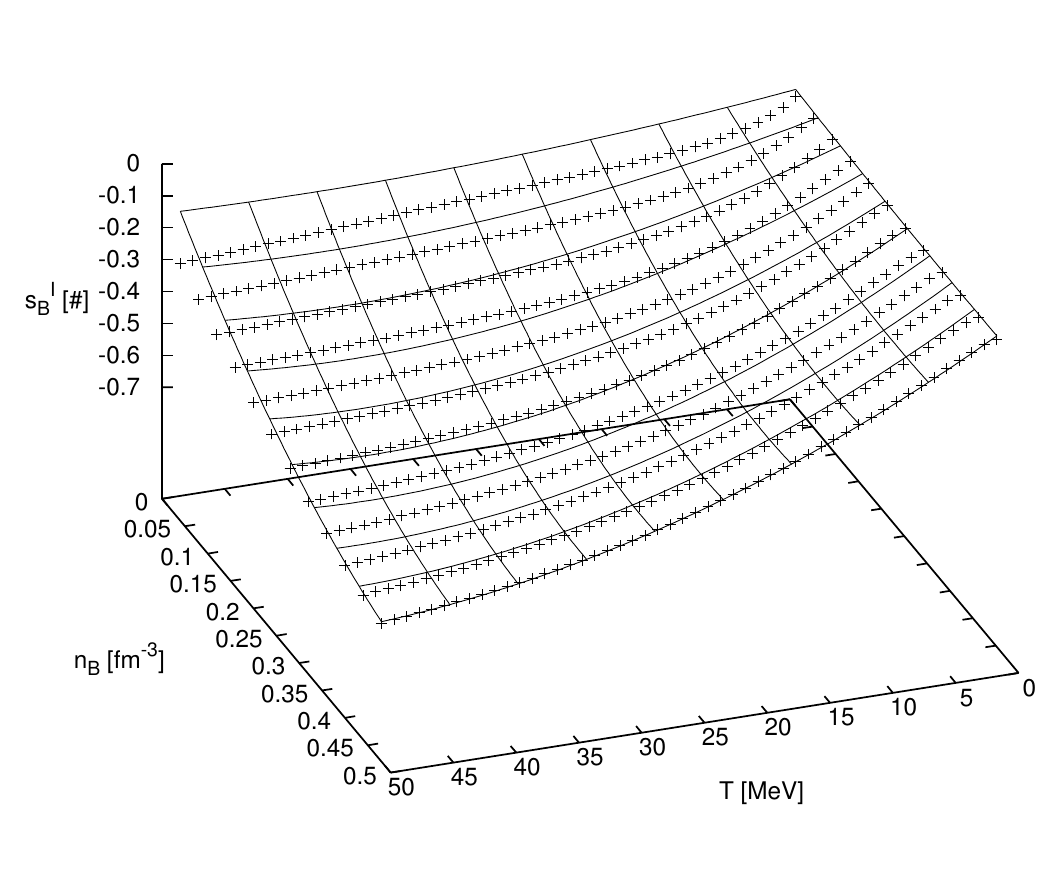}
\includegraphics[width=0.65\textwidth]{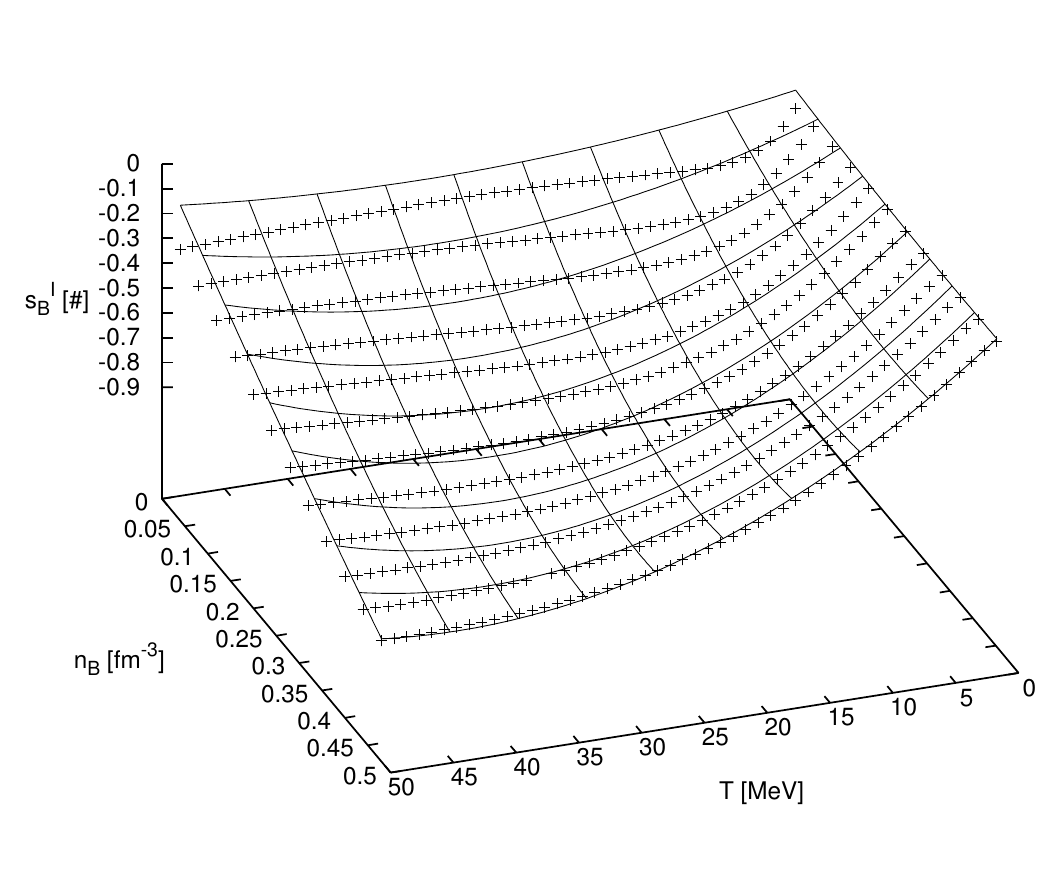}
}
\centerline{
\includegraphics[width=0.65\textwidth]{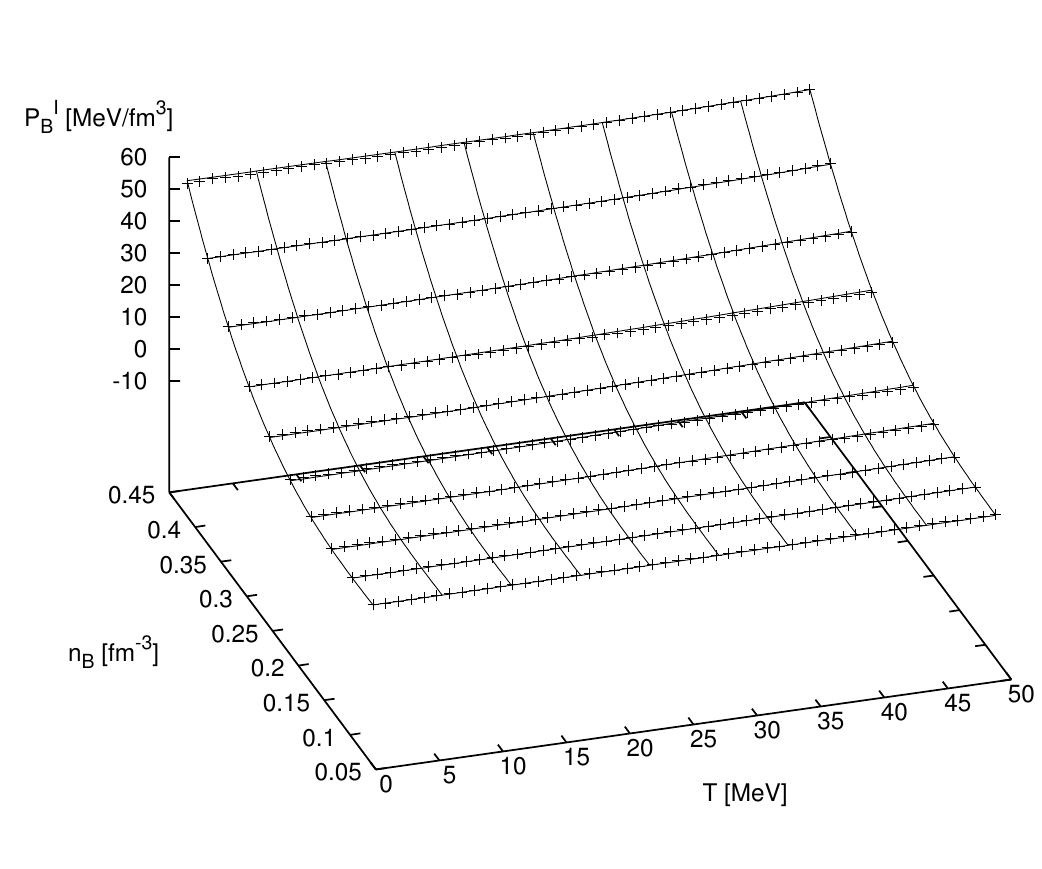}
\includegraphics[width=0.65\textwidth]{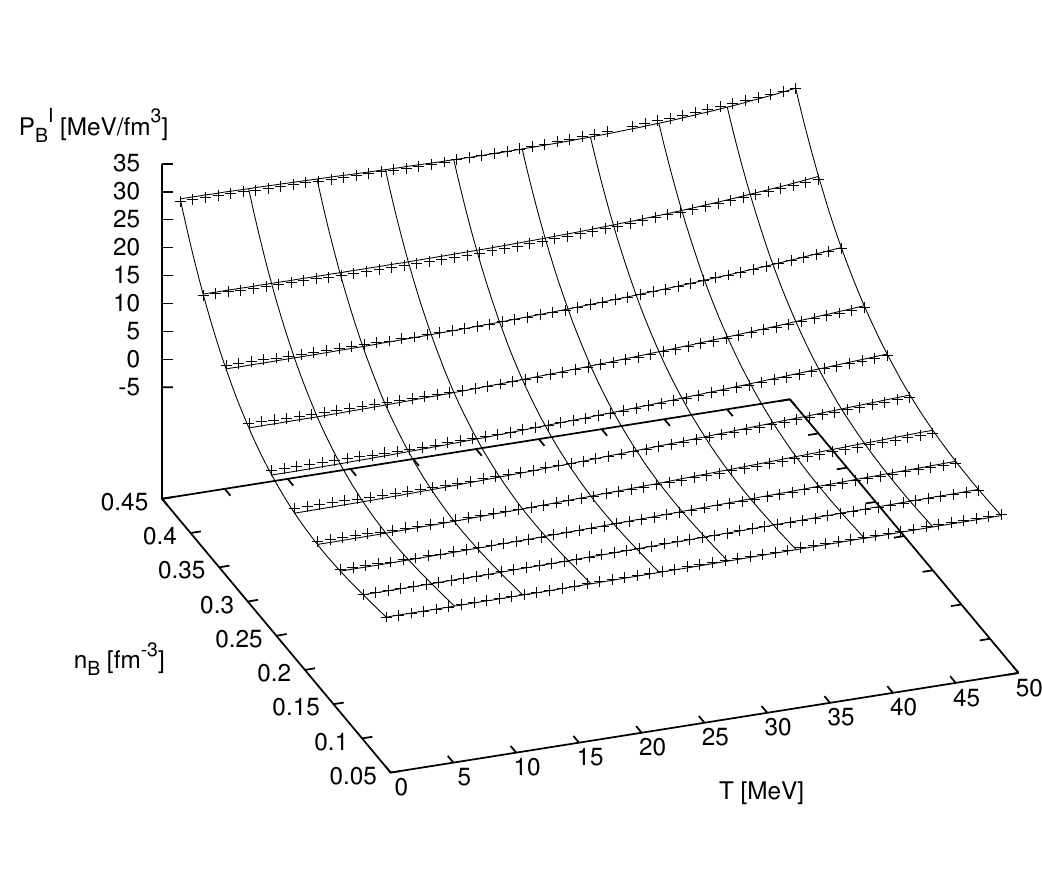}
}
\caption{As Fig.~\ref{fig:fit_GM3}, for the CBF-EI EoS.}
\label{fig:fit_CBFEI}
\end{figure}

The fit has been done with a set of points on an evenly spaced Cartesian
$11\times50\times12$ grid in $(Y_p;T;\nB)$, from
$(0;\unit[1]{MeV};\unit[0.04]{fm^{-3}})$ to
$(0.5;\unit[50]{MeV};\unit[0.48]{fm^{-3}})$, with steps of
$(0.05;\unit[1]{MeV};\unit[0.04]{fm^{-3}})$. The fit is strictly valid for $n_\mathrm{B}\in(0.04;0.48)\,\unit{fm^{-3}}$ and
$T\in(1;50)\,\unit{MeV}$, but its analytic form is suitable
to be used also for $n_\mathrm{B}<\unit[0.04]{fm^{-3}}$ and $T<\unit[1]{MeV}$. First, we have
fitted \emph{only} the interacting free energy $f^I_\mathrm{B}$, and saved the
resulting rms $\sigma_f$.  We have done the same for the interacting entropy and
pressure, obtaining $\sigma_s$ and $\sigma_P$.  Then, we have
\emph{simultaneously} fitted the interacting free energy, entropy, and pressure, giving to each fitting point $p_i$
an uniform error $\sigma_{i=\{f;s;P\}}$ that depends on which quantity that
point is describing (the free energy, the entropy, or the pressure).  The
result of the fit of the GM3 and CBF-EI EoS, obtained with
\texttt{Gnuplot}\footnote{www.gnuplot.info}, is shown in Tab.~\ref{tab:f_fit}
and Figs.~\ref{fig:fit_CBFEI} and \ref{fig:fit_GM3}. Including
in the fit also the second order derivatives, $\partial^2 f_\mathrm{B}/\partial
T^2$, $\partial^2 f_\mathrm{B}/\partial \nB^2$, and $\partial^2
f_\mathrm{B}/\partial T\partial \nB$, did not improved the accuracy.

We have checked that in the range considered in the fit, the results for the
GM3 and the CBF-EI EoSs (Tab.~\ref{tab:f_fit}) satisfy the thermodynamic stability conditions
\citep[Eqs.~(13) and (14) of][]{Swesty.1996}
\begin{align}
\left.\frac{\partial s_\mathrm{B}}{\partial T}\right|_{\nB}>{}&0,\\
\left.\frac{\partial P_\mathrm{B}}{\partial n}\right|_T>{}&0.
\end{align}
\begin{table}
\caption{Interacting baryon free energy per baryon fitting parameters, Eqs.~\eqref{eq:ffit} and
\eqref{eq:fSNM_PNM}. In the first column, we report the fitting coefficient for SNM and PNM, in
the second and third columns we report the results of the fit for the GM3 and
CBF-EI EoSs, in the fourth and last column, we report the polynomial that is
multiplied by that coefficient in the fitting formula.  In the last two rows we
report the number of points used in the fit and the reduced chi square given by \texttt{gnuplot}.
Energies are in MeV and lengths in fm. See text for details on the fit.}
\label{tab:f_fit}
\centering
\begin{tabular}{cccl}
coeff. & GM3 & CBF-EI & polynomial \\
\hline
$a_{1,\mathrm{SNM}}$ & $-402.401    $ & $-284.592   $ & $4Y_p(1-Y_p)\nB$      \\
$a_{2,\mathrm{SNM}}$ & $1290.54     $ & $676.121    $ & $4Y_p(1-Y_p)\nB^2$    \\
$a_{3,\mathrm{SNM}}$ & $-1540.52    $ & $-662.847   $ & $4Y_p(1-Y_p)\nB^3$    \\
$a_{4,\mathrm{SNM}}$ & $903.8       $ & $667.492    $ & $4Y_p(1-Y_p)\nB^4$    \\
$a_{5,\mathrm{SNM}}$ & $0.0669357   $ & $0.112911   $ & $4Y_p(1-Y_p)\nB T^2$   \\
$a_{6,\mathrm{SNM}}$ & $-0.000680098$ & $-0.00124098$ & $4Y_p(1-Y_p)\nB T^3$   \\
$a_{7,\mathrm{SNM}}$ & $-0.0769298  $ & $-0.148538  $ & $4Y_p(1-Y_p)\nB^2T^2$ \\
$a_{8,\mathrm{SNM}}$ & $0.000915968 $ & $0.00192405 $ & $4Y_p(1-Y_p)\nB^2T^3$ \\
\hline                                                                 
$a_{1,\mathrm{PNM}}$ & $-274.544    $ & $-121.362    $ & $(1-2Y_p)^2\nB$      \\
$a_{2,\mathrm{PNM}}$ & $1368.86     $ & $101.948     $ & $(1-2Y_p)^2\nB^2$    \\
$a_{3,\mathrm{PNM}}$ & $-1609.15    $ & $1079.08     $ & $(1-2Y_p)^2\nB^3$    \\
$a_{4,\mathrm{PNM}}$ & $916.956     $ & $-924.248    $ & $(1-2Y_p)^2\nB^4$    \\
$a_{5,\mathrm{PNM}}$ & $0.0464766   $ & $0.0579368   $ & $(1-2Y_p)^2\nB T^2$   \\
$a_{6,\mathrm{PNM}}$ & $-0.000388966$ & $-0.000495044$ & $(1-2Y_p)^2\nB T^3$   \\
$a_{7,\mathrm{PNM}}$ & $-0.0572916  $ & $-0.0729861  $ & $(1-2Y_p)^2\nB^2T^2$ \\
$a_{8,\mathrm{PNM}}$ & $0.00055403  $ & $0.000749914 $ & $(1-2Y_p)^2\nB^2T^3$ \\
\hline
$N$ & $19782$ & $18686$& \\
$\tilde\chi$ & $4.18$ & $2.05$ & \\
\hline
\end{tabular}
\end{table}

For the LS-bulk EoS, conversely, we use the expression for the bulk of
\citet{Lattimer+Swesty.1991},
Eqs.~\eqref{eq:fLS}, \eqref{eq:fILS}, \eqref{eq:pardelta}--\eqref{eq:parDelta}.
We remark that Eq.~\eqref{eq:fLS} is identical to Eq.~\eqref{eq:f}, and
that Eq.~\eqref{eq:fILS} can be cast in a form where it is apparent that the
dependence on $Y_p$ in the LS-bulk EoS is identical to that in
Eq.~\eqref{eq:ffit}.

To conclude this section, we provide an expression that gives the
proton and neutron chemical potentials from the baryon free energy per baryon.
Let's consider an EoS with only protons and neutrons. One can make the change of
variables
\begin{align}
\label{eq:nB}
n_\mathrm{B}(n_p,n_n)={} n_p+n_n,\\
\label{eq:Yp}
Y_p(n_p,n_n)={}\frac{n_p}{n_p+n_n},
\end{align}
such that
$f_\mathrm{B}(T,n_\mathrm{B},Y_p) = \tilde f_\mathrm{B}(T,n_p,n_n) \equiv
f_\mathrm{B}$ (we drop the tilde in the following).  The differentiation of the
free energy density in Eq.~\eqref{eq:mu2} is performed with respect to
$n_p=Y_pn_\mathrm{B}$ and $n_n=(1-Y_p)n_\mathrm{B}$; after the change of
variables we obtain
\begin{align}
\label{eq:d_dnp}
\left.\frac\partial {\partial n_p}\right|_{T,n_n} ={}& \left.\frac {\partial
\nB(n_p,n_n)}{\partial n_p}\right|_{n_n} \! \left.\frac \partial {\partial
\nB}\right|_{T,Y_p} + \left.\frac {\partial Y_p(n_p,n_n)}{\partial
n_p}\right|_{n_n} \left.\frac \partial {\partial Y_p}\right|_{T,\nB}\!,\\
\label{eq:d_dnn}
\left.\frac\partial {\partial n_n}\right|_{T,n_p}  ={}& \left.\frac {\partial
\nB(n_p,n_n)}{\partial n_n}\right|_{n_p} \left.\frac \partial {\partial
\nB}\right|_{T,Y_p} + \left.\frac {\partial Y_p(n_p,n_n)}{\partial
n_n}\right|_{n_p} \left.\frac \partial {\partial Y_p}\right|_{T,\nB}\!,
\end{align}
(to differentiate the total energy density instead of the free energy density,
one has to fix the entropy instead of the temperature).  From
Eqs.~\eqref{eq:mu2}, \eqref{eq:P1}, \eqref{eq:nB}, \eqref{eq:Yp},
\eqref{eq:d_dnp} and \eqref{eq:d_dnn} we obtain
\begin{align}
\label{eq:mup}
\mu_p={}&
f_\mathrm{B}+\frac{P_\mathrm{B}}{n_\mathrm{B}}+\left.(1-Y_p)\frac{\partial
f_\mathrm{B}}{\partial Y_p}\right|_{T,n_\mathrm{B}}\!\! =
e_\mathrm{B}+\frac{P_\mathrm{B}}{n_\mathrm{B}} + \left.(1-Y_p)\frac{\partial
e_\mathrm{B}}{\partial Y_p}\right|_{s,n_\mathrm{B}}\!,\\
\label{eq:mun}
\mu_n={}&
f_\mathrm{B}+\frac{P_\mathrm{B}}{n_\mathrm{B}}-\left.Y_p\frac{\partial
f_\mathrm{B}}{\partial Y_p}\right|_{T,n_\mathrm{B}} =
e_\mathrm{B}+\frac{P_\mathrm{B}}{n_\mathrm{B}}-\left.Y_p\frac{\partial
e_\mathrm{B}}{\partial Y_p}\right|_{s,n_\mathrm{B}}.
\end{align}

As a final remark, we notice that it is hazardous to pretend great precision on
a derivative [e.g., $g'(x)$] obtained from the differentiation of a fit
performed on a function [e.g., $g(x)$], unless there are strong theoretical
reasons to assume that fitting formula. This is the reason we have performed
the fit simultaneously on the free energy and on its first derivatives with
respect to the baryon density and the temperature\footnote{The second order
derivatives have been implicitly used in the evolutionary code described in the
next chapter, while we did not use them in the fitting procedure. However, (i)
as we have already remarked, we found out that including the second order
derivatives in the fit does not improve it, and (ii)
their exact value is physically concerning, for the determination of the
stellar quasi-normal oscillations, only in the determination of the sound
speed, $c_s^2=\partial P/\partial \epsilon$, and its value is well recovered
(see Fig.~\ref{fig:thermo_n}).  Finally, (iii) we will see in
Chapter~\ref{cha:evolution} that the results in the evolution using the GM3 EoS
obtained from the fit (Sec.~\ref{sec:totaleos}) are very similar to those
obtained with the real GM3 EoS (Sec.~\ref{sec:GM3}).}. However, we have simply
assumed a quadratic dependence on the proton fraction, without considerations
on its first derivative; and from Eqs.~\eqref{eq:mup} and \eqref{eq:mun} we see
that the derivative of the free energy with respect to the proton fraction does
appear as a contribution to the proton and neutron chemical potentials.
From Figs.~\ref{fig:Yp}, \ref{fig:mu_n} and
\ref{fig:thermo_n} it is apparent that the fit works well; however in
future it would be interesting and useful to go beyond the quadratic dependence
on the proton fraction with considerations on the behaviour of the proton and
neutron chemical potentials.

\section{Total EoS numerical implementation}
\label{sec:totaleos}

The thermodynamical quantities that we have discussed in Sec.~\ref{sec:fBfit}
refer to baryons, namely protons $p$ and neutrons $n$.  But in our star we
include also electrons $e^-$, positrons $e^+$, and the 3 species of neutrinos
($\nu_e$, $\nu_\mu$, and $\nu_\tau$) and antineutrinos ($\bar\nu_e$,
$\bar\nu_\mu$, and $\bar\nu_\tau$). In this section we discuss how one can
obtain the total EoS thermodynamical quantities from the baryon ones. Muon
and tauon neutrino chemical potentials are assumed to be zero, and we assume
beta-equilibrium [Reaction~\eqref{eq:beta}].  We do not consider photons, whose
contribution is negligible\footnote{The density and energy density for a photon
gas with $\mu_\gamma=0$ [Eq.~\eqref{eq:mugamma}] at $\unit[50]{MeV}$ is
$n_\gamma=\unit[0.004]{fm^{-3}}$ and
$\epsilon_\gamma=\unit[0.54]{MeV\,fm^{-3}}$, to be compared with the typical energy
densities of Fig.~\ref{fig:thermo_n}.}.

Usually, one uses \emph{barotropic} EoSs to describe ``cold'' neutron stars.
Baro{-}tropic means that all the thermodynamical quantities can be derived (analytically or by
table interpolation) from only one independent variable. The form of the TOV equations
(see Sec.~\ref{sec:tov}), that permit to determine the structure of the star,
suggests to employ as independent variable the pressure $P$ (from which the term
``barotropic'').  As we have discussed in Sec.~\ref{sec:thermodynamical}, a
finite temperature EoS needs more than one independent variable, and it is therefore
called \emph{non-barotropic}. The particular choice of the independent variables
depends on which use one has to make of such an EoS.  Since we want to solve
the TOV equations, it is useful to adopt the total pressure $P$ as one of the
independent variables.  The form of the diffusion equations \eqref{eq:YLfinal} and
\eqref{eq:sfinal} suggests to use as additional independent variables the total
entropy $s$ and the electron lepton fraction $Y_L\equiv Y_{L_e}$. The choice to
use as third independent variable the lepton fraction is convenient also because the
lepton number is conserved by the reactions that we consider in the evolution
and may change only because of neutrino diffusion (whereas, e.g., electron
total number is not conserved since electrons transform in electron neutrinos and
viceversa). Three variables are enough to determine all the other
thermodynamical quantities, since for the particle species we are considering
the request of beta-equilibrium and of charge conservation provide constraints related to
all the other particle number fractions [see Eq.~\eqref{eq:mu_nudYL}].  In
other words, since only electron type neutrinos have a non vanishing chemical
potential and we assume beta equilibrium, only three independent variables are
needed to determine the total EoS\footnote{In principle, one can relax the
requests of thermal and beta equilibrium for neutrinos.  However, that results
in increasing the transport equations to be solved. If one relaxes the request
of beta-equilibrium, one has to add an equation for the neutrino number
evolution.  If one relaxes both the requests of thermal and beta-equilibrium,
one has to use a multi-energy transport scheme \citep{Roberts.2012}.}.

In practice to obtain the total thermodynamical EoS we have run a Newton-Raphson
cycle using as independent variables the proton and neutron \emph{auxiliary} ``free''
chemical potentials $\mu^K_p$ and $\mu^K_n$, the temperature $T$, and the
electron chemical potential $\mu_{e^-}$.  $\mu^K_p$ and $\mu^K_n$ are the
chemical potentials of a neutron and proton free Fermi gas with the same $\nB$,
$T$ and $Y_p$ of the total interacting baryon EoS. Using the relativistic and
finite temperature free Fermi gas EoS of
\citet{Eggleton+Faulkner+Flannery.1973} and \citet{Johns+Ellis+Lattimer.1996},
we have determined $Y_p$ and $\nB$ from $\mu^K_p$, $\mu^K_n$ and $T$. Then,
using the fitting formula \eqref{eq:f}, we have determined the baryon
thermodynamical quantities, including the total chemical potentials,
$\mu_p=\mu_p^K+\mu_p^I$ and $\mu_n=\mu_n^K+\mu_n^I$ [the interacting chemical
potentials are given by Eqs.~\eqref{eq:mup} and \eqref{eq:mun}].  Requiring
beta-equilibrium [Eq.~\eqref{eq:beta1}] and assuming that the muon and tauon
neutrinos have vanishing chemical potential [Eq.~\eqref{eq:no-trapped}], we have obtained the
other lepton chemical potentials,
\begin{align}
\label{eq:beta1}
\mu_{\nu_e}={}&\mu_p-\mu_n+\mu_{e^-},\\
\label{eq:no-trapped}
\mu_{\nu_\mu}={}&\mu_{\nu_\tau}=0,\\
\mu_{\bar \nu_{\{e,\mu,\tau\}}}={}&-\mu_{\nu_{\{e,\mu,\tau\}}},\\
\mu_{e^+}={}&-\mu_{e^-},
\end{align}
and, from them, the other thermodynamical quantities.  Neutrinos have been assumed
to be massless, and we have adopted the EoS of free massless Fermions,
Eqs.~(C.1) and (C.3) of \citet{Lattimer+Swesty.1991},
\begin{align}
\label{eq:Y_nu}
Y_{\nu_i}-Y_{\bar\nu_i}={}&\frac1{6\pi^2\hbar^3c^3n_\mathrm{B}}
\mu_{\nu_i}\left(\mu_{\nu_i}^2 + \pi^2T^2\right),\\
\label{eq:eps_nu}
\epsilon_{\nu_i}+\epsilon_{\bar\nu_i}={}&\frac1{8\pi^2\hbar^3c^3}\left(\mu_{\nu_i}^4
+ 2\pi^2T^2\mu_{\nu_i}^2 +\frac7{15}\pi^4T^4\right),\\
\label{eq:P_nu}
P_{\nu_i}+P_{\bar\nu_i}={}&\frac{\epsilon_{\nu_i}+\epsilon_{\bar\nu_i}}{3},\\
\label{eq:s_nu}
s_{\nu_i}+s_{\bar\nu_i}={}&\frac1{6\hbar^3c^3n_\mathrm{B}} T\left(\mu_{\nu_i}^2
+ \frac7{15}\pi^2 T^2\right),
\end{align}
where $T\equiv T_\mathrm{matter}$ because neutrinos are in thermal equilibrium
with the matter via the nuclear processes (scattering, absorption and
emission). Since muon and tauon neutrinos have vanishing chemical potential,
their number fraction is zero everywhere (i.e., they are produced in pairs).
This is the reason we consider only the electron-type (net) lepton number,
$Y_L\equiv Y_{L_e}$.  Massive leptons have been treated as a non-interacting relativistic
Fermi gas \citep{Eggleton+Faulkner+Flannery.1973, Johns+Ellis+Lattimer.1996}.
The Newton-Raphson cycle ends when the total thermodynamical quantities converge
to the targeted ones (which we have set to be $P$, $s$, and $Y_L$, see discussion
above), with the additional request of electric neutrality. In other words, the
procedure we have just described consists in varying four independent variables
($\mu_p^K$, $\mu_n^K$, $\mu_e$, and $T$), trying to fulfill four independent
equations, three for obtaining the targeted EoS quantities ($P$, $s$, and
$Y_L$) and one for the charge neutrality. The other thermodynamical quantities
of interest, for example $\epsilon$, $\mu_{\nu_e}$, are obtained
from the above procedure as well.

\section{Comparison between EoSs}
\label{sec:comparison}

The EoS influences the PNS evolution both directly and indirectly; directly because
the thermodynamical quantities are used in solving the structure and
diffusion equations, and indirectly because to compute the diffusion
coefficients it is fundamental to account for the Fermi blocking relations
\citep{Pons+.1999} and the effective interaction
\citep{Reddy+Prakash+Lattimer.1998}, see Sec.~\ref{sec:diffusion} of this
thesis.  To illustrate the behaviour of the EoSs in different regimes, we
consider in this section three cases: (i) $Y_L=0.4$ and $s=1$ (that corresponds to the
condition at the center of the PNS at the beginning of the simulation), (ii) $Y_\nu\equiv
Y_{\nu_e}=0$ and $s=2$ (the condition present in the star at the end of the
deleptonization phase), and (iii) $Y_\nu\equiv Y_{\nu_e}=0$ and
$T=\unit[5]{MeV}$ (which is the condition in most of the star at the end of our
simulations, i.e., toward the end of the cooling phase).

\begin{figure}
\centerline{
\includegraphics[width=1.25\textwidth]{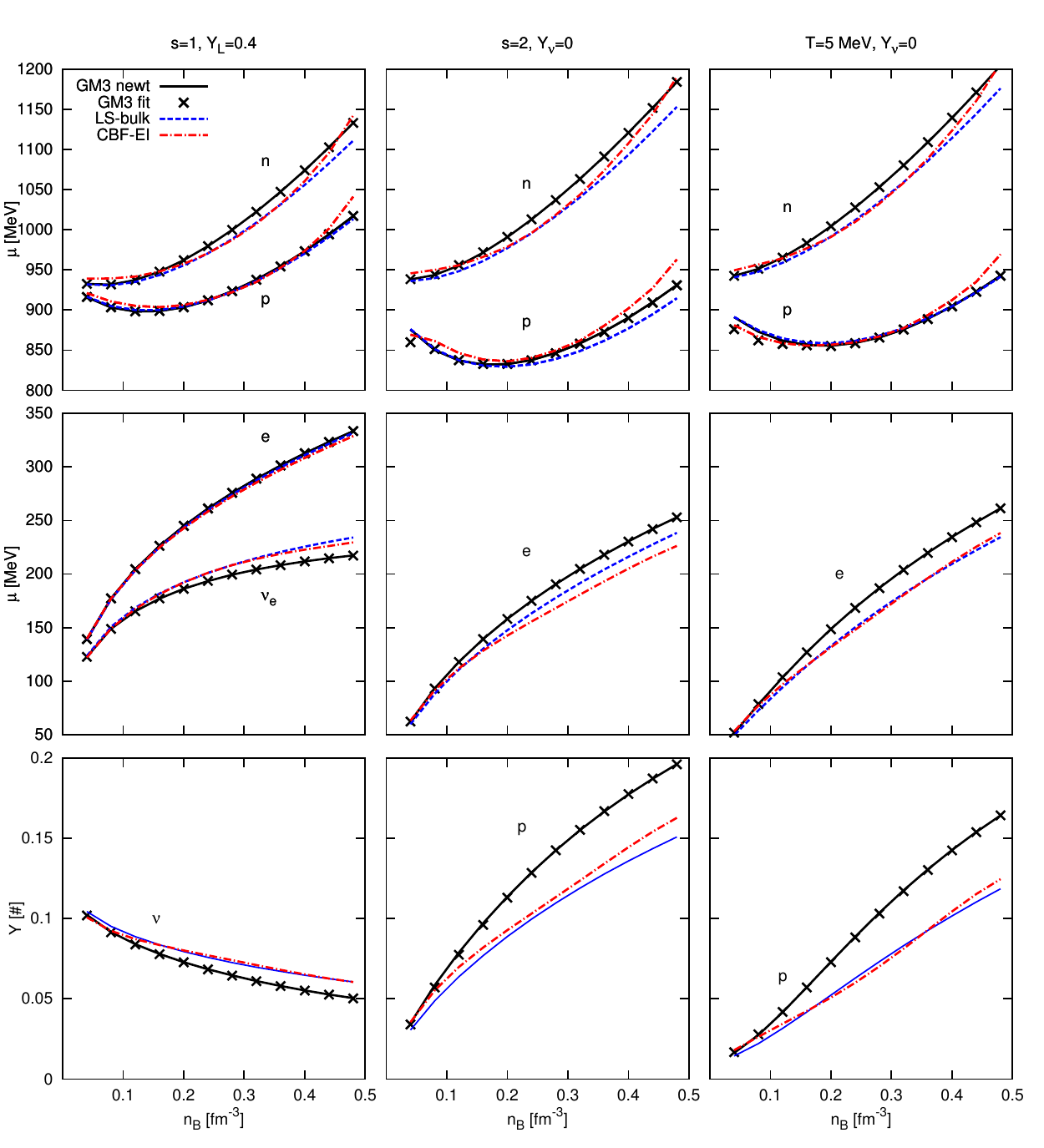}}
\caption{Comparison between the three EoSs used in
this thesis, in the three regimes described in Sec.~\ref{sec:comparison}.
In the top and middle rows, we plot the chemical potential (in
regimes (ii) and (iii) $\mu_{\nu_e}=0$, because $Y_\nu=0$).
In the lower row, we plot the particle abundances
(the abundances not plotted may be easily determined with the relations $Y_n=1-Y_p$,
$Y_e=Y_p$ and $Y_L=Y_e+Y_{\nu_e}$).
The black solid line refers to the GM3 EoS
determined by solving numerically the mean-field equations, 
the black crosses to
the GM3 EoS determined through the fit and the procedure described in
Sec.~\ref{sec:totaleos}, the blue dashed line to the LS-bulk EoS, and the red
dot-dashed line to the CBF-EI EoS.}
\label{fig:mu_n}
\end{figure}

\begin{figure}
\centerline{
\includegraphics[width=1.25\textwidth]{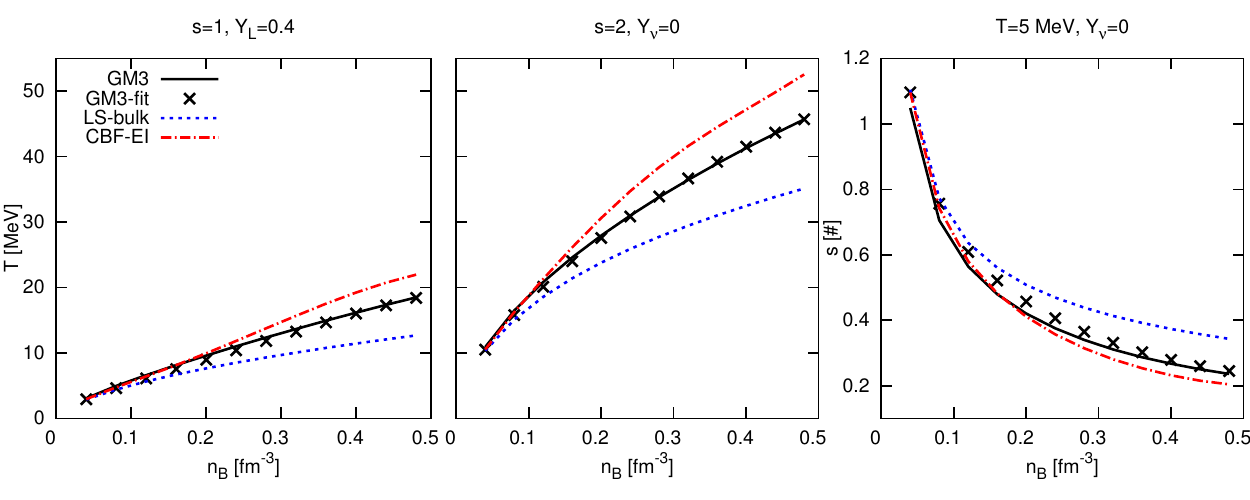}
}
\caption{Comparison between the three EoSs in the three cases described in
Sec.~\ref{sec:comparison}. In the left and central plots we
show the temperature and in the right plot we show the entropy per baryon.
Colors and line-styles are as in Fig.~\ref{fig:mu_n}.}
\label{fig:temp_entr}
\end{figure}

In Fig.~\ref{fig:mu_n} we show the chemical potentials and
the number fractions of the different species present in the PNS, in the three
regimes (i)-(iii). All the three EoSs exhibit a very similar behaviour, since
these EoSs have the same particle species and similar symmetry energy and
imcompressibility parameter.  As expected, a high electron type lepton content
[$Y_L=0.4$, regime (i)] causes a high proton fraction. This is due to the high electron fraction, combined
with the request of charge neutrality. Also higher temperatures cause a higher proton fraction (compare regimes (ii) and (iii)
in Figs.~\ref{fig:mu_n} and \ref{fig:temp_entr}).  Therefore, at the end of the deleptonization phase
[i.e., in regime (ii)], the proton fraction decreases but it is still high
enough to allow for charged current reactions.  When $Y_\nu=0$, the proton and the electron
number fractions become identical because of
charge neutrality (there are almost no positrons).

\begin{figure}
\centerline{
\includegraphics[width=1.25\textwidth]{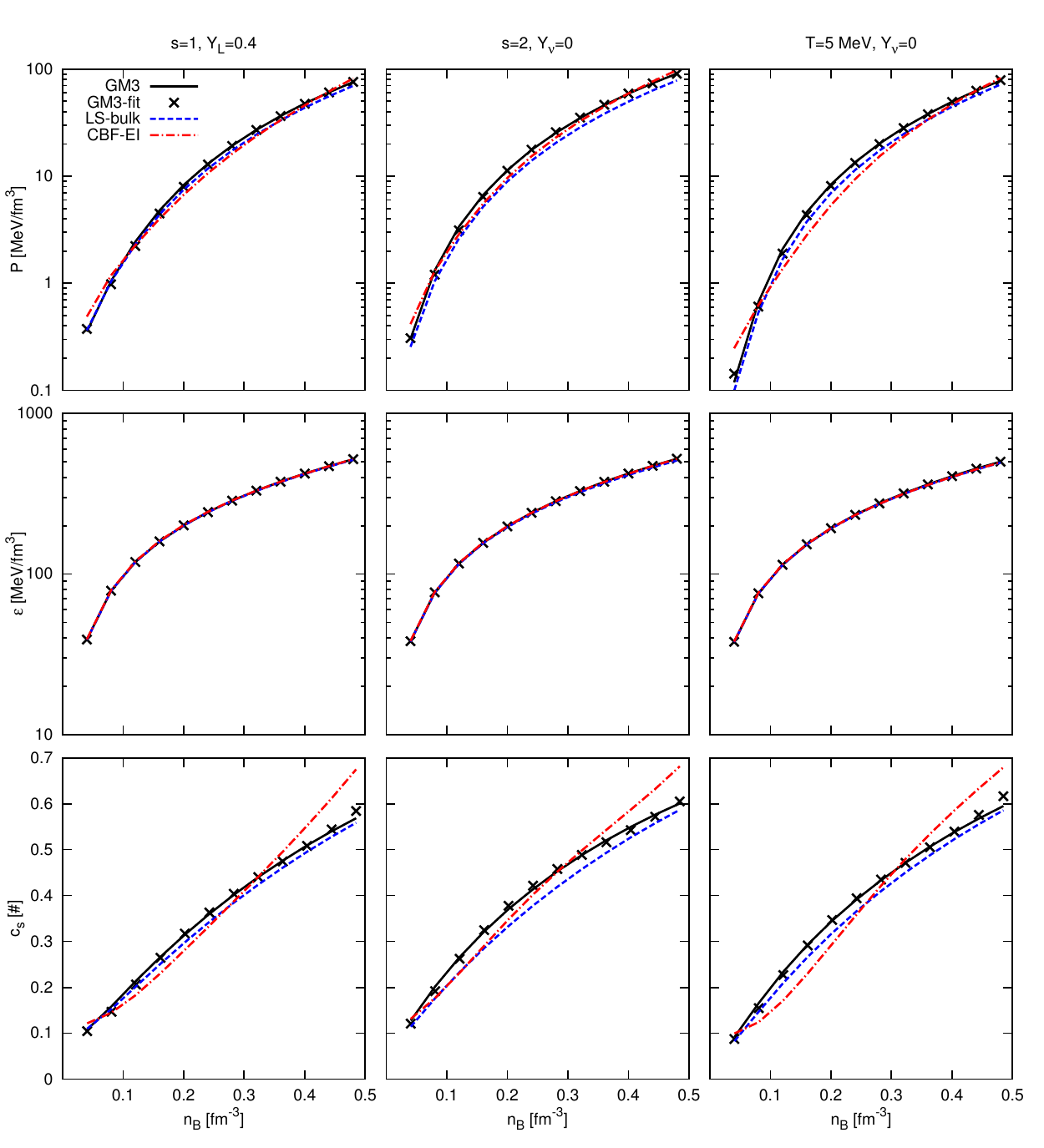}}
\caption{Comparison between the pressure (upper row), the energy density (middle row), and the
square of the speed of sound (bottom row) for the three EoSs
used in this thesis.
Colors and line-styles are as in
Fig.~\ref{fig:mu_n}.}
\label{fig:thermo_n}
\end{figure}

In Figs.~\ref{fig:temp_entr} and \ref{fig:thermo_n} we show the dependence on
the baryon number density of the temperature, entropy per baryon, pressure,
energy density, and square of the sound speed,
\begin{equation}
\label{eq:sound_speed}
c_s^2=\left.\frac{\partial P}{\partial \epsilon}\right|_{s,Y_L},
\end{equation}
for the three EoSs in the three different regimes.
The pressure and energy density in the three regimes
have a very similar dependence on the number density, since their major contribute is due to the baryon
interaction and degeneracy \citep{Pons+.1999}, rather than being thermal. At
saturation density $n_s$ (whose exact value is slightly different for the three
EoSs, but is in the range $n_s=\unit[0.15\textrm{-}0.16]{fm^{-3}}$), the sound
speed is slightly larger (smaller) for the EoS with larger (smaller)
imcompressibility parameter $K_s$. At high baryon density, the sound speed of
the CBF-EI EoS is larger than that of the LS-bulk and GMr EoSs; this is due to
a well-known problem of the many-body EoSs, which violate causality at very
high density. However, in the regime of interest for this thesis, this
unphysical behaviour can safely be neglected.
We also notice that at given entropy per
baryon $s$ and baryon density $\nB$ the LS-bulk EoS is colder than GM3 EoS,
whereas CBF-EI EoS is hotter than GM3 EoS (see
Fig.~\ref{fig:temp_entr}).  The LS-bulk EoS is colder because we do not include
thermal contributions to the interacting part of the LS-bulk EoS,
Eq.~\eqref{eq:fLS}, and therefore the entropy is given only by the kinetic
part.  On the other hand, the CBF-EI EoS is hotter because the interaction is
stronger for the CBF-EI, and therefore the (negative) entropy contribution is
larger: the mean-field EoS is ``more disordered'' than the many-body one. In
fact, the interaction lowers the total entropy, see
Figs.~\ref{fig:fit_GM3} and \ref{fig:fit_CBFEI}, see also the right panel
of Fig.~\ref{fig:temp_entr}.  Therefore, fixing the total entropy contribution,
the corresponding temperature is lower for the LS-bulk EoS, and higher for
CBF-EI EoS.

\begin{figure}
\centering
\includegraphics[width=\textwidth]{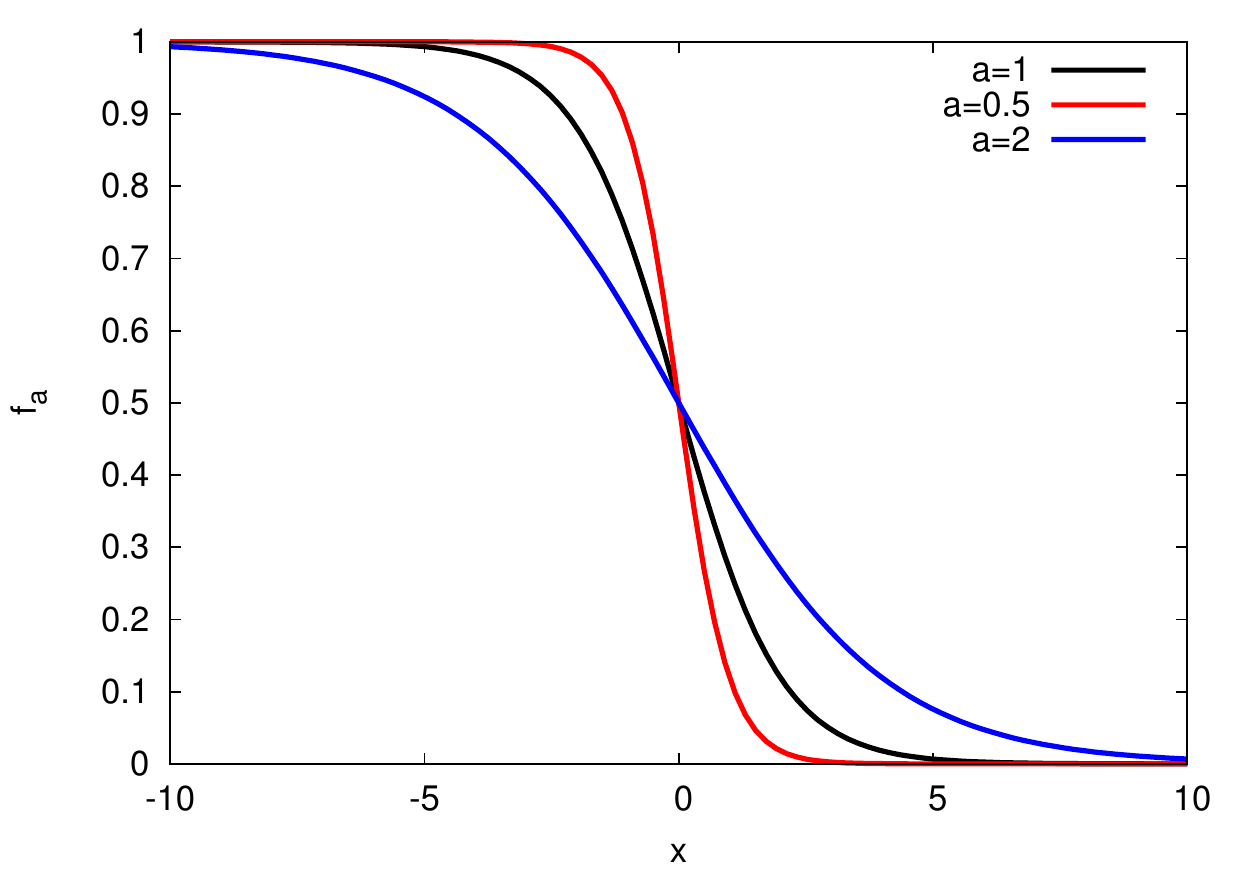}
\caption{Plot of the ``distribution function'' of
Eq.~\eqref{eq:fake_distr_func}, varying the value of the parameter $a$.}
\label{fig:fake_distr_func}
\end{figure}

\begin{figure}
\centerline{
\includegraphics[width=1.25\textwidth]{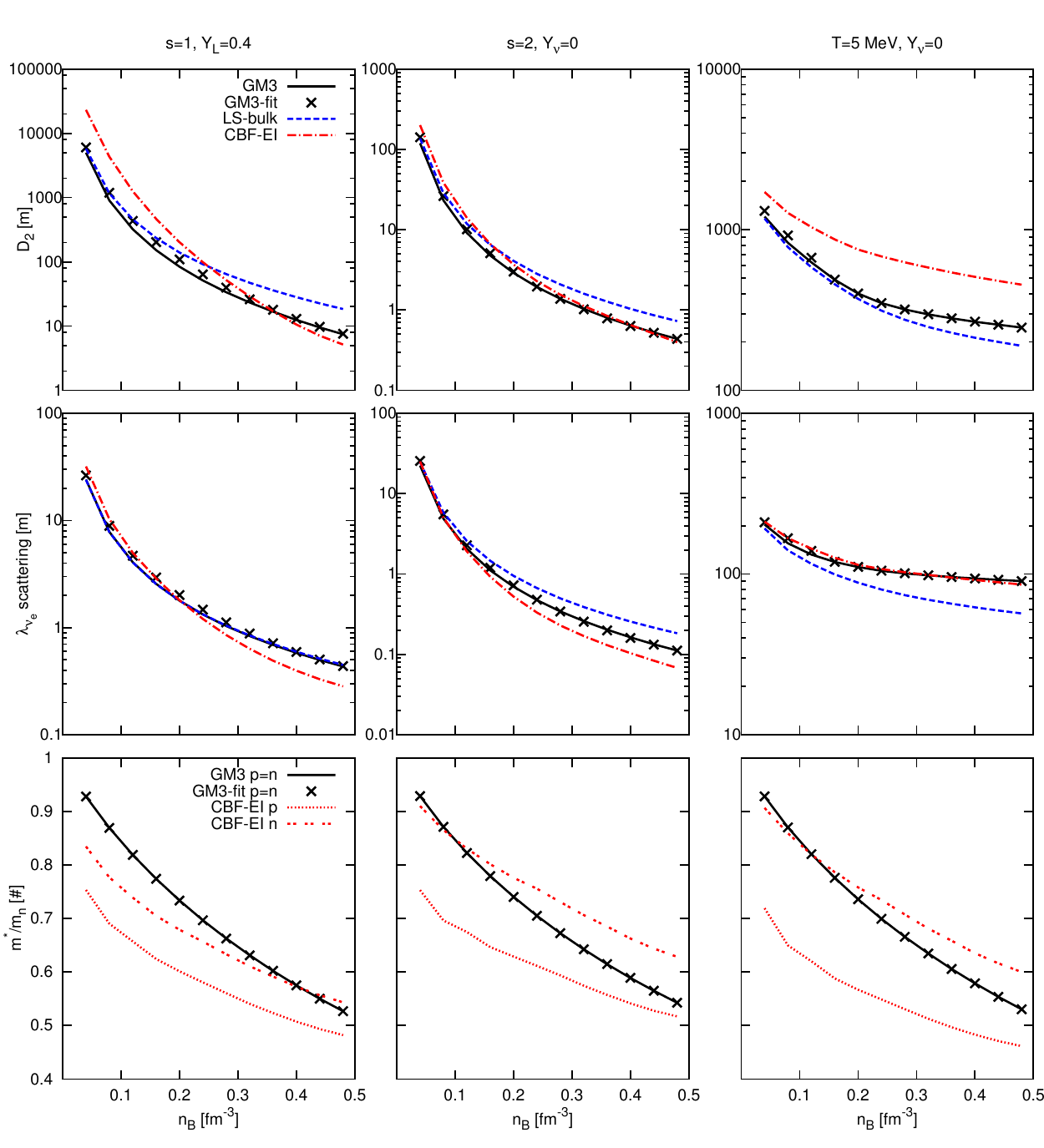}}
\caption{Comparison between the diffusion coefficient $D_2$ (upper panel), the
electron neutrino scattering mean free paths [middle panel,
the neutrino incoming energy is $E_{\nu_e}=\max(\mu_{\nu_e},\pi T)$], and 
the baryon effective masses, $m^\ast/m_n$ (lower panel) for the three EoSs 
considered in this thesis in the three cases described in
Sec.~\ref{sec:comparison}. 
For the GM3 EoS, the effective masses of proton and neutron are
identical~\cite{Glendenning.1985,Glendenning+Moszkowski.1991}.  We do not show the LS-bulk EoS effective masses, since
we have set them equal to the bare ones, $m^\ast_{\{p,n\}}/m_n=1$.
Colors and line-styles are as in
Fig.~\ref{fig:mu_n},
apart for the line-styles in the lower panel, where the CBF-EI proton (neutron) effective masses are dotted (double-dashed).}
\label{fig:diff}
\end{figure}

In Fig~\ref{fig:diff} we plot the neutrino diffusion coefficient $D_2$, the
electron neutrino scattering mean free path, and the baryon effective
masses at different regimes ---
the incident neutrino energy used to compute the neutrino mean free path is
$E_{\nu_e}=\max(\mu_{\nu_e},\pi T)$. To understand the role of interaction and
finite temperature in the neutrino diffusion, we consider their effects on the
baryon distribution function.  To illustrate that, it is useful to look at the
behaviour of the function
\begin{equation}
\label{eq:fake_distr_func} f_a(x)=\left(1+\mathrm e^{\frac xa}\right)^{-1}
\rightarrow\frac1{h^3}\left(1+\mathrm e^{\frac
{k^2}{2m^\ast T} - \frac{\mu-U-m}{T}}\right)^{-1},
\end{equation}
where $a$ plays the role of the temperature $T$ and/or effective mass $m^\ast$,
as it is clear after making a suitable change of variable that makes $f_a(x)$
the distribution function of a non-relativistic Fermion gas [right part of Eq.~\eqref{eq:fake_distr_func}].  As it is apparent
from Fig.~\ref{fig:fake_distr_func}, $f_a$ approaches a theta-function
as $a$ decreases, whereas increasing $a$ it becomes smoother.
Returning to the problem of neutrino diffusion, this means that for lower
temperatures $T$ and effective masses $m^\ast$ (i.e., lower $a$), the baryon
function becomes steeper, low-energy neutrinos can interact only with particles
near the Fermi sphere, and therefore the mean free paths and diffusion
coefficients increase. Conversely, a greater temperature and effective mass
means that the particle distribution function is smoother, low-energy neutrinos
may interact with more particles (since the Pauli blocking effect is lower),
and therefore the mean free paths and the diffusion coefficients are smaller.
The scattering mean free paths  reflect the temperature dependence of the three
EoSs: when the
matter is hotter, the scattering is more effective
(cf.~Fig.~\ref{fig:temp_entr} and the left and central plots of the middle row
of Fig.~\ref{fig:diff}). At equal temperature, the interaction is more
effective when the effective mass is greater (right plot of the lower row of
Fig.~\ref{fig:diff}, cf.~effective masses in Fig.~\ref{fig:diff}).
The behaviour of the diffusion coefficient $D_2$ (higher row of
Fig.~\ref{fig:diff}) results from a
complex interplay between scattering and absorption, for which the effective
masses and single particle potentials play an important role.  The comparison between
the diffusion coefficient $D_2$ for the three EoSs suggests that towards
the end of the cooling phase (in which the thermodynamical conditions
are roughly similar to the case (iii) described in this section), the CBF-EI
star evolves faster than the other EoSs.

\chapter{Proto-neutron star evolution}
\label{cha:evolution}
The evolution of the proto-neutron star phase, despite of being simpler to
study with respect to the precedent core collapse and core bounce phases, has
received a relatively smaller attention in the literature than the former
phases.  In fact, the main focus of numerical relativity groups has been to
develop codes that could handle the more complex and shorter (up to hundreds of
milliseconds) explosion phase.
Nevertheless, there exist codes written for the PNS phase
\citep{Burrows+Lattimer.1986, Keil+Janka.1995, Pons+.1999, Roberts.2012}.
These codes have been used to study the evolutionary dynamics of the
PNS \citep{Burrows+Lattimer.1986}, the neutrino detection from this phase
\citep{Burrows.1988, Keil+Janka.1995, Pons+.1999}, the dependence of the
evolution on the underlying EoS (\citealp{Burrows.1988, Keil+Janka.1995,
Pons+.1999, Roberts+2012} but to our knowledge the case of a many-body EoS has
never been studied), to presence of accretion \citep{Burrows.1988}, and to
determine whether and under which conditions a black hole would form in a SN
event \citep{Burrows.1988, Keil+Janka.1995, Pons+.1999}.
\citet{Roberts.2012} wrote a spectral PNS evolution code to study the
nucleosynthesis due to neutrino interaction with the SN remnant
\citep{Roberts+2012}. \citet{Miralles+Pons+Urpin.2000} and \citet{Roberts+2012}
also studied the PNS evolution with convection.

The study of the gravitational waves from quasi-normal oscillations, emitted during the
PNS phase, has been done only with the GM3 mean-field EoS by
\citet{Ferrari+Miniutti+Pons.2003}. The case of a PNS with a many-body EoS
(that of \citealp{Burgio+Schulze.2010}) has been studied by
\citet{Burgio+2011}, where the authors mimic the PNS time evolution using some
reasonable but \emph{not} self-consistently evolved radial profiles of entropy
per baryon and lepton fraction.  The main goal of this thesis is to determine how the GW
frequencies emitted during the PNS phase depends on the underlying EoS in a
\emph{self-consistent} fashion.  Since the timescale of interest for the GW
emission in the PNS phase is on the order of ten seconds \citep{Ferrari+Miniutti+Pons.2003, Burgio+2011},
we have to follow the evolution of the PNS during this period. We have therefore
written a general (i.e., that could be run with a general EoS) evolutionary
PNS code.

In this chapter we discuss the approximations, the equations, and the code used to follow the PNS evolution,
and describe the results with the three different nuclear EoSs introduced in Chapter~\ref{cha:EoS} and with three
stellar baryon masses $M_B=1.25,1.40,1.60\unit{M_\odot}$. In Sec.~\ref{sec:tov}
we discuss the equations of stellar structure, that is, the
Tolman-Oppenheimer-Volkoff equations. Then we derive the neutrino transport
equations that determine the time evolution of the PNS, first empirically in
Sec.~\ref{sec:BLEempirical}, and then rigorously in Sec.~\ref{sec:BLE}.  In
Sec.~\ref{sec:code} we describe the numerical implementation of the PNS
evolution that, even though simpler than the implementation of a SN explosion code,
is not trivial at all. The last two sections, in which we describe the
differences for the three EoSs and the three stellar masses in the PNS evolution (Sec.~\ref{sec:PNSresults})
and in the neutrino signal in a terrestrial detector
(Sec.~\ref{sec:detectors}), contain the main original results of
this chapter \citep{Camelio+2017}.

\section{Stellar structure equation (TOV)}
\label{sec:tov}
Describing the evolution of a star means that one knows the stellar structure
at each time. What we mean with ``stellar structure'' is the knowledge of
the (radial, since we work in spherical symmetry) profiles of the local and integrated physical quantities.
With ``local'' we refer to the thermodynamical quantities and to the metric functions (that are defined below),
and with ``integrated'' we refer for example to the baryonic or gravitational mass enclosed in a radius $r$.
Since, as we will show later, the evolution of a PNS is quasi-stationary, the stellar
structure is given by the so called TOV equations, which determine the structure of a static and spherical
star. In this section we discuss the TOV equations and recast them in a form which is more suitable to our case, that is, for a PNS.

To obtain the TOV equations, one first considers the most general static and
spherically symmetric metric, that may be written as
\begin{equation}
\label{eq:metric}
\mathrm ds^2=-\mathrm e^{2\phi}\mathrm dt^2+\mathrm e^{2\lambda}\mathrm dr^2 +
r^2\mathrm d\theta^2+r^2\sin^2\theta\mathrm d\varphi^2.
\end{equation}
where $\phi\equiv\phi(r)$ and $\lambda\equiv\lambda(r)$ are metric functions that depend only on
the coordinate radius $r$, since we have imposed staticity and spherical symmetry.
One then assumes that the matter may be described by a perfect fluid, that is, shear stresses
and energy transport are negligible. Since we want to study the evolution of a PNS, the latter
condition may seem contradictory. However, the quasi-stationarity of the PNS evolution
guarantees that the energy transport is negligible on a hydrodynamic timescale, which is the
timescale in which the star rearranges itself reaching structural equilibrium, and which is
therefore the timescale relevant for the TOV equations.
The perfect fluid stress-energy tensor is given by
\begin{equation}
\label{eq:Tmunu}
T^{\mu\nu}=(\epsilon+P)u^\mu u^\nu + Pg^{\mu\nu},
\end{equation}
where $\epsilon$ is the total (of all species, included the neutrinos) energy density, $P$ is the pressure, $g^{\mu\nu}$ is
the metric, and $u^\mu$ the fluid four-velocity.  Then, one usually considers
the four equations of the energy-momentum conservation,
\begin{equation}
\label{eq:Tconservation}
T^{\mu\nu}_{\hphantom{\mu\nu};\nu}=0,
\end{equation}
plus $6$ other equations from the $16$ field equations
\begin{equation}
\label{eq:fieldeqs}
G_{\mu\nu}=8\pi GT_{\mu\nu},
\end{equation}
where $G_{\mu\nu}$ is the Einstein field tensor.  After the calculations (for
details, see \citealp[e.g.][Chapter~23]{Misner+Thorne+Wheeler.book}), one
obtains inside the star
\begin{equation}
\label{eq:TOVoriginal}
\frac{\mathrm dP}{\mathrm dr} = - \frac{(\epsilon+P)(m + 4\pi r^3
P)}{r(r-2m)},
\end{equation}
where
\begin{equation}
\label{eq:MG_r}
m(r)= \int_0^r 4\pi r^2\epsilon \mathrm d r,
\end{equation}
is the total mass-energy inside radius $r$. Eqs.~\eqref{eq:TOVoriginal} and \eqref{eq:MG_r} are the TOV equations.
The relation between $\epsilon$ and $P$ is given by the EoS. However, as we
have seen in Chapter~\ref{cha:EoS}, the PNS EoS is not barotropic, that is, it
depends on more than one variable, for example it depends also on the entropy per
baryon $s$ and the lepton fraction $Y_L$, $\epsilon\equiv\epsilon(P,s,Y_L)$.  Therefore,
to integrate the TOV equations, one needs information on the stellar
thermal and composition content, that is, one needs to know the entropy and lepton
fraction profiles, $s(r)$ and $Y_L(r)$.

The other quantities that determine the stellar structure are similarly obtained,
\begin{align}
\frac{\mathrm d\phi}{\mathrm d r} = {}& \frac{m + 4\pi r^3P}{r(r-2m)},\\
\label{eq:a_R}
a(r)={}& \int_0^r 4\pi r^2 \nB\mathrm e^\lambda\mathrm dr,\\
\label{eq:lambda}
\lambda(r)={}&-\frac12\log\big(1-2m(r)/r\big),\\
m_\mathrm{B}(r)={}& m_n a(r),
\end{align}
where $a(r)$ and $m_\mathrm{B}(r)$ are the total baryon number and total baryon mass
enclosed in a sphere of radius $r$.

It is easy to cast the stellar structure equations \eqref{eq:TOVoriginal}-\eqref{eq:a_R} in the following
form,
\begin{align}
\label{eq:drda}
\frac{\mathrm dr}{\mathrm da} ={}& \frac 1{4\pi r^2 \nB \mathrm e^\lambda},\\
\label{eq:dmda}
\frac{\mathrm dm}{\mathrm da} ={}& \frac \epsilon{\nB\mathrm e^\lambda},\\
\label{eq:dphida}
\frac{\mathrm d\phi}{\mathrm da} ={}& \frac{\mathrm e^\lambda}{4\pi r^4\nB}(m + 4\pi r^3P),\\
\label{eq:dPda}
\frac{\mathrm d P}{\mathrm da} ={}& -(\epsilon+P)\frac{\mathrm e^\lambda}{4\pi r^4\nB}(m + 4\pi r^3P),
\end{align}
where $\lambda$ is given by Eq.~\eqref{eq:lambda}. We have chosen as independent variable the enclosed baryon number
$a$, that is a more natural independent variable than the radius $r$.  In fact, as we have discussed in
Chapter~\ref{cha:EoS}, all particles but neutrinos are locked during the
timescale of PNS evolution, and therefore
the total entropy $S$, lepton number $N_L$, and baryon number $N_\mathrm{B}$ contained
in the $i$-th stellar layer (which is centered in $a_i$ and whose baryon
content is $\mathrm da_i$) is conserved during the evolution, apart for the
entropy and lepton number moved from that layer to the neighbouring ones by neutrino
diffusion (see Secs.~\ref{sec:BLEempirical} and \ref{sec:BLE}). As before, to solve the TOV equations one needs information
on the stellar thermal and composition content, and therefore needs the
profiles $s(a)$ and $Y_L(a)$.

The structure equations \eqref{eq:drda}--\eqref{eq:dPda} are completed by the boundary conditions
\begin{align}
\label{eq:m0}
m(a=0)={}&0,\\
\label{eq:r0}
r(a=0)={}&0,\\
\label{eq:PA}
P(a=A)={}&P_s,\\
\label{eq:phiA}
\phi(a=A)={}&\frac12 \log(1-2M/R),
\end{align}
where $A$ is the PNS total baryon number (that is conserved
during the evolution, since we do not account for accretion or expulsion of
material from the PNS), $R=r(a=A)$ and $M=m(a=A)$ are the stellar radius and the
total gravitational mass, respectively, and $P_s$ is the surface pressure, that is, a very
low pressure that defines for numerical purposes the stellar border. The exact value of $P_s$ does not
influence the determination of the stellar structure, provided that
it is sufficiently low.

The numerical solution of the stellar structure
equations~\eqref{eq:drda}--\eqref{eq:dPda} with the boundary
conditions~\eqref{eq:m0}--\eqref{eq:phiA} is not trivial and is discussed in
Sec.~\ref{sec:code}.

\section{Neutrino transport equations: semi-empirical derivation}
\label{sec:BLEempirical}

The PNS structure changes in its first minute of life because the EoS depends on the
temperature and lepton fraction, which change due to the
neutrino diffusion. The hydrodynamical timescale is shorter than the neutrino
diffusion timescale, and hence one can describe the PNS evolution as a sequence
of quasi-stationary configurations; then, the evolutionary timescale is
determined by neutrinos. This means that, to determine the PNS evolution, one has
first to describe the neutrino diffusion process, that is, to obtain the neutrino transport equations.
In this section we present a pedagogical introduction to the transport equations,
dropping all the complications related to GR. This should give a direct physical insight that
is absent from the rigorous derivation of the neutrino transport equations, which is given in the next
section.

A mathematical model for the diffusion of a salt in a liquid was proposed for
the first time by Fick in 1855 (\citealp{Fick.1855}, see \citealp{Fick.1995}
for an English reprint of the original article), in analogy to the Fourier
mathematical model for heat transport and Ohm theory for the diffusion of
electricity in a conductor. Even though the original Fick laws are, strictly
speaking, incorrect for the study of the PNS evolution, they have a more
immediate physical interpretation; moreover, a na\"ive generalization of Fick
laws has been used in the first codes of PNS evolution
\citep{Burrows+Lattimer.1986, Keil+Janka.1995}. Here we obtain the Fick laws
with a simple argument.
We consider the diffusion of a particle species in 1D (e.g., a tube). Between
the coordinates $x-\mathrm dx/2$ and $x+\mathrm dx/2$ (a cell centered in $x$)
there are $N(x)$ particles of that species and between $x+\mathrm dx/2$ and
$x+3/2\mathrm dx$ (a cell centered in $x+\mathrm dx$) there are $N(x+\mathrm
dx)$.  If $v$ is the speed of the particles along $x$, $\mathrm dt=\mathrm
dx/v$ is the time that a particle employs to cross the distance $\mathrm dx$.
If all particles have the same speed $v$, in a time $\mathrm dt$ half of the
particles in the cell centered in $x+\mathrm dx$ move to the left (in $x$), and
half of the particles in the cell centered in $x$ move to the right (in
$x+\mathrm dx$). Then, the net number of particles that move from $x$ to
$x+\mathrm dx$ are $-\frac12 N(x+\mathrm dx) + \frac12 N(x)$, and the
corresponding net flux is
\begin{equation}
F = -\frac12 \left(\frac{N(x+\mathrm dx)}{A\mathrm dt}- \frac{N(x)}{A \mathrm
dt}\right),
\end{equation}
where $A$ is the section of the tube.  Multiplying both the numerator and the
denominator by $\mathrm dx^2$, and calling $n(x)=N(x)/A\mathrm dx$ the particle
number density in $x$, we obtain the first Fick law,
\begin{align}
\label{eq:Fsimple}
F(x) ={}& -D\frac{\mathrm dn}{dx},\\
\label{eq:Dsimple}
D={}& \frac{\mathrm dx^2}{2\mathrm dt}=\frac{v\mathrm dx}{2},
\end{align}
where $D$ is called \emph{diffusion coefficient} and it is related to the speed
$v$ of the particles.  In a time $\mathrm dt$, the number of particles in $x$
may change due to the incoming flux from $x-\mathrm dx$, the outgoing flux in
$x+\mathrm dx$, and eventually due to particle generation in $x$,
\begin{multline}
N(x,t+\mathrm dt)-N(x,t)=A\mathrm dt\big(F(x-\mathrm dx/2)\\
-F(x+\mathrm dx/2)+S(x)n_\mathrm{fluid}\mathrm dx\big),
\end{multline}
where $n_\mathrm{fluid}$ is the fluid particles number density and $S(x)$ is
the number of particle created in $x$ per each fluid particle and per unit time
$\mathrm dt$.  The Fick second law is obtained by dividing the previous
expression by $\mathrm dtA\mathrm dx$,
\begin{equation}
\frac{\partial n}{\partial t} + \frac{\partial F}{\partial x} \equiv
\frac{\partial n}{\partial t} - D\frac{\partial^2n}{\partial x^2} =
S(x)n_\mathrm{fluid},
\end{equation}
where we have assumed that the diffusion coefficient $D$ is constant along the
tube.

In three dimensions the second Fick law may be written as
(\citealp[see e.g.][]{Burrows+Lattimer.1986, Keil+Janka.1995}. At variance
with them, here we drop the GR terms to simplify the discussion)
\begin{equation}
\label{eq:Fnui_noGR}
\frac{\partial Y_i}{\partial t} + \frac1{4\pi r^2\nB} \frac{\partial
(4\pi r^2F_i)}{\partial r} = \frac{\partial Y_i}{\partial t} + \frac{\partial
(4\pi r^2F_i)}{\partial a} = S_i,
\end{equation}
where $Y_i=n_i/n_\mathrm{B}$, $F_i$, and $S_i$ are the number fraction, the
number flux, and the number source of the $i$-th particle, respectively, and $a$
is the enclose baryon number. We put the total baryon number $N_\mathrm{B}$ in
the shell $\mathrm d a$ inside the time differentiation because the baryons are
locked during the PNS evolution and the baryon number inside the shell $\mathrm
da$ is conserved. A rigorous justification of this step will be given in the
next section.

Apart for the neutrino diffusion, there is neither particle nor heat flow through
the stellar layers during the timescale of the PNS evolution. Therefore, the
energy transport equation is
\begin{equation}
\label{eq:dedtsimple}
\frac{\partial e}{\partial t} +P\frac{\partial(1/n_\mathrm{B})}{\partial t}
+\sum_i\frac{\partial(4\pi r^2 H_i)}{\partial a} = 0,
\end{equation}
where $e=E/N_\mathrm{B}$ is the total (of all kinds of particle) energy per
baryon and $H_i$ is the energy flux due to particle $i$ (in our case, neutrino
species $i$).  To understand the equation, one should realize that the first
two terms represent the change of the internal energy plus the work of
compression $P\mathrm dV/\mathrm dt$.  There is no energy source term as in the
equation for the number transport because the energy gained (lost) by one
particle is lost (gained) by the rest of the matter (there are no energy
sinks), and hence the total sum is null: the total energy can vary only because
of the net flux of energy $\sum_i H_i$ and by the work of compression.

The fluxes are proportional to the gradients of the number and energy density
of the $i$-th particle and are usually written as \citep{Burrows+Lattimer.1986,
Keil+Janka.1995}
\begin{align}
\label{eq:flux-simple}
F_i ={}& -D_i^n \frac{\partial n_i}{\partial r},\\
\label{eq:flux-simple2}
H_i={}&-D_i^\epsilon \frac{\partial \epsilon_i}{\partial r},\\
D_i^n={}&\frac{c\lambda_i^n}3,\\
\label{eq:Disimple}
D_i^\epsilon={}&\frac{c\lambda_i^\epsilon}3,
\end{align}
where $D_i^{n,\epsilon}$ are the diffusion coefficients of particle $i$, and
$\lambda_i^n$ is the spectral average of the $i$-th particle mean free path,
and $\lambda_i^\epsilon$ is the spectral average of the $i$-th particle mean
free path equivalent to the Rosseland mean free path in the photon case (cf.{}
Eqs.~\eqref{eq:flux-simple}--\eqref{eq:Disimple} with
Eqs.~\eqref{eq:Fsimple}--\eqref{eq:Dsimple} of this thesis; see also Eqs.~(A6),
(A11), and (A12) of \citealp{Keil+Janka.1995}).

Now we specialize Eqs.~\eqref{eq:Fnui_noGR} and \eqref{eq:dedtsimple} to our case.
We have set $\mu_{\nu_\mu}=\mu_{\bar\nu_\mu}=\mu_{\nu_\tau}=\mu_{\bar\nu_\tau}\equiv0$,
and therefore $Y_{L_\mu}=Y_{L_\tau}\equiv0$: only the electron neutrino fraction $Y_{\nu_e}\equiv Y_L$ may
change. Beta-equilibrium implies that $S_\nu=-S_e$, that is, for each electron neutrino
that is created, an electron is destroyed, and viceversa. Finally, as we have already discussed,
only neutrinos move through the stellar layers. Therefore, summing Eq.~\eqref{eq:Fnui_noGR} for
electrons to the same equation for neutrinos, one obtains
\begin{equation}
\label{eq:dyLdt_noGR}
\frac{\partial Y_L}{\partial t} + \frac{\partial
(4\pi r^2F_{\nu_e})}{\partial a} = 0.
\end{equation}
Finally, using Eqs.~\eqref{eq:de} and \eqref{eq:mu_nudYL}, one may rewrite
Eq.~\eqref{eq:dedtsimple} as
\begin{equation}
\label{eq:dsdtsimple}
T\frac{\partial s}{\partial t} +\mu_{\nu_e}\frac{\partial Y_L}{\partial t}
+\sum_{i\in\{e,\mu,\tau\}}\frac{\partial(4\pi r^2 H_{\nu_i})}{\partial a} = 0,
\end{equation}
since also muon and tauon neutrinos move energy and entropy through the stellar layers (but not lepton number, see discussion above).

In the next section we consistently derive the transport equation in GR;
however, it is instructive to empirically modify Eqs.~\eqref{eq:Fnui_noGR} and
\eqref{eq:dedtsimple} to include the effects of GR. This may be done by
including a redshift term for the energies and temperatures, and
including the time lapse effect in the derivatives with respect to the comoving frame, $\mathrm d\tau\rightarrow\mathrm
e^{-\phi}\mathrm d\tau$. Then, the transport equations~\eqref{eq:dyLdt_noGR} and
\eqref{eq:dsdtsimple} becomes
\begin{align}
\label{eq:bho1}
\frac{\partial Y_\mathrm{L}}{\partial t} +\frac{\partial (4\pi r^2\mathrm
e^\phi F_\nu)}{\partial a}={}&0,\\
\label{eq:bho2}
\mathrm e^\phi T\frac{\partial s}{\partial t} + \mathrm e^\phi
\mu_\nu\frac{\partial Y_\mathrm{L}}{\partial t} + \frac{\partial(4\pi r^2
\mathrm e^{2\phi} H_\nu)}{\partial a}={}&0,
\end{align}
where there are no modification to the derivatives with respect to the time $t$ measured by an observer at infinity.
Eqs.~\eqref{eq:bho1} and \eqref{eq:bho2} close the stellar structure equations,
which need the entropy and density profiles at each timestep to be solved.
The numerical solution of the transport equations is discussed in Sec.~\ref{sec:code}.

\section{Neutrino transport equations: rigorous derivation}
\label{sec:BLE}
In this section, following \citet{Pons+.1999}, we give a rigorous derivation of the neutrino transport equations
in GR, that have been obtained in a semi-empirical way in Sec.~\ref{sec:BLEempirical}.

We take the static spherical metric of Eq.~\eqref{eq:metric} and consider a
frame comoving with the matter [with our assumptions\footnote{\cite{Pons+.1999}
and \citet{Lindquist.1966} made the limit $v\ll c$ at the end of their
calculations; instead to simplicity we neglect the 3-velocity of the matter from the
beginning, $v\ll c$.
}, the four velocity of the matter is $\mathbf u=(\mathrm e^{-\phi},0,0,0)$].
The comoving basis $\{\mathbf e_a\}$ is defined by
[\citealp[][Eq.~(23.15a)]{Misner+Thorne+Wheeler.book}]
\begin{align}
\mathbf e_a \cdot \mathbf e_b ={}& \eta_{ab},\\
\mathbf e_0 \equiv{}& \mathbf u = \mathrm e^{-\phi} \partial_t,\\
\mathbf e_1 ={}& \mathrm e^{-\lambda} \partial_r,\\
\mathbf e_2 ={}& \frac1r \partial_\theta,\\
\mathbf e_3 ={}& \frac1{r\sin\theta} \partial_\varphi,
\end{align}
and therefore the non-zero tetrad components are
\begin{align}
e^t_0={}&= (e^0_t)^{-1} = \mathrm e^{-\phi},\\
e^r_1={}& = (e^1_r)^{-1} = \mathrm e^{-\lambda},\\
e^\theta_2={}& = (e^2_\theta)^{-1} = \frac1r,\\
e^\varphi_3={}& = (e^3_\varphi)^{-1} = \frac1{r\sin\theta},
\end{align}
In an orthonormal basis [\citealp[][Eqs.~(8.24b) and
(8.14)]{Misner+Thorne+Wheeler.book}],
\begin{align}
[\mathbf e_a,\mathbf e_b]={}& (e_a^\alpha\partial_\alpha e_b^\beta
-e_b^\alpha\partial_\alpha
e_a^\beta)\partial_\beta=c_{ab}^{\hphantom{ab}c}\mathbf e_c,\\
c_{abc}={}&\eta_{cd}c_{ab}^{\hphantom{ab}d},\\
\Gamma^{b}_{\hphantom{b}mn}={}& \frac{\eta^{ba}}{2} (c_{amn} + c_{anm} -
c_{mna}),
\end{align}
where the non-zero commutation coefficients $c_{abc}$ are
\begin{align}
c_{010}={}&-c_{100}=-\mathrm e^{-\lambda}\partial_r\phi,\\
c_{011}={}&-c_{101}=-\mathrm e^{-\phi}\partial_t\lambda,\\
c_{122}={}&-c_{212}=c_{133}=-c_{313}=-\frac{\mathrm e^{-\lambda}}{r},\\
c_{233}={}&-c_{323}=-\frac{\cot\theta}{r},
\end{align}
and the non zero Ricci rotation coefficients are
\begin{align}
\Gamma^1_{00}={}&\Gamma^0_{10}=\mathrm e^{-\lambda}\partial_r\phi,\\
\Gamma^0_{11}={}&\Gamma^1_{01}=\mathrm e^{-\phi}\partial_t\lambda,\\
\Gamma^1_{22}={}&-\Gamma^2_{12}=\Gamma^1_{33}=-\Gamma^3_{13}=-\frac{\mathrm
e^{-\lambda}}r\\
\Gamma^2_{33}={}&-\Gamma^3_{23}= - \frac{\cot\theta}{r}.
\end{align}
The Boltzmann equations for massless particles in GR are
[\citealp[][Eq.~(2.34)]{Lindquist.1966}]
\begin{equation}
\label{eq:BLE}
p^b\left(e_b^\beta \frac{\partial f}{\partial x^\beta} -
\Gamma^a_{bc}p^c\frac{\partial f}{\partial p^a}\right) = \left(\frac{\mathrm
df}{\mathrm d\tau}\right)_\mathrm{coll},
\end{equation}
where $f$ is the invariant neutrino distribution function, and $p^a$ is the
neutrino four momentum in the comoving basis
[\citealp[cf.][Eq.~(3.3)]{Lindquist.1966}],
\begin{equation}
\label{eq:momentum_BLE}
p^a=(E,E\mu,E\sqrt{1-\mu^2}\cos\bar\varphi,E\sqrt{1-\mu^2}\sin\bar\varphi),
\end{equation}
where $E$ is the neutrino energy, $\mu\equiv\cos\bar\vartheta$ is the cosine of
the neutrino four momentum with respect to the radial direction, and
$\bar\varphi$ is the angle of the neutrino four momentum with respect to the
$\theta$ direction.  Since in spherical symmetry $f\equiv f(t,r,E,\mu)$, we
would express the derivatives with respect to $p^a$ as derivatives with
respect to $E$ and $\mu$.  We first notice that
\begin{align}
E={}&\sqrt{p_1^2+p_2^2+p_3^2}\equiv p_0,\\
\mu={}&\frac{p_1}{\sqrt{p_1^2+p_2^2+p_3^2}},\\
\frac{\partial f}{\partial p^a}\mathrm dp^a = {}& \frac{\partial f}{\partial E}\mathrm dE+\frac{\partial f}{\partial \mu} \mathrm d\mu,\\
\mathrm dE={}&\alpha\{\mu\mathrm dp_1+\sqrt{1-\mu^2}\cos\bar\varphi\mathrm dp_2
+\sqrt{1-\mu^2}\sin\bar\varphi \mathrm dp_3\}\notag\\
&+(1-\alpha)\mathrm dp_0,\\
\mathrm d\mu={}&\frac{1-\mu^2}{E} \mathrm dp_1
-\frac{\mu\sqrt{1-\mu^2}\cos\bar\varphi}{E}\mathrm dp_2
-\frac{\mu\sqrt{1-\mu^2}\sin\bar\varphi}{E}\mathrm dp_3 ,
\end{align}
where $\alpha$ is an arbitrary parameter that we have included to stress the
freedom we have in expanding the differential of $\mathrm dE$. This freedom is
due to the reduced number of variables ($E$ and $\mu$ instead of $p^a$,
$a=0\ldots3$) on which the distribution function explicitly depends after the
change of variable (the following calculations do not depend on the exact value
of $\alpha$, and we shall set it equal to zero to simplicity). Then,
\begin{align}
\frac{\partial f}{\partial p^0}={}&(1-\alpha)\frac{\partial f}{\partial E},\\
\frac{\partial f}{\partial p^1}={}& \alpha\mu \frac{\partial f}{\partial E} +
\frac{1-\mu^2}E \frac{\partial f}{\partial \mu},\\
\frac{\partial f}{\partial p^2}={}& \alpha\sqrt{1-\mu^2}\cos\bar\varphi
\frac{\partial f}{\partial E} - \mu\frac{\sqrt{1-\mu^2}}{E}\cos\bar\varphi
\frac{\partial f}{\partial \mu},\\
\frac{\partial f}{\partial p^3}={}& \alpha\sqrt{1-\mu^2}\sin\bar\varphi
\frac{\partial f}{\partial E} - \mu\frac{\sqrt{1-\mu^2}}{E}\sin\bar\varphi
\frac{\partial f}{\partial \mu},
\end{align}
and, after lengthy but straightforward calculations, Eq.~\eqref{eq:BLE}
becomes
[\citealp[cf.][Eq.~(3.8a)]{Lindquist.1966}]
\begin{multline}
\label{eq:BLE2}
\mathrm e^{-\phi}E\partial_t f +\mathrm e^{-\lambda}E\mu\partial_rf
-E^2\left(\mu\mathrm e^{-\lambda}\partial_r\phi +\mu^2\mathrm
e^{-\phi}\partial_t\lambda\right)\partial_E f\\
+E(1-\mu^2)\left(\frac{\mathrm e^{-\lambda}}r - \mathrm
e^{-\lambda}\partial_r\phi - \mu\mathrm e^{-\phi}\partial_t\lambda \right)
\partial_\mu f =\left(\frac{\mathrm df}{\mathrm d\tau}\right)_\mathrm{coll}.
\end{multline}
We remark that in Eq.~\eqref{eq:BLE2} the arbitrary parameter $\alpha$ has disappeared.

It is customary to work with the angular moments of Eq.~\eqref{eq:BLE2}
\citep{Lindquist.1966, Thorne.1981}.  Since we are interested in the zeroth and
first moment of the Boltzmann-Lindquist equation, we apply to
Eq.~\eqref{eq:BLE2} the operator
\begin{equation}
\frac 12\int_{-1}^{+1} \mu^i\mathrm d\mu^i\qquad i=\{0;1\},
\end{equation}
obtaining two equations [\citealp[cf.][Eqs.~(6)--(7)]{Pons+.1999}],
\begin{multline}
\label{eq:moment0}
\mathrm e^{-\phi}E\partial_tM_0 +\mathrm e^{-\lambda}E\partial_rM_1
-E^2(\mathrm e^{-\lambda}\phi'\partial_EM_1+ \mathrm
e^{-\phi}\dot\lambda\partial_EM_2)\\
+E\mathrm e^{-\lambda}\left(\frac1r -\phi'\right)2M_1 +E\mathrm
e^{-\phi}\dot\lambda(M_0-3M_2)=Q_0,
\end{multline}
\begin{multline}
\label{eq:moment1}
\mathrm e^{-\phi}E\partial_tM_1 +\mathrm e^{-\lambda}E\partial_rM_2
-E^2(\mathrm e^{-\lambda}\phi'\partial_EM_2+ \mathrm
e^{-\phi}\dot\lambda\partial_EM_3)\\
+E\mathrm e^{-\lambda}\left(\frac1r - \phi' \right)(3M_2-M_0) +2E\mathrm
e^{-\phi}\dot\lambda(M_1-2M_3)=Q_1,
\end{multline}
where $\phi'=\partial_r\phi$, $\dot\lambda=\partial_t\lambda$, and
\begin{align}
\label{eq:Mi}
M_i={}&\frac12 \int_{-1}^{+1}\mu^i f\mathrm d\mu,\\
Q_i={}& \frac12 \int_{-1}^{+1}\mu^i \left(\frac {\mathrm df}{\mathrm
d\tau}\right)_\mathrm{coll}\mathrm d\mu,
\end{align}
are the moments of the distribution function and of the source term.  Since we
are interested in energy averaged equations, we apply to Eq.~\eqref{eq:moment0}
the operators
\begin{align}
&4\pi\int_0^\infty \mathrm dE E,\\
&4\pi\int_0^\infty \mathrm dE E^2,
\end{align}
obtaining
\begin{align}
\label{eq:finalmente1}
\mathrm e^{-\phi}\partial_t n_\nu +\mathrm e^{-\lambda}\left(\partial_rF_\nu+
\phi'F_\nu+\frac2rF_\nu\right) + \mathrm e^{-\phi}\dot\lambda n_\nu ={}& S_N,\\
\label{eq:finalmente2}
\mathrm e^{-\phi}\partial_t \epsilon_\nu +\mathrm
e^{-\lambda}\left(\partial_rH_\nu+ 2\phi'H_\nu+\frac2rH_\nu\right) + \mathrm
e^{-\phi}\dot \lambda(\epsilon_\nu+P_\nu)={}& S_E,
\end{align}
where\footnote{\label{ftn:1} The difference in the factors between our
definition of Eqs.~\eqref{eq:nnudef}--\eqref{eq:SEdef} and that of
\citet[][cf.~their Eqs.~(11)--(12)]{Pons+.1999} is due to their choice
to set $\hbar=1$. Compare also with
\citet[][Eq.~(3.9)]{Lindquist.1966}.}
\begin{align}
\label{eq:nnudef}
n_\nu={}&4\pi\int_0^\infty \mathrm dE M_0E^2,\\
\label{eq:Fnudef}
F_\nu={}&4\pi\int_0^\infty \mathrm dE M_1E^2,\\
S_N={}&4\pi\int_0^\infty \mathrm dE Q_0E,\\
\epsilon_\nu={}&4\pi\int_0^\infty \mathrm dE M_0E^3,\\
\label{eq:Hnudef}
H_\nu={}&4\pi\int_0^\infty \mathrm dE M_1E^3,\\
P_\nu={}&4\pi\int_0^\infty \mathrm dE M_2E^3,\\
\label{eq:SEdef}
S_E={}&4\pi\int_0^\infty \mathrm dE Q_0E^2.
\end{align}
Using the baryon number conservation equation in GR,
\begin{equation}
\label{eq:baryon_GR}
(nu^a)_{;a} = \mathrm e^{-\phi}\partial_t\nB+\mathrm
e^{-\phi}\dot\lambda\nB=0,
\end{equation}
one obtains
\begin{align}
\label{eq:finalmente3}
\mathrm e^{-\phi}\left(\partial_t n_\nu-\frac{n_\nu}{\nB}\partial_t\nB\right) +
\frac{\mathrm e^{-\lambda}\mathrm e^{-\phi}}{4\pi r^2}\partial_r( 4\pi r^2
\mathrm e^\phi F_\nu) ={}& S_N,\\
\label{eq:finalmente4}
\mathrm e^{-\phi}\left(\partial_t
\epsilon_\nu-\frac{\epsilon_\nu+P_\nu}{\nB}\partial_t\nB\right) + \frac{\mathrm
e^{-\lambda}\mathrm e^{-2\phi}}{4\pi r^2}\partial_r( 4\pi r^2 \mathrm e^{2\phi}
H_\nu) ={}& S_E,
\end{align}
and, multiplying Eqs.~\eqref{eq:finalmente3} and \eqref{eq:finalmente4} by
$\mathrm e^\phi/\nB$, and using Eq.~\eqref{eq:drda},
\begin{align}
\label{eq:finalmente5}
\partial_t Y_\nu + \partial_a( 4\pi r^2 \mathrm e^\phi F_\nu) ={}& \mathrm
e^{\phi}\frac{S_N}{\nB},\\
\label{eq:finalmente6}
\partial_t e_\nu-\frac{P_\nu}{\nB^2}\partial_t\nB + \mathrm
e^{-\phi}\partial_a( 4\pi r^2 \mathrm e^{2\phi} H_\nu) ={}& \mathrm
e^{\phi}\frac{S_E}{\nB},
\end{align}
where $Y_\nu=n_\nu/\nB$ is the neutrino fraction and $e_\nu=\epsilon_\nu/\nB$
the neutrino energy per baryon.  As we have already discussed
(Secs.~\ref{sec:totaleos} and \ref{sec:BLEempirical}), we assume that muon and
tauon neutrinos have vanishing chemical potentials, and hence the only non
vanishing neutrino number is the electron one, $Y_{\nu_e}$.  Since we
additionally assume beta-equilibrium, for each neutrino that is absorbed
(emitted), there is an electron that is emitted (absorbed), and analogously for
the antineutrinos/positrons. Moreover, electrons are locked in the timescales
of PNS evolution, and therefore only neutrinos diffuse through the star.  Then,
\begin{equation}
\label{eq:electronsGR}
\partial_t Y_e=-\mathrm e^{\phi}\frac{S_N}{\nB}.
\end{equation}
Summing Eqs.~\eqref{eq:finalmente3} and \eqref{eq:electronsGR}, one finally
obtains
\begin{equation}
\label{eq:dYL_dt}
\frac{\partial Y_{\mathrm L}}{\partial t}=- \frac{\partial(4\pi r^2\mathrm
e^\phi F_\nu)}{\partial a}.
\end{equation}
Similarly, the energy gained (lost) by neutrinos is lost (gained) by the rest
of the matter (in our case, protons, neutrons, electrons, and positrons),
\begin{equation}
\label{eq:matterGR}
\partial_t e_\mathrm{matter}-\frac{P_\mathrm{matter}}{\nB^2}\partial_t\nB =
-\mathrm e^{\phi}\frac{S_E}{\nB}.
\end{equation}
Summing Eqs.~\eqref{eq:finalmente4} and \eqref{eq:matterGR}, and using the
thermodynamical relations~\eqref{eq:de} and \eqref{eq:mu_nudYL}, we finally
obtain
\begin{equation}
\label{eq:ds_dt}
T\frac{\partial s}{\partial t}+\mu_{\nu_e}\frac{\partial Y_{\mathrm L}}
{\partial t} = - \sum_{i\in\{e;\mu;\tau\}}\mathrm e^{-\phi} \frac{\partial(4\pi
r^2\mathrm e^{2\phi}H_{\nu_i})}{\partial a},
\end{equation}
where $s$ is the total (matter plus neutrinos) entropy per baryon, and we have
summed over all neutrino species because even though muon and tauon neutrinos have
been assumed to have a (fixed and constant) vanishing chemical potential and
then their respective lepton numbers are fixed and vanishing. In fact, they are not
locked in the star and therefore can move energy through the stellar layers.
We remark that Eqs.~\eqref{eq:dYL_dt} and \eqref{eq:ds_dt} are the same
equations that we have previously obtained with a semi-empirical argument,
Eqs.~\eqref{eq:bho1} and \eqref{eq:bho2}.

To obtain the relativistic expression for the neutrino number and energy
fluxes, we have to make an assumption on the distribution function that allows
to close the set of equations~\eqref{eq:moment0} and \eqref{eq:moment1}.  This
assumption is the so called \emph{diffusion approximation}, that is, we expand
the distribution function dependence on the angle made by the neutrino
direction with the radial direction ($\bar\vartheta$, and
$\mu\equiv\cos\bar\vartheta$) in Legendre polynomials and we retain only the
first two moments,
\begin{equation}
f(E,\mu)=f_0(E)+\mu f_1(E),
\end{equation}
where $f_0$ is the (ultra-relativistic) neutrino distribution function (see
Appendix~\ref{app:eos}).  Then, the moments of the distribution function
[Eq.~\eqref{eq:Mi}] become
\begin{align}
M_0={}& f_0,\\
\label{eq:M1def}
M_1={}& \frac13 f_1,\\
M_2={}& \frac13 f_0,\\
M_3={}& \frac15 f_1,
\end{align}
and the first moment of the Boltzmann-Lindquist equation
[Eq.~\eqref{eq:moment1}] becomes (since the transport is driven by spatial
gradients, we neglect time derivatives)
\begin{equation}
Q_1=\frac E3 \mathrm e^{-\lambda}
\left(\partial_rf_0-E\phi'\partial_Ef_0\right).
\end{equation}
It is possible to show \citep{Pons+.1999, Reddy+Prakash+Lattimer.1998} that
\begin{equation}
\label{eq:Q1find}
Q_1= -\frac{E}{3}f_1(E)\lambda_\mathrm{tot}^{-1}(E),
\end{equation}
where $\lambda_\mathrm{tot}(E)$ is the total mean free path (due to scattering,
absorption, emission, and their inverse processes, see Sec.~\ref{sec:diffusion}) of a neutrino with
energy $E$, and therefore
\begin{equation}
\label{eq:f1find}
f_1= -\lambda_\mathrm{tot}(E)\mathrm e^{-\lambda} \left(\partial_rf_0-E
\phi'\partial_Ef_0\right).
\end{equation}
Since
\begin{align}
\label{eq:drf0}
\partial_r f_0 ={}& + (2\pi\hbar)^3 f_0 (1-f_0) \left(\partial_r \eta + \frac
E{T^2} \partial_rT\right),\\
\label{eq:dEf0}
\partial_E f_0 ={}& - (2\pi\hbar)^3 f_0 (1-f_0) \frac1T,\\
\label{eq:eta}
\eta\equiv{}&\frac{\mu_{\nu_e}}{T},
\end{align}
from Eqs.~\eqref{eq:Fnudef}, \eqref{eq:Hnudef}, \eqref{eq:M1def},
\eqref{eq:f1find}--\eqref{eq:eta}, we finally obtain
\citep{Pons+.1999}
\begin{align}
\label{eq:Fnu}
F_{\nu_i}={}&-\frac{\mathrm e^{-\lambda}\mathrm e^{-\phi}T^2}{6\pi^2\hbar^3}
\Bigg(D_3^i\frac{\partial(T\mathrm e^\phi)}{\partial r}+(T\mathrm
e^\phi)D_2^i\frac{\partial\eta}{\partial r}\Bigg),\\
\label{eq:Hnu}
H_{\nu_i}={}&-\frac{\mathrm e^{-\lambda}\mathrm e^{-\phi}T^3}{6\pi^2\hbar^3}
\Bigg(D_4^i\frac{\partial(T\mathrm e^\phi)}{\partial r}+(T\mathrm
e^\phi)D_3^i\frac{\partial\eta}{\partial r}\Bigg),\\
\label{eq:Ddef}
D_n^i={}& \int_0^\infty \mathrm d(E/T) (E/T)^n
f^i_0(E)\big(1-f^i_0(E)\big) \lambda_\mathrm{tot}^{-1}(E),
\end{align}
where $F_\nu$ and $H_\nu$ are the neutrino number and energy fluxes,
respectively, the index $i=\{e;\mu;\tau\}$ refers to the lepton family, $f_0^i$
is the Fermi-Dirac distribution function of the $i$-th neutrino at equilibrium
 and the integration is performed on the neutrino
energy divided by the temperature, $E/T$.
As already stated, only electron neutrinos move lepton
number through the stellar layers, whereas all
neutrinos move energy.  This fact, together with the assumption of muon and tauon
null chemical potential, permits to rewrite the transport equations in a
simplified way \citep{Pons+.1999},
\begin{align}
\label{eq:Fnufinal}
F_\nu={}&-\frac{\mathrm e^{-\lambda}\mathrm e^{-\phi}T^2}{6\pi^2\hbar^3}
\Bigg(D_3\frac{\partial(T\mathrm e^\phi)}{\partial r}+(T\mathrm
e^\phi)D_2\frac{\partial\eta}{\partial r}\Bigg),\\
\label{eq:Hnufinal}
H_\nu={}&-\frac{\mathrm e^{-\lambda}\mathrm e^{-\phi}T^3}{6\pi^2\hbar^3}
\Bigg(D_4\frac{\partial(T\mathrm e^\phi)}{\partial r}+(T\mathrm
e^\phi)D_3\frac{\partial\eta}{\partial r}\Bigg),\\
\label{eq:YLfinal}
\partial_t Y_\mathrm{L} +{}&  \frac{\partial 4\pi\mathrm e^\phi r^2
F_\nu}{\partial a}=0,\\
\label{eq:sfinal}
T\partial_t s +{}& \mu_{\nu_e}\partial_t Y_\mathrm{L} + \mathrm e^{-\phi}\frac{\partial
4\pi\mathrm e^{2\phi} r^2 H_\nu}{\partial a}=0,\\
\label{eq:D2final}
D_2={}&D_2^{\nu_e}+D_2^{\bar\nu_e},\\
\label{eq:D3final}
D_3={}&D_3^{\nu_e}-D_3^{\bar\nu_e},\\
\label{eq:D4final}
D_4={}&D_4^{\nu_e}+D_4^{\bar\nu_e}+4D_4^{\nu_\mu},
\end{align}
where we have neglected the differences between muon and tauon neutrinos (and
antineutrinos) in the microphysical processes.

Eqs.~\eqref{eq:Fnufinal}--\eqref{eq:D4final} have to be completed by boundary
conditions. For the center of the star, these are the request of null fluxes,
$F_\nu(r=0)=0$ and $H_\nu(r=0)=0$; for the stellar border, one has to do an
assumption on the leaving neutrinos. Usually one assumes that the number and
energy fluxes at the stellar border are proportional to the neutrino number and
energy densities. The numerical solution of
Eqs.~\eqref{eq:Fnufinal}--\eqref{eq:D4final} is discussed in the next section.

A final remark is in order. We notice that the ``semi-empirical'' definition of
the fluxes~\eqref{eq:flux-simple}--\eqref{eq:flux-simple2} is equivalent (apart
for the metric coefficients) to Eqs.~\eqref{eq:Fnufinal}--\eqref{eq:Hnufinal},
if we adopt for $n_\nu$ and $\epsilon_\nu$ the expressions~\eqref{eq:Y_nu} and
\eqref{eq:eps_nu}, provided that the diffusion coefficients are\footnote{I am
indebted to J.~A.~Pons for pointing out this to me. See also Eq.~(25) of \citet{Burrows+Lattimer.1986}.}
\begin{align}
D_2\propto{}&3\eta^2+\pi^2,\\
D_3\propto{}&3(\eta^3+\pi^2\eta),\\
D_4\propto{}&3\left(\eta^4+2\pi^2\eta^2 +\frac75\pi^4\right).
\end{align}

\section{Numerical implementation}
\label{sec:code}
The equations that describe the PNS structure and evolution, Eqs.~\eqref{eq:drda}--\eqref{eq:dPda}
and \eqref{eq:Fnufinal}--\eqref{eq:sfinal}, are relatively simple. However, their numerical
implementation is not trivial at all. In this section we discuss the subtleties
involved.

The PNS has been evolved using a sequence of predictor-corrector steps,
that is, one separately solves
the structure and the transport equations, and then iterate to
evolve the star. This method is equivalent to use a full implicit scheme \citep{Pons+.1999}.
The iterative procedure is easier to understand by inspecting the skeleton of the code in top
language,

\begin{verbatim}
# 1) iterating (predictor-corrector step) is equivalent
#    to solve the structure and transport equations
#    simultaneously
# 2) structure() solves the TOV equation (relaxation method)
# 3) transport() evolves the thermodynamical profiles
#    (implicit scheme)
main:
   load initial profiles (a, YL, s)
   (P, phi, r, m) = structure(YL, s, a)
   # evolve the star
   for each time:
      # iterate until convergence of YL' and s'
      repeat:
         # determine the optimal timestep dt
         repeat:
            YL''=YL'
            s''=s'
            (YL', s') = transport(P, phi, r, m, YL, s, dt)
            temp = max(sum|s'-s''|/sum|s''|,
                       sum|YL'-YL''|/sum|YL''|)
            if(eps/2 < temp < eps*2):
               exit
            else:
               dt = dt * eps / temp
         (P, phi, r, m) = structure(YL', s', a)
         temp = max(sum|s'-s|/sum|s|, sum|YL'-YL|/sum|YL|)
         if(temp < tol):
            exit
      YL=YL'
      s=s'
      eventually, print the evolved profiles
\end{verbatim}

Using the original stellar profiles $P(a)$, $m(a)$, $r(a)$ and $\phi(a)$, we
have (i) first solved the transport equations, determining the evolved lepton
fraction and entropy profiles, $s'(a)$ and $Y_L'(a)$.  Then, (ii) using these evolved
profiles, we have determined the new stellar structure, $P'(a)$, $m'(a)$, $r'(a)$ and
$\phi'(a)$. Using these evolved stellar profiles, but the original lepton fraction and entropy profiles $s(a)$ and $Y_L(a)$,
we have (ib) solved the transport equations obtaining the evolved $s''(a)$ and $Y_L''(a)$ profiles, from which
we have obtained (iib) the evolved stellar profiles $P''(a)$, $m''(a)$, $r''(a)$ and $\phi''(a)$.
We have iterated the steps (i)-(ii) until convergence of the evolved lepton fraction and entropy profiles.

In addition to the iteration between the structure and transport part
(predictor-corrector step), we have also iterate to find the optimal timestep,
that is, the timestep such that the maximum fractional change in the entropy and lepton fraction profiles
is greater than \texttt{eps/2} but smaller than \texttt{2*eps}.

The usual procedure to solve the TOV equation is to fix the central pressure
$P_c$ and then integrate Eq.~\eqref{eq:TOVoriginal} and \eqref{eq:MG_r} up to
the final pressure $P_s$, whose exact value does not influence the
determination of the stellar properties, provided that it is small enough. In
this way, one obtains the total radius, gravitational mass, baryon mass, and the
thermodynamical and metric profiles of a star with central pressure $P_c$. However, this procedure can not
be applied to our case, since at each timestep one does not know the central
pressure $P_c$ and conversely the total baryon mass of the PNS is fixed and
constant during the evolution (since we do not include effects of accretion or
mass ejection from the PNS), that is, one has to solve the structure equations
with the boundary conditions~\eqref{eq:m0}--\eqref{eq:phiA}.  A simple way to
achieve this is to use the \emph{shooting method}, that is, one first solves
the structure equations fixing a tentative central pressure $P'_c$ and obtains a
total baryon mass $M_\mathrm{B}'$, then iteratively varies the central pressure
until the right total baryon mass is reached within the given error.  This
method has the advantage to be conceptually simple, however (i) it is
numerically quite slow (ii) it may fail to converge when it is used in
combination with the Newton-Raphson procedure to determine the EoS.  This is due
to the fact that the intrinsic variability of the Newton-Raphson procedure does
not allow to obtain an arbitrarily small accuracy in the mass determination. To
circumvent this, one may (a) increase the precision of the Newton-Raphson
procedure (but this lowers the chances of Newton-Raphson convergence), or (b) get the EoS quantities by table interpolation (but to get
the EoS by table interpolation in a hydrodynamical code the second order
derivatives must be continue, see \citealp{Swesty.1996}, and for an EoS with
three independent variables this is quite complex, see discussion in
Sec~\ref{sec:fBfit}).  We have therefore solved the structure equations with
a \emph{relaxation method} \citep[see][Sec.~17.3]{Press+book}, as
done also in \citet{Pons+.1999}.  The relaxation method consists in iteratively
modify a tentative solution of a given differential equation on a grid, until
the differential equation is satisfied within a given error. To fix the ideas, we take a set
of $N$ ordinary differential equations
\begin{equation}
\label{eq:differential}
\frac {\mathrm d \mathbf y}{\mathrm d x} = \mathbf g(x,\mathbf y),
\end{equation}
where the $N$ functions $y_{i}$, with $i=1,\ldots,N$, are defined on a grid of $M$
points $x_j$, with $j=1,\ldots,M$ (we call $y_{i,j}$ the $j$-th point of the $i$-th
function).  If the differential equations are not fulfilled by the tentative
solution $\mathbf y$, then\footnote{Apart for the first two grid points
[$\mathbf E_1(\mathbf y_2,\mathbf y_1)$], we take the mean value of $\mathbf g$
between the points $x_j$ and $x_{j-1}$, instead of taking the value of $\mathbf
g$ in $(x_j+x_{j+1})/2$, because otherwise the system of equations on the grid
of $M$ points decouples in two subsystems on two subgrid of $M/2$ points.
However, in
$\mathbf E_1$ we have considered $\mathbf g(x_{1+1/2},\mathbf y_{1+1/2})$,
since in $x_1\equiv 0$ the TOV equation is singular.  We could have taken a
very small but finite $x_1$ instead (J.~A.~Pons, private
communication).
}
\begin{equation}
\label{eq:residual}
\mathbf E_j(\mathbf y_j,\mathbf y_{j-1}) = \mathbf y_j-\mathbf y_{j-1} -
(x_j-x_{j-1})\frac{\mathbf g(x_j,\mathbf y_j)+\mathbf g(x_{j-1},\mathbf
y_{j-1})}{2} \neq 0,
\end{equation}
(Eq.~\eqref{eq:residual} is valid for an internal point,
$j=2,\ldots,M$, however it may be easily generalized for the
boundary conditions, i.e.{} $j=1,M+1$) and the improved solution
$\mathbf y'$ of the differential equations $\mathbf g$,
\begin{equation}
\label{eq:ressol}
\mathbf y'=\mathbf y+\Delta \mathbf y,
\end{equation}
is given by the condition
\begin{equation}
\label{eq:residual2}
\mathbf E_j(\mathbf y_j+\Delta \mathbf y_j, \mathbf y_{j-1}+\Delta \mathbf
y_{j-1})\simeq 0.
\end{equation}
In other words, one must invert the system
\begin{equation}
\label{equation}
\sum_{k=1}^N\frac{\partial E_{i,j}}{\partial y_{k,j}} \Delta y_{k,j} +
\sum_{k=1}^N\frac{\partial E_{i,j-1}}{\partial y_{k,j-1}} \Delta y_{k,j-1} =
-E_{i,j},
\end{equation}
for $i=1,\ldots,N$ and $j=2,\ldots,M$ (plus the boundary conditions, $j=1,M+1$,
that are similarly handled) to find $\Delta y_{i,j}$ and the improved solution
$y'_{i,j}$. This procedure is repeated until convergence \citep[for details,
see][Sec.~17.3]{Press+book}. We have used the relaxation method to solve the
set of structure equations~\eqref{eq:drda}--\eqref{eq:dPda}, with the boundary
conditions~\eqref{eq:m0}--\eqref{eq:phiA}, on a grid of $124$ points.  The EoS
has been determined with a Newton-Raphson cycle.

The neutrino transport equations are diffusive equations, whose prototype is in
the form \citep[see][Secs.~19.1 and 19.2]{Press+book}
\begin{equation}
\label{eq:PDE_example}
\frac{\partial y}{\partial t}-D\frac{\partial^2y}{\partial x^2}=0.
\end{equation}
Numerically solving a diffusion equation, that is a parabolic equation, is not
easy at all.  The simpler numerical algorithm one can think of,
\begin{align}
\label{eq:explicit}
\frac{f^{t+\mathrm dt}_i-f^t_i}{\mathrm dt} ={}&
D\frac{f^t_{i+1}-2f^t_i+f^t_{i-1}}{\mathrm dx^2},\\
f^{t+\mathrm dt}_i={}&\left(\delta_{ij} + \frac{D\mathrm dt}{\mathrm
dx^2}(\delta_{i+1,j}-2\delta_{i,j}+\delta_{i-1,j})\right)f^t_j,
\end{align}
is called \emph{forward Euler} and is not stable, unless one chooses a very low
time-step $\mathrm dt$.  Indeed, a stability analysis shows that the
algorithm~\eqref{eq:explicit} is stable only if \citep[][Sec.~19.2]{Press+book}
\begin{equation}
\label{eq:courant}
\frac{2D\mathrm dt}{\mathrm dx^2} \le 1.
\end{equation}
This condition can be understood with the following argument, called
\emph{Courant condition}.  Eq.~\eqref{eq:explicit} implies that, from the point
of view of the algorithm, the point $f_i^{t+\mathrm d t}$ depends only on the
information present in the points $f_{i-1}^t$, $f_i^t$, and $f_{i+1}^t$.  Since
Eq.~\eqref{eq:PDE_example} determines the physical propagation velocity of the
information [$v\simeq 2D/\mathrm dx$, cf.~Eq.~\eqref{eq:Dsimple}], choosing a
timestep too large one should also consider the information present in the other
neighbouring points $f_{i+2}^t$, $f_{i-2}^t$, etc. Not doing that results in a
numerical instability. In order to preserve causality, then, the timestep must
be smaller than (the order of) the distance between two consecutive grid points
times the dispersion velocity, Eq.~\eqref{eq:courant}.  The forward Euler
algorithm~\eqref{eq:explicit} is an \emph{explicit} algorithm, because to
obtain the value of the function at the next timestep $f^{t+\mathrm dt}$ one
makes use of the values of the function at the previous timestep, $f^t$.
Conversely, in an \emph{implicit} algorithm the right hand of
Eq.~\eqref{eq:PDE_example} is evaluated at the next timestep,
\begin{align}
\label{eq:implicit}
\frac{f^{t+\mathrm dt}_i-f^t_i}{\mathrm dt} ={}& D\frac{f^{t+\mathrm
dt}_{i+1}-2f^{t+\mathrm dt}_i+f^{t+\mathrm dt}_{i-1}}{\mathrm dx^2},\\
\label{eq:implicit2}
f^t_i={}& \left(\delta_{ij} - \frac{D\mathrm dt}{\mathrm
dx^2}(\delta_{i+1,j}-2\delta_{i,j}+\delta_{i-1,j})\right)f^{t+\mathrm dt}_j,
\end{align}
and the value of the function at the next timestep $f^{t+\mathrm dt}$ is
determined by inverting Eq.~\eqref{eq:implicit2}.  A stability analysis applied
to the implicit algorithm shows that it is unconditionally stable, that is, it
converges for whatever timestep $\mathrm dt$ one chooses. Of course, the smaller
$\mathrm dt$, the more accurate the solution, that is, a numerical
convergence does not imply a convergence to the true physical solution
\citep[][Sec.~19.2]{Press+book}.

To specialize the implicit algorithm to our case, we notice that (i) on the
right hand side $f\equiv(\mathrm e^\phi T,\eta)$, (ii) on the left hand side
the time derivatives are performed to $Y_L\equiv Y_L(T,\eta)$ and $s\equiv
s(T,\eta)$. Moreover, the diffusion coefficients are (iii) not constant and
(iv) they depend on the EoS, $D_i\equiv D_i(T,\eta)$. It is easy to account for
(i) and (ii), we just have to consider two equations instead of one and to
consider the convenient thermodynamical derivatives.  To handle (iii), we just
determine the value of the diffusion coefficients on a \emph{staggered} grid of
$M-1$ points, whose knots $a_{i+1/2}$ are the enclosed baryon mass in a radius
\begin{equation}
r_{i+1/2}=\frac{r_{i+1}+r_{i}}{2},
\end{equation}
see \citet[][Sec.~19.2]{Press+book}. The conceptually more difficult difference
between the neutrino diffusion Equations~\eqref{eq:dYL_dt}--\eqref{eq:ds_dt}
and the idealized diffusion Equation~\eqref{eq:PDE_example} is encoded in (iv),
that is, the diffusion coefficients depend on the value of $T$ and $\eta$, and
therefore we should evaluate them at the next timestep, $D^{t+\mathrm dt}$.
However, this would greatly increase the complexity of the numerical
implementation; we therefore (as done also by \citealp{Pons+.1999}) do not
include implicitly the dependence on the EoS in the diffusion coefficients, but
evaluate them explicitly, that is, at the current timestep, $D^t$.  The
discretization of the neutrino transport
equations~\eqref{eq:dYL_dt}--\eqref{eq:ds_dt} is
\begin{multline}
\label{eq:implicit_YT}
\left(\left.\frac{\partial (\mathrm e^\phi T)}{\partial Y_L}\right|_j^t
\delta_{ij} - \tilde D^{YT}_{j+1/2}(\delta_{i+1,j}-\delta_{i,j}) + \tilde
D^{YT}_{j-1/2}(\delta_{i,j} -\delta_{i-1,j})\right)(\mathrm e^\phi
T)^{t+\mathrm dt}_j\\
=\left.\frac{\partial (\mathrm e^\phi T)}{\partial Y_L}\right|_i^t (\mathrm
e^\phi T)^t_i,
\end{multline}
\begin{multline}
\label{eq:implicit_Yeta}
\left(\left.\frac{\partial \eta}{\partial Y_L}\right|_j^t \delta_{ij} - \tilde
D^{Y\eta}_{j+1/2}(\delta_{i+1,j}-\delta_{i,j}) + \tilde
D^{Y\eta}_{j-1/2}(\delta_{i,j} -\delta_{i-1,j})\right)\eta^{t+\mathrm dt}_j\\
=\left.\frac{\partial \eta}{\partial Y_L}\right|_i^t \eta^t_i,
\end{multline}
\begin{multline}
\label{eq:implicit_sT}
\left(\left[\frac{\partial (\mathrm e^\phi T)}{\partial Y_L} + \frac{\partial
(\mathrm e^\phi T)}{\partial s}\right]_j^t \delta_{ij} - \tilde
D^{sT}_{j+1/2}(\delta_{i+1,j}-\delta_{i,j})\right.\\
\left.+ \tilde D^{sT}_{j-1/2}(\delta_{i,j} -\delta_{i-1,j})\right)(\mathrm
e^\phi T)^{t+\mathrm dt}_j = \left[\frac{\partial (\mathrm e^\phi T)}{\partial
Y_L} + \frac{\partial (\mathrm e^\phi T)}{\partial s}\right]^t_i (\mathrm
e^\phi T)^t_i,
\end{multline}
\begin{multline}
\label{eq:implicit_seta}
\left(\left[\frac{\partial \eta}{\partial Y_L} + \frac{\partial \eta}{\partial
s}\right]_j^t \delta_{ij} - \tilde
D^{s\eta}_{j+1/2}(\delta_{i+1,j}-\delta_{i,j})\right.\\
\left.+ \tilde D^{s\eta}_{j-1/2}(\delta_{i,j}
-\delta_{i-1,j})\right)\eta^{t+\mathrm dt}_j = \left[\frac{\partial
\eta}{\partial Y_L} + \frac{\partial \eta}{\partial s}\right]_i^t \eta^t_i,
\end{multline}
where $\tilde D^{YT}$, $\tilde D^{Y\eta}$, $\tilde D^{sT}$, and $\tilde
D^{s\eta}$ are diffusion coefficients conveniently redefined to simplify the
equations, explicitly (i.e., at the current time $t$) evaluated on the staggered
grid.
Eqs.~\eqref{eq:implicit_YT}--\eqref{eq:implicit_seta} are valid for internal
points of the grid.  At the center of the star, we assume that the fluxes are
null,
\begin{align}
F_\nu(0)=0,\\
H_\nu(0)=0,
\end{align}
and at the border, that the fluxes are proportional to the neutrino number and
energy densities,
\begin{align}
F_\nu(A)=\alpha n_\nu(A),\\
H_\nu(A)=\alpha \epsilon_\nu(A),
\end{align}
where $A$ is the total baryon number of the star, $A\equiv a(R)$, and $\alpha\in(0,1)$ an
arbitrary parameter.  The
diffusion approximation (Sec.~\ref{sec:BLE}) causes the fluxes computed via
Eqs.~\eqref{eq:dYL_dt} and \eqref{eq:ds_dt} to exceed the black-body limit in
regions of small optical depth (i.e., near the stellar border). To stabilize
the code in this region, we apply the flux limiter $3\Lambda(x)$ of
\citet{Levermore+Pomraning1981},
\begin{align}
\label{eq:flux_limiter}
3\Lambda(x)={}&\frac3x\left(\coth(x)-\frac1x\right),\\
F_\nu\rightarrow{}& 3\Lambda(F_\nu/n_\nu)F_\nu,\\
H_\nu\rightarrow{}& 3\Lambda(H_\nu/\epsilon_\nu)H_\nu.
\end{align}

Our simulations start at $\unit[200]{ms}$ from the core bounce, when the
evolution may be considered quasi-stationary.  As \citet{Pons+.1999}, our
initial entropy and lepton fraction profiles are obtained from the
core-collapse simulations of \citet{Wilson+Mayle.1989}, conveniently rescaled
with the total PNS baryon mass, which has been fixed at the beginning of our
simulations.  We have done this since we expect that the masses in the inner
core roughly contains a constant entropy and lepton fraction
\citep{Pons+.1999}. The rescaling applied to the entropy and lepton fraction profiles
is the following
\begin{align}
s(a,t=\unit[200]{ms})={}&\frac{M_\mathrm{B}}{M'_\mathrm{B}}s'\!\left(a',t=\unit[200]{ms}\right),\\
Y_L(a,t=\unit[200]{ms})={}&\frac{M_\mathrm{B}}{M'_\mathrm{B}}Y_L'\!\left(a',t=\unit[200]{ms}\right),\\
a={}&\frac{M_\mathrm{B}}{M'_\mathrm{B}}a',
\end{align}
where the prime refers to the reference profiles of~\citet{Wilson+Mayle.1989}.

The code described in this section has been written from scratch in
\texttt{FORTRAN90}, with some auxiliary \texttt{FORTRAN77} subroutine from
\citet{Press+book} and some other auxiliary subroutines provided by J.~A.~Pons.

\section{Results: PNS evolution}
\label{sec:PNSresults}

\begin{figure}
\centerline{
\includegraphics[width=1.3\textwidth]{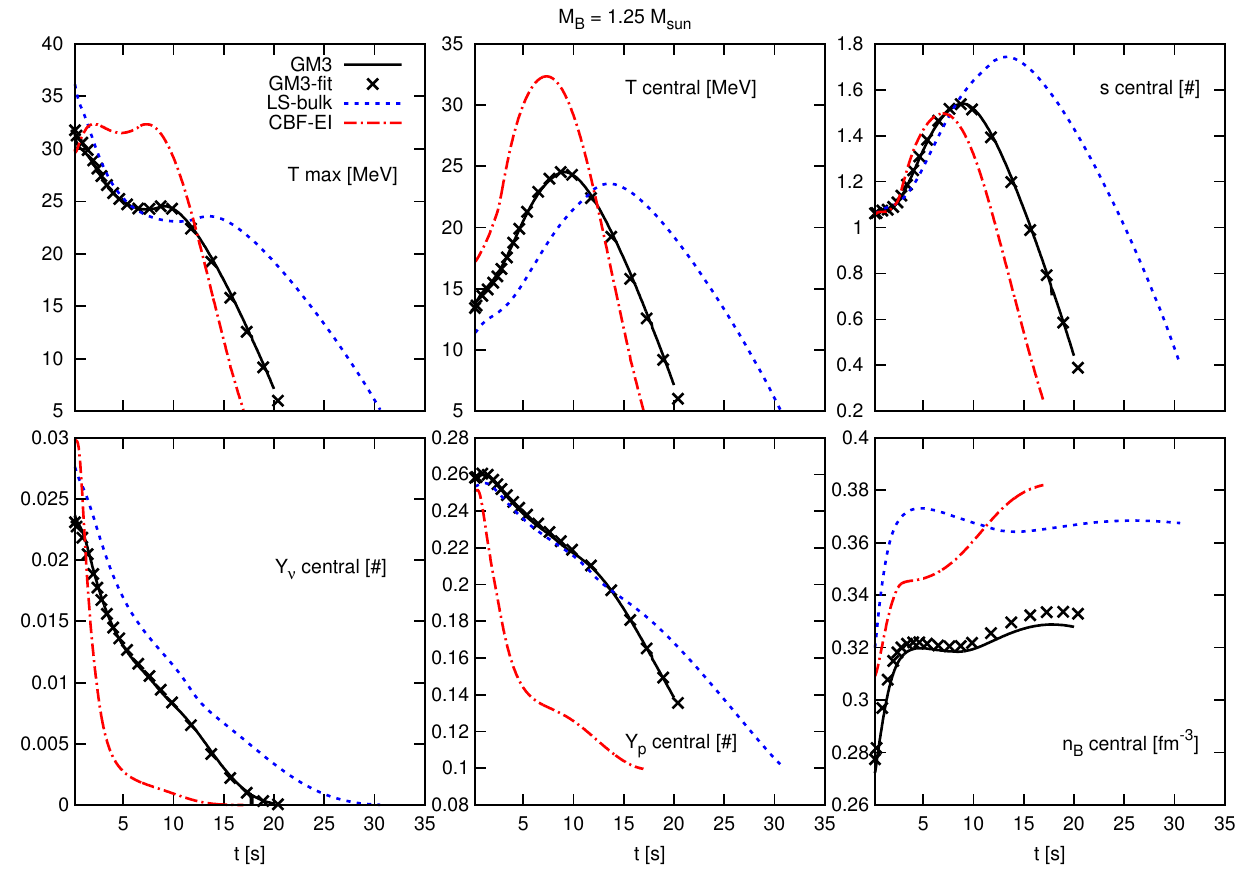}}
\caption{Time dependence of the maximum and central temperature (left and
central upper plots), central entropy per baryon (right upper plot), neutrino
and proton fraction (left and central lower plots) and central baryon
density (right lower plot), for a star with total baryon mass
$M_\mathrm{B}=\unit[1.25]{M_\odot}$ evolved using the three EoSs. Colors and
line-styles are as in Fig.~\ref{fig:mu_n}.}
\label{fig:central_1.25}
\end{figure}

\begin{figure}
\centerline{
\includegraphics[width=1.3\textwidth]{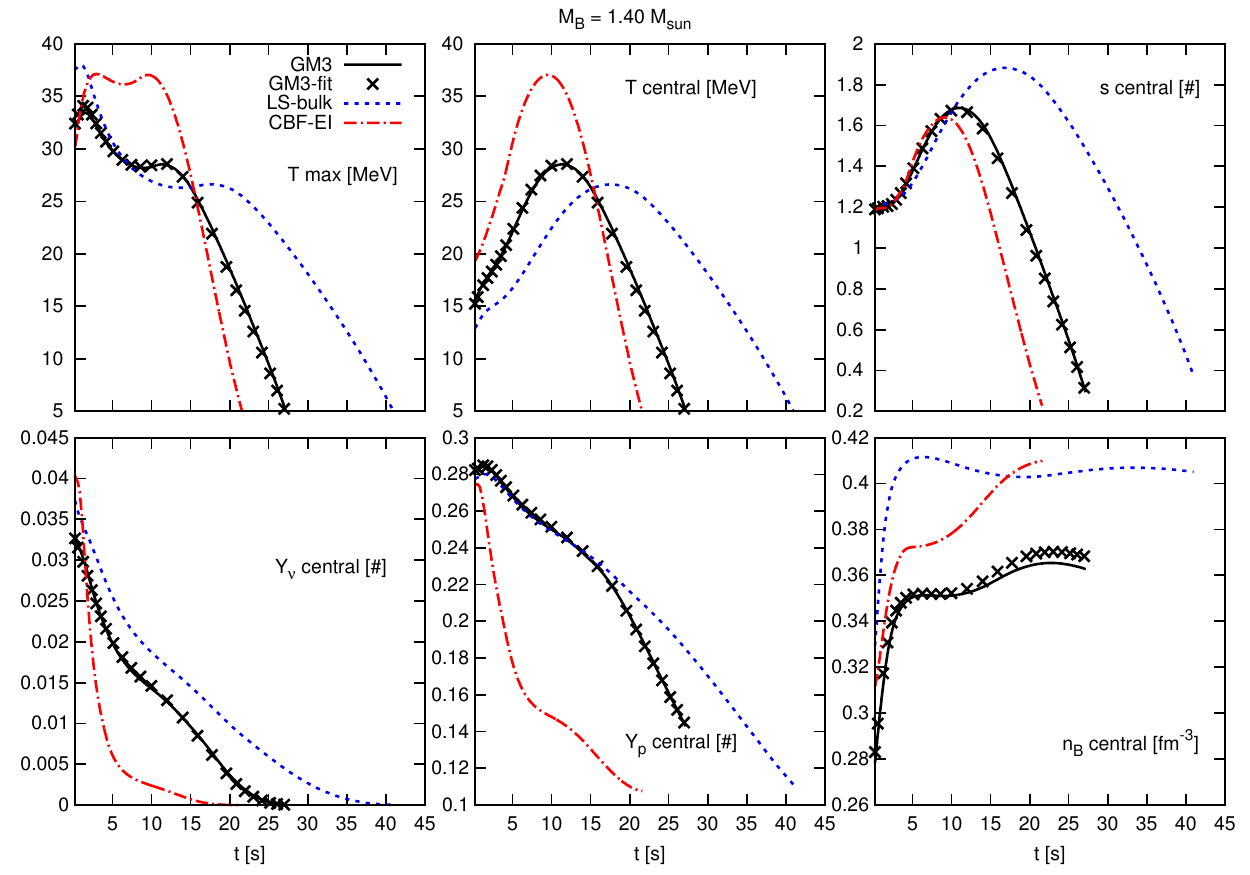}}
\caption{As in Fig~\ref{fig:central_1.25}, but for a star with $M_\mathrm{B}=\unit[1.40]{M_\odot}$.}
\label{fig:central_1.40}
\end{figure}

\begin{figure}
\centerline{
\includegraphics[width=1.3\textwidth]{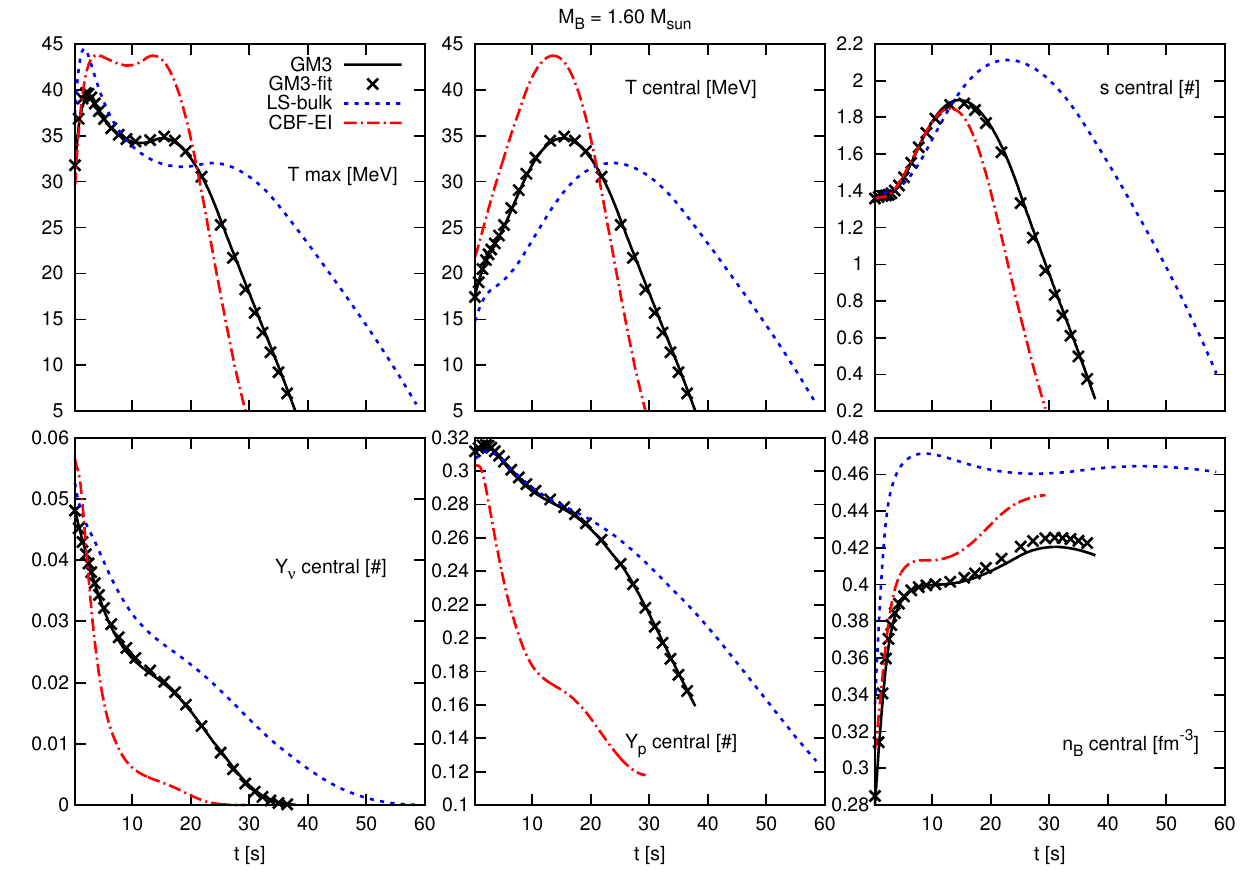}}
\caption{As in Fig~\ref{fig:central_1.25}, but for a star with $M_\mathrm{B}=\unit[1.60]{M_\odot}$.}
\label{fig:central_1.60}
\end{figure}

\begin{figure}
\centerline{
\includegraphics[width=1.25\textwidth]{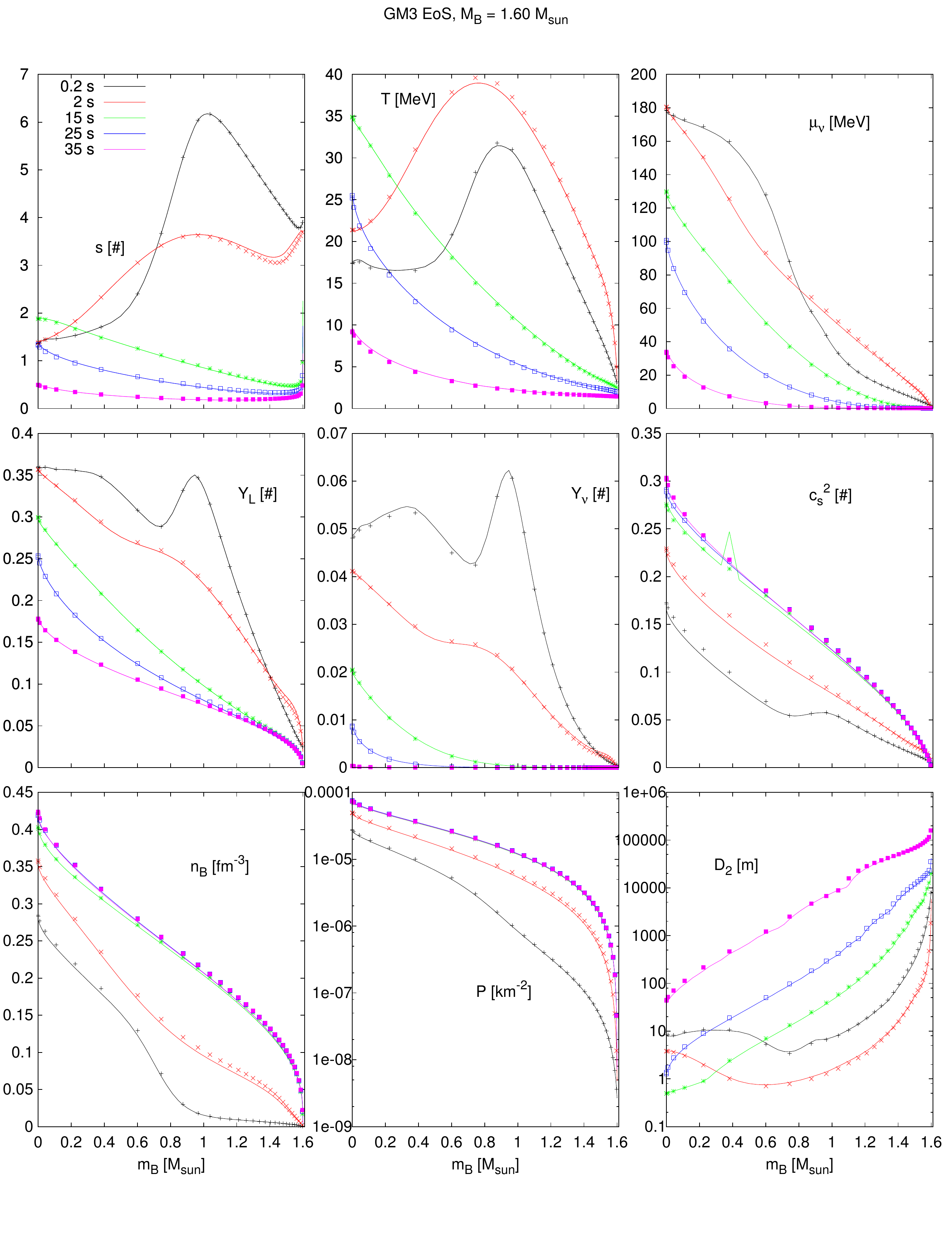}}
\vspace{-1cm}
\caption{Stellar profiles at different times of the PNS evolved with the GM3 EoS. The solid lines correspond to the
GM3 EoS determined with the methods explained in Sec.~\ref{sec:GM3}, the dots to the GM3 EoS determined by fitting the
interacting contribution, see Secs.~\ref{sec:fBfit} and \ref{sec:totaleos}. From left to right and from top to bottom,
we plot: the entropy per baryon, the temperature, the neutrino chemical potential, the lepton and neutrino fraction,
the square of the speed of sound, the baryon density, the pressure and the diffusion coefficient $D_2$.}
\label{fig:evolution_GM3}
\end{figure}

\begin{figure}
\centerline{
\includegraphics[width=1.25\textwidth]{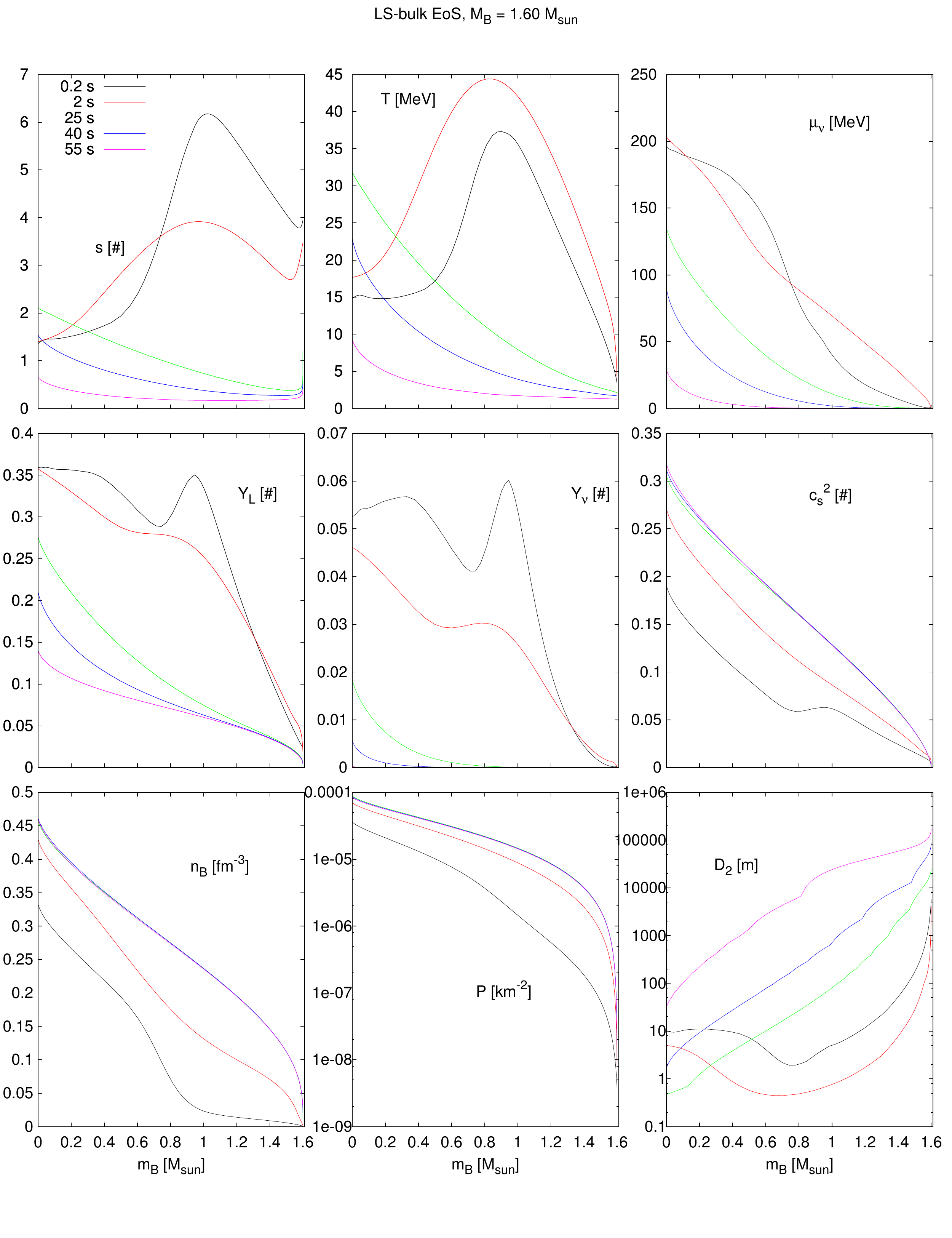}}
\caption{As in Fig.~\ref{fig:evolution_GM3}, for the LS-bulk EoS.}
\label{fig:evolution_bulk}
\end{figure}

\begin{figure}
\centerline{
\includegraphics[width=1.25\textwidth]{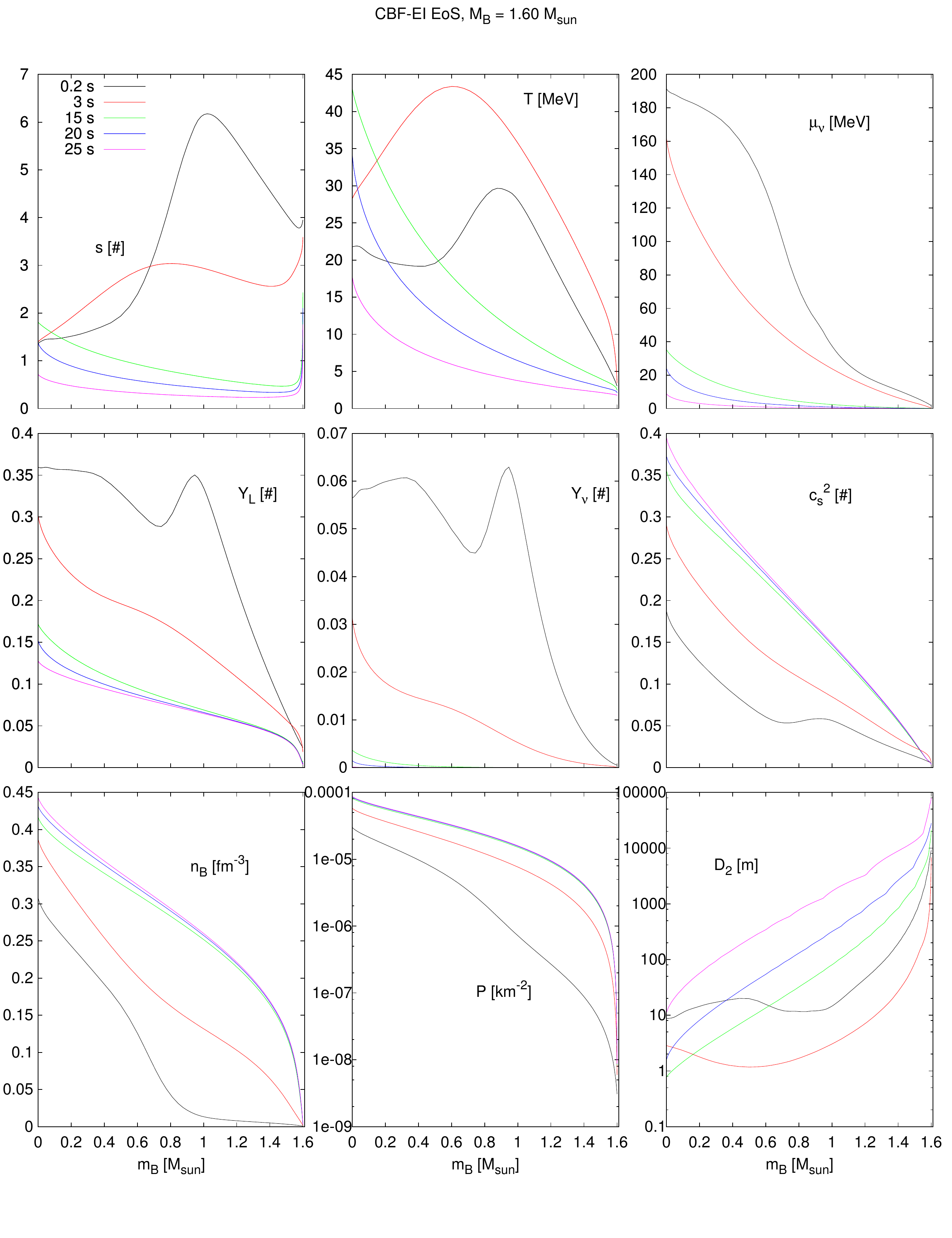}}
\caption{As in Fig.~\ref{fig:evolution_GM3}, for the CBF-EI EoS.}
\label{fig:evolution_CBFEI}
\end{figure}

In this section we discuss the EoS and mass dependence of the evolution of a
PNS. The EoSs we consider are those introduced in Chapter~\ref{cha:EoS}. 
In Table~\ref{tab:central} we summarize the timescale of the evolution
and the maximum central temperature for the three EoSs and three total baryon masses
$M_\mathrm{B}=\unit[(1.25;1.40;1.60)]{M_\odot}$. In
Figs.~\ref{fig:central_1.25}--\ref{fig:central_1.60} we show the evolution of
the central and maximum temperature, central entropy per baryon, central
neutrino and proton fraction, and central baryon density for the same
cases of Table~\ref{tab:central}.
In Figs.~\ref{fig:evolution_GM3}--\ref{fig:evolution_CBFEI} we show some
snapshots of the stellar profiles (as a function of the enclosed baryon mass
$m_\mathrm{B}$) of the entropy per baryon, temperature, neutrino chemical potential,
lepton fraction, neutrino fraction, sound speed, baryon number density,
pressure, and diffusion coefficient $D_2$ for the three EoSs, for a star with
total baryonic mass $M_\mathrm{B}=\unit[1.60]{M_\odot}$.  The times of the
snapshots have been chosen to approximately describe the same periods of the
stellar evolution; in fact the timescale of the evolution is different for each
case considered.

As is shown in Figs.~\ref{fig:central_1.25}--\ref{fig:central_1.60}, the
qualitative behaviour of the stellar evolution is the same for the three EoSs
and the three stellar masses, even though the timescales and the thermodynamical
profiles are quantitatively different.  At the beginning of the evolution,
which is 200 ms from core bounce, the PNS has a (relatively) low entropy core
and a high entropy envelope.  The neutrino chemical potential initially is very
high in the center of the star; the process of neutrino diffusion transfers
this degeneracy energy from neutrinos to the matter and this causes the heating
of the PNS core. Moreover, on timescales of about 10 s, the star contracts
from about $30$ km to its final radius of about $12$--$13$ km. The region which
is affected the most from this contraction is the envelope, and in fact the
contraction causes a consistent heating of the PNS envelope.  At the same time,
the steep negative neutrino chemical potential in the envelope causes a
deleptonization of the envelope.  The neutrinos leave the star, bringing with
them energy. The joint effect of the envelope heating caused by contraction and
cooling caused by neutrino emission is apparent in the behaviour of the maximum
stellar temperature: before the central temperature reaches its maximum, the
maximum stellar temperature increases, reaches a maximum value, and then
decreases.  During the progress of the evolution, the maximum stellar
temperature position moves inward, until reaches the center of the star. Since
the neutrino diffusion causes an heating of the PNS core (because as already
explained the neutrino degeneracy energy is converted in thermal energy), it is
not surprising that after some seconds the maximum stellar temperature increases
again.  This initial phase, during which the central temperature increases, lasts for several
seconds and has been referred to as \emph{Joule heating} phase in previous
works.  We may place the end of this phase at the end of the central
temperature increase, that corresponds with the time at which the maximum
temperature position reaches the center of the star. After the end of the Joule
heating, there is a general cooling of the star as the deleptonization
proceeds.  \citet{Burrows+Lattimer.1986} and \citet{Keil+Janka.1995} found that
the end of the Joule-heating phase coincides with the end of deleptonization,
whereas \citet{Pons+.1999}, with the GM3 EoS and with a more sensible
treatment of the neutrino opacities, found that the deleptonization is
longer than the Joule-heating phase.  As \citet{Pons+.1999} we find that
the deleptonization precedes the end of cooling for all the cases considered,
however in the case of the CBF-EI we find that the most of neutrinos have been
radiated by the end of the Joule-heating phase, that is, by the time at which
the central temperature reaches its maximum value
(Figs.~\ref{fig:central_1.25}--\ref{fig:central_1.60}, see Table~\ref{tab:central}).

Our results for the $M_\mathrm{B}=\unit[1.60]{M_\odot}$ PNS with the GM3 EoS are
in qualitative agreement with those of \citet{Pons+.1999}. In particular, the duration
of the Joule-heating phase is in good agreement (cf.{} Fig.~\ref{fig:central_1.60} of
this thesis with Fig.~17 in \citealp{Pons+.1999}); however we find lower stellar temperatures and
a shorter cooling phase than
that found in \citet{Pons+.1999}. We think that the quantitative differences\footnote{The differences amount
in about $10\%$ in the value of the central temperature maximum
and of the deleptonization time, and in less than
$2\%$ for the time of the end of Joule-heating phase, compare Tab.~\ref{tab:central}
and Fig.~17 of \citealp{Pons+.1999}.} between
our results and those of \citet{Pons+.1999} are due to some differences in the
initial profiles and in the details of the treatment of the diffusion process.

The PNS radius approaches its limiting value before the end of the cooling
phase, while the gravitational mass has not completely reached its limiting
value at the end of our simulation.  This is due to the fact that we end our
simulations when the central temperature is still relatively high,
$T_c=\unit[5]{MeV}$.  However, at this point the remnant energy to be
emitted is small, and in fact the final luminosity is at least two orders of
magnitude smaller than its initial value. We notice that the total
gravitational mass at the end of the evolution is quite similar for the three
EoSs.  This is due to the fact that all considered EoSs are nucleonic EoSs, and
therefore have a similar stiffness and stellar concentration (i.e., $M/R$),
that largely determines the gravitational binding energy (and
then the total gravitational mass). However, the gravitational mass for the
three EoSs is not exactly the same, and this results in a different total
neutrino emission, see Sec.~\ref{sec:detectors}.

For each EoS, the evolutionary timescales are smaller for a smaller stellar
baryon mass. This is due to the way we have rescaled the initial entropy per
baryon and lepton fraction profiles with the total stellar baryon mass, but
also to the fact that a lower stellar mass is equivalent to lower densities and
then to longer neutrino mean free paths.  We also notice that a lower stellar
mass corresponds to lower temperatures. This again depends on the
initial entropy profiles and on the different densities present in the star,
see Fig.~\ref{fig:temp_entr}: for a given entropy per baryon and lepton (or
neutrino) fraction, lower densities correspond to lower temperatures. To
overcome the uncertainties related to the initial configuration, in future we
should use as initial profiles the ending configuration of core-collapse simulation of a star with the
correct baryonic mass (see \citealp{Pons+.1999} for a study on how the initial conditions affect
the PNS evolution).

The fact that the LS-bulk evolution is slower than the GM3 one, which in turn is
slower than that of the CBF-EI EoS, may well be explained by the fact that in
the many-body CBF-EI EoS nuclear correlations are stronger than in the
mean-field GM3 EoS, in which in turn are stronger than in the LS-bulk EoS (where
the baryon masses are equal to the bare ones).  A
smaller neutrino cross section is a consequence of a greater baryon correlation
(Sec.~\ref{sec:comparison}).  This effect is relevant even at the
mean-field level, where one adopts the description of the baryon spectra in term
of effective masses and single-particle potentials to obtain the diffusion
coefficients. For example, the fact that the proton effective mass is
significantly smaller than the neutron one in the CBF-EI framework (Fig.~\ref{fig:diff}) is a
consequence of the tensor correlations which are stronger in the n-p channel
than in the n-n or p-p channels.

To check this interpretation, that is, that the different timescales are mainly
due to the details of the microphysics (i.e., the baryon spectra and hence the
neutrino mean free paths and diffusion coefficients), we have run a simulation
of a $M_\mathrm{B} =\unit[1.60]{M_\odot}$ PNS with the LS-bulk and CBF-EI EoSs,
\emph{but} with the diffusion coefficients of the GM3 EoS. As expected, we find
out that the LS-bulk timescale is reduced with respect to that of a
self-consistent simulation (i.e., using the LS-bulk diffusion coefficients),
and the CBF-EI timescale is increased with respect to that of a self-consistent
simulation.  Of course, the timescales and the evolutionary profiles found in
this non-consistent manner are not equal to those corresponding to the GM3 EoS,
the differences due to the details of the EoS. For example, each EoS has a different thermal
content and neutrino degeneracy, and different thermodynamical derivatives
that determine how the stellar profiles change while energy and leptons diffuse
through the star. Both the EoS and the neutrino mean free paths are important in determining the PNS
evolution.

\begin{table}
\caption{Significant quantities describing the PNS evolution for the three EoSs
described in this thesis and for three stellar baryon masses. The first column
contains the name of the EoS, the second column contains the stellar baryon
mass, the third and fourth columns contain the maximum central temperature and the
corresponding time (the latter approximately corresponds to the end of the
Joule-heating phase), respectively, the fifth column contains the time at which the central
neutrino fraction becomes equal to $Y_\nu=0.005$ (this is an indication on the
duration of the deleptonization phase), and the sixth column contains the time
at which our simulation ends (namely, when the central temperature becomes
equal to $T=\unit[5]{MeV}$). All simulations start at $t_\mathrm{start}=\unit[0.2]{s}$.}
\label{tab:central}
\centering
\begin{tabular}{cccccc}
EOS & $M_\mathrm{B}$ [$M_\odot$]& $T_\mathrm{max}$ [MeV] & $t_\mathrm{Jh}$ [s] & $t_\mathrm{del}$ [s] & $t_\mathrm{end}$ [s] \\
\hline
GM3     & 1.25 & 24.6 &  9.0 & 13.1 & 20.8 \\
GM3     & 1.40 & 28.7 & 11.2 & 18.6 & 27.1 \\
GM3     & 1.60 & 34.9 & 15.2 & 27.9 & 37.7 \\
LS-bulk & 1.25 & 23.6 & 13.5 & 17.5 & 30.6 \\
LS-bulk & 1.40 & 26.6 & 17.6 & 26.3 & 41.0 \\
LS-bulk & 1.60 & 32.1 & 23.8 & 41.4 & 59.2 \\
CBF-EI  & 1.25 & 32.3 & 7.31 & 3.46 & 17.0 \\
CBF-EI  & 1.40 & 37.0 & 9.55 & 5.65 & 21.6 \\
CBF-EI  & 1.60 & 43.7 & 13.6 & 11.7 & 29.4
\end{tabular}
\end{table}

\begin{figure}
\centerline{
\includegraphics[width=1.3\textwidth]{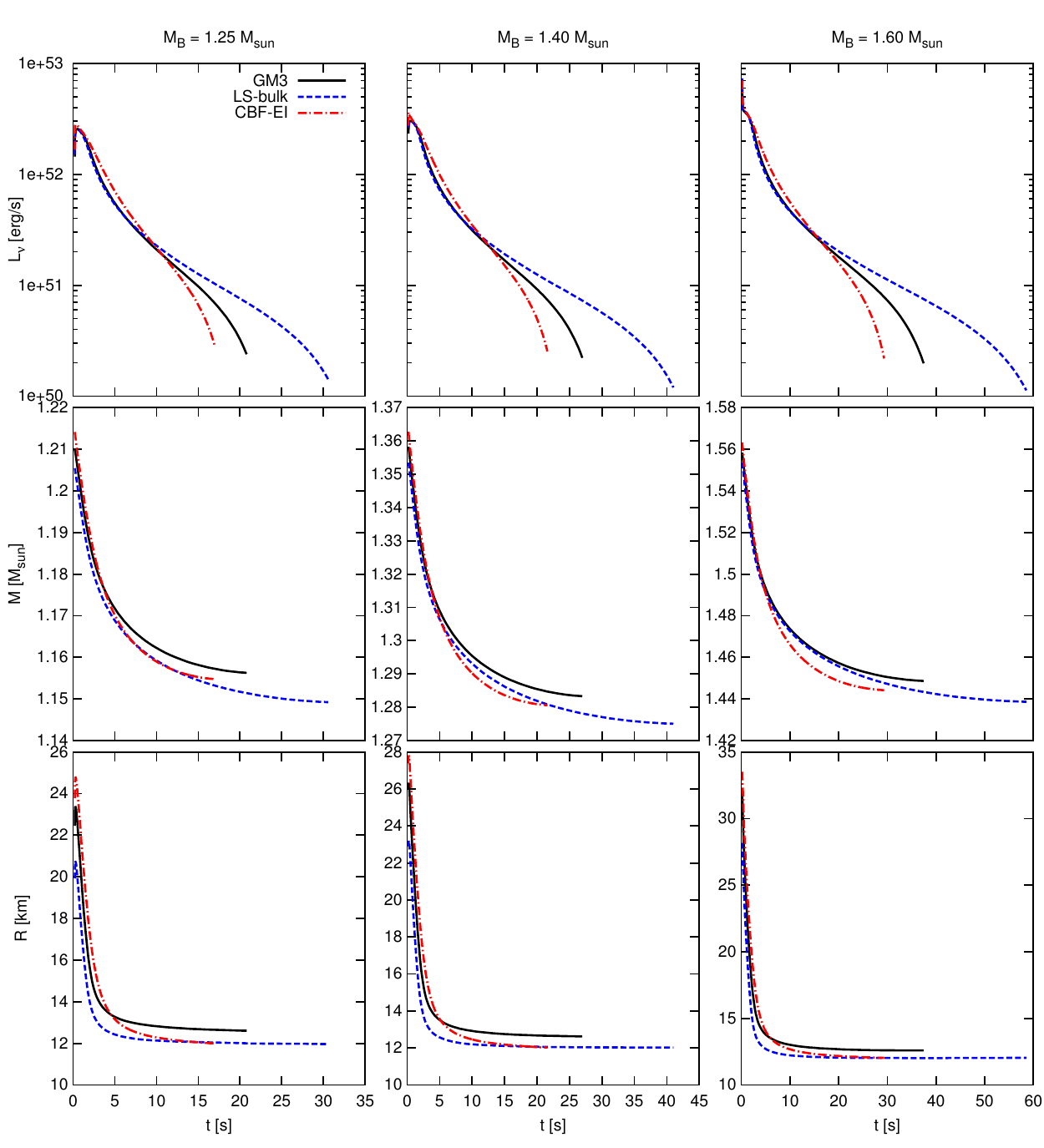}}
\caption{Time dependence of the total neutrino luminosity (upper panels),
gravitational mass (middle panels), and stellar radius (lower panels) of a PNS
evolved with the three EoSs considered in this thesis and the baryon stellar
masses $M_\mathrm{B}=\unit[(1.25,1.40,1.60)]{M_\odot}$.  Colors and line-styles
as in Fig.~\ref{fig:mu_n}. The gravitational masses at
the end of the simulations are: for $M_\mathrm{B}=1.25M_\odot$,
$M_\mathrm{GM3}=1.1554 M_\odot$, $M_\mathrm{LS-bulk}=1.1492 M_\odot$,
$M_\mathrm{CBF-EI}=1.1548M_\odot$;
 for $M_\mathrm{B}=1.40M_\odot$,
$M_\mathrm{GM3}=1.2824 M_\odot$, $M_\mathrm{LS-bulk}=1.2750 M_\odot$,
$M_\mathrm{CBF-EI}=1.2806M_\odot$;
and  for $M_\mathrm{B}=1.60M_\odot$,
$M_\mathrm{GM3}=1.4478 M_\odot$, $M_\mathrm{LS-bulk}=1.4386 M_\odot$,
$M_\mathrm{CBF-EI}=1.4442M_\odot$.
}
\label{fig:L_M_R}
\end{figure}

\section{Results: neutrino signal in terrestrial detectors}
\label{sec:detectors}

In 1987 a supernova (SN1987a) has been observed in the Large Magellanic Cloud
\citep{Kunkel+1987}. Together with the electromagnetic signal, 19 neutrinos
were detected by the Cherenkov detectors Kamiokande~II \citep{Hirata+1987} and
IMB \citep{Bionta+1987}, and 5 neutrinos at the Baksan underground
scintillation telescope \citep{Alekseev+1987}. These neutrinos have been
observed on a timescales of ten seconds, and are therefore thought to have been
emitted during the PNS phase. However, they were too few to accurately
constrain the emitted neutrino spectrum and its time dependence (see
e.g.~\citealp{Lattimer+Yahil.1989}) and to give unambiguous answers about the
proto-neutron star physics \citep{Burrows.1988, Lattimer+Yahil.1989,
Keil+Janka.1995, Pons+.1999}. Today, with the current detectors, a SN event
such that of the 1987 would generate $\sim10^4$ neutrino detection
\citep{Ikeda+2007}, that would permit to accurately discriminate between the
different physical scenarios that could occur during the PNS evolution. It is
therefore fundamental to determine how the underlying EoS modifies the observed
PNS neutrino signal. In this section we consider in particular the
Super-Kamiokande III detector \citep{Hosaka+2006, Ikeda+2007}.

Our code has some limitations in reconstructing the emitted spectrum: we assume
(i) beta-equilibrium, (ii) a Fermi distribution for all neutrino species, and
(iii) a vanishing chemical potential for the muon and tauon neutrinos
everywhere in the star.  Assumptions (i) and (ii) are reasonable in the
interior of the star, and lose accuracy near the stellar border, where the
diffusion approximation breaks down and in practice the fluxes are always
flux-limited (see Appendix~\ref{sec:betaequilibrium}). To obtain a precise
description of the neutrino emitted spectrum, one has to employ multi-flavour
multi-group evolutionary codes \citep[e.g.,][]{Roberts.2012}, that also account for
neutrino leakage near the stellar border.  This is outside the aims of our
work; however our approximations are reasonable as far as one is interested
in total quantities, in particular the total neutrino luminosity $L_\nu$,
which is equal to minus the gravitational mass variation rate,
\begin{equation}
\label{eq:Lnu_Mdot}
L_\nu=\mathrm e^{2\phi}4\pi R^2H_\nu(R)=-\frac{\mathrm dM}{\mathrm dt},
\end{equation}
where $H_\nu(R)$ is the neutrino energy luminosity at the stellar border
(Figs.~\ref{fig:central_1.25}--\ref{fig:central_1.60}, lower-left plot).

We determine the formula to estimate the signal in terrestrial detectors following the discussion in
\citet{Burrows.1988} and applying a slight modification introduced by
\citet{Pons+.1999}, and we specify our results for the Super-Kamiokande III detector
\citep{Hosaka+2006, Ikeda+2007}.  The main reaction that occurs in a water
detector like Super-Kamiokande is the electron antineutrino absorption on
protons, $\bar\nu_e + p\rightarrow n+e^+$ [\citealp[][Eq.~(1)]{Ikeda+2007}], whose
cross section for an antineutrino of energy $E$ is
[\citealp[][Eq.~(3c)]{Burrows.1988}]
\begin{align}
\label{eq:cross_SK}
\sigma_p\simeq{}& \tilde \sigma_0(E-\Delta_m)\sqrt{(E-\Delta_m)^2-m_e^2},\\
\tilde \sigma_0 ={}& \unit[0.941\times10^{-43}]{cm^2MeV^{-2}},
\end{align}
where we have neglected weak-magnetism corrections, $\Delta_m$ is the
neutron-proton mass difference, and $m_e$ is the electron mass. We remark that
since $\tilde \sigma_0$ \emph{has not} the dimensions of a cross section (there
is an additional factor $\unit{MeV^{-2}}$), we have explicitly put a tilde over it.

For each incoming neutrino, the detection probability is given by
\begin{equation}
\label{eq:Pnu_SK}
P_{\bar\nu_e}=N_p \frac{4\pi \int_{E_\mathrm{th}}^\infty E^2
f(E,\mathrm e^{\phi_\nu}T_\nu,\mathrm e^{\phi_\nu}\mu_{\bar\nu_e}) \sigma_p(E) W(E) \mathrm d E} {4\pi \int_0^\infty E^2
f(E,\mathrm e^{\phi_\nu}T_\nu,\mathrm e^{\phi_\nu}\mu_{\nu_e})\mathrm dE},
\end{equation}
where $E_\mathrm{th}$ is the incoming neutrino energy threshold (to cut off the
low-energy neutrino background that is a noise for high-energy
SN and PNS neutrinos, \citealp{Ikeda+2007}), $N_p$ is the number of free protons in the detector (i.e., the hydrogen atoms),
$f$ is the electron antineutrino distribution function at the neutrinosphere, and $W(E)$ is the
efficiency of the detector at incoming neutrino energy $E$. $\mathrm e^{\phi_\nu}$, $T_\nu$, and $\mu_{\bar\nu_e}$
are the redshift, temperature, and antineutrino chemical potential at the
neutrinosphere, that is the sphere inside the PNS at whose radius $R_\nu$
neutrinos decouple from matter (therefore, $\mathrm e^{\phi_\nu} T_\nu$ and $\mathrm e^{\phi_\nu}\mu_{\bar\nu_e}$ are
the temperature and the chemical potential at the neutrinosphere, seen by an
observer at infinity), and the
denominator is the antineutrino number density $n_{\bar\nu_e}$ at the neutrinosphere\footnote{\label{ftn:2} It may be thought that
the denominator of Eq.~\eqref{eq:Pnu_SK} is the electron antineutrino density at the detector. However, this is not the case:
in fact $f$ \emph{is not} the distribution function at the detector, since
neutrinos do not follow a Fermi-Dirac distribution at the detector. In
fact, they are not in thermal equilibrium with the matter anymore, and their
density is so low that the Pauli blocking is ineffective. What has happened is
that the neutrino distribution freezes when neutrinos decoupled from matter at
the neutrinosphere. To obtain their distribution function at the detector one
should also account for their dilution while they move away from the PNS.
This factor does not appear in Eq.~\eqref{eq:Pnu_SK} since it is factorized by
an identical factor at the numerator.}. The number flux of antineutrinos
reaching the detector is given by
\begin{equation}
\label{eq:dNnu_SK}
\frac{\mathrm dN_{\bar\nu_e}}{\mathrm dt} = \frac{4\pi R^2\mathrm e^\phi F_{\bar\nu_e}}{4\pi D^2},
\end{equation}
where $D$ is the distance of the detector from the PNS and $4\pi R^2 \mathrm
e^\phi F_{\bar\nu_e}$ is the total (integrated over the sphere) number flux of
electron antineutrinos at a radius $R\ge R_\nu$, seen by an observer at
infinity. Our code determines better the total energy luminosity than the
total number flux of antineutrinos and than the neutrinosphere radius (see discussion
above), and therefore we make the replacements (at variance with
\citealp{Burrows.1988}, and together with \citealp{Pons+.1999}) $4\pi
R^2\mathrm e^{\phi(R)} F_{\bar\nu_e}\rightarrow L_{\bar\nu_e}$ and
$n_{\bar\nu_e}\rightarrow\epsilon_{\bar\nu_e}$, where $L_{\bar\nu}$ is the
total luminosity of the electron antineutrino and $\epsilon_{\bar\nu_e}$ its
energy density (this latter computed at the neutrinosphere, see footnote
\ref{ftn:2}).  Putting together Eqs.~\eqref{eq:cross_SK}, \eqref{eq:Pnu_SK},
and \eqref{eq:dNnu_SK}, and assuming a vanishing electron antineutrino chemical
potential $\mu_{\bar\nu_e}=0$, we obtain
\begin{align}
\frac{\mathrm d\mathcal N}{\mathrm dt} ={}& \frac{\tilde\sigma_0 \tilde n_p
\mathcal M}{4\pi D^2}\mathrm e^{\phi_\nu} T_\nu L_{\bar \nu_e}\frac{G_W(\mathrm
e^{\phi_\nu}T_\nu,E_\mathrm{th})}{6\operatorname{F_3}(0)},\\
6\operatorname{F_3}(0)={}&\int_0^\infty \frac{x^3}{1+\mathrm e^x} \mathrm dx =\frac{7\pi^4}{120},\\
G_W(T,E_\mathrm{th})={}&\int_{E_\mathrm{th}/T}^\infty
\frac{x^2\left(x-\frac\Delta T\right)\sqrt{\left(x-\frac \Delta
T\right)^2-\left(\frac {m_e} T\right)^2}}{1+\mathrm e^{x}} W(xT)\mathrm dx,
\end{align}
where $\tilde n_p\simeq\unit[6.7\times10^{31}]{kton^{-1}}$ is the number of
free protons (i.e., hydrogen atoms) per \emph{unit water mass} of
the detector (this is the reason we put a tilde over it, since it is not
defined in the same way we have defined a ``number density'' up to now),
$\mathcal M$ is the water mass of the detector ($N_p=\mathcal M\tilde n_p$),
$F_3$ is an ordinary Fermi integral and $G_W$ is a modified and truncated Fermi
integral.

We take Super-Kamiokande III as reference detector, and therefore $\mathcal
M\simeq\unit[22.5]{ktons}$ \citep{Ikeda+2007}, $E_\mathrm{th}=\unit[7.5]{MeV}$,
and $W$ is reported in \citet[][Fig.~3]{Hosaka+2006} and is unity for
$E>E_\mathrm{th}$. We consider a galactic PNS, $D\simeq\unit[10]{kpc}$, and
assume the neutrinosphere to be at the radius at which the (total, of all flavours)
neutrino energy flux becomes one third of the (total, of all flavours) neutrino
energy density, $H_\nu/\epsilon_\nu=1/3$. Finally, we take the electron
antineutrino energy to be one sixth of the total, $L_{\bar\nu_e}=L_\nu/6$, since
(i) at the neutrinosphere all neutrino type chemical potentials are very small and
therefore their energy density is very similar [the temperature term dominates, see Eq.~\eqref{eq:eps_nu}]
and (ii) we do not account for neutrino oscillations (which should enhance the
flux by about $10\%$, \citealp{Ikeda+2007}).

\begin{figure}
\centerline{
\includegraphics[width=1.3\textwidth]{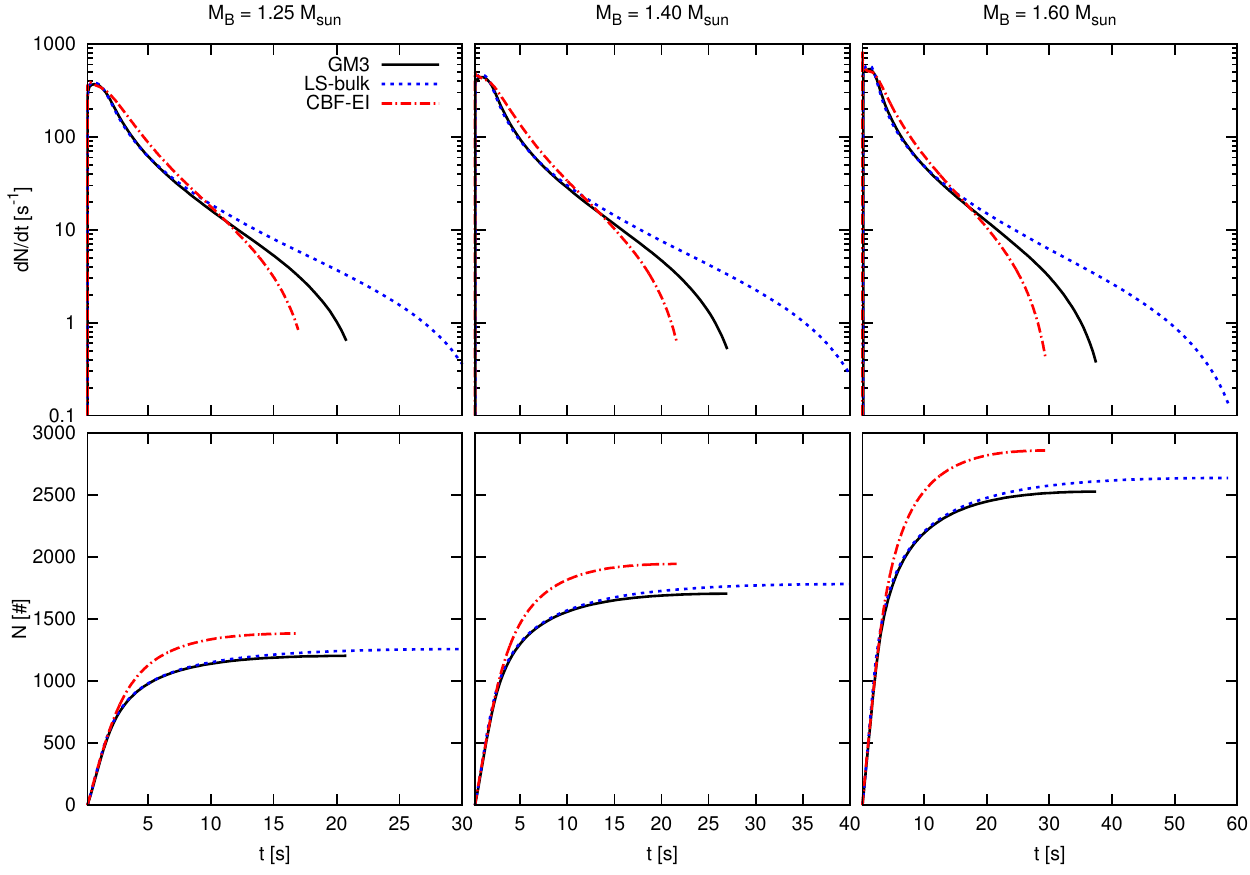}}
\caption{Signal in the Super-Kamiokande III Cherenkov detector, for the three
EoSs considered in this thesis. On the top row, electron antineutrino detection
rate; on the bottom row, electron antineutrino cumulative detection. On the left column,
we consider a star with $M_\mathrm{B}=\unit[1.25]{M_\odot}$, on the central column
$M_\mathrm{B}=\unit[1.40]{M_\odot}$, and on the right column $M_\mathrm{B}=\unit[1.60]{M_\odot}$.}
\label{fig:detector}
\end{figure}

The neutrino signal rate and total signal for the three EoSs are shown in Fig.~\ref{fig:detector}.
Since the binding energy of the cold neutron star of the three EoSs
we consider is very similar (see Figs.~\ref{fig:central_1.25}--\ref{fig:central_1.60}), the total energy
emitted by neutrinos during the PNS evolution is very similar too. On the other
hand, the rate of antineutrino emission and the temperature at the neutrinosphere varies according to the underlying
EoS. Therefore, there is an EoS signature on the cumulative antineutrino detection, and in particular for the CBF-EI,
whose signal is noticeably larger than the other EoSs, even though its gravitational binding energy at the end of the evolution
is between those of the LS-bulk and GM3 EoSs, see Figs.~\ref{fig:central_1.25}--\ref{fig:central_1.60}.
This is due to the fact that the higher temperatures of the CBF-EI EoS cause a smoother antineutrino distribution function at
the neutrinosphere, and hence more antineutrinos have an energy greater than the threshold $E_\mathrm{th}$ at the detector.

The different evolutionary timescales for the different EoSs and stellar masses
correspond to different signal timescales, that may be easily inferred from the
antineutrino detection rate.  The antineutrino detection rates are
qualitatively very similar between the three EoSs and the three stellar masses.
During the first ten seconds the LS-bulk and GM3 stars have very similar
detection rates and then diverge, the LS-bulk star having a longer evolution
than the GM3 star. The CBF-EI star, instead, has the peculiarity of maintaining
a higher antineutrino emission rate during the Joule-heating phase
(approximately, during the first ten seconds), which is due to the faster
deleptonization that we have already described in Sec.~\ref{sec:PNSresults}.

To summarize, the analysis of the antineutrino detection rate and cumulative
detection allows to determine the timescale of the evolution and to
discriminate between the underlying EoS and the total stellar baryon mass.  As
a final remark, we point out again that, for a given mass, the total
antineutrino detected do not depend uniquely on to the gravitational binding
energy: for example the CBF-EI has a higher total antineutrino detection than
the LS-bulk EoS, but a lower gravitational binding energy.

\chapter[GWs from stellar perturbation theory]{Gravitational waves from stellar perturbation theory}
\label{cha:GW}

The detection of the first gravitational wave signal \citep{Abbot+2016},
emitted by two merging black holes, has hopefully started the era of
``gravitational wave astronomy'', that promises to open a new observational
window on our universe.  The detection of a GW signal from a SN event would
permit to explore the explosion process and the PNS formation, allowing to
constrain the PNS physics. However, this would need an extended numerical study
of the possible gravitational waveforms. \citet{Ferrari+Miniutti+Pons.2003}
have studied, using stellar perturbation theory, the gravitational emission
during the first tens of seconds of the PNS life, evolved
consistently with the mean-field GM3 EoS, also allowing for a hadron-quark
transition in the core.  Later, \citet{Burgio+2011} have determined the
quasi-normal modes frequencies of a PNS modeled with the many-body EoS of
\citet{Burgio+Schulze.2010}, mimicking the PNS evolution with some reasonable
thermal and composition profiles. Also \cite{Sotani:2016uwn} have studied the
gravitational wave emission (from proto-neutron stars with LS and Shen EoSs) using ad-hoc thermal profiles that mimic the
evolution. To our knowledge, there are no studies of the
GW emission from a PNS consistently evolved with a many-body EoS.  Our aim is
therefore to fill this gap, also providing a study of the dependence of
the oscillation modes on the underlying EoS and stellar baryon mass.

In Chapter~\ref{cha:evolution} we have discussed the PNS evolution with some
different high-temperature EoSs, fixing different values of the total PNS
baryon mass.  In this chapter, we apply the relativistic theory of stellar
perturbation to determine the frequencies at which a non-rotating PNS
oscillates and emits gravitational waves, according to its underlying physics.
In Sec.~\ref{sec:GW.Eqs} we show how to find the frequencies of the
quasi-normal modes and discuss their classification; in
Sec.~\ref{sec:GW.results} we describe the code we have used to determine the
stellar oscillations and compare the GW emitted signal for different EoSs and
PNS masses.  The results discussed in Sec.~\ref{sec:GW.results} are an original
contribution of this thesis, and are also presented in \citet{Camelio+2017}.

\section{Quasi-normal modes of a relativistic star}
\label{sec:GW.Eqs}

The Newtonian theory of stellar perturbations \citep[e.g.][Chapter~27]{Cox+Giuli.book}
has been generalized to relativistic stars by \citet{Thorne+Campolattaro.1967},
and subsequently by \citet{Chandrasekhar+Ferrari.1990}. Using this theory one
can determine the frequencies at which the star oscillates, radiating
gravitational waves. These damped oscillations are the so-called
\emph{quasi-normal modes} of the star. In this section we briefly summarize the
theory of stellar perturbation and of the quasi-normal modes and refer the
interested reader to the cited literature for the details.

First of all, one makes the customary separation of the metric of the
oscillating star $g_{\mu\nu}$ in a static and spherically symmetric background
part $g_{\mu\nu}^0$ (obtained solving the TOV equations) and a perturbative
part $h_{\mu\nu}$ which is a function of all coordinates,
\begin{equation}
\label{eq:perturbation}
g_{\mu\nu}(t,r,\theta,\varphi)=g^0_{\mu\nu}(r,\theta)+h_{\mu\nu}(t,r,\theta,\varphi),\qquad\mbox{with
}|h_{\mu\nu}|\ll |g^0_{\mu\nu}|.
\end{equation}
Since the star is perturbed, a fluid element that would be placed at $x^\mu$ in
the equilibrium configuration moves to $x^\mu+\xi^\mu(x^\nu)$, being $\xi^\mu$
the fluid displacement.  At this point it is useful to introduce the concepts
of \emph{Eulerian} and \emph{Lagrangian} perturbations of a particular
quantity.  To fix the ideas, we consider the pressure $P$, but it is trivial to
generalize the discussion to other quantities, like the energy density or the
baryon number density.  An Eulerian perturbation of the pressure $\delta P$ at
a point $x^\mu$ is defined by the variation of the pressure at $x^\mu$,
\begin{equation}
\label{eq:perturbation_eulerian}
\delta P(x^\mu) = P(x^\mu) - P^0(x^\mu),
\end{equation}
where $P$ and $P^0$ are the pressure of the perturbed and unperturbed star,
respectively.  The Lagrangian perturbation of the pressure $\Delta P$ at a
point $x^\mu$ is defined as the variation of the pressure of the
\emph{displaced} fluid element that in the unperturbed star was in $x^\mu$,
that is,
\begin{equation}
\label{eq:perturbation_lagrangian}
\Delta P(x^\mu) = P(x^\mu+\xi^\mu) - P^0(x^\mu).
\end{equation}
In other words, the Lagrangian perturbation is the variation seen by an
observer comoving with the perturbation.  It is easy to convert an Eulerian
perturbation into a Lagrangian one,
\begin{equation}
\label{eq:eulerian_to_lagrangian}
\Delta P(x^\mu) \simeq \delta P(x^\mu) + \frac{\partial P^0(x^\mu)}{\partial
x^\nu}\xi^\nu(x^\mu).
\end{equation}

Choosing a suitable frame for which $\xi^0=0$, the stellar perturbation is
defined by the unknown functions $\xi^i(x^\alpha)$ ($i=\{1;2;3\}$) and
$h_{\mu\nu}(x^\alpha)$.  The crucial point of the stellar perturbation theory
is that the displacement vector $\xi^i$ and the metric perturbation tensor
$h_{\mu\nu}$ are expanded in vector and tensor spherical harmonics,
respectively.  These were introduced by \citet{Regge+Wheeler.1957}, and we
will follow their convention.  Each harmonic basis element is determined by
two indexes, that we customary call $l=0,\ldots,\infty$ (``the total angular
momentum'') and $ m$ (``the $z$ component of the angular momentum'', not to be
confused with the enclosed gravitational mass).  Since our background metric is
spherically symmetric (i.e., the star is non rotating), there is degeneracy in
the $ m$ index, and then the resulting equations for two different values of $
m$ are equivalent. To simplify the discussion, and without loss of generality,
we will assume $ m=0$.  The index $l$ permits to classify the behaviour of the
spherical harmonics under the parity operation,
$\mathbf{r}\rightarrow-\mathbf{r}$.  A harmonic that transforms as $(-1)^l$
under parity is called ``even'' (or ``polar'', or ``electric''), whereas if it
transforms as $(-1)^{l+1}$ is called ``odd'' (or ``axial'', or ``magnetic'').
Scalar harmonics are even, vector and tensor harmonics may be either even or
odd.  Making a suitable choice of the gauge, one can put to zero four components of
the metric perturbation tensor $h_{\mu\nu}$.  Many different choices of gauge
are possible, and we will use the Regge-Wheeler gauge \citep{Regge+Wheeler.1957}.

Since the fluid is perturbed, the stress-energy tensor is perturbed too and one
has to solve the perturbed Einstein equations and the perturbed continuity
equations,
\begin{align}
\label{eq:perturbed_einstein}
\delta G_{\mu\nu}={}&8\pi\delta T_{\mu\nu},\\
\label{eq:perturbed_continuity}
\delta(T_{\hphantom{\mu\nu};\nu}^{\mu\nu})={}&0,
\end{align}
where $G_{\mu\nu}$ is the Einstein tensor, $T_{\mu\nu}$ is the stress-energy
tensor of the fluid composing the star, and $\delta$ is the Eulerian
perturbation.  At this point, one could Fourier-transform the equations and
solve the perturbed equations for assigned values of the frequency $\omega$. We
shall look for complex frequency solutions, i.e.{}
\begin{equation}
\label{eq:omega_complex}
\omega=2\pi\nu+\frac\imath\tau,
\end{equation}
where $\nu$ is the (real) frequency of the perturbation and $\tau$ is the
damping time. If the imaginary part of the complex frequency is zero, we would
have \emph{normal} modes, that is, stellar oscillation modes with amplitude
constant in time (they do not grow nor damp). This is what happens in the
Newtonian case, if one neglects viscosity forces. However, in GR, the imaginary
part of the complex frequency is not zero due to gravitational wave emission,
and therefore the stellar modes are called \emph{quasi-normal modes} (QNMs).

Since the background metric is spherically symmetric, the perturbation
equations for different angular momentum $l$ and different parity are
decoupled.  In particular, there are two distinct sets of equations, one for
the even components and one for the odd ones. This allows a further
simplification.  The fluid quantities, that is, the pressure $P$, the energy
density $\epsilon$, and the baryon density $\nB$, are scalar quantities, and
therefore do not appear in the odd-parity equations, since scalar harmonics are
even. Purely gravitational modes, the so called \emph{w modes}, are obtained
from the odd equations, but since we are mainly interested in how the matter
influences the gravitational wave emission, we shall not consider odd
perturbations.

By expanding $h_{\mu\nu}$ in tensor spherical harmonics it is possible to show
that the perturbed metric can be written as \citep{Thorne+Campolattaro.1967,
Detweiler+Lindblom.1985}
\begin{multline}
\label{eq:metric_oscillations}
\mathrm ds^2=-\mathrm e^{2\phi} (1+r^l H_0Y_{l m}\mathrm e^{\imath\omega
t+\imath m\varphi})\mathrm dt^2 -2\imath\omega r^{l+1} H_1Y_{l m}\mathrm
e^{\imath\omega t+\imath m\varphi}\mathrm dt\mathrm dr\\
+ \mathrm e^{2\lambda} (1-r^lH_0Y_{l m}\mathrm e^{\imath\omega t+\imath
m\varphi})\mathrm dr^2 + r^2(1-r^lKY_{l m}\mathrm e^{\imath \omega t+\imath
m\varphi})\mathrm d\Omega,
\end{multline}
and the displacement vector as
\begin{align}
\label{eq:displacement_r}
\xi^r={}&r^{l-1}\mathrm e^{-\lambda}W(r,\omega) Y_{l m}(\theta,\varphi) \mathrm
e^{\imath\omega t+\imath m\varphi},\\
\label{eq:displacement_theta}
\xi^\theta={}&-r^{l-2} V(r,\omega)\partial_\theta Y_{l m}(\theta,\varphi) \mathrm
e^{\imath\omega t+\imath m\varphi},\\
\label{eq:displacement_phi}
\xi^\varphi={}&-\frac{r^{l-2}{V(r,\omega)}}{\sin^2\theta}\partial_\varphi
Y_{lm}(\theta,\varphi) \mathrm e^{\imath\omega t+\imath m\varphi},
\end{align}
where $Y_{lm}$ is the scalar spherical harmonic corresponding to indexes $l$
and $m$, and the functions $H_0$, $H_1$, $K$, $W$, $V$ depend on the radial
coordinate $r$ and on the frequency $\omega$. They are the solution of a
coupled set of equations obtained by inserting the metric
\eqref{eq:metric_oscillations} into Eqs.~\eqref{eq:perturbed_einstein} and
\eqref{eq:perturbed_continuity}; setting the index $m=0$ these equations are ($m(r)$ is
the gravitational mass)
\begin{multline}
\label{eq:osc_start}
\left(3m(r)+\frac{(l+2)(l-1)}2r+4\pi r^3P\right)H_0=8\pi r^3 \mathrm e^{-\phi}
X\\
+\left(\frac12(l+2)(l-1)r-\omega^2 r^3\mathrm e^{-2\phi}-\frac{\mathrm
e^{2\lambda}}{r}\big(m(r)+4\pi r^3P\big)\big(3m(r)-r+4\pi r^3P\big)\right)K\\
-\left(\frac{l(l+1)}2\big(m(r)+4\pi r^3P\big)-\omega^2r^3\mathrm
e^{-2(\phi+\lambda)}\right)H_1,
\end{multline}
\begin{align}
X={}& \omega^2 (\epsilon+P)\mathrm e^{-\phi}V-\frac{\mathrm
e^{\phi-\lambda}}{r} \frac{\mathrm dP}{\mathrm dr} W
+\frac{\epsilon+P}{2}\mathrm e^\phi H_0,\\
\frac{\mathrm dH_1}{\mathrm dr}={}& -\left(l + 1 + \frac{2m(r)}{r}\mathrm
e^{2\lambda}+ 4\pi r^2\mathrm e^{2\lambda}(P-\epsilon)
\right)\frac{H_1}{r}\notag\\
{}&+\frac{\mathrm e^{2\lambda}}{r}\Big(H_0+K-16\pi(\epsilon+P)V\Big),\\
\frac{\mathrm dK}{\mathrm dr}={}&\frac{H_0}r + \frac{l(l+1)}{2r}H_1 -
\left(\frac{l+1}r-\frac{\mathrm d\phi}{\mathrm dr}\right)K
-8\pi\frac{\epsilon+P}{r}\mathrm e^\lambda W,\\
\frac{\mathrm dW}{\mathrm dr}={}&-\frac{l+1}{r}W + r\mathrm
e^\lambda\left(\frac{\mathrm e^{-\phi}}{(\epsilon+P)c_s^2}X
-\frac{l(l+1)}{r^2}V+\frac{H_0}2+K\right),\\
\frac{\mathrm dX}{\mathrm dr}={}&-\frac{l}{r}X + (\epsilon+P)\mathrm
e^\phi\left\{\left(\frac1r-\frac{\mathrm d\phi}{\mathrm dr}\right)
\frac{H_0}2\right.\notag\\
{}&+\left(r\omega^2\mathrm e^{-2\phi}+\frac{l(l+1)}{2r}\right)\frac{H_1}{2} +
\left(3\frac{\mathrm d\phi}{\mathrm dr}-\frac1r\right)\frac K2
-\frac{l(l+1)}{r^2}\frac{\mathrm d\phi}{\mathrm dr}V\notag\\
\label{eq:osc_end}
{}&\left.- \frac1r \left[4\pi(\epsilon+P)\mathrm e^\lambda+\omega^2\mathrm
e^{\lambda-2\phi} 
- r^2\frac{\mathrm d}{\mathrm dr}\left(\frac{\mathrm
  e^{-\lambda}}{r^2}\frac{\mathrm d\phi}{\mathrm dr}\right)\right]W\right\},
\end{align}
where $\phi(r)$, $\lambda(r)$ [the metric functions defined in
Eq.~\eqref{eq:metric}], $m(r)$, $P(r)$, and $\epsilon(r)$ are the unperturbed
quantities determined solving the TOV equations, the  $c_s^2$ is the square of the 
speed of sound,
\begin{equation}
\label{eq:cs2}
c_s^2=\left.\frac{\partial P}{\partial \epsilon}\right|_{s,Y_L},
\end{equation}
and the auxiliary function $X(r,\omega)$ is proportional to the Lagrangian
perturbation of the pressure,
\begin{equation}
\label{eq:X_osc}
\Delta P=-r^l\mathrm e^{-\phi} X(r).
\end{equation}

To summarize, Eqs.~\eqref{eq:osc_start}--\eqref{eq:osc_end} allow to determine
the stellar perturbation \emph{in the interior of the star}, for a fixed
harmonic angular momentum $l$ and for a complex pulsation $\omega$.  Since
these equations are singular at $r=0$, to obtain physical solutions (i.e.,
regular at $r=0$), one has to expand the functions near the center, obtaining
\begin{align}
H_1(r)={}&H_1(0)+O(r^2),\\
K(r)={}&K(0)+O(r^2),\\
W(r)={}&W(0)+O(r^2),\\
X(r)={}&X(0)+O(r^2),\\
\label{eq:osc_cond1}
H_1(0)={}&\frac{2lK(0)+16\pi(\epsilon_c+P_c)W(0)} {l(l+1)},\\
\label{eq:osc_cond2}
X(0)={}& (\epsilon_c+P_c)\mathrm e^\phi\left(\left(\frac{4\pi}3
(\epsilon_c+3P_c)-\frac {\omega^2}l \mathrm e^{-2\phi}\right)W(0) + \frac 12
K(0)\right),
\end{align}
where $\epsilon_c=\epsilon(0)$ and $P_c=P(0)$ are the energy density and the
pressure at the center.  After imposing conditions \eqref{eq:osc_cond1} and
\eqref{eq:osc_cond2}, it is possible to show that the system has two
independent solutions.  In order to find them, one has to set the value of $W$
and $K$ at $r=0$; we set for instance, following
\citet{Detweiler+Lindblom.1985},
\begin{align}
W(0)={}&1,\\
K(0)={}&\pm (\epsilon_c+P_c).
\end{align}
At this point, the System~\eqref{eq:osc_start}--\eqref{eq:osc_end} has to be
integrated for the two independent solutions (we put a tilde and a hat over the
perturbation quantities to identify the two independent solutions), with the
condition that at the surface the Lagrangian perturbation of the pressure
[Eq.~\eqref{eq:X_osc}] vanishes, that is, $X(R)=0$. In this way, the solution
is uniquely determined:
\begin{align}
X(r)={}&\hat X(r)-\frac{\hat X(R)}{\tilde X(R)}\tilde X(r),\\
H_1(r)={}&\hat H_1(r)-\frac{\hat X(R)}{\tilde X(R)}\tilde H_1(r),
\end{align}
and similarly for the other perturbation functions.  To determine which values
of $\omega$ correspond to a quasi-normal mode, one has to impose that there is
no ingoing gravitational radiation at infinity; therefore the perturbation
equations for the stellar exterior have to be solved too.

To do that, using the solution of the perturbation equations which we have
integrated inside the star, we compute the values of the Zerilli function
$Z(r)$ and of its first derivative at the surface of the star, where the fluid
perturbations vanish. Indeed, at $r\ge R$ the perturbation equations reduce to
a single wave equation, found by \citep{Zerilli.1970b},
\begin{align}
\frac{\mathrm d^2 Z(r)}{\mathrm d r^{\ast
2}}=&{}\left(V(r)-\omega^2\right)Z(r),\\
V(r)=&{}\left(1-\frac{2M}{r}\right) \frac{2n^2(n+1)r^3+6n^2Mr^2+18nM^2r +
18M^3}{r^3(nr+3M)^2},\\
n=&{} (l-1)(l+2)/2,\\
r^\ast={}&r+2M\log(r/2M-1),\\
Z(R)=&{}\frac{R^{l+2}}{nR+3M}\big[K(R)-(1-2M/R)H_1(R)\big],\\
\left.\frac{\mathrm dZ(r)}{\mathrm
dr^\ast}\right|_{r=R}=&{}\frac{R^l}{(nR+3M)^2}\Big\{\big[3M(nR+M)-nR^2\big]K(R)\notag\\
&{}+(1-2M/R)\big[3M(nR+M)+n(n+1)R^2\big]H_1(R)\Big\},
\end{align}
where $r^\ast$ is the tortoise coordinate and $M=m(R)$ is the total gravitational
mass of the star.  With these initial conditions, the Zerilli equations can be
integrated from $R$ to radial infinity. When $r\to \infty$, the potential tends
to zero and the solution can be written as a superposition of ingoing and
outgoing waves,
\begin{equation}
\lim_{r\to\infty}Z(r)= Z_+(r) + Z_-(r) =A_+(\omega)\mathrm e^{\imath \omega(t+
r^\ast)} + A_-(\omega) \mathrm e^{\imath \omega (t-r^\ast)},
\end{equation}
from which it is apparent that the ingoing wave part is $Z_+$ and the outgoing
wave part is $Z_-$.  The quasi-normal modes in which we are interested
correspond to purely outgoing waves, for which $A_+=0$. To determine the QNM
frequencies, one should first set the value of the complex frequency $\omega$,
integrate the perturbation equations inside the star, convert the perturbation
functions in the Zerilli function at the stellar surface, solve the Zerilli
equations until a very large radius, and determine the values of $A_+$ and
$A_-$ for that frequency $\omega$.  At this point one iteratively varies the
complex frequency $\omega$ until the condition $A_+=0$ is reached.

An efficient way to determine whether the condition $A_+=0$ holds, is to
transform the Zerilli equations into the Regge-Wheeler ones and apply the
continued fraction method (for details on this method, see
\citealp{Leins+Nollert+Soffel.1993}, see also Appendix~B of
\citealp{Sotani+Tominaga+Maeda.2002} for a clear description of the continued
fraction method). In this way one can avoid to solve the Zerilli equation;
however this method may be applied only if $r>4M$.  Therefore, if the stellar
radius is such that $R<4M$, one should first integrate the Zerilli equations
from $r=R$ to $r>4M$, and then apply the continued fraction method. Anyway, the
stellar configurations we are interested in are always such that $R>4M$.

The quasi-normal modes are classified by their properties (as in Newtonian
gravity, \citealp{Cowling.1941}), in particular by which force dominates in
restoring the equilibrium of a displaced fluid element.  For a non-rotating
relativistic star, the quasi-normal modes are
\begin{itemize}
\item{g-modes:} if the star has thermal or composition gradients (as in the PNS
case), there exists a class of modes, the \emph{gravity modes} (also called
\emph{buoyancy-driven modes} or \emph{g-modes}), whose main restoring force is
the gravity through the buoyancy. These modes have at least one radial node,
frequencies smaller than the fundamental one,
$\nu_f>\nu_{g_1}>\nu_{g_2}>\ldots$, and very long damping times.
\citet{Ferrari+Miniutti+Pons.2003} found that during the first second of PNS
evolution the first g-mode frequency is about $\nu_{g_1}\simeq \unit[900]{kHz}$
and its damping time about tens of seconds.  After the first second, the first
g-mode frequency decreases and its damping time increases. \citet{Burgio+2011}
confirmed these results. To quantify the effect of the thermal and composition
gradients, it is useful to define the Schwarzschild discriminant
\citep[e.g.,][]{Thorne.1966},
\begin{equation}
\label{eq:S_discriminant}
S(r)=\frac{\mathrm dP}{\mathrm dr}\left(1-\frac{c_s^2}{c_0^2}\right),
\end{equation}
where $c_s^2$ is defined in Eq.~\eqref{eq:cs2} and $c_0^2$ is
\begin{equation}
\label{eq:c02}
c_0^2=\frac{\mathrm dP/\mathrm dr}{\mathrm d\epsilon/\mathrm dr}.
\end{equation}
In a cold star without composition gradients, $c_s=c_0$ and the g-modes are all
degenerate at zero frequency.  To study the stability against convection, we
first notice that, since $\mathrm dP/\mathrm dr<0$, if $S(r)>0$ then $c_0<c_s$,
and therefore $\mathrm d\epsilon/\mathrm dP>\partial \epsilon/\partial P$
(where the ``$\mathrm d$'' refers to variations along the radial profile, and
the partial differentiation is made at constant $s$ and $Y_L$).  We now take a
small volume of fluid and adiabatically displace it by a small $\Delta r>0$,
that is, we keep $s$ and $Y_L$ constant in the small displaced volume. The
pressure of the small volume $P_V$ will be equal to the pressure of the
surrounding matter, $P_V\equiv P(r+\Delta r)=P(r) + \Delta P$, with $\Delta
P<0$ (since $\Delta r>0$). However, since we have displaced it adiabatically,
its energy density will differ from that of the surrounding matter,
\begin{equation}
\epsilon_V\equiv\epsilon(r)+\Delta P\frac{ \partial \epsilon}{\partial
P}>\epsilon(r+\Delta r) \equiv \epsilon(r)+\Delta P \frac{\mathrm
d\epsilon}{\mathrm d P}.
\end{equation}
Since the small displaced volume is \emph{heavier} than the surrounding matter,
it will be pushed down by the Archimedes force (i.e., by buoyancy), its initial
position will be restored, and the perturbation is stable against convection.
Viceversa, if $S(r)<0$, the perturbation is unstable against convection. Off
course, if there are no entropy nor composition gradients along the stellar
profile, $\partial P/\partial \epsilon \equiv \mathrm d P/\mathrm d \epsilon$
and $S(r)\equiv 0$.
\item{p-modes:} for the \emph{pressure or p- modes} the main restoring force is
the pressure gradient.  For these modes the fluid radial displacement has one
or more nodes, that is, $\xi^r$ changes sign at least once.  In a cold star,
the first (i.e., it has one radial node) p-mode has frequencies
$\nu_{p_1}\simeq 5$--$\unit[10]{kHz}$ or higher and damping times
$\tau_{p_1}\simeq1$--$\unit[10]{s}$. The other p-modes have higher frequencies,
$\nu_f<\nu_{p_1}<\nu_{p_2}<\ldots$.
\item{f-mode:} is intermediate between the g- and the p- modes; the solution of
the perturbation equations has no radial nodes (that is, $\xi^r$ and therefore
$W(r)$ never change sign at fixed time). It is called \emph{fundamental mode}
or \emph{f-mode}, and its main restoring force is the pressure. In fact, it
is actually a pressure mode with no radial nodes, and therefore some papers refer to it
with the symbol $p_0$. It corresponds to an expansion/contraction of the entire
star, with frequency $\nu_f$. In a cold star, $\nu_f$ is about
$1.5$--$\unit[2.5]{kHz}$ and its damping time $\tau_f\simeq\unit[0.1]{s}$.
As already mentioned, the frequency of the f-mode is intermediate between those of
the g- and p-modes, $\ldots<\nu_{g_1}<\nu_f<\nu_{p_1}<\ldots$\,\,.
\item{w-modes:}  the so-called \emph{gravitational-wave} or \emph{w-} modes are
very weakly coupled to fluid motions, and are therefore (almost) pure spacetime
modes.  w-modes have at least one radial node, and the number of nodes allows
to label them. In a cold star, the first w-mode has a frequency around
$\nu_{w_1}\simeq 8$--$\unit[10]{kHz}$ and damping times
$\tau_{w_1}\simeq10^{-5}$--$\unit[10^{-4}]{s}$.  These modes, that exist also
for the odd-parity perturbations (see discussion above), produce a negligible
fluid displacement, and therefore we will not consider them in the following
discussion.
\end{itemize}

\section{Results}
\label{sec:GW.results}
In this section we describe the code which finds the PNS quasi-normal modes.
Moreover, we shall discuss and compare the time evolution of the QNM
frequencies of a PNS, consistently evolved with the three EoSs described in
Chapter~\ref{cha:EoS} and with different stellar masses.

To determine the quasi-normal mode frequencies at a given time $t$ of the
stellar evolution, we have first evolved the PNS as explained in
Chapter~\ref{cha:evolution}, and saved the stellar profiles at chosen values of
time $t$, in particular the pressure $P(r,t)$, the energy density
$\epsilon(r,t)$, the baryon density $\nB(r,t)$, and the square of the sound
speed, $c^2_s(r,t)$. Then, we have determined the
effective EoS at a given time: we have first inverted the pressure monotonic
dependence on the radius, $r= r(P,t)$, and then we have determined the other
thermodynamical quantities by means of this relation,
$\bar\epsilon_t(P)=\epsilon(r(P,t),t)\rightarrow\epsilon(P)$, and similarly for
$\nB(P)$ and $c_s^2(P)$.  We have then fixed a value for the complex frequency
$\omega$ and solved the perturbation
equations~\eqref{eq:osc_start}--\eqref{eq:osc_end}, together with the TOV
equations to determine the background metric, using as independent variable the
logarithm of the pressure (the stellar quantities depend more smoothly on the
logarithm of the pressure than on the radius).  Then, at the stellar surface we
construct the initial values of the quantities needed to integrate the
perturbation equations outside the star and apply the continued fraction method
to determine whether the condition $A_+=0$ holds.  We have then varied the
complex frequency $\omega$ with a Newton-Raphson cycle until the condition
$A_+=0$ is satisfied.  The complex value of the frequency for which this
condition is satisfied is a quasi-normal mode frequency, and from it we obtain
the pulsation frequency (real part) and the damping time (imaginary part) [see
Eq.~\ref{eq:omega_complex}].

\begin{figure}
\centering
\includegraphics[width=0.8\textwidth]{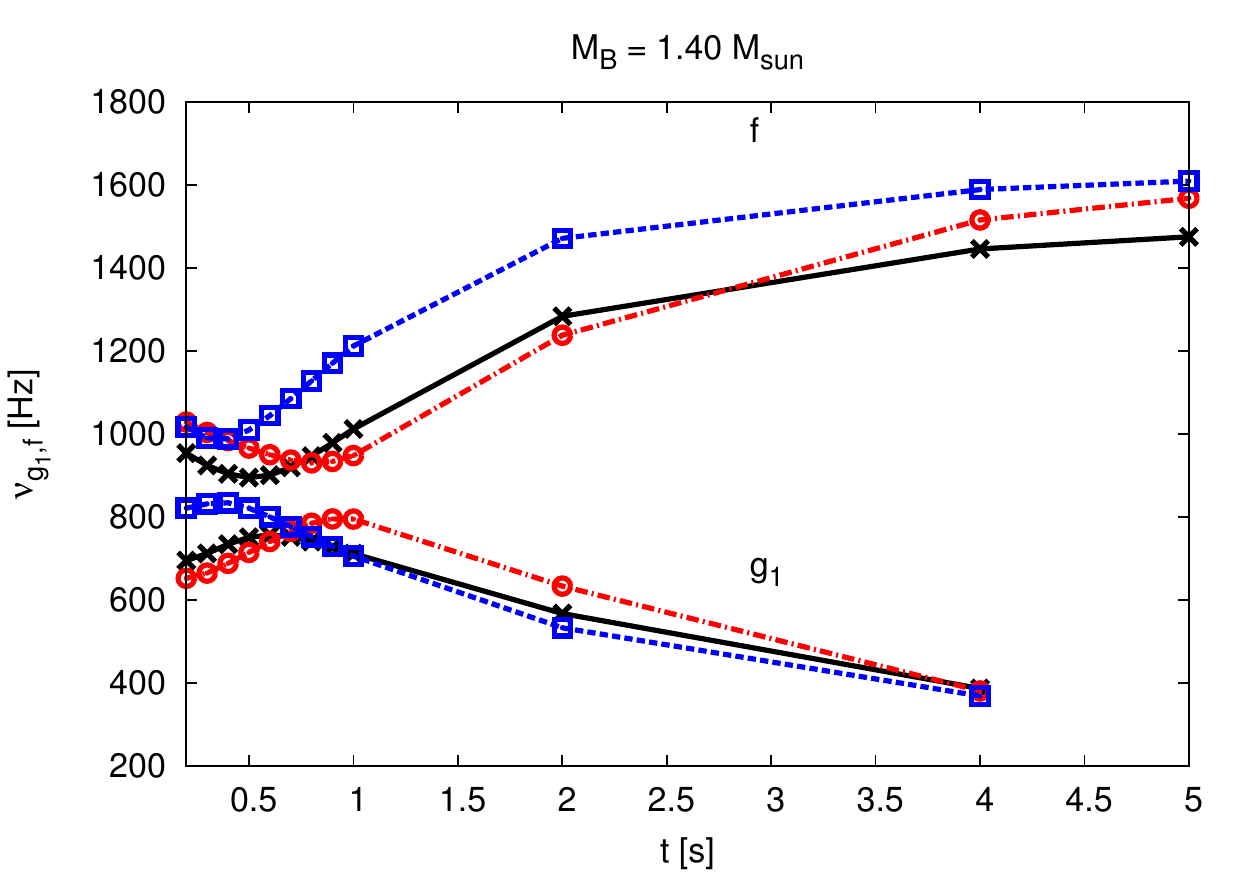}
\includegraphics[width=0.8\textwidth]{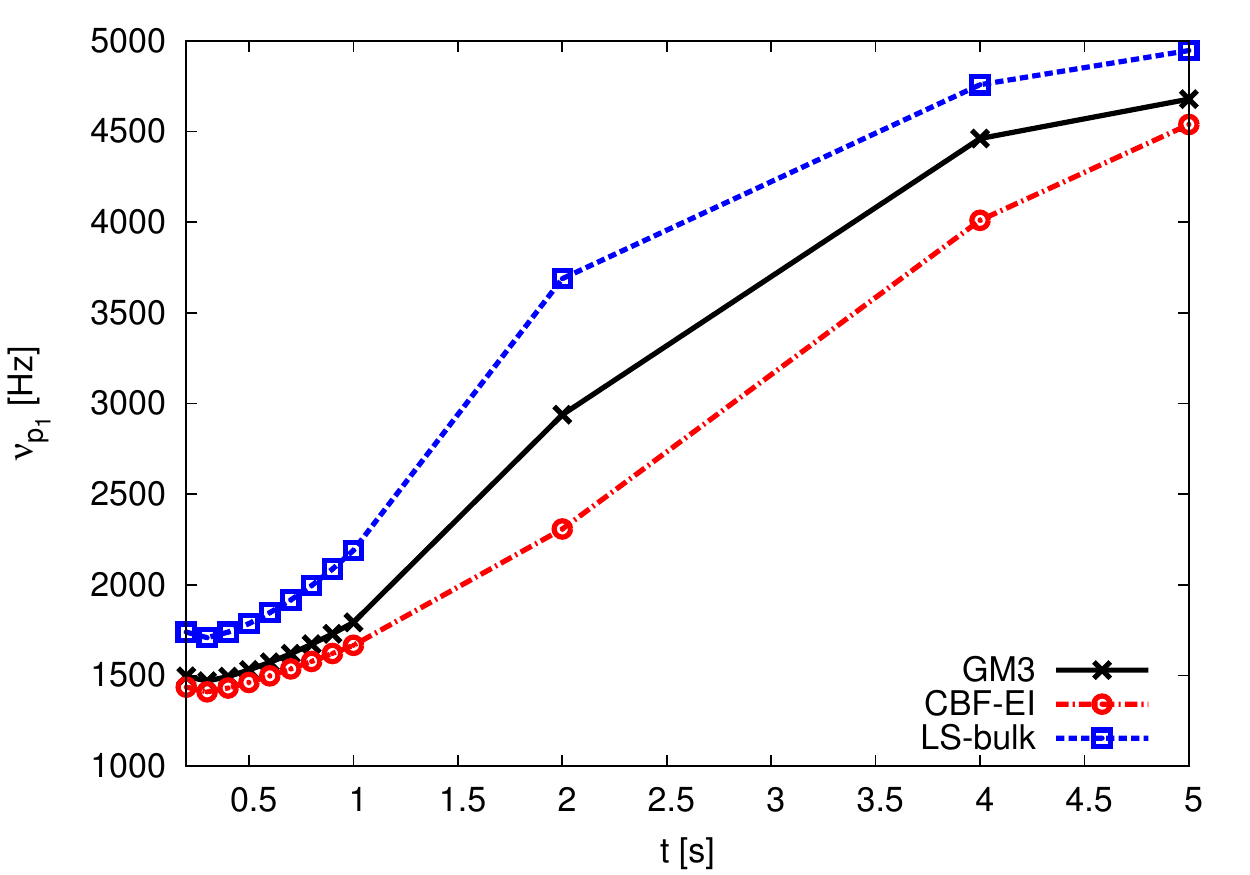}
\includegraphics[width=0.8\textwidth]{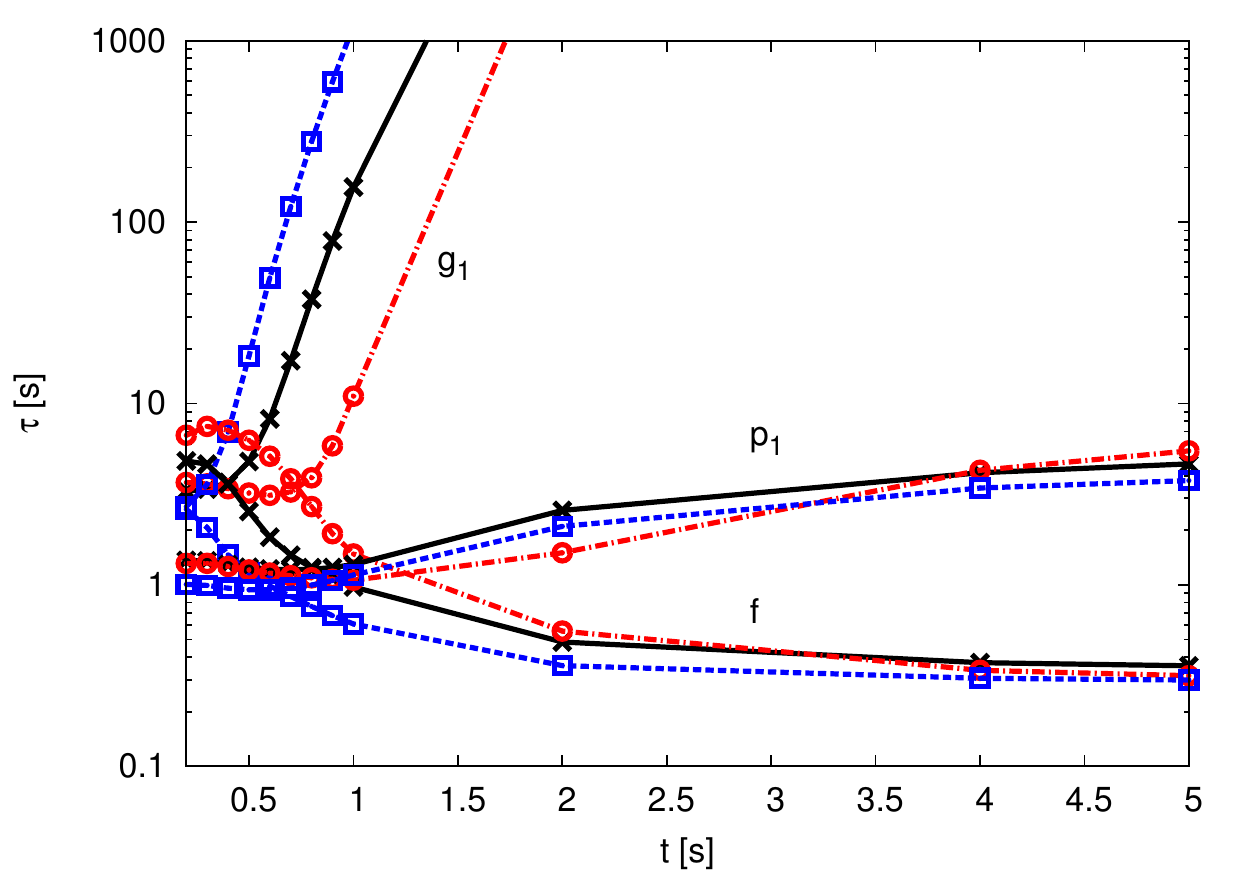}
\caption{Time dependence of the PNS quasi-normal mode frequencies and damping times for the three EoSs
and for $M_\mathrm{B}=\unit[1.40]{M_\odot}$.}
\label{fig:QNM_1.40}
\end{figure}

In Tabs.~\ref{tab:GM3_1.25}--\ref{tab:LSbulk_1.60} we report the QNM frequencies and damping times during the evolution
for the cases considered. In Fig.~\ref{fig:QNM_1.40} we show the time dependence of the QNM frequencies
and damping times for the three EoSs and for the $M_\mathrm{B}=\unit[1.40]{M_\odot}$ star.
Irrespective to the EoS considered, our results are in qualitative agreement
with those of \citet{Ferrari+Miniutti+Pons.2003} and \citet{Burgio+2011}.  The
$g_1$, $f$, and $p_1$ modes have initial frequencies clustered around 1 kHz,
and after the first $\sim \unit[1]{s}$ they rapidly migrate toward their ending
values (those of the cold star). During the first second, the $g_1$ frequency
approaches the $f$ one, but they do never cross and are always distinguishable
(since they have different number of radial nodes of the perturbation
functions). The $g_1$ frequency increases, reaches a
maximum before $\unit[1]{s}$, and then decreases, whereas the $f$ behaviour is
opposite (the $f$ frequency reaching its minimum slightly before the $g_1$
reaches its maximum).

In order to be competitive
in extracting energy from the PNS \citep{Ferrari+Miniutti+Pons.2003},
the timescales of the gravitational wave emission (i.e., the damping times)
should be smaller than the evolutionary timescale. This
means that the QNM damping times should be smaller than $\sim \unit[10]{s}$.
Moreover, the smaller is the QNM damping time, the greater is the energy
radiated. In fact, the amplitude of a QNM with frequency $\nu$ and damping time
$\tau$ is \citep{Ferrari+Miniutti+Pons.2003}
\begin{equation}
h(t)=h_0\mathrm e^{-(t-t_0)/\tau}\sin[2\pi\nu(t-t_0)],
\end{equation}
where $h_0$ is the initial amplitude and $t_0$ the initial time.  Then, since
the QNM energy is proportional to the square of the amplitude,
$E_\mathrm{QNM}\propto h^2$, $E_\mathrm{QNM}\propto\mathrm e^{-2(t-t_0)/\tau}$,
and the GW luminosity is \citep{Burgio+2011}
\begin{equation}
L_\mathrm{GW}=-\dot E_\mathrm{QNM}\simeq \frac{2E_\mathrm{QNM}}{\tau}.
\end{equation}
Therefore, the QNM with smaller damping times are more effective in extracting
energy from the PNS.  In a cold star, the damping times of the $f$ mode are
generally smaller than those of the $p_1$ mode, typical values being $\tau_f\simeq\unit[0.1]{s}$ and
$\tau_{p_1}\simeq\unit[5]{s}$. This means that
in a cold star the $f$ mode is the most effective in extracting energy.  During the first second of the PNS
life, the situation is completely different: the $p_1$ mode has damping times
$\tau_{p_1}\simeq\unit[1]{s}$ \emph{smaller} than that of the $f$ mode
$\tau_f\simeq\unit[10]{s}$ and it could be \emph{more effective} in extracting
energy than the fundamental mode. Moreover, during the first second, the $g_1$
mode has a damping time $\tau_{g_1}\simeq \unit[10]{s}$ that is \emph{smaller
or comparable} to the timescales of the neutrino diffusive processes, and may
therefore be excited. At later times the damping times rapidly become those of
a cold NS.

Having a damping time shorter than the evolutionary timescale (in our case,
tens of seconds) is a necessary condition to detect the QNMs, but it is not
enough\footnote{
Recent 3-D simulations of the early explosion phase of
core-collapse supernovae and of the following accretion
phase~\cite{Andresen:2016pdt,Kuroda:2016bjd}
show that other phenomena than stellar oscillations may
contribute to gravitational wave emission; for instance,  standing accretion shock
instability and convection, which are shown to be associated to stochastic oscillations,
and to unstable $g$-modes, different from the stable $g$-modes considered in this thesis.
For a review see also~\cite{Kotake:2011yv}.
}.  In fact, the QNM detectability strongly depends on how much energy
goes in a specific mode. We cannot determine this value because our
evolutionary code is 1D and therefore we have to resort to estimates or to the
results of 2D or 3D core-collapse codes. A SN event is expected to radiate
$\sim 10^{-8}$--$\unit[10^{-7}]{M_\odot}$ in GW emission
\citep{Dimmelmeier+Font+Muller.2002}, and therefore we expect that a QNM can be
detected only by a galactic supernova \citep{Andersson+2011, Andresen+2016}.

\citet{Ferrari+Miniutti+Pons.2003} studied only a PNS with baryon mass
$M_\mathrm{B}=\unit[1.60]{M_\odot}$ and two EoSs, the mean-field GM3 EoS and a
hadron-quark EoS; \citet{Burgio+2011} studied a PNS with
$M_\mathrm{B}=\unit[1.50]{M_\odot}$ and the many-body EoS of
\citet{Burgio+Schulze.2010}.  We have considered three stellar baryon masses
(1.25, 1.40, and 1.60 $\unit{M_\odot}$) and the three nucleonic EoSs described
in Chapter~\ref{cha:EoS}, between which the GM3 EoS.  Our results for the
$\unit[1.60]{M_\odot}$ star with the GM3 EoS are in agreement with the results
for the equivalent ``model A'' of \citet{Ferrari+Miniutti+Pons.2003}. In
particular, the $f$ and $p_1$ initial and final frequencies are in quantitative
agreement within 2\%; the $g_1$ initial, maximum, and final frequencies and the
$f$ minimum frequency are in agreement within $\sim10\%$.  The time of the
$g_1$ maximum / $f$ minimum is shifted by about $0.4$/$0.3$ s, and the
maximum/minimum are more accentuated in our case.  The QNM damping times are in
reasonable agreement with those of \citet{Ferrari+Miniutti+Pons.2003} (note
that in \citealp{Ferrari+Miniutti+Pons.2003} there is a typo, the damping times
are $2\pi$ times the true ones).  We think that the quantitative differences
between our results and those of \citet{Ferrari+Miniutti+Pons.2003} are due to
some differences in the initial profiles and in the details of the treatment of
the diffusive processes.

It may be noted that our results are qualitatively different from those
of~\cite{Burgio+2011} and~\cite{Sotani:2016uwn}, where the QNMs show a monotonic
increase of the $f$- and $p$-modes, and a monotonic decrease of the $g$-mode.
We think that this is due to the fact that a consistent evolution of the
PNS is  crucial to describe the behaviour of the QNMs.

\begin{figure}
\centering
\includegraphics[width=\textwidth]{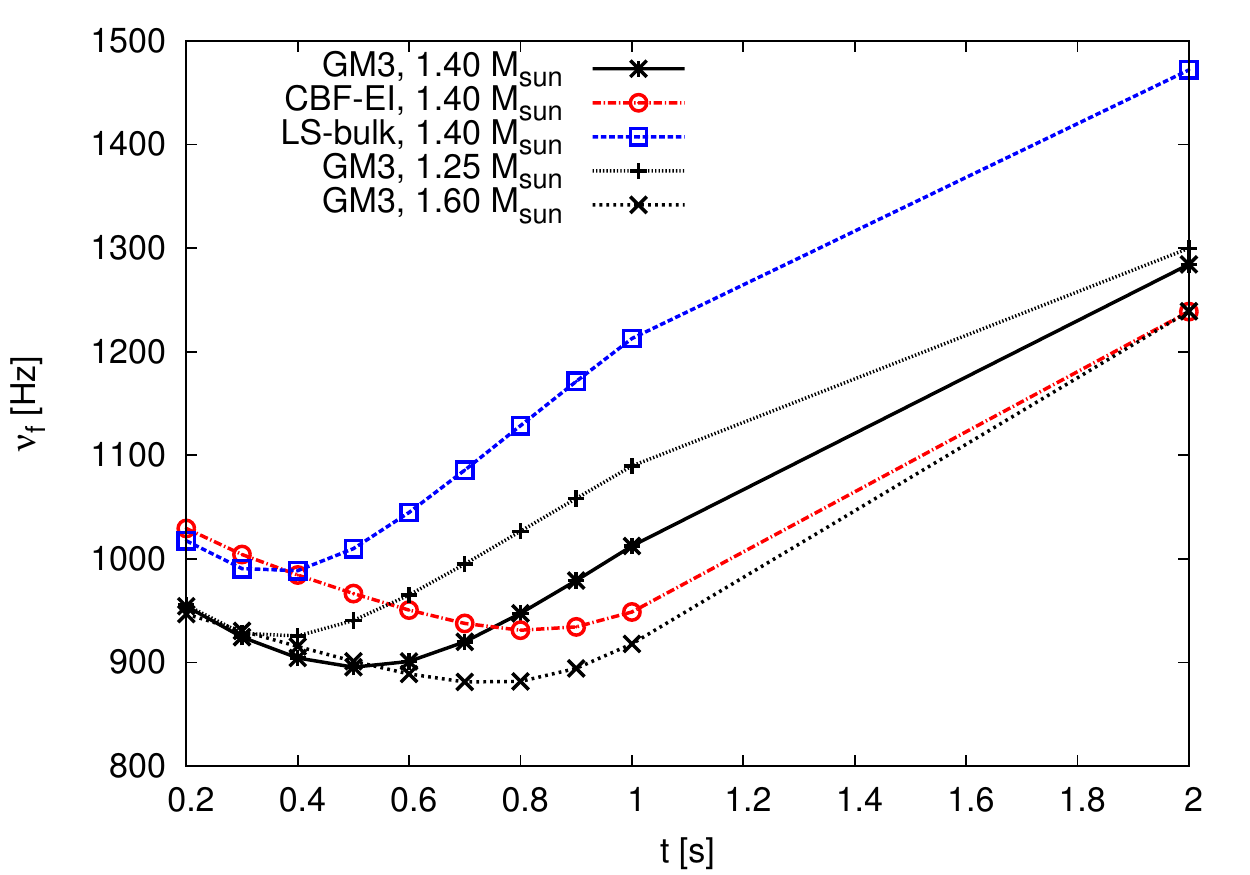}
\includegraphics[width=\textwidth]{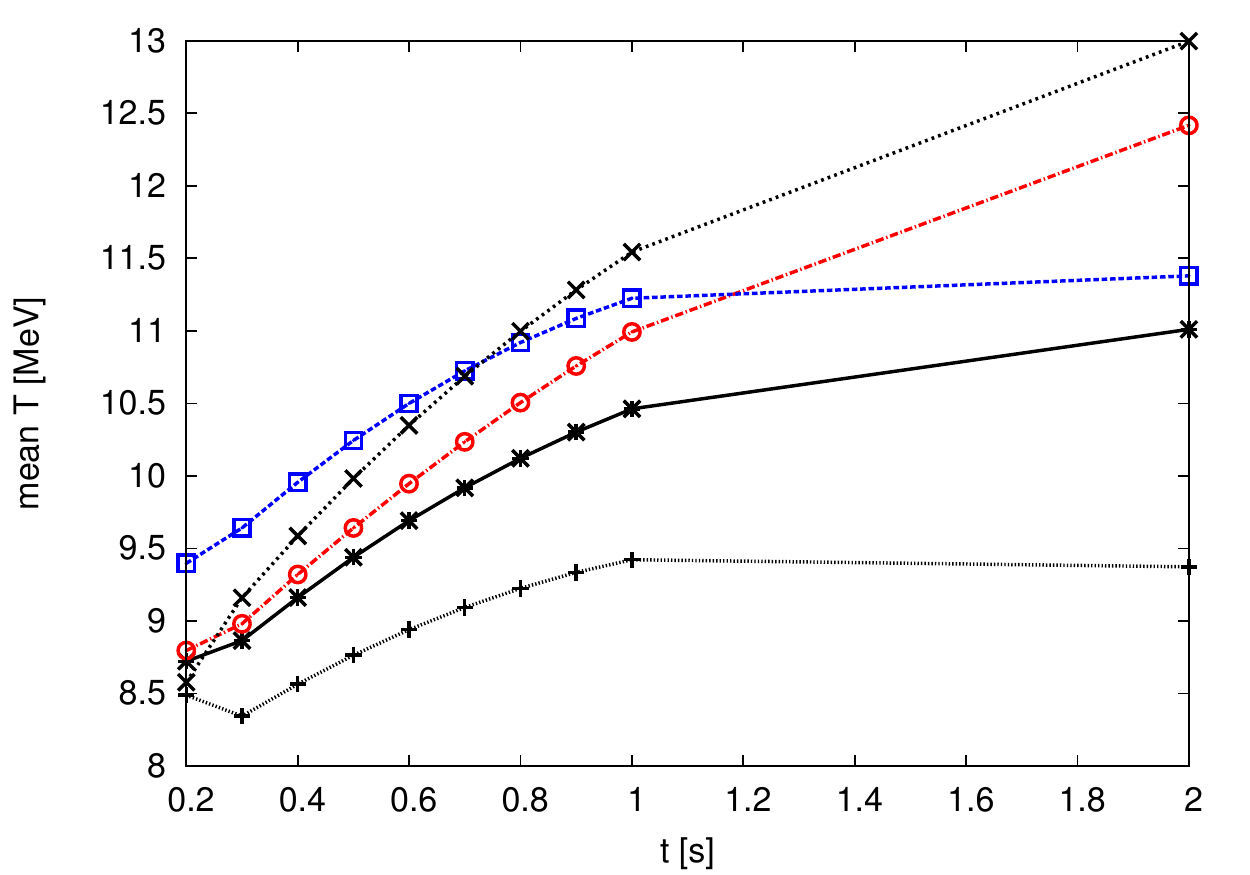}
\caption{Fundamental mode frequency and mean stellar temperature for some stellar configurations from $\unit[0.2]{s}$ to $\unit[2]{s}$.}
\label{fig:QNM_2s}
\end{figure}

\begin{figure}
\centering
\includegraphics[width=0.8\textwidth]{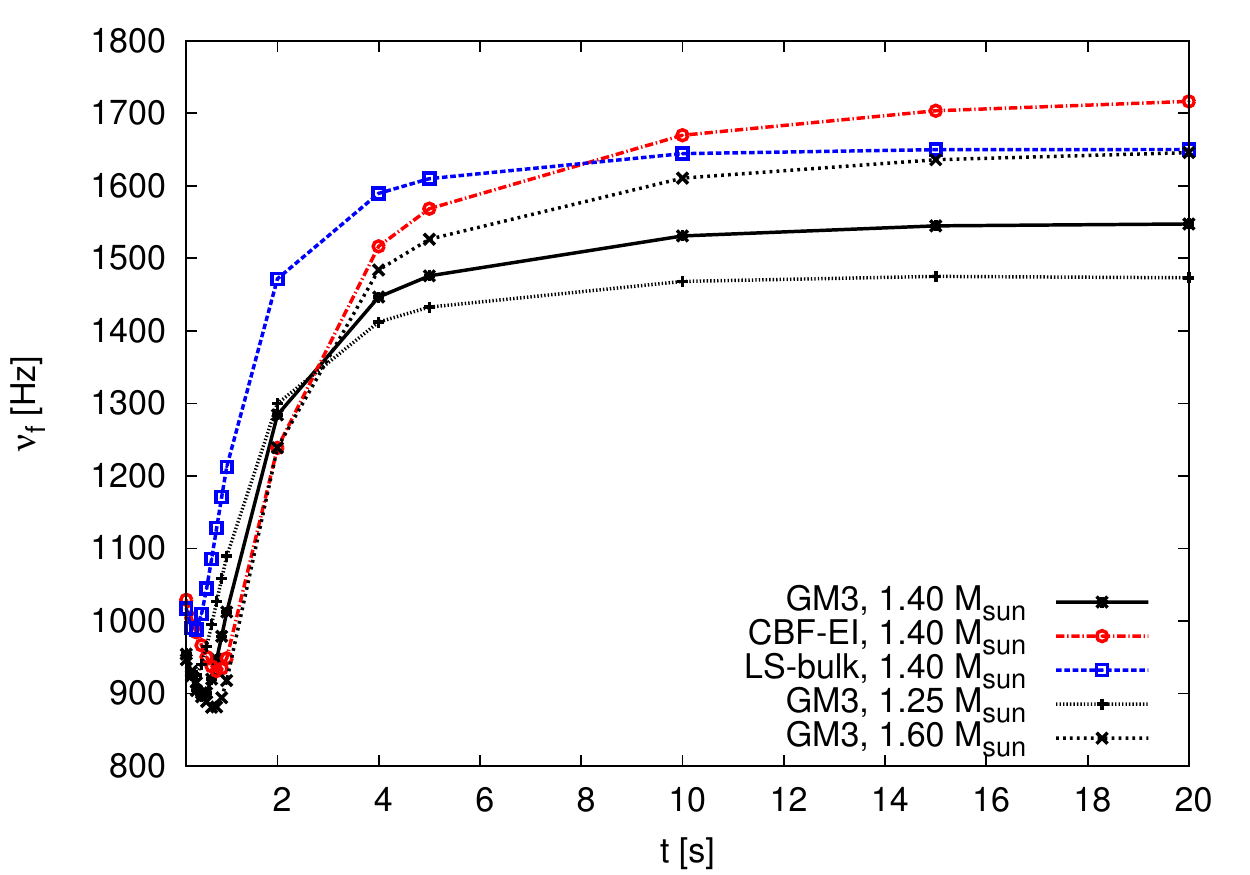}
\includegraphics[width=0.8\textwidth]{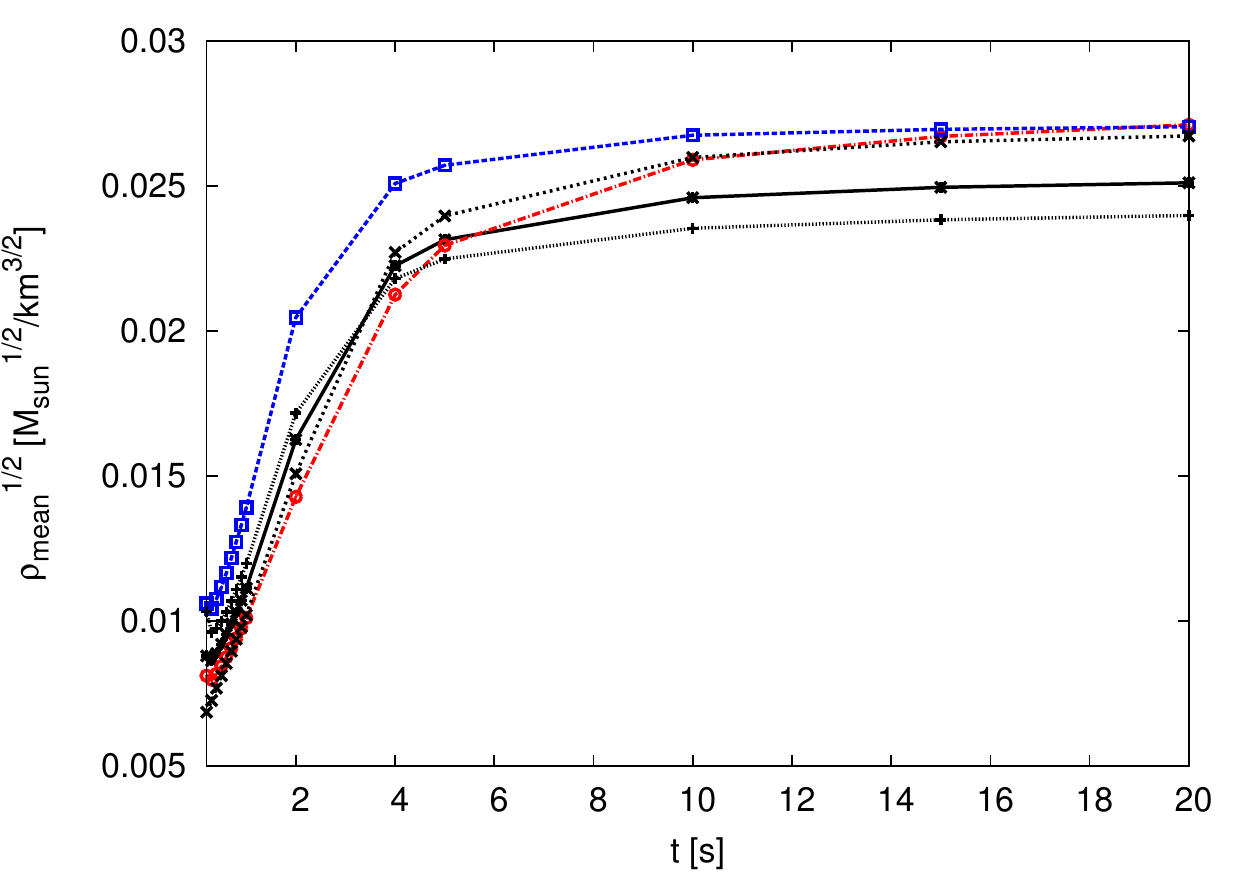}
\includegraphics[width=0.8\textwidth]{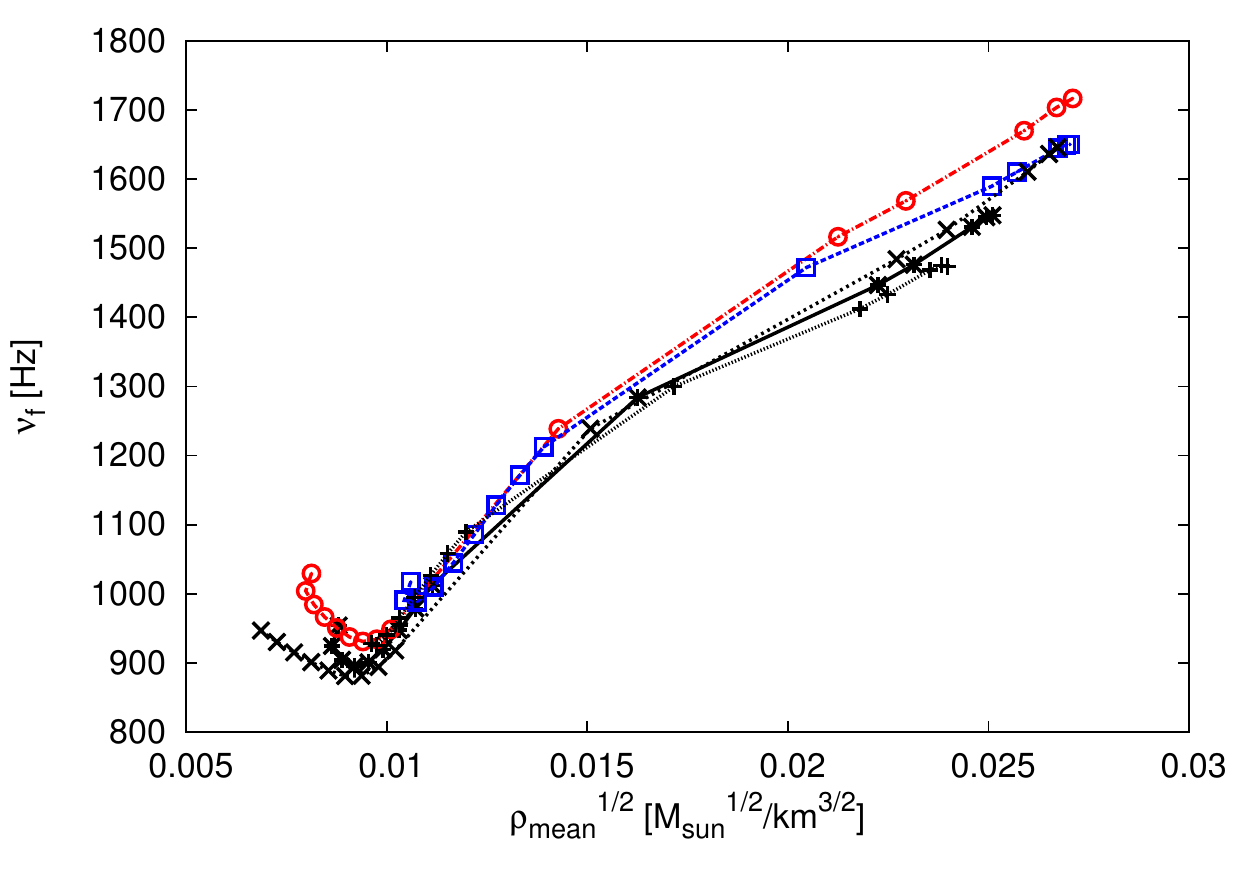}
\caption{In the top plot we show the fundamental frequency evolution, in the middle plot the root square of the mean density $\sqrt{M/R^3}$
evolution, and in the bottom plot the fundamental density vs the root square of
the mean density (the evolutionary tracks begin at the bottom left of the plot and end
at $t=\unit[20]{s}$ at the top right).}
\label{fig:QNM_20s}
\end{figure}

\begin{figure}
\centering
\includegraphics[width=\textwidth]{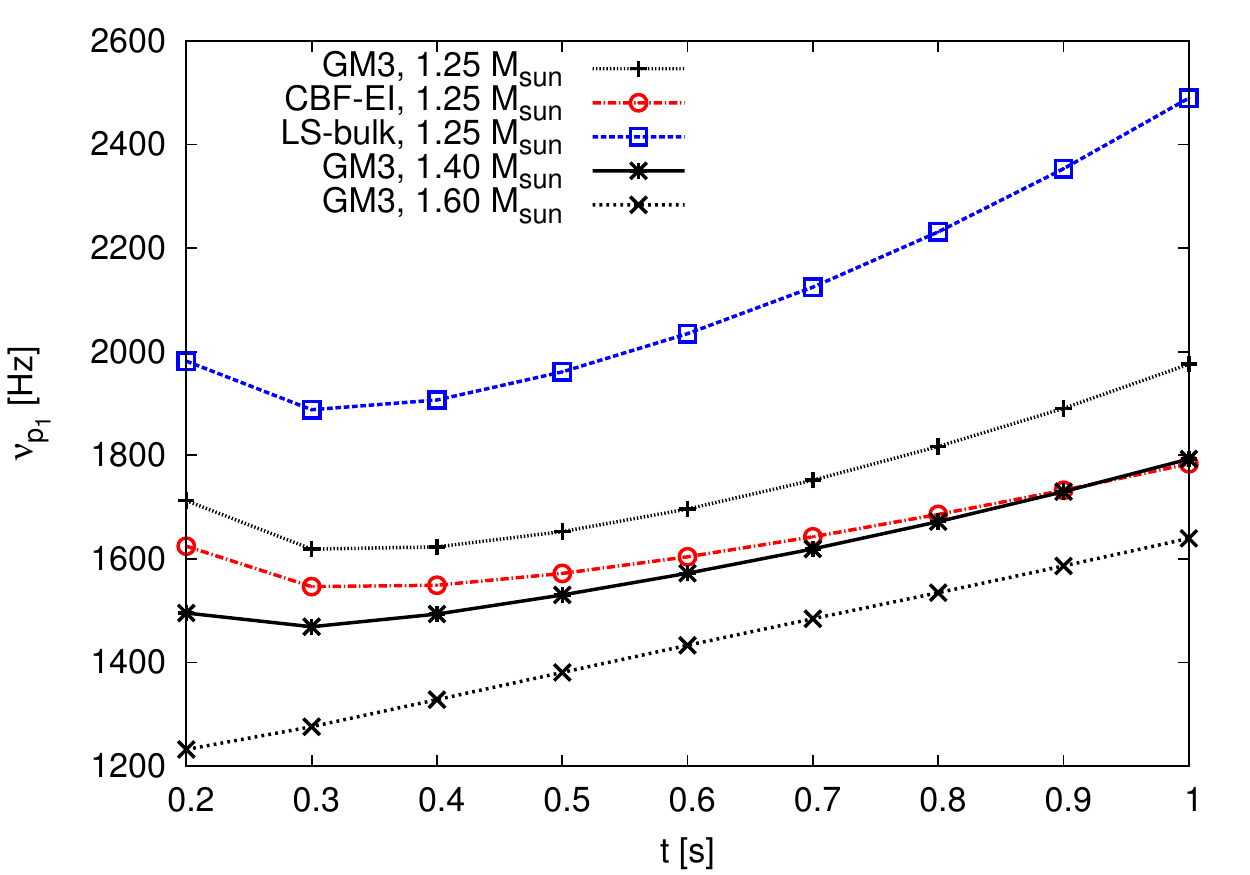}
\caption{First pressure mode frequency for some stellar configurations from $\unit[0.2]{s}$ to $\unit[1]{s}$.}
\label{fig:QNM_1s}
\end{figure}

\begin{figure}
\centering
\includegraphics[width=\columnwidth]{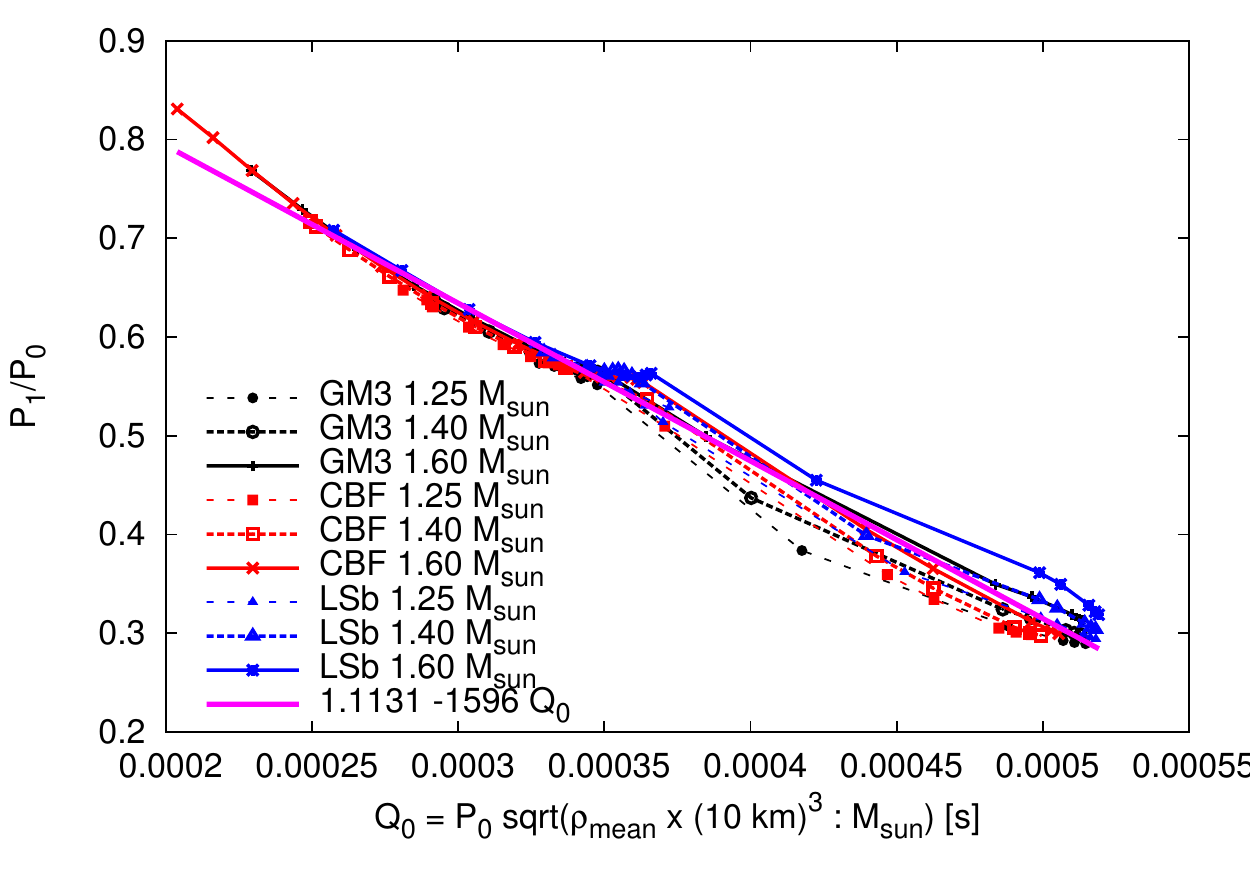}
\caption{Ratio between the first p-mode period and the fundamental period as a
function of the quantity $Q_0$, for a PNS in the configurations studied in this
thesis. The points at the top-left correspond to late evolutionary times, whereas
the initial configurations are in the bottom-right.}
\label{fig:P1oP0vsQ0}
\end{figure}

In Fig.~\ref{fig:QNM_1.40} we plot the evolution of the QNM
frequencies and damping times for a star with baryon mass $\unit[1.40]{M_\odot}$.
In Figs.~\ref{fig:QNM_2s}--\ref{fig:QNM_1s} we show the evolution
of the fundamental frequency, the mean stellar temperature, the square root of the
mean stellar density, and the $p_1$-mode frequency, for some configurations.
In Figs.~\ref{fig:QNM_20s} and \ref{fig:P1oP0vsQ0} we show the dependence of the
QNM frequencies on some global properties of the star.
In Tables~\ref{tab:GM3_1.25}--\ref{tab:LSbulk_1.60} we report the QNM frequencies and
damping times, and the stellar gravitational masses and radii for all the configurations
considered and at different time snapshots.
We remark the following
interesting features.
\begin{itemize}
\item The QNM frequencies show a signature of the underlying EoS and of the
stellar baryon mass, even though the frequencies are not dramatically different,
despite of the different theories that have been adopted to obtain a given EoS.  For a
given time, the QNM frequencies of the different configurations may vary as much as
$\sim 100$--$\unit[200]{Hz}$ according to the underlying EoS and total stellar
mass, see Fig.~\ref{fig:QNM_1.40} and Tabs.~\ref{tab:GM3_1.25}--\ref{tab:LSbulk_1.60}.
\item After some seconds, the frequency of the fundamental mode scales with the
square root of the mean stellar density (Fig.~\ref{fig:QNM_20s}), but during the first second the
fundamental mode behaviour deviates from the expected scaling, Fig.~\ref{fig:QNM_2s}.  In this initial
period, the PNS temperature is very high, and this is probably the cause of the
discrepancy. However, we have not succeed in finding a clear dependence of the
fundamental mode frequency with the stellar thermal content. We will return on
this problem in future, since the scaling of the QNM
frequencies with the stellar properties is important to better understand the
gravitational wave emission in the PNS phase.
\item In non-relativistic variable stars (as Chepheids), the ratio of the periods
$P_1/P_0=\nu_f/\nu_{p_1}$ of the first overtone (that corresponds in the
language of stellar oscillations in GR to the first p-mode) and of the
fundamental mode is a function of the quantity $Q_0=P_0\sqrt{\bar
\rho/\rho_\odot}$, where $P_0=\nu_f^{-1}$ and
$\rho_\odot=\unit[2.97\times10^{-18}]{M_\odot/km^3}$ is the mean Sun density
(see e.g.~\citealp{Christy.1966}). We have fitted the ratio
$P_1/P_0\equiv\nu_f/\nu_{p_1}$ with a linear dependence on
$Q_0\propto\sqrt{\bar\rho}/\nu_f$, obtaining
\begin{equation}
\label{eq:fitGW}
\frac{P_1}{P_0}=1.1131(\pm0.0066) - 1596(\pm17)\frac{P_0}{\unit[1]{s}}\sqrt{\bar
\rho \frac{\unit[10^3]{km}}{\unit[1]{M_\odot}}}.
\end{equation}
The result of the fit is shown in Fig.~\ref{fig:P1oP0vsQ0}; the corresponding
reduced chi-square is rather low: $\tilde\chi=3.2\times10^{-4}$. This is an
indication that, even in PNSs, the ratio $P_1/P_0$ could be an universal
property, independent of the masses and EoSs of the PNS.
\item At $\unit[20]{s}$, the QNM frequencies of the PNS with the LS-bulk and GM3 EoSs have
already reached the cold star values; instead that of the PNS with the CBF-EI EoS has not converged yet,
see the upper plot of Fig.~\ref{fig:QNM_20s}.
This is due to the fact that at those times the CBF-EI PNS configuration has not yet
relaxed to the cold NS, see the middle plot of Fig.~\ref{fig:QNM_20s}.
\item Lowering the baryonic mass, the qualitative
evolution of the $p$ mode changes (see Fig.~\ref{fig:QNM_1s}): in the $\unit[1.25]{M_\odot}$ star the
frequency of the $p_1$ mode initially decreases, reaches a minimum, and then
increases, in a way qualitatively similar to the behaviour of the $f$ mode. This feature is
present for all EoSs considered, and it is present also in the
$\unit[1.40]{M_\odot}$ star, where however it is less pronounced.
\end{itemize}

\begin{table}
\caption{QNMs for a $M_\mathrm{B}=\unit[1.25]{M_\odot}$ star evolved with the GM3 EoS.
The column content, from left to right, is: time of the snapshot (in s), frequency (in Hz) and
damping time (in s)
of the $g_1$, $f$, and $p_1$ modes, stellar gravitational mass (in $M_\odot$), and stellar radius (in km).}
\label{tab:GM3_1.25}
\centering
\begin{tabular}{ccccccccc}
$t$ & $\nu_{g_1}$ & $\tau_{g_1}$ & $\nu_f$ & $\tau_f$ & $\nu_{p_1}$ & $\tau_{p_1}$ & $M$ & $R$ \\
\hline
0.2 & 784.1 & 4.01 & 955.3 & 2.49 & 1712. & 1.57 & 1.2064 & 22.444 \\
0.3 & 776.6 & 5.91 & 928.2 & 2.18 & 1619. & 1.59 & 1.2051 & 23.530 \\
0.4 & 768.0 & 11.5 & 925.7 & 1.75 & 1623. & 1.56 & 1.2036 & 23.320 \\
0.5 & 752.9 & 25.9 & 940.5 & 1.48 & 1653. & 1.55 & 1.2021 & 22.917 \\
0.6 & 734.4 & 58.2 & 965.2 & 1.31 & 1696. & 1.57 & 1.2006 & 22.430 \\
0.7 & 715.1 & 123. & 995.0 & 1.17 & 1752. & 1.62 & 1.1992 & 21.894 \\
0.8 & 695.8 & 247. & 1027. & 1.05 & 1817. & 1.71 & 1.1977 & 21.349 \\
0.9 & 676.7 & 472. & 1058. & .952 & 1890. & 1.83 & 1.1962 & 20.812 \\
1.0 & 658.2 & 873. & 1090. & .869 & 1976. & 1.98 & 1.1948 & 20.272 \\
2.0 & 504.3 & $1.5\times10^5$ & 1300. & .534 & 3386. & 3.96 & 1.1828 & 15.893 \\
4.0 & 326.8 & $2.8\times10^5$ & 1412. & .446 & 4593. & 5.96 & 1.1697 & 13.502 \\
5.0 & -     & -    & 1433. & .434 & 4756. & 6.46 & 1.1660 & 13.215 \\
10. & -     & -    & 1468. & .415 & 5017. & 6.77 & 1.1562 & 12.778 \\
15. & -     & -    & 1475. & .413 & 5074. & 6.66 & 1.1520 & 12.659 \\
20. & -     & -    & 1473. & .415 & 5091. & 6.88 & 1.1502 & 12.600 \\
\hline
\end{tabular}
\end{table}
\begin{table}
\caption{As Tab.~\ref{tab:GM3_1.25}, for a $M_\mathrm{B}=\unit[1.25]{M_\odot}$ star with the CBF-EI EoS.}
\label{tab:CBFEI_1.25}
\centering
\begin{tabular}{ccccccccc}
$t$ & $\nu_{g_1}$ & $\tau_{g_1}$ & $\nu_f$ & $\tau_f$ & $\nu_{p_1}$ & $\tau_{p_1}$ & $M$ & $R$ \\
\hline
0.2 & 756.2 & 3.13 & 1036. & 4.41 & 1625. & 1.36 &1.2118 & 23.807 \\
0.3 & 753.2 & 3.35 & 1001. & 4.94 & 1546. & 1.38 &1.2102 & 24.808 \\
0.4 & 767.0 & 3.47 & 976.8 & 4.45 & 1549. & 1.34 &1.2085 & 24.628 \\
0.5 & 783.5 & 3.82 & 959.2 & 3.53 & 1572. & 1.30 &1.2068 & 24.238 \\
0.6 & 796.5 & 4.88 & 950.3 & 2.60 & 1604. & 1.27 &1.2052 & 23.765 \\
0.7 & 800.8 & 7.85 & 952.9 & 1.93 & 1643. & 1.25 &1.2035 & 23.251 \\
0.8 & 795.4 & 15.4 & 967.7 & 1.55 & 1686. & 1.25 &1.2019 & 22.723 \\
0.9 & 782.6 & 32.2 & 990.7 & 1.33 & 1732. & 1.26 &1.2004 & 22.200 \\
1.0 & 766.1 & 66.4 & 1019. & 1.18 & 1783. & 1.28 &1.1988 & 21.679 \\
2.0 & 568.0 & $2.5\times10^4$ & 1296. & .556 & 2542. & 2.32 & 1.1856 & 17.255 \\
4.0 & 377.7 & $3.7\times10^5$ & 1512. & .385 & 4206. & 6.46 & 1.1695 & 13.684 \\
5.0 & - & - & 1553. & .365 & 4643. & 7.75 &1.1647&13.113 \\
10. & - & - & 1635. & .330 & 5355. & 8.57 &1.1535&12.240 \\
15. & - & - & 1659. & .321 & 5507. & 8.04 &1.1499&12.012 \\
20. & - & - & 1664. & .319 & 5568. & 8.00 &1.1492&11.903 \\
\end{tabular}
\end{table}
\begin{table}
\caption{As Tab.~\ref{tab:GM3_1.25}, for a $M_\mathrm{B}=\unit[1.25]{M_\odot}$ star with the LS-bulk EoS.}
\label{tab:LSbulk_1.25}
\centering
\begin{tabular}{ccccccccc}
$t$ & $\nu_{g_1}$ & $\tau_{g_1}$ & $\nu_f$ & $\tau_f$ & $\nu_{p_1}$ & $\tau_{p_1}$ & $M$ & $R$ \\
\hline
0.2 & 870.3 & 8.57 & 1047. & 1.21 & 1982. & 1.23 &1.2020 & 19.925 \\
0.3 & 839.0 & 22.0 & 1045. & 1.12 & 1888. & 1.27 &1.2006 & 20.780 \\
0.4 & 805.6 & 62.2 & 1068. & 1.01 & 1907. & 1.27 &1.1992 & 20.506 \\
0.5 & 772.9 & 161. & 1102. & .911 & 1961. & 1.30 &1.1976 & 20.047 \\
0.6 & 742.2 & 380. & 1139. & .821 & 2035. & 1.37 &1.1961 & 19.533 \\
0.7 & 713.6 & 829. & 1176. & .743 & 2125. & 1.46 &1.1946 & 19.003 \\
0.8 & 686.9 & $1.7\times10^3$ & 1213. & .676 & 2231. & 1.59 &1.1931 & 18.467\\
0.9 & 662.0 & $3.4\times10^3$ & 1248. & .622 & 2353. & 1.74 &1.1917 & 17.948\\
1.0 & 638.7 & $6.5\times10^3$ & 1279. & .579 & 2490. & 1.92 &1.1903 & 17.446\\
2.0 & 473.5 & $5.9\times10^6$ & 1457. & .415 & 4027. & 3.22 &1.1789 & 13.939\\
4.0 & 321.8 & $3.3\times10^5$ & 1538. & .372 & 4864. & 4.89 &1.1673 & 12.557\\
5.0 & 278.9 & $5.9\times10^5$ & 1552. & .366 & 5010. & 5.28 &1.1638 & 12.378\\
10. & - & - & 1572. & .359 & 5267. & 6.01 & 1.1540 & 12.091 \\
15. & - & - & 1575. & .359 & 5328. & 6.15 & 1.1493 & 12.025 \\
20. & - & - & 1575. & .360 & 5352. & 6.22 & 1.1465 & 11.986 \\
\end{tabular}
\end{table}
\begin{table}
\caption{As Tab.~\ref{tab:GM3_1.25}, for a $M_\mathrm{B}=\unit[1.40]{M_\odot}$ star with the GM3 EoS.}
\label{tab:GM3_1.40}
\centering
\begin{tabular}{ccccccccc}
$t$ & $\nu_{g_1}$ & $\tau_{g_1}$ & $\nu_f$ & $\tau_f$ & $\nu_{p_1}$ & $\tau_{p_1}$ & $M$ & $R$ \\
\hline
0.2 & 695.5 &       3.28 & 954.3 & 4.83 & 1495. & 1.36 & 1.3553 & 25.961 \\
0.3 & 712.6 &       3.38 & 924.5 & 4.62 & 1469. & 1.35 & 1.3536 & 26.289 \\
0.4 & 734.6 &       3.67 & 904.4 & 3.61 & 1494. & 1.29 & 1.3518 & 25.782 \\
0.5 & 751.3 &       4.78 & 895.6 & 2.54 & 1530. & 1.25 & 1.3500 & 25.174 \\
0.6 & 757.2 &       8.21 & 901.1 & 1.83 & 1572. & 1.22 & 1.3482 & 24.560 \\
0.7 & 752.5 &       17.2 & 919.9 & 1.46 & 1619. & 1.21 & 1.3464 & 23.937 \\
0.8 & 741.5 &       37.5 & 947.4 & 1.24 & 1672. & 1.22 & 1.3447 & 23.312 \\
0.9 & 727.4 &       78.7 & 979.1 & 1.09 & 1730. & 1.25 & 1.3429 & 22.704 \\
1.0 & 712.1 &       156. & 1012. & .972 & 1793. & 1.29 & 1.3412 & 22.119 \\
2.0 & 567.7 & $3.1\times10^4$ & 1284. & .484 & 2938. & 2.57 & 1.3260 & 17.122 \\
4.0 & 387.1 & $4.6\times10^6$ & 1447. & .372 & 4461. & 4.15 & 1.3084 & 13.827 \\
5.0 & -     &       -    & 1476. & .357 & 4678. & 4.64 & 1.3033 & 13.450 \\
10. & -     &       -    & 1531. & .335 & 5050. & 5.14 & 1.2893 & 12.872 \\
15. & -     &       -    & 1545. & .330 & 5142. & 4.98 & 1.2827 & 12.725 \\
20. & -     &       -    & 1548. & .330 & 5175. & 4.95 & 1.2791 & 12.661 \\
\end{tabular}
\end{table}
\begin{table}
\caption{As Tab.~\ref{tab:GM3_1.25}, for a $M_\mathrm{B}=\unit[1.40]{M_\odot}$ star with the CBF-EI EoS.}
\label{tab:CBFEI_1.40}
\centering
\begin{tabular}{ccccccccc}
$t$ & $\nu_{g_1}$ & $\tau_{g_1}$ & $\nu_f$ & $\tau_f$ & $\nu_{p_1}$ & $\tau_{p_1}$ & $M$ & $R$ \\
\hline
0.2 & 652.7 &            3.67 & 1029. & 6.68 & 1436. & 1.31 & 1.3613 & 27.434 \\
0.3 & 665.1 &            3.60 & 1004. & 7.48 & 1410. & 1.31 & 1.3591 & 27.740 \\
0.4 & 688.6 &            3.39 & 984.5 & 7.13 & 1430. & 1.26 & 1.3570 & 27.266 \\
0.5 & 715.0 &            3.20 & 966.5 & 6.27 & 1462. & 1.21 & 1.3550 & 26.671 \\
0.6 & 741.6 &            3.12 & 950.6 & 5.11 & 1498. & 1.16 & 1.3529 & 26.044 \\
0.7 & 766.0 &            3.25 & 937.7 & 3.84 & 1536. & 1.12 & 1.3510 & 25.421 \\
0.8 & 785.3 &            3.89 & 931.2 & 2.70 & 1578. & 1.09 & 1.3490 & 24.805 \\
0.9 & 795.7 &            5.84 & 934.4 & 1.91 & 1621. & 1.07 & 1.3471 & 24.204 \\
1.0 & 795.4 &            11.0 & 948.7 & 1.48 & 1666. & 1.06 & 1.3453 & 23.622 \\
2.0 & 633.8 & $5.4\times10^3$ & 1239. & .556 & 2308. & 1.50 & 1.3290 & 18.681 \\
4.0 & 381.0 & $3.7\times10^5$ & 1516. & .337 & 4010. & 4.28 & 1.3079 & 14.254 \\
5.0 &  -    &            -    & 1568. & .315 & 4539. & 5.48 & 1.3013 & 13.520 \\
10. &  -    &            -    & 1670. & .280 & 5459. & 7.33 & 1.2844 & 12.420 \\
15. &  -    &            -    & 1703. & .270 & 5674. & 6.92 & 1.2778 & 12.146 \\
20. &  -    &            -    & 1717. & .266 & 5761. & 6.69 & 1.2752 & 12.018 \\
\end{tabular}
\end{table}
\begin{table}
\caption{As Tab.~\ref{tab:GM3_1.25}, for a $M_\mathrm{B}=\unit[1.40]{M_\odot}$ star with the LS-bulk EoS.}
\label{tab:LSbulk_1.40}
\centering
\begin{tabular}{ccccccccc}
$t$ & $\nu_{g_1}$ & $\tau_{g_1}$ & $\nu_f$ & $\tau_f$ & $\nu_{p_1}$ & $\tau_{p_1}$ & $M$ & $R$ \\
\hline
0.2 & 821.7 &            2.70 & 1017. & 2.61 & 1739. & 1.00 & 1.3505 & 22.909 \\
0.3 & 832.1 &            3.58 & 990.7 & 2.07 & 1708. & .993 & 1.3488 & 23.135 \\
0.4 & 834.6 &            6.95 & 988.7 & 1.46 & 1739. & .958 & 1.3470 & 22.658 \\
0.5 & 821.9 &            18.2 & 1010. & 1.16 & 1786. & .939 & 1.3451 & 22.081 \\
0.6 & 800.7 &            49.2 & 1045. & .986 & 1846. & .938 & 1.3433 & 21.461 \\
0.7 & 776.9 &            122. & 1086. & .862 & 1916. & .957 & 1.3415 & 20.833 \\
0.8 & 752.7 &            278. & 1128. & .761 & 1996. & .996 & 1.3396 & 20.219 \\
0.9 & 729.1 &            595. & 1171. & .677 & 2087. & 1.05 & 1.3379 & 19.615 \\
1.0 & 706.3 & $1.2\times10^3$ & 1213. & .608 & 2190. & 1.13 & 1.3361 & 19.035 \\
2.0 & 533.6 & $2.5\times10^5$ & 1472. & .359 & 3688. & 2.10 & 1.3214 & 14.668 \\
4.0 & 368.9 & $6.3\times10^7$ & 1590. & .305 & 4757. & 3.42 & 1.3058 & 12.757 \\
5.0 & -     &            -    & 1610. & .298 & 4946. & 3.76 & 1.3012 & 12.532 \\
10. & -     &            -    & 1644. & .288 & 5299. & 4.41 & 1.2877 & 12.165 \\
15. & -     &            -    & 1650. & .288 & 5394. & 4.58 & 1.2807 & 12.082 \\
20. & -     &            -    & 1650. & .289 & 5431. & 4.64 & 1.2764 & 12.042 \\
\end{tabular}
\end{table}
\begin{table}
\caption{As Tab.~\ref{tab:GM3_1.25}, for a $M_\mathrm{B}=\unit[1.60]{M_\odot}$ star with the GM3 EoS.}
\label{tab:GM3_1.60}
\centering
\begin{tabular}{ccccccccc}
$t$ & $\nu_{g_1}$ & $\tau_{g_1}$ & $\nu_f$ & $\tau_f$ & $\nu_{p_1}$ & $\tau_{p_1}$ & $M$ & $R$ \\
\hline
0.2 & 548.5 &            4.72 & 946.8 & 10.6 & 1232. & 1.41 & 1.5571 & 32.104 \\
0.3 & 587.0 &            4.05 & 930.4 & 9.24 & 1276. & 1.32 & 1.5546 & 30.898 \\
0.4 & 626.4 &            3.53 & 915.4 & 7.84 & 1329. & 1.22 & 1.5522 & 29.722 \\
0.5 & 664.1 &            3.18 & 901.1 & 6.38 & 1381. & 1.14 & 1.5498 & 28.653 \\
0.6 & 699.0 &            3.04 & 889.0 & 4.82 & 1433. & 1.07 & 1.5475 & 27.675 \\
0.7 & 728.6 &            3.24 & 881.2 & 3.35 & 1484. & 1.02 & 1.5453 & 26.797 \\
0.8 & 749.6 &            4.27 & 881.7 & 2.22 & 1535. & .983 & 1.5431 & 25.993 \\
0.9 & 758.7 &            7.45 & 894.3 & 1.57 & 1586. & .958 & 1.5410 & 25.236 \\
1.0 & 757.3 &            15.6 & 917.9 & 1.25 & 1640. & .946 & 1.5388 & 24.525 \\
2.0 & 636.9 & $5.4\times10^3$ & 1239. & .464 & 2481. & 1.46 & 1.5195 & 18.838 \\
4.0 & 458.1 & $6.7\times10^5$ & 1484. & .305 & 4246. & 2.61 & 1.4949 & 14.259 \\
5.0 & 397.8 & $10^6$          & 1526. & .289 & 4531. & 3.03 & 1.4877 & 13.735 \\
10. &  -    &            -    & 1611. & .263 & 5050. & 3.69 & 1.4675 & 12.955 \\
15. &  -    &            -    & 1636. & .256 & 5210. & 3.62 & 1.4572 & 12.750 \\
20. &  -    &            -    & 1646. & .254 & 5261. & 3.51 & 1.4509 & 12.665 \\
\end{tabular}
\end{table}
\begin{table}
\caption{As Tab.~\ref{tab:GM3_1.25}, for a $M_\mathrm{B}=\unit[1.60]{M_\odot}$ star with the CBF-EI EoS.}
\label{tab:CBFEI_1.60}
\centering
\begin{tabular}{ccccccccc}
$t$ & $\nu_{g_1}$ & $\tau_{g_1}$ & $\nu_f$ & $\tau_f$ & $\nu_{p_1}$ & $\tau_{p_1}$ & $M$ & $R$ \\
\hline
0.2 & 522.7 &            7.45 & 995.1 & 20.4 & 1197. & 1.41 & 1.5637 & 33.618 \\
0.3 & 549.3 &            5.01 & 986.9 & 15.3 & 1231. & 1.36 & 1.5606 & 32.492 \\
0.4 & 581.2 &            4.17 & 978.8 & 12.3 & 1273. & 1.29 & 1.5577 & 31.376 \\
0.5 & 614.4 &            3.62 & 969.0 & 10.4 & 1318. & 1.21 & 1.5551 & 30.341 \\
0.6 & 647.5 &            3.21 & 957.7 & 8.95 & 1362. & 1.14 & 1.5525 & 29.389 \\
0.7 & 680.3 &            2.90 & 945.4 & 7.56 & 1408. & 1.08 & 1.5500 & 28.496 \\
0.8 & 711.9 &            2.69 & 933.3 & 6.14 & 1453. & 1.02 & 1.5476 & 27.677 \\
0.9 & 741.4 &            2.62 & 922.2 & 4.67 & 1497. & .978 & 1.5452 & 26.910 \\
1.0 & 767.2 &            2.80 & 914.3 & 3.28 & 1542. & .941 & 1.5430 & 26.187 \\
1.1 & 786.1 &            3.58 & 913.0 & 2.20 & 1588. & .911 & 1.5407 & 25.510 \\
1.2 & 794.6 &            5.98 & 921.9 & 1.55 & 1634. & .889 & 1.5386 & 24.860 \\
1.3 & 792.6 &            12.2 & 941.4 & 1.22 & 1682. & .874 & 1.5364 & 24.234 \\
1.4 & 783.8 &            26.1 & 967.9 & 1.04 & 1732. & .867 & 1.5343 & 23.633 \\
1.8 & 725.5 &            365. & 1095. & .691 & 1957. & .910 & 1.5263 & 21.458 \\
1.9 & 709.4 &            648. & 1127. & .635 & 2022. & .939 & 1.5244 & 20.962 \\
2.0 & 693.2 & $1.1\times10^3$ & 1158. & .586 & 2090. & .975 & 1.5226 & 20.486 \\
4.0 & 453.8 & $5.5\times10^5$ & 1522. & .291 & 3758. & 2.55 & 1.4944 & 14.944 \\
5.0 & -     &               - & 1590. & .267 & 4348. & 3.37 & 1.4853 & 14.006 \\
10. & -     &               - & 1714. & .233 & 5492. & 5.69 & 1.4600 & 12.639 \\
15. & -     &               - & 1755. & .224 & 5801. & 5.88 & 1.4485 & 12.295 \\
20. & -     &               - & 1776. & .220 & 5926. & 5.64 & 1.4425 & 12.143 \\
\end{tabular}
\end{table}
\begin{table}
\caption{As Tab.~\ref{tab:GM3_1.25}, for a $M_\mathrm{B}=\unit[1.60]{M_\odot}$ star with the LS-bulk EoS.}
\label{tab:LSbulk_1.60}
\centering
\begin{tabular}{ccccccccc}
$t$ & $\nu_{g_1}$ & $\tau_{g_1}$ & $\nu_f$ & $\tau_f$ & $\nu_{p_1}$ & $\tau_{p_1}$ & $M$ & $R$ \\
\hline
0.2 & 662.9 &            3.07 & 1020. & 6.83 & 1440. & .997 & 1.5519 & 28.240 \\
0.3 & 709.0 &            2.72 & 997.3 & 5.89 & 1494. & .919 & 1.5495 & 27.040 \\
0.4 & 753.6 &            2.52 & 976.8 & 4.56 & 1555. & .846 & 1.5471 & 26.001 \\
0.5 & 792.9 &            2.64 & 962.0 & 3.06 & 1618. & .787 & 1.5448 & 25.029 \\
0.6 & 819.9 &            3.71 & 959.7 & 1.86 & 1680. & .745 & 1.5424 & 24.138 \\
0.7 & 827.0 &            8.06 & 977.4 & 1.25 & 1743. & .717 & 1.5401 & 23.325 \\
0.8 & 818.3 &            21.6 & 1012. & .988 & 1809. & .701 & 1.5379 & 22.547 \\
0.9 & 802.1 &            55.8 & 1055. & .838 & 1879. & .699 & 1.5356 & 21.818 \\
1.0 & 782.9 &            132. & 1102. & .728 & 1956. & .708 & 1.5334 & 21.124 \\
2.0 & 606.4 & $5.3\times10^4$ & 1473. & .315 & 3237. & 1.24 & 1.5140 & 15.752 \\
4.0 & 426.3 & $1.3\times10^5$ & 1657. & .245 & 4586. & 2.17 & 1.4924 & 12.975 \\
5.0 & 374.9 & $9.2\times10^6$ & 1687. & .237 & 4825. & 2.44 & 1.4860 & 12.678 \\
10. &  -    &            -    & 1742. & .225 & 5304. & 3.03 & 1.4670 & 12.201 \\
15. &  -    &            -    & 1755. & .223 & 5452. & 3.19 & 1.4568 & 12.081 \\
20. &  -    &            -    & 1757. & .224 & 5510. & 3.25 & 1.4500 & 12.032 \\
\end{tabular}
\end{table}

\chapter{Rotation}
\label{cha:rotation}
In the previous chapters, we have considered a non-rotating proto-neutron star.
However, when a supernova explodes, the hot, lepton-rich remnant (the proto-neutron star) is also presumably rapidly
rotating.  In the early stages of its
evolution, the PNS cools down and loses its high lepton content, while its radius and  rotation
rate decrease. In this phase, a huge amount of energy and of angular momentum is
released, mainly through neutrino emission \citep{Burrows+Lattimer.1986,Keil+Janka.1995,Pons+.1999}.
A fraction of this energy is expected to be emitted in the gravitational wave
channel; indeed, as a consequence of the violent collapse, non-radial oscillations can be
excited, making PNSs promising sources for present and future gravitational
detectors \citep{Ferrari+Miniutti+Pons.2003,Ott.2009,Burgio+2011,Fuller+2015}.
In addition, since the star rotates, if its shape deviates from axisymmetry it
emits gravitational waves also due to rotation.

The study of the rotation rate evolution of a PNS during the
quasi-stationary, Kelvin-Helmholtz phase is important because it allows to link the
supernova explosion simulations with the observed properties of the young
pulsars population. In fact, current SN simulations (see
e.g.~\citealp{Thompson+Quataert+Burrows.2005,Ott+2006,Hanke+2013,Couch+Ott.2015,Nakamura+2014}) predict
that the PNS initial rotation period may be as small as few ms, whereas young
pulsars have been observed with rotational periods greater than about 100 ms (see \citealp{Miller+Miller.2014}, and references therein).

The evolution of rotating PNSs has been studied in \cite{Villain+.2004}.
\citet{Villain+.2004} have accounted for rotation effectively, employing the thermodynamical profiles of \cite{Pons+.1999}
(i.e., those of a non-rotating PNS evolved with the GM3 EoS) as effective time-dependent EoSs.
These effective EoSs have been used to construct the rotating configurations
with the non-linear BGSM code \citep{Gourgoulhon+1999}. A similar approach has been followed
in \cite{Martinon+2014}, which used  the profiles of \cite{Pons+.1999}
and \cite{Burgio+2011}. The main limitations of these works are the
following.
\begin{itemize}
\item The evolution of the PNS rotation rate is due not only to the change of the moment of inertia (i.e., to the
contraction), but also to the angular momentum change due to neutrino emission \citep{Epstein.1978}.  This was
neglected in \citet{Villain+.2004}, and described with a heuristic formula in \citet{Martinon+2014}.
\item As we shall discuss in this chapter, when the PNS profiles describing a non-rotating
star are treated as an effective EoS, one can obtain
configurations which are unstable to radial perturbations, unless particular care is taken in modelling
the effective EoS.
\end{itemize}

In this chapter, we study the spin evolution of the PNS in its first tens of
seconds of life, using the entropy and lepton fraction profiles obtained with
the quasi-stationary evolution of a spherically symmetric PNS
(Chapter~\ref{cha:evolution}) with the GM3 EoS, the same adopted by
\citet{Villain+.2004}.  In Sec.~\ref{sec:rotation_HT} we describe the
Hartle-Thorne equations, that we use to obtain the slowly-rotating PNS
profiles.  In Sec.~\ref{sec:rotation_eff} we describe our model to
effectively include rotation in the PNS evolution; our approach is different from that
used in \cite{Villain+.2004}. In particular, we
model the evolution of the angular momentum (due to neutrino emission) with the
Epstein's formula \citep{Epstein.1978}. In Sec.~\ref{sec:rotation_results} we describe
our results on the time dependence of the angular momentum and in Sec.~\ref{sec:rotation_GW} we discuss the gravitational wave
emission which is associated with this process.
The results of this chapter have been published in \citet{Camelio+2016}.

\section{Slowly-rotating neutron star: the Hartle-Thorne equations}
\label{sec:rotation_HT}

In this section, we briefly describe the equations of the perturbative Hartle-Thorne approach. For
further details we refer the reader to \citet{Hartle.1967,Hartle.1968,Hartle.1973} and to the Appendix
of \cite{Benhar+2005}.

In the Hartle-Thorne approach, the rotating star is described as a stationary
perturbation of a spherically symmetric background, for small values of the
angular velocity $\Omega=2\pi\nu$ (as seen by an observer at infinity), that is, for $\nu\ll\nu_{ms}$ ($\nu_{ms}$ is
the mass-shedding frequency, at which the star starts losing mass at the
equator, see Sec.~\ref{sec:rotation_results}).  As shown
in~\cite{Martinon+2014}, this ``slow rotation'' approximation is reasonably
accurate for rotation rates up to $\sim0.8$ of the mass-shedding limit,
providing values of mass, equatorial radius and moment of inertia which differ
by $\lesssim0.5\%$ from those obtained with fully relativistic, nonlinear
simulations.  In our approach we assume uniform rotation; PNSs are expected to
have a significant amount of differential rotation at birth
\citep{Janka+Monchmeyer.1989} which, however, is likely to be removed by
viscous mechanisms, such as, for instance, magnetorotational instability
\citep{Mosta+2015}, in a fraction of a second.  Our approach should be
considered as a first step towards a more detailed description of rotating
PNSs, in which we shall include differential rotation.

The spacetime metric, up to third order in $\Omega$, can be written as
\begin{multline}
\label{metricHartle}
\mathrm ds^2 = -e^{2\phi(r)} \big[ 1 + 2h_0(r) + 2h_2(r)P_2(\mu) \big] \mathrm dt^2 \\
+ e^{2\lambda (r)}\left[ 1 + \frac{ 2m_0(r) + 2m_2(r)P_2(\mu) }{
r - 2M(r) } \right]\mathrm dr^2 
+ r^2 [ 1 + 2k_2(r)P_2(\mu) ]\\
\times\left\{\mathrm d\theta^2 + \sin^2\theta \left[\mathrm d\phi - \left(\omega (r) +
w_1(r) + w_3(r)P^{'}_3(\mu)\right) \mathrm dt \right]^2 \right\},
\end{multline}
where $\mu=\cos\theta$ and $P_n(\mu)$ is the Legendre polynomial of order $n$,
the prime denoting the derivative with respect to $\mu$.  The perturbations of
the non-rotating star are described by the functions $\omega$ [of $O(\Omega)$],
$h_0$, $m_0$, $h_2$, $m_2$, $k_2$ [of $O(\Omega^2)$], and $w_1$, $w_3$ [of
$O(\Omega^3)$].  The energy-momentum tensor is
\begin{equation}
T^{\mu\nu} = (\mathcal E + \mathcal P)u^{\mu}u^{\nu} + \mathcal Pg^{\mu\nu}
\end{equation}
where $g_{\mu\nu}$, $u^{\mu}$ are the metric and four-velocity in the rotating
configuration, and $\mathcal E$ and $\mathcal P$ are the energy density and pressure in the rotating star. An element of
fluid, at position $(r,\theta)$ in the non-rotating star, is displaced by
rotation to the position
\begin{equation}
\label{displ1}
\bar{r}= r + \xi(r,\theta),
\end{equation}
where $\xi(r,\theta)=\xi_0(r) + \xi_2(r)P_2(\mu) +O(\Omega^4)$ is the Lagrangian displacement.
The energy and pressure Eulerian perturbations (for the difference between
Eulerian and Lagrangian perturbation, see discussion in Sec.~\ref{sec:GW.Eqs})
are
\begin{align}
\delta P={}&\big(\epsilon(r)+P(r)\big)\big(\delta p_0(r)+\delta p_2(r)P_2(\mu)\big).\\
\delta\epsilon={}&\frac{\mathrm d\epsilon/\mathrm dr}{\mathrm dP/\mathrm dr}\delta P,
\end{align}
where $\epsilon$ and $P$ are the energy density and pressure in the
non-rotating star and the perturbations depend on the functions $\delta p_0(r)$
and $\delta p_2(r)$.  The background spacetime is described by the TOV
equations, see Sec.~\ref{sec:tov}.

In the Hartle-Thorne approach, one assumes that if the fluid element of the
non-rotating star has pressure $P$ and energy density $\epsilon$, the
displaced fluid element of the rotating star has the same values of pressure
and energy density.  In other words, the Lagrangian perturbations of the energy
density and the pressure vanish (see Eq.~(6) of \citealp{Hartle.1967}),
$\Delta \epsilon=\Delta P=0$; the modification of these quantities is only due to the
displacement \eqref{displ1}
\begin{align}
\label{epspert}
\delta \epsilon(r,\theta) ={}& - \frac{\mathrm d\epsilon}{\mathrm dr}\xi(r,\theta),\\
\label{Ppert}
\delta P(r,\theta) ={}& - \frac{\mathrm dP}{\mathrm dr}\xi(r,\theta).
\end{align}
We remark that as long as we neglect terms of $O(\Omega^4)$,
$\delta \epsilon(r,\theta)\simeq\delta \epsilon(\bar r,\theta)$.

Einstein's equations, expanded in powers of $\Omega$ and in Legendre polynomials, can be written as a set of ordinary
differential equations for the perturbation functions.
The spacetime perturbation to first order in $\Omega$ is described by the function $\omega(r)$, which is responsible for the
dragging of inertial frames; it satisfies the equations
\begin{align}
\frac{\mathrm d\chi}{\mathrm dr}={}&\frac{u}{r^4}-\frac{4\pi r^2(\epsilon+P)\chi}{r-2M},\\
\frac{\mathrm du}{\mathrm dr}={}&\frac{16\pi r^5(\epsilon+P)\chi}{r-2M},
\label{eqpai}
\end{align}
where $M$ is the gravitational mass of the non-rotating star (obtained with the TOV equations),
$\varpi=\Omega-\omega$, $j(r)= e^{-\phi}\sqrt{1-2M/r}$, $\chi=j\varpi$ and $u=r^4jd\varpi/dr$. The angular
momentum $J$ is obtained by matching the interior with the exterior solution $\chi(r)=\Omega-2J/r^3$, $u(r)=6J$ at
$r=R$. The moment of inertia, at zero-th order in the rotation rate, is $I=J/\Omega$.
The perturbations to second order in $\Omega$ are described by the metric functions $h_l(r)$, $m_l(r)$ ($l=0,2$),
$k_2(r)$, and by the fluid pressure perturbations $\delta p_l$. The $l=0$ perturbations satisfy the equations
\begin{align}
\frac{\mathrm d}{\mathrm dr}\left(\delta p_0+h_0-\frac{\chi^2r^3}{3(r-2M)}\right)={}&0,\\
\delta p_2+h_2-\frac{\chi^2r^3}{3(r-2M)}={}&0,
\end{align}
\begin{align}
\frac{\mathrm dm_0}{\mathrm dr}={}&4\pi r^2\frac{\mathrm d\epsilon}{\mathrm dP}[\delta p_0(\epsilon+P)]+
\frac{u^2}{12r^4}+\frac{8\pi r^5(\epsilon+P)\chi^2}{3(r-2M)},\\
\frac{\mathrm d\delta p_{0}}{\mathrm dr}={}&\frac{u^2}{12r^4(r-2M)}-\frac{m_0(1+8\pi r^2P)}{(r-2M)^2}
-\frac{4\pi(\epsilon+P)r^2\delta p_0}{r-2M}\notag\\
{}&+\frac{2r^2\chi}{3(r-2M)}\left[\frac{u}{r^3}
+\frac{(r-3M-4\pi r^3P)\chi}{r-2M}\right].
\end{align}

Matching the interior and the exterior solutions at $r=R$ ($R$ is the radius of the non-rotating star), it is possible to
compute the monopolar stellar deformation, and then the correction to the
gravitational mass due to stellar rotation, $\delta M=m_0(R)+J^2/R^3$. The
baryonic mass correction $\delta M_B=\delta m_B(R)$ is found by solving the
equation
\begin{equation}
\frac{\mathrm d\delta m_B}{\mathrm dr}=4\pi r^2e^{\lambda}\left[\left(1+\frac{m_0}{r-2m}+\frac{1}{3}
r^2\varpi^2e^{-2\phi}\right)\epsilon
+\frac{\mathrm d\epsilon/\mathrm dr}{\mathrm dP/\mathrm dr}(\epsilon+P)\delta p_0\right].
\end{equation}
The $l=2$ perturbations satisfy the equations
\begin{align}
\frac{\mathrm dv_2}{\mathrm dr}={}&-2\frac{\mathrm d\phi}{\mathrm dr}h_2+\left(\frac{1}{r}+
\frac{\mathrm d\phi}{\mathrm dr}\right)\left[
\frac{8\pi r^5(\epsilon+P)\chi^2}{3(r-2M)}+\frac{u^2}{6r^4}\right],\\
\frac{\mathrm dh_2}{\mathrm dr}={}&\left[-2\frac{\mathrm d\phi}{\mathrm dr}+
\frac{r}{r-2M}\left(2\frac{\mathrm d\phi}{\mathrm dr}\right)^{-1}\left(8\pi(\epsilon+P)-
\frac{4M}{r^3}\right)\right]h_2\nonumber\\
{}&-\frac{4v_2}{r(r-2M)}\left(2\frac{\mathrm d\phi}{\mathrm dr}\right)^{-1}+
\frac{u^2}{6r^5}\left[
\frac{\mathrm d\phi}{\mathrm dr}r-\frac{1}{r-2M}\left(2\frac{\mathrm d\phi}{\mathrm dr}\right)^{-1}\right]\nonumber\\
{}&+\frac{8\pi r^5(\epsilon+P)\chi^2}{3(r-2M)}\left[
\frac{\mathrm d\phi}{\mathrm dr}r+
\frac{1}{r-2M}\left(2\frac{\mathrm d\phi}{\mathrm dr}\right)^{-1}\right],
\end{align}
where $v_2=k_2+h_2$.  Matching the interior and the exterior solutions, it is
possible to determine the quadrupole deformation of the PNS and then the
quadrupole moment.

The equations for the peturbations at $O(\Omega^3)$, $w_l(r)$ ($l=1,3$), have a
similar structure but they are longer and are not reported here; we refer the
reader to \citet{Hartle.1973} and the Appendix of \citet{Benhar+2005}. They
yield the octupole moment, the third-order corrections to the angular momentum
and the second-order corrections to the moment of inertia.

For each value of the central pressure $P_c$ (or, equivalently, of the central
energy density $\epsilon_c$) and of the rotation rate $\Omega$, the numerical
integration of the perturbation equations yields the perturbation functions,
and then the values of the multipole moments of the star (in particular, the
gravitational mass of the rotating star $M^\mathrm{rot}$ and the angular
momentum $J$), and of its baryonic mass $M^\mathrm{rot}_\mathrm{B}$. These quantities
can be written as $M^\mathrm{rot}=M+\delta M$, $J=\delta J$,
$M^\mathrm{rot}_\mathrm{B}=M_\mathrm{B}+\delta M_B$, etc., where the quantities with superscript $^\mathrm{rot}$
refer to the rotating star.  Given a non-rotating star
with central pressure $P_c$, gravitational mass $M$, and baryon mass $M_\mathrm{B}$, the rotating star (with spin
$\Omega$) with the same central pressure has a baryon mass $M_\mathrm{B}+\delta
M_B$, which is generally larger than $M_\mathrm{B}$. Therefore, a rotating star with the
same baryon mass $M_\mathrm{B}$ as the non-rotating one, has necessarily a smaller value
of the central pressure, $P_c+\delta P_c$, with $\delta P_c <0$ (this is not
surprising: when a star is set into rotation, its central pressure decreases).

\section{Including the rotation in an effective way}
\label{sec:rotation_eff}
In this section we describe how we account for slow rotation in the
evolution of the PNS.

In order to integrate the structure equations of the PNS we need to assign an
equation of state which is non-barotropic, thus we also need to know the
profiles of entropy and lepton fraction throughout the star (see discussion in
Secs.~\ref{sec:totaleos} and \ref{sec:tov}). As discussed in Chapter~\ref{cha:evolution}, these profiles
are obtained by our evolutionary code for a spherical, non-rotating PNS at
selected values of time.

The non-rotating profiles can be used to compute the structure of a rotating
PNS in different ways. A possible approach is the following.  Let us consider
a spherical PNS with baryon mass $M_B$  at a given evolution time $t$. The
numerical code discussed in Chapter~\ref{cha:evolution} provides the functions
$P(a)$, $\epsilon(a)$, $s(a)$, $Y_L(a)$, where we remind that $a$ is the
enclosed baryon number.  If we replace the inverse function of $P(a)$ into the
non-barotropic EoS, we obtain an ``effective barotropic EoS'',
$\tilde\epsilon(P)=\epsilon(P,s(a(P)),Y_L(a(P))$, which can be used to solve
the TOV equations for the spherical configuration to  which we  add the
perturbations due to rotation, according to Hartle's procedure.  Since the
rotating star must have the same baryon mass as the spherical star, one can
proceed  as follows: (i) solve the TOV equations for a spherically symmetric
star with central pressure $P_c+\delta P_c$; (ii) solve the perturbation
equations for a chosen value of the rotation rate, to determine the actual
baryon mass of the rotating star with that central pressure; (iii) iterate
these two steps modifying $\delta P_c$  until the baryon mass coincides with
the assigned  value $M_B$.  This approach was used in~\cite{Villain+.2004},
where the rotating star was modeled solving the fully non-linear Einstein
equations.

However, this procedure has some relevant drawbacks.  Indeed, during the first
second after bounce the star is very weakly bound, and it may happen that the
procedure above yields $\delta P_c>0$, which indicates that these
configurations are in the unstable branch of the mass-radius diagram.  We think
that this is caused by the nonphysical treatment (effective, as a barotropic
EoS) of the thermodynamical profile.  This problem did not occur in the
simulations of~\cite{Villain+.2004} because the authors considered a different,
stable branch of the mass-radius curve corresponding to the ``effective'' EoS
$\tilde\epsilon(P)$, at much lower densities. Indeed, for $t\lesssim0.5$ s, at
the center of the star they had $n_B\sim\unit[10^{-2}]{fm^{-3}}$ (i.e.,
rest-mass density $\rho\sim\unit[10^{13}]{g/cm^{3}}$), which corresponds to the
outer region of the star modeled in~\cite{Pons+.1999}.  When the central
density is so low, only a small region of the star is described by the GM3 EoS;
the rest is described by the low-density EoS used to model the PNS envelope,
which does not yield unstable configurations.

Since we want to model the PNS consistently with the evolutionary models, we
decided to implement the non-rotating profiles in an alternative way.  As in
the previous approach, we consider the spherical configuration obtained by the
evolution code at time $t$, with central density $P_c$ and baryon mass
(constant during the evolution) $M_B$. To describe the rotating star, we use
the GM3 EoS $\epsilon=\epsilon(P,s,Y_L)$; since we are restricting our analysis
to slowly rotating stars, the  entropy and lepton fraction profiles $s(a)$ and
$Y_L(a)$  of the non-rotating star are a good approximation for those of the
rotating star.  We follow the steps discussed before: (i) solve the TOV
equations for a star with central pressure $P_c+\delta P_c$; at each value of
$a$, the energy density is $\epsilon(P,s(a),Y_L(a))$; (ii) solve Hartle's
perturbation equations, finding the baryon mass of the star rotating to a given
rate with this reduced central pressure and find the correction to the baryon
mass due to rotation; (iii) iterate the first two steps, finding $\delta P_c$
such that the baryon mass of the rotating star is $M_B$.  We remark that the
energy density of the rotating star in step (ii) is related to that of the
non-rotating star in step (i) by the Hartle-Thorne prescription described above
Eq.~\eqref{epspert}.  Since we are using an appropriate non-barotropic EoS, the
instability discussed above disappears, and the central pressure of the
rotating star is, as expected, always smaller than that of the non-rotating
star with the same baryon mass. However, the central density is high enough for
the most of the star to be described by the GM3 EoS.

\begin{figure}
\centerline{
\includegraphics[width=1.3\textwidth]{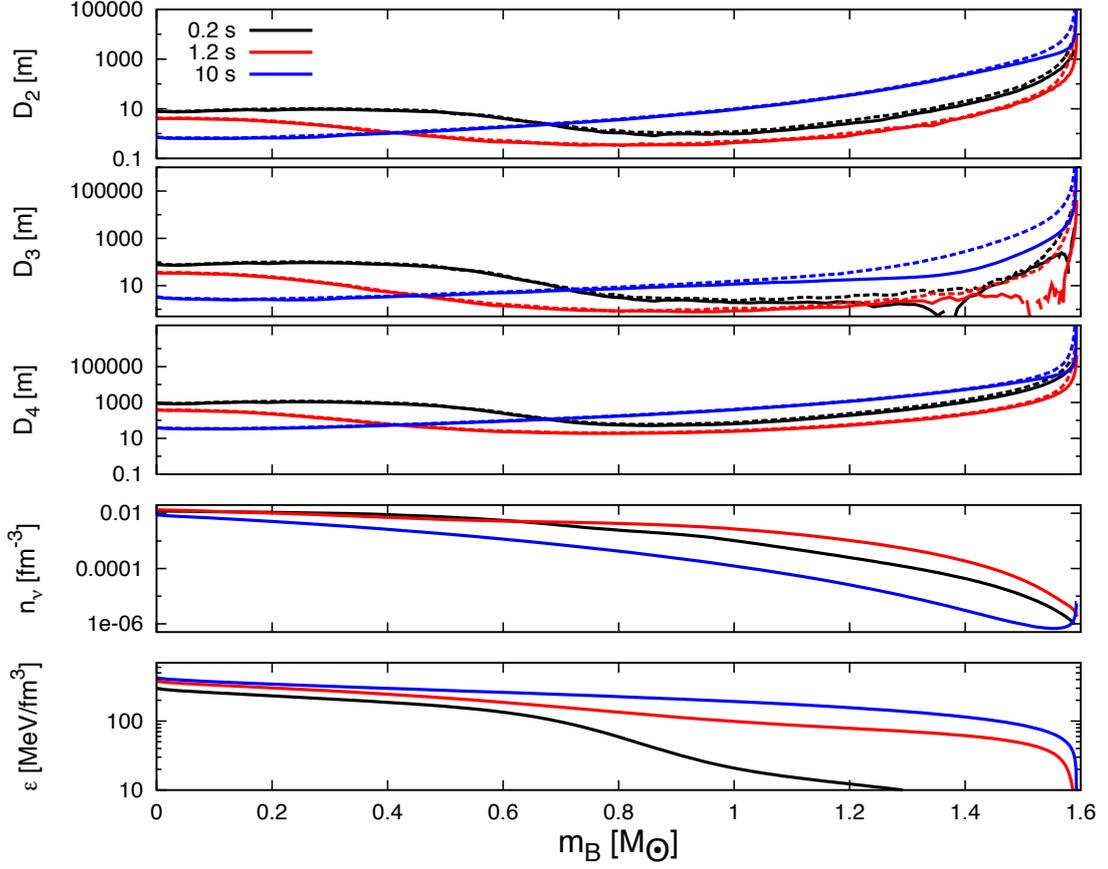}}
\caption{Neutrino diffusion coefficients $D_n$ ($n=2,3,4$), as functions of the
enclosed baryon mass, computed using the density and thermodynamical profiles
of the non-rotating (solid line) and rotating (dashed line) configurations, at
$t=0.2$ s, $t=1.2$ s and $t=10$ s (upper and middle panels).  Profiles of
neutrino number density and energy density (lower panels).  We assume
$M_B=1.6\,M_{\odot}$ and that the angular momentum is the maximum allowed
$J_{in}=J_{max}$ (for $J_{in}>J_{max}$, the PNS reaches the mass-shedding limit
during its evolution, see Sec.~\ref{sec:rotation_results}).}
\label{fig:compare}
\end{figure}
We stress again that we are using the numerical solution of the transport
equations for a {\it non-rotating} PNS, to build quasi-stationary
configurations of a {\it rotating} PNS. Therefore, we are neglecting the effect
of rotation on the time evolution of the PNS.  To be consistent, we should have
integrated the transport equations appropriate for a rotating star, which are
much more complicated.  Since these approximations affect the timescale of the
stellar evolution, we would like to estimate how faster, or slower, the
rotating star loses its thermal and lepton content with respect to the
non-rotating one.  Since the evolution timescale is governed by neutrino
diffusion processes, at each time step of the non-rotating PNS evolution, we
have computed and compared the neutrino diffusion coefficients $D_n$ [see
Eqs.~\eqref{eq:Fnu} and \eqref{eq:Hnu}] for non-rotating and rotating
configurations.  The latter have been obtained by replacing the profiles
($P(a)$, $\epsilon(a)$, etc.) of a non-rotating PNS with those of a rotating
PNS (computed as discussed above in this Section).  In the upper and middle
panels of Fig.~\ref{fig:compare} we plot $D_2, D_3$ and $D_4$ as functions of
the enclosed baryon mass $m_B=m_n a$ ($m_n$ is the neutron mass), for the
non-rotating (solid line) and rotating (dashed line) configurations, at $t=0.2$
s, $t=1.2$ s and $t=10$ s.  In the lower panels we plot the neutrino number
density and the total energy density at the same times.  We set
$M_B=1.6\,M_{\odot}$ and that the initial angular momentum, $J_{in}$, is equal
to the maximum angular momentum $J_{max}$, above which mass-shedding sets in
(see Sec.~\ref{sec:rotation_results} for further details). We see that  the
diffusion coefficients of the rotating configurations are  larger than those of
the non-rotating star.  For $m_B\lesssim 1~M_\odot$ the relative difference
$\vert D_n^\mathrm{rot}-D_n^\mathrm{non~rot}\vert/\vert D_n^\mathrm{non~rot}\vert$ is always smaller
than $\sim10-20\%$, and becomes  smaller than a few percent  after the first
few seconds.  In the outer region $m_B\gtrsim 1~M_\odot$ and early times, the
relative difference seems larger, in particular for the coefficient $D_3$, but
this has no effect for two reasons: first, as shown in the two lower panels of
Fig.~\ref{fig:compare}, both the neutrino number density and  the total energy
density are much  smaller than in the inner core; therefore, even though the
diffusion coefficients of the rotating star are larger than those of the
non-rotating one, few neutrinos are trapped in this region and transport
effects do not contribute significantly to the overall evolution; second, the
differences become large in the semi-transparent region, when the mean free
path becomes comparable to (or larger than) the distance to the star surface.
In this region the diffusion approximation breaks down and in practice the
diffusion coefficients are always numerically limited by the flux-limiter (see
Sec.~\ref{sec:code}).

From the above discussion we can conclude that the
rotating star loses energy and lepton number through neutrino emission
faster than the non-rotating one. This effect is larger at the beginning of the
evolution, that is for $t \lesssim 2$ s, and is of the order of $\sim10-20\%$, 
but becomes negligible at later times. Consequently our rotating star cools
down and contracts over a timescale which, initially, is $\sim10-20\%$ shorter
than that of the corresponding non-rotating configuration.

Once the equations describing the rotating  configuration
are solved for each value of the evolution time $t$ and for an assigned
value of the rotation  rate
$\Omega$,  the  solution of these equations allows one to  compute the
multipole moments of the rotating star,  including the angular momentum $J$. Conversely
we can choose, at  each value of $t$, the value of the angular momentum, and
determine, using a shooting method, the corresponding value of the rotation
rate.

If we want to describe the early evolution of a rotating PNS, we need a
physical prescription for  the time dependence of $J$.  For instance, we may
assume that the angular momentum is constant, as in \citet{Villain+.2004} (see
also \citealp{Goussard+Haensel+Zdunik.1997, Goussard+Haensel+Zdunik.1998}).
However, in the first minute of a PNS life, neutrino emission carries away
$\sim10\%$ of the star gravitational mass \citep{Lattimer+Prakash.2000}, and
also a significant fraction of the total angular momentum \citep{Janka.2004}.
To our knowledge, the most sensible estimate of the neutrino angular momentum
loss in PNSs has been done in \citet{Epstein.1978}
\begin{equation}
\frac{dJ}{dt}=-\frac{2}{5}qL_\nu R^2\Omega, \label{epst}
\end{equation}
where $R$ is the radius of the star, $L_\nu=-\mathrm dM/\mathrm dt$ is the
neutrino energy flux, and $q$ is an efficiency parameter, which depends on the
features of the neutrino transport and emission. If neutrinos escape without
scattering, $q=1$; if, instead, they have a very short mean free path, they are
diffused up to the surface, and then are emitted with $q=5/3$. As discussed in
\cite{Epstein.1978} (see also
\citealp{Kazanas.1977,Mikaelian.1977,Henriksen+Chau.1978}), $q=5/3$ should be
considered as an {\it upper limit} of the angular momentum loss by neutrino
emission. A more recent, alternative study \citep{Dvornikov+Dib.2009} indicates
an angular momentum emission smaller than this limit. In the following, we
shall consider Epstein's formula with $q=5/3$, and this  has to be meant as an
upper limit. We also mention that a simplified expression based on Epstein's
formula for the angular momentum loss in PNSs has been derived
in~\cite{Janka.2004} and used in~\cite{Martinon+2014}.

We mention that in \citet{O_Connor+Ott.2010} the neutrino transport equations
for a rotating star in general relativity have been solved by using an
alternative approach.  In this approach (which is believed to be accurate for
slowly rotating stars, \citealp{O_Connor+Ott.2010}) the structure and transport
equations for a spherically symmetric star are modified by adding a centrifugal
force term, to include the effect of rotation.

\section{Results: spin evolution of the proto-neutron star}
\label{sec:rotation_results}

\begin{figure}
\centerline{
\includegraphics[width=0.9\textwidth,angle=-90]{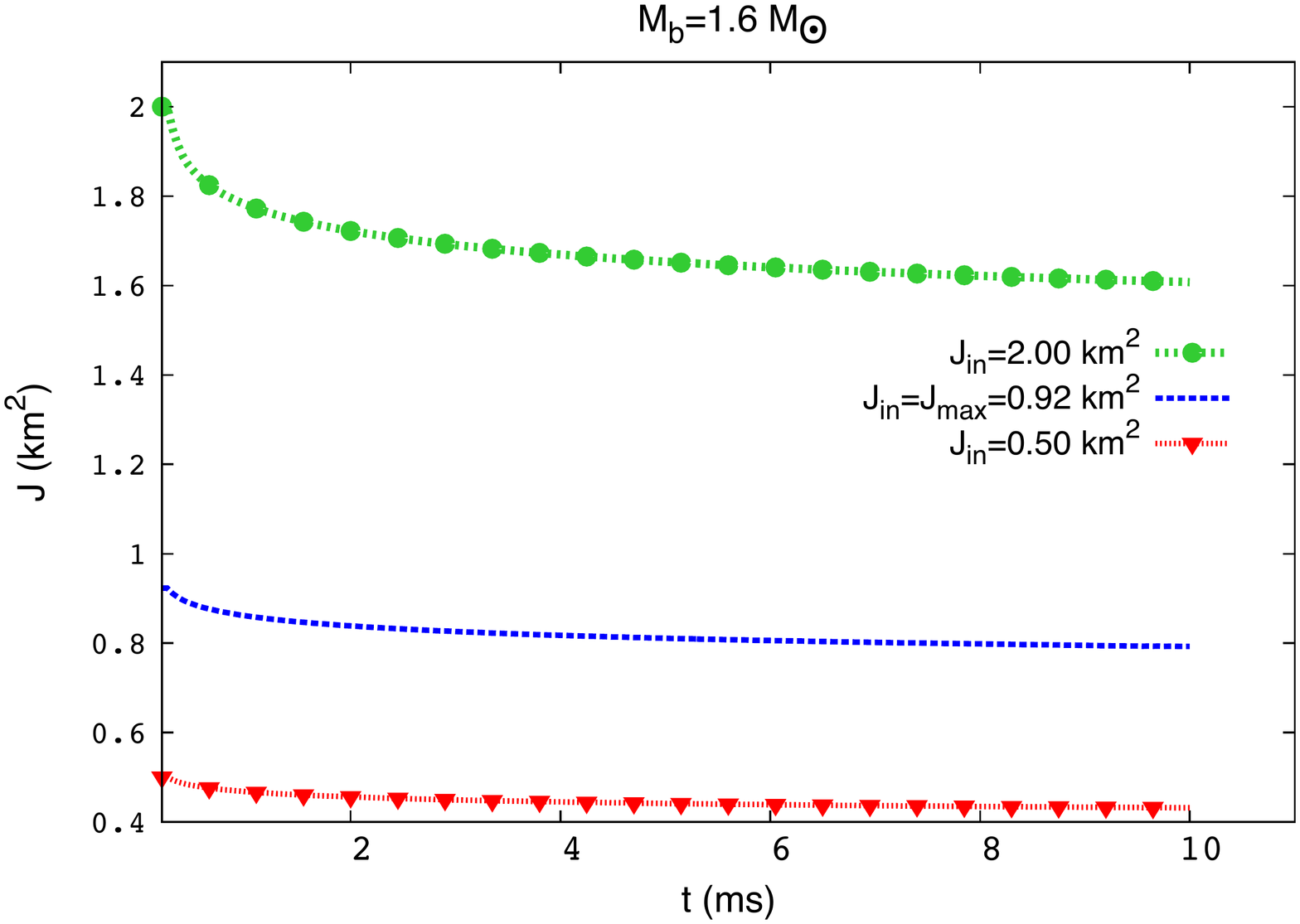}}
\caption{Angular momentum evolution due to neutrino losses, for a PNS with
baryonic mass $M_B=1.6\,M_\odot$ and initial angular momentum
$J_{in}=(2.02,\,3.71,\,8.08)\times10^{48}$erg s.}
\label{fig:Jt}
\end{figure}

In Figure~\ref{fig:Jt} we show how the angular momentum changes according to
Epstein's formula~\eqref{epst} as the PNS evolves. We assume $q=5/3$ and
baryonic mass $M_B=1.6\,M_\odot$. We consider different values of the angular
momentum $J_{in}$ at the beginning of the quasi-stationary phase ($t=0.2$ s
after the bounce): $J_{in}=\unit[2.02\times10^{48}]{erg\,s}$,
$J_{in}=\unit[3.71\times10^{48}]{erg\,s}$ and
$J_{in}=\unit[8.08\times10^{48}]{erg\,s}$. We find that, in the
first ten seconds after bounce, $13\%$ of the initial angular momentum is
carried away by neutrinos if $J_{in}=\unit[2.02\times10^{48}]{erg\,s}$ or
$J_{in}=\unit[3.71\times10^{48}]{erg\,s}$; $20\%$ of the initial angular
momentum is carried away if $J_{in}=\unit[8.075\times10^{48}]{erg\,s}$.
As mentioned above, $q=5/3$ should be considered as an upper bound; for smaller
values of $q$, the rate of angular momentum loss would be smaller.

\begin{figure}
\centerline{
\includegraphics[width=0.9\textwidth,angle=-90]{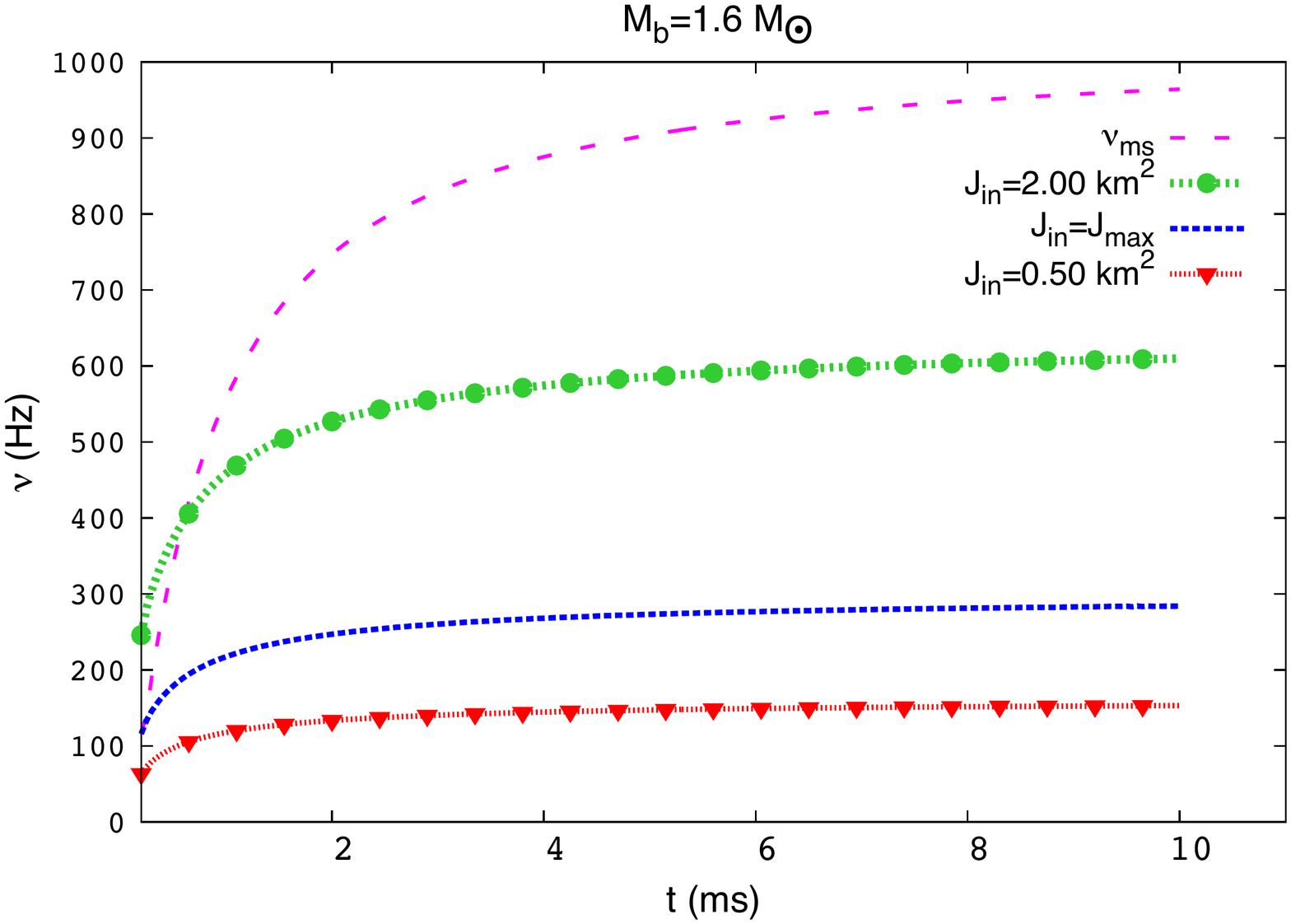}}
\caption{Evolution of the PNS rotation rate, corresponding to the angular momentum profiles
shown in Figure~\ref{fig:Jt}.}
\label{fig:nut}
\end{figure}

The corresponding evolution of the PNS rotation frequency is shown in
Figure~\ref{fig:nut}. In the same Figure we also show the mass-shedding
frequency $\nu_{ms}$ determined using a numerical fit derived in
\citet{Doneva.2013} from fully relativistic, non-linear integration of
Einstein's equations:
\begin{equation}
\nu_{ms}(Hz)=a\sqrt{\frac{M/M_\odot}{R/1{\rm km}}}+b\label{fitms}
\end{equation}
where $a=45862$ Hz and $b=-189$ Hz. We remark that the coefficients $a,b$ of
this fit do not depend on the EoS. We see that if
$J_{in}=\unit[8.08\times10^{48}]{erg\,s}$, the curves of $\nu(t)$ and of
$\nu_{ms}(t)$ cross during the quasi-stationary evolution; before the crossing,
the PNS spin is larger than the mass-shedding limit.  This means that a PNS
with such initial angular momentum would lose mass.  If we require the initial
rotation rate to be smaller than the mass-shedding limit, we must impose
$J_{in} \le J_{max}\equiv\unit[3.72\times10^{48}]{erg\,s}$.  We remark that the
value of $J_{max}$ is not affected by the efficiency of angular momentum loss
$q$: if $q<5/3$, $J_{max}$ has the same value, but the rotation rate grows more
rapidly than in Figure~\ref{fig:nut}.

It is interesting to note that, since $\nu_{ms}$ has a steeper increase than $\nu(t)$, even when the bound
$\nu\le\nu_{ms}$ is saturated at the beginning of the quasi-stationary phase the frequency becomes much smaller than the
mass shedding frequency at later times. This is an a-posteriori confirmation that the slow rotation approximation is
appropriate to study newly born PNSs.  For $t>10$ s, the PNS radius does not change significantly, and the star starts
to spin down due to electromagnetic and gravitational emission. However this spin down timescale is much longer than the
timescale of the quasi-stationary evolution we are considering; therefore it is unlikely that after this early phase the
PNS rotation rate is larger than $\sim 300$ Hz (i.e., that its period is smaller than
$\sim3$ ms), unless some spin-up mechanism (such as e.g. accretion) sets in. A less efficient angular momentum loss
($q<5/3$) would moderately increase this final value, but the general picture would remain the same.

It is worth noting that models of
pre-supernova stellar evolution \citep{Heger+Woosley+Spruit.2005} predict a similar range of the PNS rotation rate and angular
momentum. Among the models considered in \citet{Heger+Woosley+Spruit.2005}, the only one with $J>J_{max}$ (and rotation period smaller
than $3$ ms) is expected to collapse to a black hole.  Other works \citep{Thompson+Quataert+Burrows.2005, Ott+2006} have shown that
if the progenitor has a rotation rate sufficiently large, the PNS resulting from the core-collapse can have periods of
few ms; our results suggest that this scenario is unlikely, unless there is a significant mass loss in the early
Kelvin-Helmoltz phase.

\section{Results: gravitational wave emission}
\label{sec:rotation_GW}

If the evolving PNS is born with some degree of  asymmetry, it emits
gravitational waves. Assuming that the star rotates about a principal axis of
the moment of inertia tensor, that is, that there is no precession\footnote{Free
precession requires the existence of a rigid crust \citep{Jones+Andersson.2000}, thus it
should not occur in the first tens of seconds of the PNS life, when the crust has not
formed yet \citep{Suwa.2013}.}, gravitational waves are emitted at twice the orbital frequency
$\nu$, with amplitude
\citep{Zimmermann+Szedenits.1979,Thorne.book.1987,Bonazzola+Gourgoulhon.1996,Jones.2001}
\begin{align}
h_0\simeq{}&\frac{4G(2\pi\nu)^2I_3\varepsilon}{c^4 d},\label{gwamplitude}\\
\varepsilon={}&\frac{I_1-I_2}{I_3},
\end{align}
where $G$ is the gravitational constant, $c$ the speed of light, $d$ the distance from the detector,
and the deviation from axisymmetry is described by the {\it ellipticity}
$\varepsilon$. 
$I_1$, $I_2$, and $I_3$ are the principal moments of inertia of the
PNS and $I_3$ is assumed to be aligned with the rotation axis.
For old neutron stars, the loss of energy through gravitational waves
is compensated by a decrease of rotational energy,
which contributes to the spin-down of the star (the main contribution to the spin-down 
being  that of the magnetic field). 

In the case of a newly born PNS, the situation is different.
As the star contracts, due to the processes related to neutrino production and
diffusion, its rotation rate increases. If the PNS has a finite ellipticity,
it emits gravitational waves, whose amplitude and frequency also
increase as the star spins up. The timescale of this process is of the order of
tens of seconds. In our model, for simplicity we shall assume that the PNS
ellipticity remains constant over this short time interval.

Unfortunately, the ellipticity of a PNS is  unknown. In cold,
old NSs it is expected to be, at most, as large as
$\sim10^{-5}-10^{-4}$ \citep{Haskell+Jones+Andersson.2006,Ciolfi+Ferrari+Gualtieri.2010} (larger values are
allowed for EoSs including exotic matter
phases, \citealp{Horowitz+Kadau.2009,Johnson_McDaniel+Owen.2013}).
For newly born PNSs, it may be larger, but we have no hint on its actual value.
To our knowledge, current numerical simulations of core-collapse do not provide
estimates of the PNS ellipticity. We remark that although there is observational
evidence of large asymmetries in supernova
explosions \citep{Wang+2003,Leonard+2006}, there is no evidence that
they can be inherited by the PNS.  In the following, we shall assume
$\varepsilon=10^{-4}$, but this should be considered as a fiducial value: the
gravitational wave amplitude (which is linear in $\varepsilon$) can be easily
rescaled for different values of the PNS ellipticity.

\begin{figure}
\centering
\includegraphics[width=\textwidth]{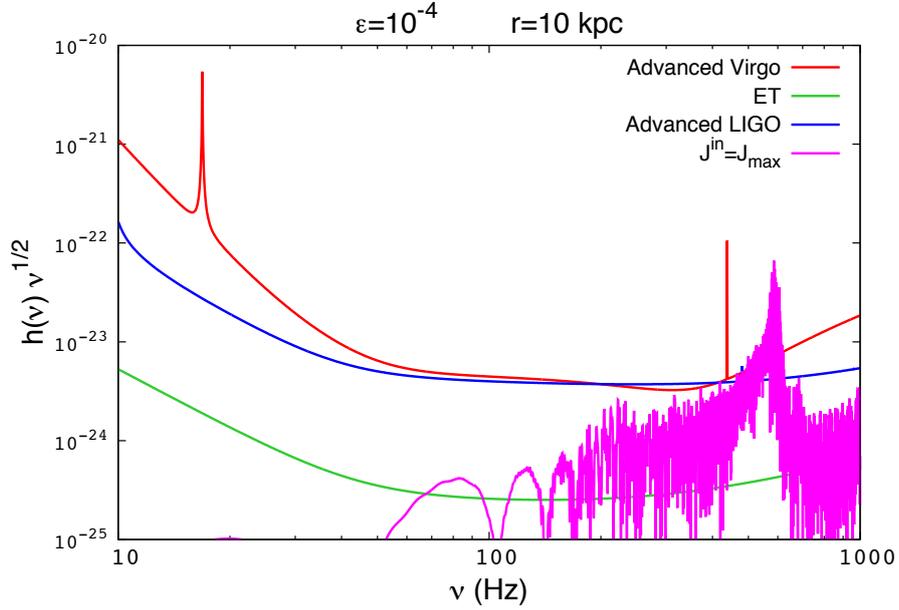}
\caption{
The strain amplitude $\tilde h(f)\sqrt{f}$
of the gravitational wave signal emitted by a PNS with 
$\epsilon=10^{-4}$, $J_{in}=J_{max}$, at a distance $r=10$ kpc,
is compared with the noise curves of Advanced Virgo, Advanced LIGO and ET.
}
\label{fig:h}
\end{figure}

If the PNS has a finite ellipticity $\varepsilon$ it emits
gravitational waves with frequency $f(t)=2\nu(t)$ and amplitude given by
Eq.~\eqref{gwamplitude}, with $\nu\equiv \nu(t)$ and $I_3\equiv I_3(t)$.
As the spin rate $\nu(t)$ increases, both the frequency and the amplitude of the
gravitational wave increase; therefore, the signal is a sort of ``chirp''; this
is different from the chirp emitted by neutron star binaries before coalescence,
because the amplitude increases at a much milder rate and the frequency migrates at a
rate which depends on the evolutionary timescale.
In Figure~\ref{fig:h} we show the strain amplitude $\tilde
h(f)\sqrt{f}=\sqrt{f}\sqrt{(\tilde h_+(f)^2+\tilde h_\times(f)^2)/2}$, 
where $\tilde h_{+,\times}(f)$ are the Fourier transform of the two polarization
of the gravitational wave signal
\begin{align}
h_+={}& h_0\frac{1+\cos^2i }{2}\cos\big(2\pi f(t) t\big),\\
h_\times={}& h_0\cos i\sin\big(2\pi f(t) t\big),
\end{align}
and $i$ is the angle between the rotation axis and the line of sight.
In Fig.~\ref{fig:h} the signal strain amplitude,  computed assuming optimal
orientation, $J_{in}=J_{max}$, $\varepsilon=10^{-4}$, and a distance of
$r=10$ kpc,  is
compared with the sensitivity curves of Advanced Virgo\footnote{https://inspirehep.net/record/889763/plots},
Advanced LIGO\footnote{https://dcc.ligo.org/LIGO-T0900288/public}, and of the
third generation detector ET\footnote{http://www.et-gw.eu/etsensitivities}.
We see that the signal is marginally above noise
for the advanced detectors, but it is definitely above the noise curve for ET.
This signal would be seen by Advanced Virgo with a signal-to-noise ratio
$\mathrm{SNR}=1.4$, and by Advanced LIGO with $\mathrm{SNR}=2.2$, too low to extract it
from the detector noise; however, since the
signal-to-noise ratio scales linearly with the ellipticity, a star born with
$\varepsilon=10^{-3}$ would be detected with $\mathrm{SNR}=14$ and $\mathrm{SNR}=22$ by Advanced
Virgo and LIGO, respectively.
The third generation detectors like ET would detect the signal coming from a
galactic PNS born with $\varepsilon=10^{-4}$ with a very large signal-to-noise
ratio, i.e.  $\mathrm{SNR}=22$. If the source is in the Virgo cluster ($d=\unit[15]{Mpc}$), the
ellipticity of the PNS should be as large as $5\cdot 10^{-2}$ to be seen by ET with
$\mathrm{SNM}=8$.

\chapter{Conclusions and outlook}
\label{cha:conclusions}

In this thesis we have studied the evolution and the gravitational wave
emission of a proto-neutron star in the Kelvin-Helmoltz phase, that is the
period of the neutron star life subsequent to the supernova explosion, until
the star becomes transparent to neutrinos.  This phase lasts for tens of
seconds and its beginning may be placed at about 200 ms after core bounce. The
evolution can be modeled as a sequence of quasi-stationary configurations in
which the matter is almost in beta-equilibrium \citep{Burrows+Lattimer.1986}.
To perform this study, we have written a new general relativistic, one-dimensional, energy-averaged,
and flux-limited PNS evolutionary code which evolves a general EoS
consistently.  In particular, we have considered three nucleonic equation of
states and three stellar masses, and we have determined the neutrino cross
sections self-consistently with the corresponding EoS. The EoSs considered are
all nucleonic (without hyperons) and are obtained (i) by the extrapolation of
the nuclear matter properties determined by terrestrial experiments \citep[the
LS-bulk EoS,][]{Lattimer+Swesty.1991} (ii) by the nuclear relativistic
mean-field theory \citep[the GM3 EoS,][]{Glendenning+Moszkowski.1991} (iii) by
the nuclear non-relativistic many-body and effective interaction theory
(the CBF-EI EoS, \citealp{Benhar+Lovato2017}; \citeauthor[][forthcoming]{Lovato+prep}).  We have
determined the frequencies of the quasi-normal oscillation modes for the
different EoSs and stellar masses \citep{Camelio+2017} and
we have studied the angular momentum evolution and the gravitational wave
emission of the proto-neutron star in the case of the GM3 mean-field EoS,
including rotation in an effective way \citep{Camelio+2016}.

The main improvements with respect to previous works introduced by our study are
the following.
\begin{itemize}
\item We have developed and tested a new fitting formula for the interacting part of the baryon
free energy (i.e., neutrons plus protons), which is valid for high density matter, finite temperature, and intermediate proton fractions.
We used this fitting formula to derive the other thermodynamical quantities.  This formula is suitable
to be used in evolutionary codes.
\item We have computed the neutrino cross sections for the many-body theory EoS
of \citet{Benhar+Lovato2017}.  They have been computed at
the mean-field level \citep{Reddy+Prakash+Lattimer.1998}, that is, the
interaction between baryons has been accounted for modifying the baryon
energy spectra by means of density-, temperature-, and composition-dependent
effective masses and single particle energies.
\item We used these neutrino cross sections to evolve the PNS with the
many-body EoS in a consistent way.  To our knowledge, this is the
first time that a PNS with a many-body EoS has been evolved with
consistently determined neutrino opacities. From this  evolution, we
have determined the stellar quasi-normal modes.
\item We have adopted a new method to account for rotation in an effective way, namely we have used a
time-dependent effective EoS which depends on the enclosed baryon mass, instead of on the pressure like
\citet{Villain+.2004}. Our method has the advantage to avoid nonphysical 
instabilities in the procedure adopted to include rotation.
\item We have consistently accounted for angular momentum loss by neutrino emission, using the Epstein formula
\citep{Epstein.1978}. Earlier, the angular momentum was assumed to be constant \citep{Villain+.2004}
or it was modeled with a heuristic formula \citep{Martinon+2014}.\\
\end{itemize}

The main results of this thesis are the following.
\begin{itemize}
\item The PNS evolution depends on the adopted EoS. In particular, for the
many-body EoS CBF-EI the PNS cooling is faster than that with the mean-field
EoS GM3, which in turn is faster than that with the extrapolated EoS LS-bulk.
We explain this with the fact that in the CBF-EI many-body theory the interaction
between baryons is stronger than in the GM3 mean-field theory (in fact, the baryon
masses are smaller in the CBF-EI than in the GM3 EoS).  In the extrapolated EoS
LS-bulk the effective baryon masses have been assumed equal to the bare ones,
and the result is that, for what regards the computation of the
neutrino cross sections, this EoS is ``less interacting'' than the others.
\item The deleptonization of the PNS with the CBF-EI EoS is almost completed at
the end of the Joule-heating phase (similarly to what was found in the first
PNS numerical studies by \citealp{Burrows+Lattimer.1986} and
\citealp{Keil+Janka.1995}), whereas the deleptonization for the GM3 and LS-bulk
EoSs proceeds during the cooling phase (as found by \citealp{Pons+.1999}).
\citet{Pons+.1999} explained the difference in the duration of the
deleptonization with the over-simplifications in the treatment of the neutrino
opacities in the first works of \citet{Burrows+Lattimer.1986} and
\citet{Keil+Janka.1995}.  However, we compute the neutrino cross sections for
the CBF-EI and the other EoSs with the same procedure of \citet{Pons+.1999}.
Therefore, the faster deleptonization is a feature due also to the EoS
properties and not only to the treatment of neutrino opacities.
\item The total number of electron antineutrinos detected depends on the
gravitational binding energy but is not completely determined by it. In
particular, the CBF-EI EoS has more antineutrino detected than the LS-bulk EoS,
even though its binding energy is smaller. This is due to the fact that the CBF-EI
has higher temperatures than the other EoSs, hence the electron antineutrino
distribution function at the neutrinosphere is smother and more antineutrino
have energies larger than the detector energy threshold at the detector.  This
result remarks the importance of an accurate modeling of the PNS evolution in
order to extract information on the PNS physics from the neutrino signal.
\item 
We show that during the first second, the frequencies at which the PNS
oscillates emitting  gravitational waves have a non monotonic behaviour.
The fundamental mode frequency decreases, reaches a minimum and then increases 
toward the value corresponding to the  cold neutron star which forms at
the end of the evolution. The frequency of the first $g$ mode increases, reaches
a maximum and then decreases to the asymptotic zero limit, that of the mode $p_1$ has  
a less pronounced minimum at earlier times with respect to the $f$ mode. We show
that this behaviour, already noted in~\cite{Ferrari+Miniutti+Pons.2003} for the EoS GM3
for the $f$ and $g_1$ modes,
is a generic feature when the PNS evolution is consistently described, and that
the timescale depends on the EoS. Indeed the time needed to reach the minimum
(maximum) for the $f$- ($g_1$-) mode can differ by as much as half a second 
for the EoS we consider.

During the first second, the damping time of all modes  is shorter than the
neutrino diffusive timescale ($\sim \unit[10]{s}$);  therefore gravitational
wave emission may be competitive in subtracting energy from the star.  This
remains true at later time only for the fundamental mode and for the first $p$
mode. However, the damping time of the $f$ mode is much shorter, thus we should
expect that after the first few seconds gravitational waves will be emitted
mainly at the corresponding frequency.
\item The QNM frequencies depend not only on the EoS, but also on the stellar baryon mass. In
particular, we find that for a lower mass, at the beginning the $p_1$-mode has 
higher frequency; for instance, for the $\unit[1.25]{M_\odot}$ star it approaches
2 kHz.
\item We have found a relation between the fundamental and first p-mode
frequencies and the mean stellar density [Eq.~\eqref{eq:fitGW}] which is valid
during all PNS phases for the cases considered in this thesis. This may be an
universal property of PNSs, independent of the mass and the EoS.
\item During this early evolution, the star spins up due to contraction.  By
requiring that the initial rotation rate does not exceed the mass-shedding
limit (i.e., the limit at which the PNS start to lose material), we have
estimated the maximum rotation rate at the end of the spin-up phase. For a PNS
of $M_\mathrm{B}=\unit[1.6]{M_\odot}$ we  find  that  one minute after bounce
the star would rotate at  $\nu_\mathrm{max}\lesssim \unit[300]{Hz}$,
corresponding to a rotation period of
$\tau_\mathrm{min}\gtrsim\unit[3.3]{ms}$.
\item If the PNS is born with a finite ellipticity $\epsilon$, while spinning
up it emits gravitational waves at twice the rotation frequency. This signal
increases both in frequency and amplitude, like the ``chirp'' signal of a
binary neutron star before coalescence.  However, the PNS signal is different
from those of a binary NS because its amplitude increases at a much
milder rate and the frequency migrates at a rate which depends on the
evolutionary timescale. We find that for a galactic supernova, if
$\epsilon=10^{-3}$ this signal could be detected by Advanced LIGO/Virgo with a
signal-to-noise ratio $\gtrsim 14$.  To detect farther sources, third
generation detectors like ET would be needed.\\
\end{itemize}

The work of this thesis may be improved in many direction. 
\begin{itemize}
\item We have developed a high density, finite temperature, arbitrary proton fraction fitting formula
for the interacting part of the baryon free energy, for a nucleonic EoS with only neutrons
and protons. It would be interesting to develop an analogous fitting formula for the hyperon case.
\item We have considered an interacting gas of baryons, without considering the formation of alpha particle,
pasta phases nor crystalline structures. For a hot PNS this approximation is valid for most of the star,
but while it cools down we expect that an increasing fraction of the star is interested by the formation of
alpha particles and nuclei (see Appendix~\ref{sec:gas}). While we do not expect dramatic changes (see Appendix~\ref{sec:gas}
for an \emph{a posteriori} justification of our assumption),
it would be interesting to consider also these lower density phases.
\item We have assumed beta-equilibrium everywhere in the star, as \citet{Keil+Janka.1995}.
This is a good approximation (see Appendix~\ref{sec:betaequilibrium}, and also
\citealp{Burrows+Lattimer.1986, Pons+.1999}), however in future it would be more consistent to account
for an eventual deviation from beta-equilibrium.
\item We have assumed a vanishing muon and tauon neutrino chemical potentials, as \citet{Burrows+Lattimer.1986,
Keil+Janka.1995, Pons+.1999}. This is a good approximation (see \citealp{Burrows+Lattimer.1986, Keil+Janka.1995}),
however to consistently account for the presence of muons (which are present in the PNS, even though in lower quantity than electrons)
we should in future consider also a finite muon lepton fraction and therefore a finite muon chemical potential.
Therefore, we should consider the process of moun neutrino transport and possibly also of tauon neutrino transport.
\item The neutrino cross sections are fundamental in the determination of the PNS evolution, and in this thesis
we have found that the baryon interaction strongly influences them. Therefore, to obtain a more accurate evolution,
it is important to accurately describe the neutrino diffusive processes. Many improvements may be done to the
neutrino cross section treatment of this thesis, for example including the random phase approximation 
and the weak magnetism correction.
Consistently computed neutrino cross
sections in the many-body theory for finite temperature and high density matter
would be welcome too.
\item Our initial profiles are obtained from the ending configuration of an old
SN simulation \citep{Wilson+Mayle.1989}, and moreover we have quite arbitrarily rescaled
the entropy per baryon and the lepton fraction to consider PNSs with different
masses. As we have discussed, this brings a significant amount of uncertainties
in the PNS evolution (see \citealp{Pons+.1999} for a discussion on this issue).
In the future, we should consistently link our PNS evolution
with the core-collapse phase, that is, we should consistently adopt as initial profiles the ending snapshots of
more modern SN simulations.
\item We have not considered accretion \citep{Burrows.1988, Takiwaki+Kotake+Suwa.2014,
Melson+Janka+Marek.2015, Muller.2015} and convection
\citep{Miralles+Pons+Urpin.2000, Roberts+2012} in our simulations. Both processes may produce observable effects,
and it would be interesting to include them consistently in future simulations.
\item It is known that in the first seconds the QNM frequencies deviate from the scaling laws valid for cold NSs
for a general EoS. It would be interesting to find how the QNM frequencies and damping times scale with the PNS
properties in this early phase.
\item In estimating the gravitational wave emission from a rotating PNS we have assumed a value of the
PNS initial ellipticity $\epsilon$. However,
we remark that  the actual value of PNS ellipticities is unknown, and depends on
the details of the supernova core collapse.
Accurate numerical simulations of supernova explosion addressing this
issue are certainly needed to  provide a quantitative estimate of
the range of $\epsilon$.
\item 
In our approach the effects of the PNS rotation are
consistently included in the structure equations, but they are neglected when
solving the neutrino transport equations. We estimate that due to this
approximation, we overestimate the evolution timescale at early times of, at
most, $\sim10-20\%$.  Moreover, we have assumed uniform rotation. It would
be interesting to improve our code to allow for differential rotation
and to treat the neutrino diffusion process consistently with
rotation.\\
\end{itemize}

The investigation of the proto-neutron star phase deepens our
understanding of the stellar physics and of the largely unknown high density and
high temperature microphysics.  This thesis is a further step in the study of
this important phase of the neutron star life.

\appendix

\chapter{Some non-interacting Fermion EoSs}
\label{app:eos}

In this Appendix we present the analytic form of some non-interacting, single-particle Fermion EoSs.
We explicit the fundamental constants $c$ and $h$, and set to unity the Boltzmann constant, $k_\mathrm{B}=1$.
It is easy to reinstate the Boltzmann constant, one just need to make the substitution
$T\rightarrow k_\mathrm{B} T$.

\section{General facts}
\label{sec:EoS_e}

The Fermion distribution function is
\begin{equation}
\label{eq:fdirac}
f(\mathbf{k})=\frac1{h^3}\frac g{1+\mathrm e^{\frac{\mathcal E(\mathbf{k})-\mu}{T}}},
\end{equation}
which has the dimension of $\mbox{number}/(\mbox{length}\times
\mbox{momentum})^3=1/\mbox{action}^3$. Here, $\mathbf k$ is the particle momentum, $\mathcal E$ is the single-particle spectrum, $\mu$ is
the particle chemical potential, $T$ is the temperature, and $g$ is the particle degeneracy ($g=2$ for
electrons, muons, and tauons and $g=1$ for neutrinos).
We include the mass in the chemical potential, that is, we include the rest mass energy
in the single-particle spectrum. The relativistic expression
of the single-particle spectrum is
\begin{equation}
\label{eq:Erel}
\mathcal E(\mathbf{k})=c\sqrt{k^2+m^2c^2},
\end{equation}
where $m$ is the particle mass. Since for the free particle the dependence of the energy spectrum
on the particle momentum does not depends on the momentum direction, we will drop the dependence on the direction from
here on, $\mathbf k\rightarrow k$. The Fermi momentum is defined as the momentum of a particle with energy equal
to its chemical potential, $\mathcal E(k_F)\equiv\mu$,
\begin{equation}
\label{eq:kFrel}
k_F=\sqrt{\mu^2/c^2-m^2c^2}.
\end{equation}

The distribution function [Eq.~\eqref{eq:fdirac}] admits two different limits:
(i) the ultra-degenerate limit, for which $\mu-mc^2\gg T$,
\begin{equation}
\label{eq:fultradeg}
f^\mathrm{UD}(k)=\frac g{h^3} \Theta\big(\mathcal E(k)-\mu\big),
\end{equation}
where $\Theta$ is the theta-function, and (ii) the non-degenerate limit, for which $\mu-mc^2\ll T$,
\begin{equation}
\label{eq:fnondeg}
f^\mathrm{ND}(k)=\frac g{h^3} \mathrm e^{-\frac{\mathcal E(k)-\mu}T}.
\end{equation}
We remark that the ultra-degenerate limit is equivalent to keep the limit $T\to0$, and
therefore is also called zero temperature limit.

The relativistic spectrum [Eq.~\eqref{eq:Erel}] has two different limits:
(i) the non-relativistic limit, for which $mc\gg k$,
\begin{align}
\label{eq:Enonrel}
\mathcal E^\mathrm{NR}(k)={}&\frac{k^2}{2m}+mc^2,\\
\label{eq:kFnonrel}
k^\mathrm{NR}_F={}&\sqrt{2m(\mu-mc^2)},
\end{align}
and (ii) the ultra-relativistic limit, for which $k \gg mc$,
\begin{align}
\label{eq:Eultrarel}
\mathcal E^\mathrm{UR}(k)={}&kc,\\
\label{eq:kFultrarel}
k^\mathrm{UR}_F={}&\frac \mu c.
\end{align}
In the non-degenerate case, one can employ the non-relativistic limit if $T\ll mc^2$, and the ultra-relativistic limit
if $T\gg mc^2$; in the ultra-degenerate case, one can employ the non-relativistic limit if $\mu-mc^2\ll mc^2$, and the ultra-relativistic limit
if $\mu-mc^2\gg mc^2$.

The thermodynamical quantities of interest are
\begin{align}
\label{eq:n}
n(\mu,T)={}&4\pi\int_0^\infty f(k)k^2\mathrm dk,\\
\label{eq:eps}
\epsilon(\mu,T)={}&4\pi\int_0^\infty \mathcal E(k) f(k)k^2\mathrm dk,\\
\label{eq:P}
P(\mu,T)={}&\frac{4\pi}3\int_0^\infty kvf(k)k^2\mathrm dk,\\
\label{eq:sigma}
\sigma(\mu,T)={}&-4\pi\int_0^\infty \big[f\log f +(1-f) \log(1-f) \big] k^2 \mathrm dk,
\end{align}
where $n$ is the particle density, $\epsilon$ the energy density, $P$ the pressure,
$\sigma\equiv n_\mathrm{B}s$ the entropy density, and the velocity $v$ is defined by the relation
\begin{equation}
\label{eq:v}
v=\frac{\partial \mathcal E(k)}{\partial k}.
\end{equation}

Before solving the Integrals~\eqref{eq:n}--\eqref{eq:sigma} in the different limits, we remark that
\begin{itemize}
\item in the non-relativistic case,
\begin{equation}
\label{eq:vnonrel}
v^\mathrm{NR}=\frac km,
\end{equation}
and therefore
\begin{equation}
\epsilon^\mathrm{NR}=\frac32 P^\mathrm{NR} + n^\mathrm{NR}mc^2;
\end{equation}
\item in the ultra-relativistic case,
\begin{equation}
\label{eq:vultrarel}
v^\mathrm{UR}=c,
\end{equation}
and therefore
\begin{equation}
\epsilon^\mathrm{UR}=3P^\mathrm{UR};
\end{equation}
\item in the ultra-degenerate case, since the distribution function is proportional to the theta function,
\begin{equation}
\label{eq:sultradeg} \sigma^\mathrm {UD}=0.
\end{equation}
This was expected, since the ultra-degenerate limit is equivalent to the limit $T\to 0$, for which the entropy vanishes;
\item the entropy density may also be obtained from the general thermodynamical Relation~\eqref{eq:fundamental},
\begin{equation}
\label{eq:fundamental2}
\sigma(\mu,T)=\frac{\epsilon(\mu,T)+P(\mu,T)-n(\mu,T)\mu}{T},
\end{equation}
that holds for a system which is scale-invariant, see Sec.~\ref{sec:thermodynamical}.
\end{itemize}

In the following sections we consider the analytic expressions for the
thermodynamical quantities in some relevant limits of the single-particle free
Fermion gas. To consider a gas composed by more than one particle species, one
can use the method described in Sec.~\ref{sec:totaleos}. For a fit of the most
interesting and most general case, that is, the degenerate relativistic case
[that corresponds to the distribution function \eqref{eq:fdirac} and the energy
spectrum \eqref{eq:Erel}], we refer the reader to
\citet{Eggleton+Faulkner+Flannery.1973}, and \citet{Johns+Ellis+Lattimer.1996}.

\section{Ultra-degenerate, non-relativistic}
\label{sec:UDNR}

\begin{align}
\label{eq:n_UDNR}
n={}&\frac{4\pi g}{h^3}\frac{k_\mathrm{F}^3}{3},\\
\label{eq:kF_UDNR}
k_F={}&h\left(\frac{3n}{4\pi g}\right)^{1/3},\\
\label{eq:eps_UDNR}
\epsilon={}&\frac{4\pi g}{h^3}\frac{k_\mathrm{F}^5}{10m}+nmc^2,\\
\label{eq:P_UDNR}
P={}&\frac{4\pi g}{h^3}\frac{k_\mathrm{F}^5}{15m},\\
\label{eq:sigma_UDNR}
\sigma={}&0.
\end{align}

\section{Ultra-degenerate, ultra-relativistic}
\label{sec:UDUR}

\begin{align}
\label{eq:n_UDUR}
n={}&\frac{4\pi g}{h^3}\frac{k_\mathrm{F}^3}{3},\\
\label{eq:kF_UDUR}
k_F={}&h\left(\frac{3n}{4\pi g}\right)^{1/3},\\
\label{eq:eps_UDUR}
\epsilon={}&\frac{\pi gc}{h^3}k_\mathrm{F}^4,\\
\label{eq:P_UDUR}
P={}&\frac{\pi gc}{3h^3}k_\mathrm{F}^4,\\
\label{eq:sigma_UDUR}
\sigma={}&0.
\end{align}

\section{Non-degenerate, non-relativistic}
\label{sec:NDNR}

\begin{align}
\label{eq:n_NDNR}
n={}&\frac{4\pi g}{h^3\mathrm e^{\frac{mc^2-\mu}{T}}}\int_0^\infty
k^2\mathrm e^{-\frac{k^2}{2mT}}\mathrm
dk=\left[x^2\equiv\frac{k^2}{2mT}\right]\notag\\
={}&\frac{4\pi g(2mT)^{3/2}}{h^3\mathrm e^{\frac{mc^2-\mu}{T}}} \int_0^\infty
x^2\mathrm e^{-x^2}\mathrm
dx\notag\\
={}&\frac{g(2\pi mT)^{3/2}}{h^3}\mathrm e^{\frac{\mu-mc^2}{T}},\\
\label{eq:eps_NDNR}
\epsilon={}&\frac{4\pi gT(2mT)^{3/2}}{h^3\mathrm e^{\frac{mc^2-\mu}{T}}} \int_0^\infty
x^4\mathrm e^{-x^2}\mathrm
dx + nmc^2\notag\\
={}&\frac32 nT +nmc^2,\\
\label{eq:P_NDNR}
P={}&nT,\\
\label{eq:s_NDNR}
\sigma={}&\frac52n +\frac{mc^2-\mu}{T}n\simeq\frac52n,
\end{align}
where we have used
\begin{align}
\int_0^\infty x^2\mathrm e^{-x^2}\mathrm dx={}&\frac{\sqrt{\pi}}4,\\
\int_0^\infty x^4\mathrm e^{-x^2}\mathrm dx={}&\frac{3\sqrt{\pi}}8.
\end{align}

\section{Non-degenerate, ultra-relativistic}
\label{sec:NDUR}

\begin{align}
\label{eq:n_NDUR}
n={}&\frac{4\pi g\mathrm e^{\mu/T}}{h^3}\int_0^\infty
k^2\mathrm e^{-ck/T}\mathrm
dk=\left[x\equiv\frac{ck}{T}\right]\notag\\
={}&\frac{4\pi g\mathrm e^{\mu/T}}{h^3} \left(\frac Tc\right)^3\int_0^\infty
x^2\mathrm e^{-x}\mathrm
dx\notag\\
={}&\frac{8\pi gT^3}{(hc)^3}\mathrm e^{\mu/T},\\
\label{eq:eps_NDUR}
\epsilon={}&\frac{4\pi g\mathrm e^{\mu/T}}{h^3} \left(\frac Tc\right)^4c\int_0^\infty
x^3\mathrm e^{-x}\mathrm
dx=\frac{24\pi g\mathrm e^{\mu/T}}{(hc)^3} T^4\notag\\
={}&3 nT,\\
\label{eq:P_NDUR}
P={}&nT,\\
\label{eq:sigma_NDUR}
\sigma={}&4n -\frac{\mu}{T}n\simeq4n,
\end{align}
where we have used
\begin{align}
\int_0^\infty x^2\mathrm e^{-x}\mathrm dx={}&2,\\
\int_0^\infty x^4\mathrm e^{-x}\mathrm dx={}&6.
\end{align}

\section{Degenerate, non-relativistic}
\label{sec:degNR}

\begin{align}
\label{eq:n_degNR}
n={}&\frac{4\pi g}{h^3}\int_0^\infty \frac{k^2}{1+\mathrm e^{\frac{k^2}{2mT}+\frac{mc^2-\mu}T}}\mathrm dk\notag\\
={}&\Big[x^2\equiv \frac{k^2}{2mT}\Big]=\frac{4\pi g (2mT)^{3/2}}{h^3}\int_0^\infty\frac{x^2}{1+\mathrm e^{x^2-\frac{\mu-mc^2}{T}}}\mathrm dx\notag\\
={}&-\frac {g(2\pi mT)^{3/2}}{h^3}\operatorname{Li_{3/2}}\left(-\mathrm e^{\frac{\mu-mc^2}T}\right),\\
\label{eq:eps_degNR}
\epsilon={}&\frac{4\pi g(2mT)^{5/2}}{2mh^3}\int_0^\infty\frac{x^4}{1+\mathrm e^{x^2-\frac{\mu-mc^2}{T}}}\mathrm dx+nmc^2\notag\\
={}&-\frac32\frac{g(2\pi mT)^{3/2}T}{h^3}\operatorname{Li_{5/2}}\left(-\mathrm e^{\frac{\mu-mc^2}T}\right)+nmc^2,\\
\label{eq:P_degNR}
P={}&-\frac{g(2\pi mT)^{3/2}T}{h^3}\operatorname{Li_{5/2}}\left(-\mathrm e^{\frac{\mu-mc^2}T}\right),\\
\label{eq:sigma_degNR}
\sigma={}&-\frac {g(2\pi mT)^{3/2}}{h^3}\left[
\frac52\operatorname{Li_{5/2}}\left(-\mathrm e^{\frac{\mu-mc^2}T}\right)\right.\notag\\
{}&\left.+\frac{mc^2-\mu}T\operatorname{Li_{3/2}}\left(-\mathrm e^{\frac{\mu-mc^2}T}\right)\right],
\end{align}
where we have used the relations
\begin{align}
\int_0^\infty \frac{x^2} {\mathrm e^{x^2-a} + 1} \mathrm dx={}&
-\frac{\sqrt\pi}{4} \operatorname{Li_{3/2}}(-\mathrm e^a),\\
\int_0^\infty \frac{x^4} {\mathrm e^{x^2-a} + 1} \mathrm dx={}&
-\frac{3\sqrt\pi}{8}\operatorname{Li_{5/2}}(-\mathrm e^a),
\end{align}
and $\operatorname{Li_{3/2}}$ and $ \operatorname{Li_{5/2}}$ 
are polylogarithmic functions.

\section{Degenerate, ultra-relativistic}
\label{sec:degUR}

\begin{align}
\label{eq:n_degUR}
n={}&\left[x\equiv \frac {ck}{T}\right]= \frac{4\pi gT^3}{(h c)^3} \int_0^\infty \frac{x^2} {\mathrm
e^{x-\mu/T}+1} \mathrm d x\notag\\
={}& -\frac{8\pi gT^3}{(h c)^3} \operatorname{Li_3}\big(-\mathrm
e^{\mu/T}\big),\\
\label{eq:eps_degUR}
\epsilon={}& \frac{4\pi gT^4}{(h c)^3} \int_0^\infty \frac{x^3} {\mathrm
e^{x-\mu/T} +1} \mathrm d x\notag\\
={}& - \frac{24 \pi g T^4}{(h c)^3}
\operatorname{Li_4}\big(-\mathrm e^{\mu/T}\big),\\
\label{eq:P_degUR}
P={}& - \frac{8 \pi g T^4}{(h c)^3}
\operatorname{Li_4}\big(-\mathrm e^{\mu/T}\big),\\
\label{eq:s_degUR}
\sigma={}&\frac{8\pi g T^2}{(h c)^3} \Big(\mu
\operatorname{Li_3}\big(-\mathrm e^{\mu/T}\big) -4T \operatorname{Li_4} \big(-\mathrm
e^{\mu/T} \big)\Big),
\end{align}
where we used the relations
\begin{align}
\int_0^\infty \frac{x^2} {\mathrm e^{x-a} + 1} \mathrm dx={}&
-2\operatorname{Li_3}(-\mathrm e^a),\\
\int_0^\infty \frac{x^3} {\mathrm e^{x-a} + 1} \mathrm dx={}&
-6\operatorname{Li_4}(-\mathrm e^a),
\end{align}
where $\operatorname{Li}_n(x)$ is the polylogarithmic function \citep{Lewin.1981}.

Using the relations [\citealp[][Eqs.~(6.6), (7.81), and (7.82)]{Lewin.1981}]
\begin{align}
\label{eq:Li3}
\operatorname{Li_3}(-x)-\operatorname{Li_3}(-1/x) ={}& -\frac {\pi^2}6 \log x
-\frac16 \log^3x,\\
\label{eq:Li4}
\operatorname{Li_4}(-x)+\operatorname{Li_4}(-1/x) ={}& -\frac {7\pi^4}{360}
-\frac1{24} \log^4 x - \frac{\pi^2}{12} \log^2x,
\end{align}
one obtains the net thermodynamical quantities for a free Fermion gas of
particles and antiparticles in chemical equilibrium with respect to pair
production \cite[][Appendix~C]{Lattimer+Swesty.1991}:
\begin{align}
\label{eq:n_net}
n-\bar n={}&\frac{8\pi g}{6(hc)^3} \mu\left(\mu^2 +\pi^2T^2\right),\\
\label{eq:eps_net}
\epsilon+\bar\epsilon={}&\frac{\pi g}{(hc)^3}\left(\mu^4 +
2\pi^2T^2\mu^2 +\frac7{15}\pi^4T^4\right),\\
\label{eq:P_net}
P+\bar P={}&\frac{\epsilon+\bar \epsilon}{3},\\
\label{eq:sigma_net}
\sigma+\bar \sigma={}&\frac{4\pi^3 g}{3(hc)^3} T\left(\mu^2 + \frac7{15}\pi^2
T^2\right),
\end{align}
where the bar refers to the anti-particle quantities.

As a final remark, we notice that the Properties~\eqref{eq:Li3} and \eqref{eq:Li4} permit
to simplify the numerical evaluation of the polylogarithmic functions. In fact
\citep[see e.g.~Subsec.~A.4.2 of][]{Lewin.1981} one could tabulate the
polylogarithm function between $-1$ and $0$, and then use Eqs.~\eqref{eq:Li3}
and \eqref{eq:Li4} to determine the value of the polylogarithm from $-\infty$
to $0$. In case of need, another simple relation permits to determine the value of the
polylogarithm between $0$ and $1$ \citep[][Subsec.~A.4.2]{Lewin.1981}. This
trick may be applied only to integer order polylogarithmic functions; in fact the
relations equivalent to Eqs.~\eqref{eq:Li3} and \eqref{eq:Li4} for
semi-integer polylogarithmic functions, like $\operatorname{Li_{3/2}}$ or
$\operatorname{Li_{5/2}}$, are much more complicated.

\chapter{Code checks}
\label{app:code_checks}

In this appendix we show the checks of the accuracy of the code (Sec.~\ref{sec:conservation}),
of the assumption of beta-equilibrium (Sec.~\ref{sec:betaequilibrium}), and of the
assumption that matter is composed by an interacting Fermion gas of baryons (Sec.~\ref{sec:gas}).
For simplicity, we show the results of a PNS evolved with the CBF-EI EoS and with total baryon mass $M_\mathrm{B}=\unit[1.60]{M_\odot}$,
but the results for the other EoSs and the other baryon masses are similar.

\section{Energy and lepton number conservation}
\label{sec:conservation}

\begin{figure}
\centering
\includegraphics[width=\textwidth]{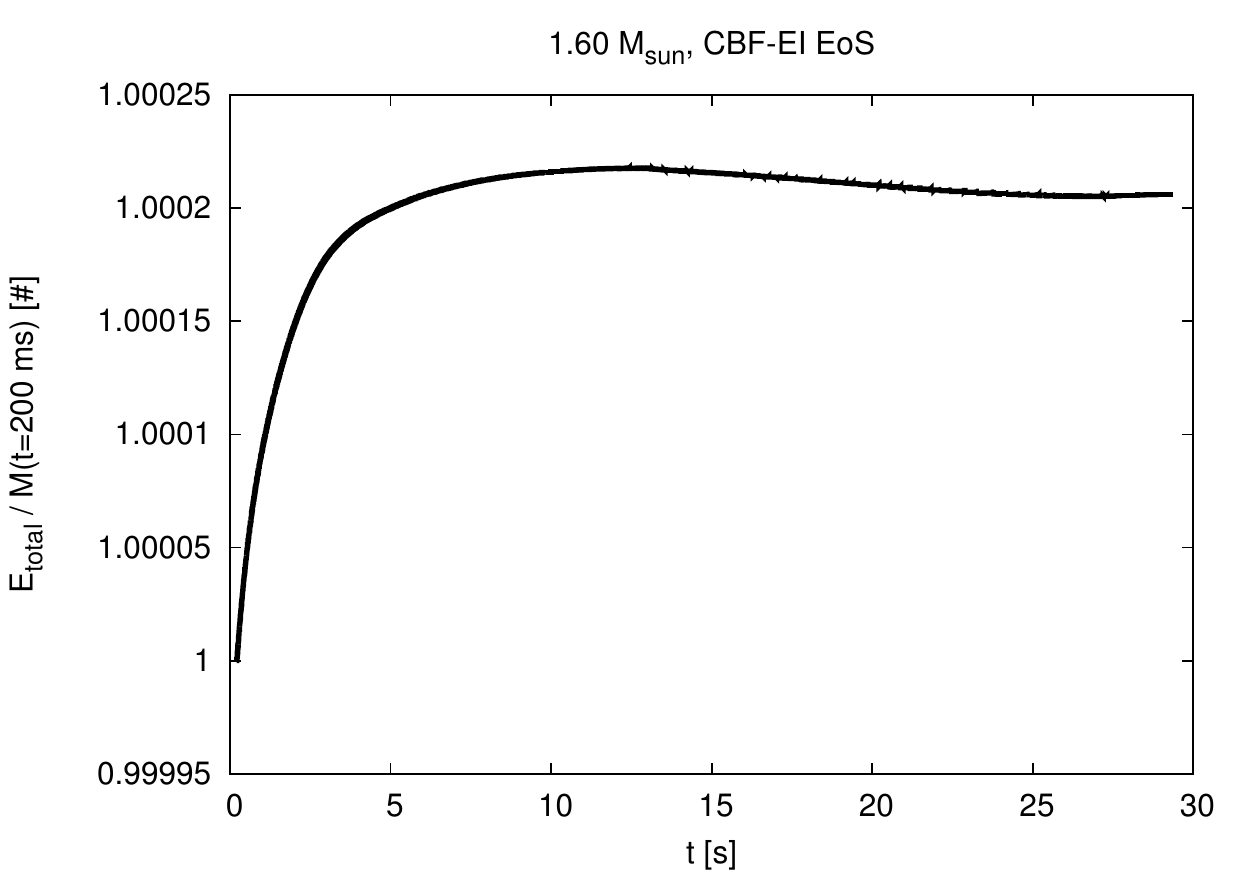}
\includegraphics[width=\textwidth]{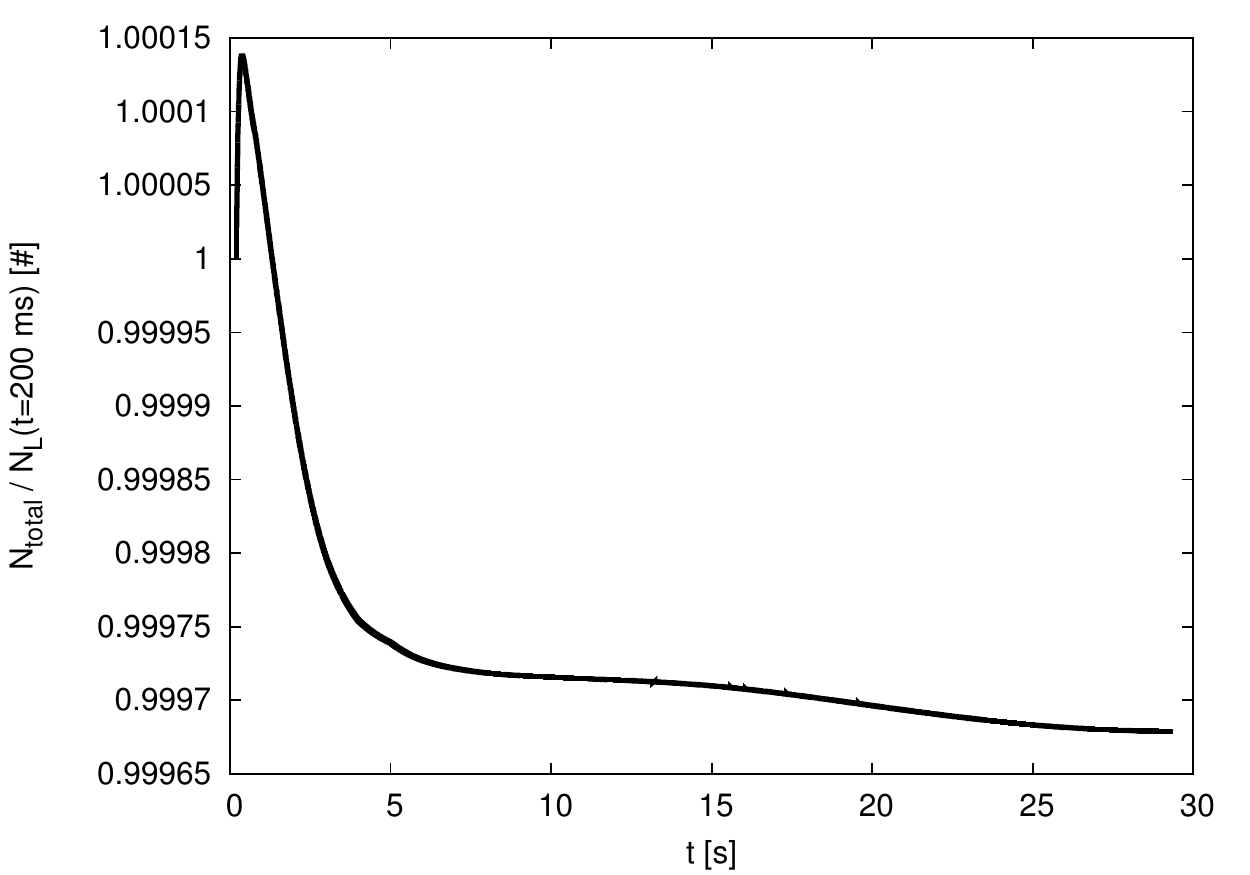}
\caption{Total energy $E_\mathrm{total}$ and total lepton number $N_\mathrm{total}$
conservation for a PNS with the CBF-EI EoS and
$\unit[1.60]{M_\odot}$ baryon mass, normalized with the stellar initial energy and
lepton number (our simulations start at $\unit[200]{ms}$, see Sec.~\ref{sec:code}). The timestep is changed during the
evolution in such a way that the relative variation in a timestep of the profiles of entropy per baryon and lepton fraction
is approximately equal to $10^{-4}$.
}
\label{fig:conservation}
\end{figure}

\begin{figure}
\centering
\includegraphics[width=\textwidth]{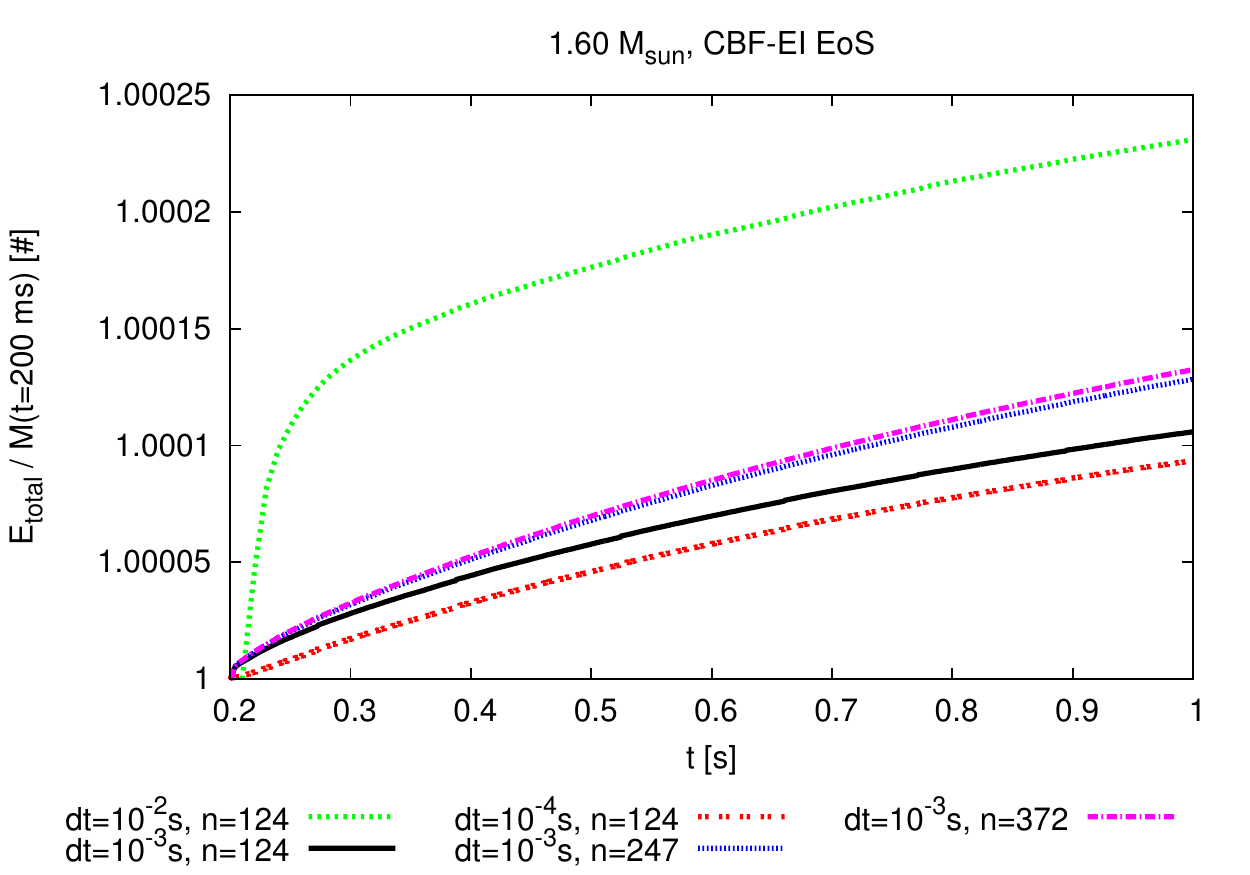}
\includegraphics[width=\textwidth]{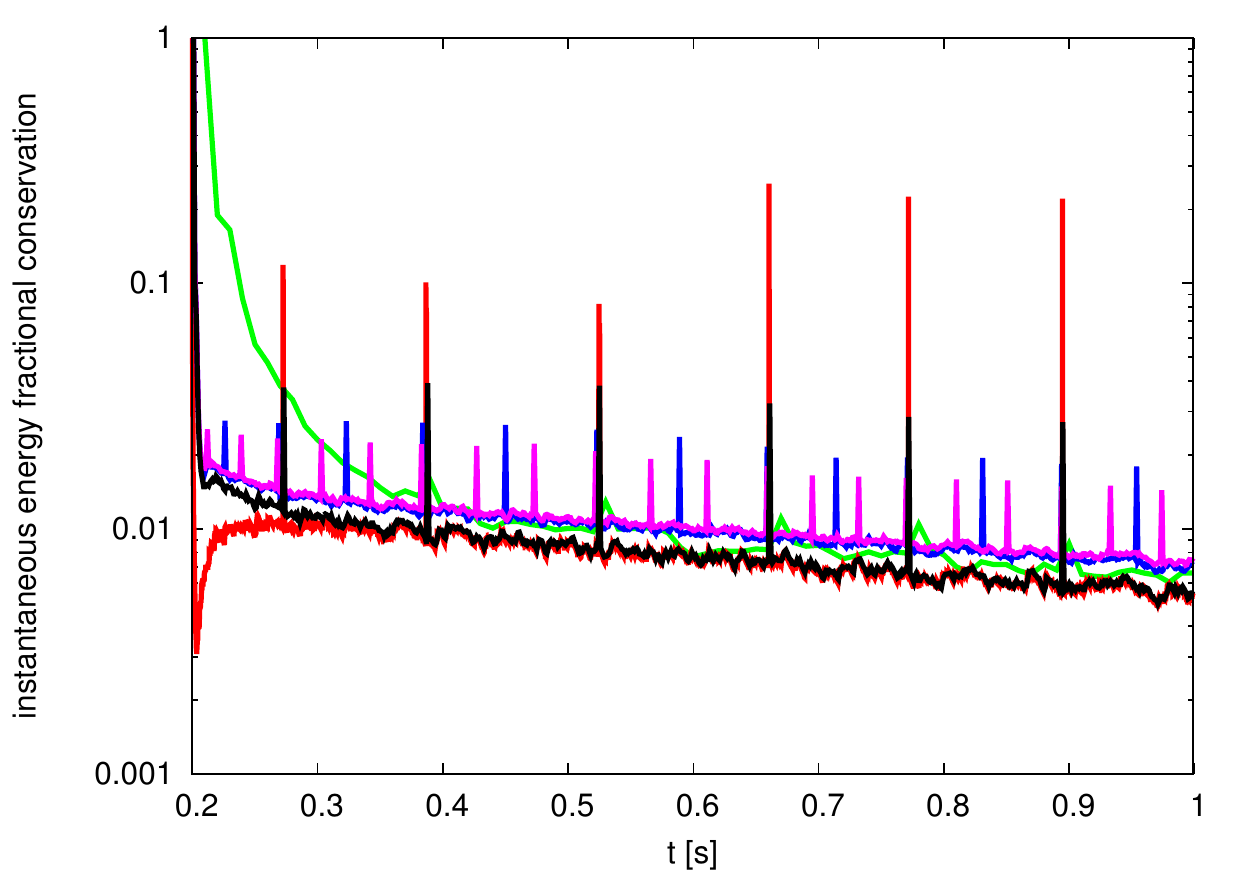}
\caption{Total energy
$E_\mathrm{total}$ (normalized with the stellar initial energy) and
instantaneous energy fractional conservation for a PNS with the CBF-EI EoS and
$\unit[1.60]{M_\odot}$ baryon mass. The timestep \texttt{dt} is kept fixed
during the evolution and \texttt{n} is the number of grid points.
The plots begin at $\unit[200]{ms}$ because this is the initial
time of our simulations (see Sec.~\ref{sec:code}).
}
\label{fig:conservation_bis}
\end{figure}

The total energy and other quantum numbers (i.e., the baryon number) are
conserved in every physical process. Our code enforces the conservation of the
total baryon number $A=M_\mathrm{B}/m_n$, but as it evolves, the PNS loses
energy and lepton number since neutrinos are allowed to escape from the star.
Since the total energy of a star (matter \emph{plus} neutrinos)
in spherical symmetry is given by its gravitational mass $M$, the total energy
of the system (stellar energy \emph{plus} energy of the emitted neutrinos) is given by
\begin{equation}
E_\mathrm{total}=M(t)+\int_{\unit[200]{ms}}^t L_\nu(t)\mathrm dt,
\end{equation}
where $L_\nu$ is defined in Eq.~\eqref{eq:Lnu_Mdot}.
Similarly, for the electron lepton number,
\begin{align}
N_\mathrm{total}={}&N_L(t) +
\int_{\unit[200]{ms}}^t4\pi R^2\mathrm e^{\phi(R)}F_\nu(R)\mathrm dt,\\
N_L ={}&
\int_0^{A} Y_L(a) \mathrm da,
\end{align}
where $N_L$ is the total number of electronic leptons in the star, and $F_\nu$ is
the electron neutrino number flux (we do not account for the other lepton numbers since
we do not include muons and tauons in the EoSs and moreover
$\mu_{\nu_\mu}=\mu_{\nu_\tau}=0$).

Since the conservation of $E_\mathrm{total}$ and
$N_\mathrm{total}$ has not been enforced, they provide a test for our
simulations. From Fig.~\ref{fig:conservation} and from the top plot of Fig.~\ref{fig:conservation_bis},
it is clear that they are conserved better than about $0.03\%$
during the evolution.

In Fig.~\ref{fig:conservation_bis} we show, for different \emph{fixed}
timesteps and for different grid \emph{dimensions}, the total and instantaneous
energy fractional conservation from 0.2 s to 1 s. The instantaneous energy fractional
conservation is defined as
\begin{equation}
\mathrm{i.e.f.c.}=\frac{|\dot M+L_\nu|}{L_\nu}.
\end{equation}
We see that reducing the timestep the energy conservation is improved. The
instantaneous energy fractional conservation as a function of time shows
regular spikes, whose number doubles (triples) if we double (triple) the grid
points, and whose magnitude is approximately inversely proportional to the
timestep.  
We explain these spikes with the non-linearity of the transport equations (\cite{Press+book},
Sec.~19.1). In fact,
the temperature and the neutrino degeneracy appear inside and outside the
gradients in the transport equations [Eqs.~\eqref{eq:Fnufinal}--\eqref{eq:sfinal}].
As a consequence, the power in the Fourier space is accumulated in the shorter
wavelengths and is finally released in the longer wavelengths of the solution.
This explains why the frequency of the peaks changes with the grid spacing.
The spikes of the instantaneous energy conservation have magnitudes which
increase when the timestep is lowered, since one is dividing over a smaller
timestep an approximately constant energy jump, $[M(t+\mathrm dt)-M(t)]/\mathrm
dt$.
These spikes do not undermine the overall conservation of the energy and lepton
number and the PNS evolution, see Fig.~\ref{fig:conservation} and upper plot of
Fig.~\ref{fig:conservation_bis}.

\section{Beta-equilibrium assumption}
\label{sec:betaequilibrium}

Our code (as \citealp{Keil+Janka.1995}) assumes beta-equilibrium, Eqs.~\eqref{eq:beta-mu}.
This approximation is valid if the timescale of the beta-equilibrium is shorter than the
dynamical timescale. We estimate the beta-equilibrium timescale using Eqs.~(16) and (17a) of
\citet{Burrows+Lattimer.1986},
\begin{align}
t_\mathrm{beta}={}&\frac{1}{D_n},\\
D_n={}&1.86\times 10^{-2} Y_pT^5[S_4(\eta_e)-S_4(\eta_\nu)]\frac{1-\mathrm e^{-\Delta/T}}{1-\mathrm e^{-\eta_e+\eta_\nu}} \unit{\frac{neutrinos}{baryon\cdot s}},\\
S_4(y)={}&\frac{y^5}5+2\pi^2\frac{y^3}{3} + 7\pi^4\frac{y}{15},
\end{align}
where $D_n$ is the net rate of production of electron-neutrinos, $\eta=\mu/T$ is the degeneracy parameter,
and $\Delta=0$ in the case of beta-equilibrium (we refer the reader to
\citealp{Burrows+Lattimer.1986} for more details). Since we have assumed
beta-equilibrium, we put $1-\exp(-\Delta/T)\equiv1$ to estimate the corresponding timescale. This means that the value of the beta-equilibrium
timescale is not fully consistent.

We estimate the dynamical timescale with the formula
\begin{equation}
t_\mathrm{dyn}=R\frac{n_\nu(r)}{F_\nu(r)},
\end{equation}
where $n_\nu(r)$ and $F_\nu(r)$ are the neutrino number density and number flux, respectively (which depend on the radial coordinate $r$),
and $R$ is the stellar radius (notice that $F_\nu/n_\nu$ has the dimension of a velocity).

\begin{figure}
\centering
\includegraphics[width=\textwidth]{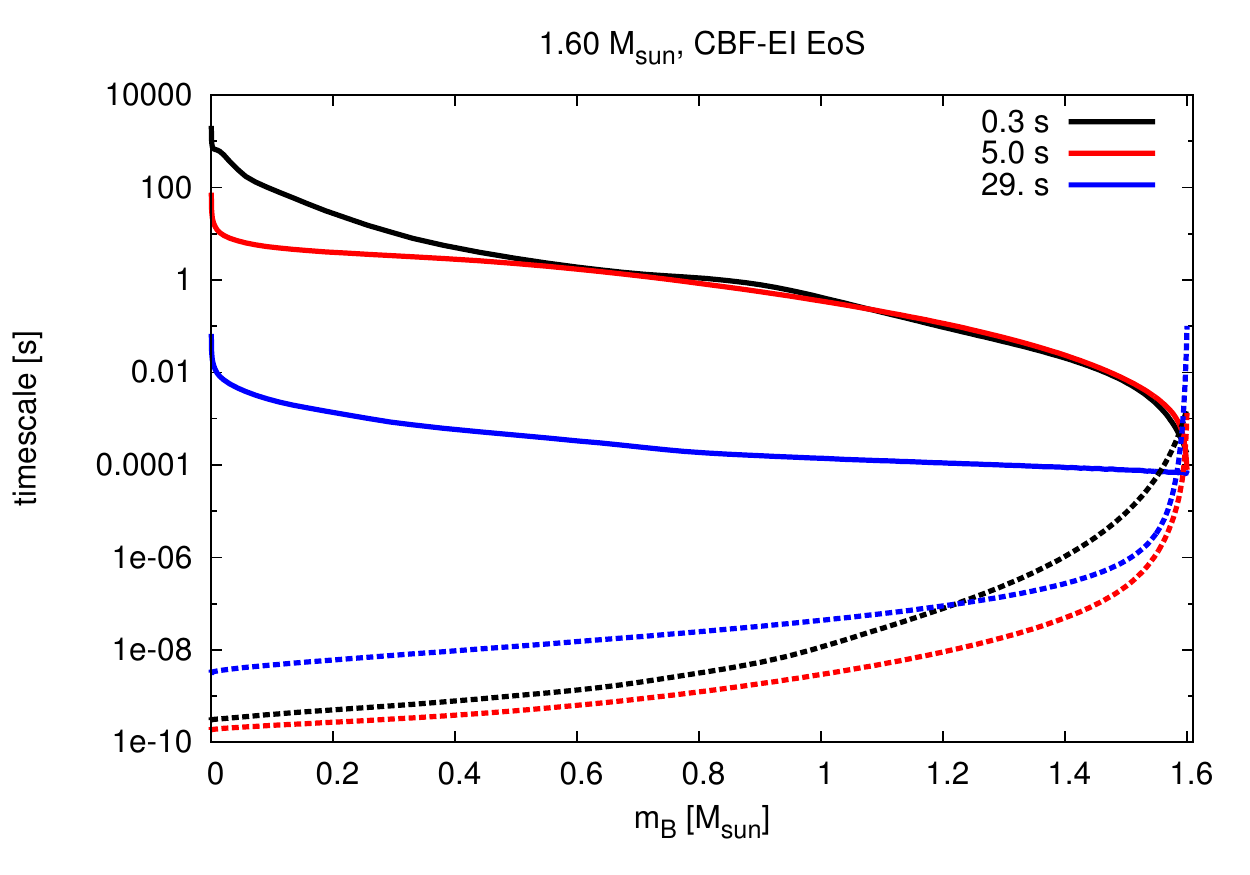}
\caption{Dynamical timescale (solid lines) and beta-equilibrium timescale (dashed lines) profiles at different
times for a PNS with the CBF-EI EoS and $\unit[1.60]{M_\odot}$ stellar mass.}
\label{fig:timescales}
\end{figure}

In Fig.~\ref{fig:timescales} we plot the dynamical and beta-equilibrium
timescales for a PNS with the CBF-EI EoS and
$M_\mathrm{B}=\unit[1.60]{M_\odot}$.  The beta-equilibrium is valid in most of
the star, apart for a thin shell near the
border.  Towards the end of the evolution the dynamical timescale seems to
reduce, and this is conterintuitive. In fact, we have associated the dynamical
timescale with the neutrino timescale. This is not true towards the end of the
evolution, since as the PNS becomes optically thin the neutrinos decouple from
the matter and the diffusion approximation breaks down.  At that point, the
neutrino timescale drops, but the stellar dynamics is actually frozen.

\section{Interacting nucleon gas assumption}
\label{sec:gas}

\begin{figure}
\centering
\includegraphics[width=\textwidth]{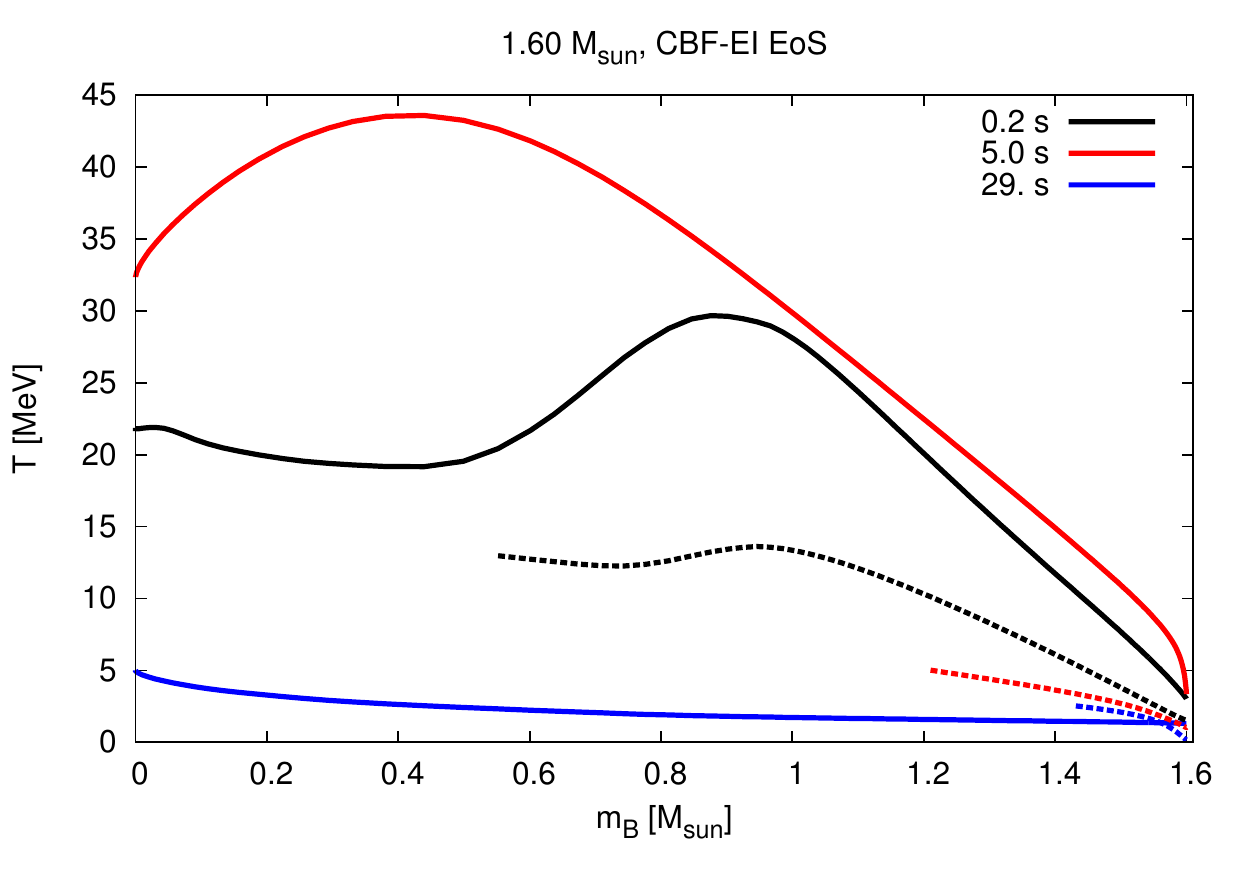}
\caption{Stellar temperature (solid lines) and critical temperature (dashed
lines) for the formation of alpha particles at different times, for a PNS with the CBF-EI EoS and
$\unit[1.60]{M_\odot}$ baryon mass.  When the
baryon density reaches the nuclei density, $\unit[0.155]{fm^{-3}}$, alpha
particles could not form and we do not plot the critical temperature anymore.}
\label{fig:Tc}
\end{figure}

In this thesis we have not consider the formation of any kind of crust or
envelope, that is, the EoS baryon part is made by an interacting
gas of protons and neutrons.  However, at low temperature and baryon density,
the matter is not constitute only by a gas of baryons.
The alpha particles (i.e., Helium nuclei) are the first species that appear
decreasing the temperature and the density.  The critical temperature at which
alpha particles begin to form, that is, the
lowest temperature at which protons and neutrons are present only as an
interacting gas, depends on the baryon density and the proton fraction.
Eq.~(2.31) of \citet[][]{Lattimer+Swesty.1991} is an estimate of this critical temperature,
\begin{equation}
\label{eq:Tcritical}
T_c(Y_p)=87.76\left(\frac{K_s}{\unit[375]{MeV}}\right)^{1/2}\left(\frac{\unit[0.155]{fm^{-3}}}{n_s}\right)^{1/3}Y_p(1-Y_p)\unit[\,]{MeV},
\end{equation}
where $n_s$ and $K_s$ are the saturation density and the imcompressibility
parameter at saturation density of symmetric nuclear matter (see
Sec.~\ref{sec:LSbulk}). Eq.~\eqref{eq:Tcritical} is valid for
$n_\mathrm{B}<n_s$, otherwise no alpha particles may form.  In
Fig.~\ref{fig:Tc} we report the profiles of the critical temperature and the
PNS temperature for different snapshots of a PNS
with the CBF-EI EoS and with $M_\mathrm{B}=\unit[1.60]{M_\odot}$; the results
for the other EoSs and baryon masses are similar. As expected, the assumption of
a proton-neutron interacting gas is valid at the beginning of the
simulation and loses accuracy towards the end of the evolution,
when it is not valid only in the outermost layers.

\chapter{On the neutrino inverse reactions}
\label{app:inverse}
The relevant nuclear processes in the neutrino diffusion during the PNS
evolution are\footnote{In this thesis we do not include nucleon-nucleon Bremsstrahlung,
whose effects have been recently studied by \cite{Fischer.2016}.}
: the neutrino scattering on neutrons, protons, and electrons, the neutrino
absorption on neutrons, the antineutrino absorption on protons, and their inverse processes. 
Such reactions are accounted in Eqs.~\eqref{eq:reaction_start}--\eqref{eq:reaction_end}.
The Reactions~\eqref{eq:reaction_start}--\eqref{eq:reaction_end} are accounted
for in \citet{Reddy+Prakash+Lattimer.1998}, in \citet{Pons+.1999}, and in our
work at the tree level, see Figs.~\ref{fig:feynman} and \ref{fig:feynman2}.
These processes are the neutrino scattering on neutrons, protons, and
electrons, the neutrino absorption on neutrons, the antineutrino absorption on
protons, \emph{and} their inverse processes.  In this appendix we explain how
to account for the inverse process in the determination of the neutrino
cross-section.

\section{Direct reaction}
\label{sec:direct}
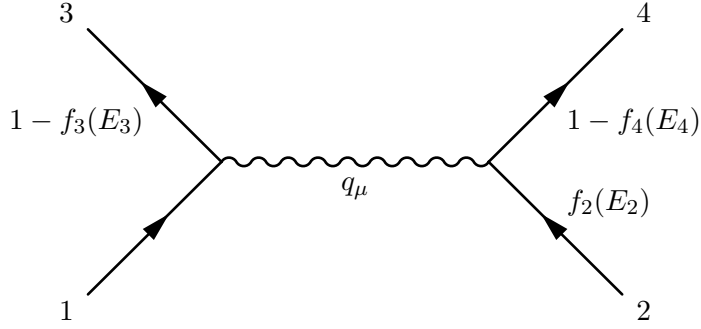
\begin{figure}
\centering
\begin{fmffile}{feynman}
\fmfframe (3,3)(3,3){
\begin{fmfgraph*}(250,100)
\fmfleft{i1,i2}
\fmfright{o1,o2}
\fmflabel{1}{i1}
\fmflabel{3}{i2}
\fmflabel{2}{o1}
\fmflabel{4}{o2}
\fmf{fermion}{i1,v1}
\fmf{fermion,label=$1-f_3(E_3)$,label.side=left}{v1,i2}
\fmf{fermion,label=$f_2(E_2)$,label.side=right}{o1,v2}
\fmf{fermion,label=$1-f_4(E_4)$,label.side=right}{v2,o2}
\fmf{photon,label=$q_\mu$,label.side=right}{v1,v2}
\end{fmfgraph*}
}
\end{fmffile}
\caption{Feynman diagram of the direct reaction: the incoming particles are 1 and 2, the outgoing particles are 3 and 4.
In the computation of the direct cross section [Eq.~\eqref{eq:eq82RPL98}], one has to account for the distribution function of particle 2
and the Pauli-blocking effect for particles 3 and 4.}
\label{fig:feynman}
\end{figure}

In Fig.~\ref{fig:feynman} we show the Feynman diagram corresponding to the
direct reaction, where one considers an incoming neutrino
(particle 1) with a given energy $E_1$ that interacts with another incoming
particle (particle 2). In the case of scattering, the outgoing particle 3 is a
neutrino, instead
in the case of absorption it is an electron or a positron.
To determine the neutrino cross-section
$\sigma_j^{\nu_i}(\omega)\equiv\sigma(E_1)$ in Eq.~\eqref{eq:mean_free_path} for the direct reactions,
that is, the cross-section of a neutrino $\nu_i$ with energy $\omega\equiv E_1$
in the \emph{direct} reaction $j$, we make use of Eq.~(82) of
\citet{Reddy+Prakash+Lattimer.1998},
\begin{align}
\label{eq:eq82RPL98}
\frac{\sigma(E_1)}V ={}& \frac{G_F^2}{2\pi^2E_1^2}\int_{-\infty}^{E_1} \mathrm
dq_0\frac{1-f_3(E_3)}{1-\exp\left[\frac{-q_0-(\mu_2-\mu_4)} {T}\right]}\notag\\
{}&\times\int_{|q_0|}^{E_1+E_3}\mathrm dq qq_\mu^2[AR_1
+R_2+BR_3].
\end{align}
As in \citet{Reddy+Prakash+Lattimer.1998}, we have set $c=h=1$, $G_F$ is the Fermi weak coupling constant, $f_i(E_i)$ is the
distribution function of the particle $i$ with energy $E_i$ and chemical potential $\mu_i$,
$q^\mu=p_1^\mu-p_3^\mu$ is the transferred four-momentum, $A=(4E_1E_3+q_\mu^2)/(2q^2)$, $B=E_1+E_3$, and $R_1$, $R_2$, and $R_3$ are the
response functions \citep{Reddy+Prakash+Lattimer.1998}.  The integration limits for $q_0$, namely $-\infty$
and $E_1$, are due to the fact that $q_0=E_1-E_3$ and $E_3>0$. The integration
limits for $q$ can be obtained from $q^2=E_1^2+E_3^2-2E_1E_3\cos\theta_{13}$,
where $\theta_{13}$ is the angle between $\mathbf p_1$ and $\mathbf p_3$.  We
write the upper limit of the integration of $q$ as $E_1+E_3$ instead of as
$2E_1-q_0$ as in \citet{Reddy+Prakash+Lattimer.1998}, because this will
simplify the discussion of the inverse reaction.
It is useful to rewrite Eq.~\eqref{eq:eq82RPL98} as
\begin{align}
\label{eq:rewrite_eq82}
\frac{\sigma^\mathrm{d}(E_1)}V ={}& \int_{-\infty}^{E_1} \mathrm
dq_0\big[1-f_3(E_3)\big] \tilde W(q_0,2,4),\\
\label{eq:tildeW}
\tilde W(q_0,2,4) ={}&
\frac{G_F^2}{2\pi^2E_1^2}\left(1-\exp\left[\frac{-q_0-(\mu_2-\mu_4)}
{T}\right]\right)\notag\\
{}&\times\int_{|q_0|}^{E_1+E_3}\mathrm dq qq_\mu^2[AR_1
+R_2+BR_3],
\end{align}
where the superscript ``d'' means that it is the cross-section of the direct
reaction and we put a tilde in the conveniently defined function $\tilde W$
because it is not a transition rate, and
it depends on the properties of particles 2 and 4 (their masses, chemical
potentials, and effective spectra, see \citealp{Reddy+Prakash+Lattimer.1998}).

\section{Inverse reaction}
\label{sec:inverse}
\begin{figure}
\centering
\begin{fmffile}{feynman2}
\fmfframe (3,3)(3,3){
\begin{fmfgraph*}(250,100)
\fmfleft{i1,i2}
\fmfright{o1,o2}
\fmflabel{1}{i1}
\fmflabel{3}{i2}
\fmflabel{2}{o1}
\fmflabel{4}{o2}
\fmf{fermion}{v1,i1}
\fmf{fermion,label=$f_3(E_3)$,label.side=left}{i2,v1}
\fmf{fermion,label=$1-f_2(E_2)$,label.side=right}{v2,o1}
\fmf{fermion,label=$f_4(E_4)$,label.side=right}{o2,v2}
\fmf{photon,label=$q_\mu$,label.side=right}{v2,v1}
\end{fmfgraph*}
}
\end{fmffile}
\caption{Feynman diagram of the inverse reaction: the incoming particles are 3 and 4, the outgoing particles are 1 and 2.
In the computation of the inverse cross section, one has to account for the distribution function of particles 3 and 4
and the Pauli-blocking effect for particle 2.}
\label{fig:feynman2}
\end{figure}
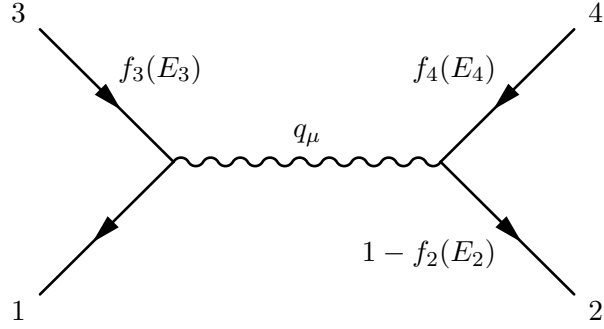

Eq.~\eqref{eq:rewrite_eq82} describes the direct process, that is, it is the
cross section of an incoming neutrino with energy $E_1$.  However, in
Eq.~\eqref{eq:mean_free_path} one needs to account also for the inverse
process, namely one should include also the cross section of
an outgoing neutrino with energy $E_1$, represented in Fig.~\ref{fig:feynman2}.
In the inverse reaction the outgoing neutrino (particle 1) with a given energy
$E_1$ is the result of the interaction of the incoming particles 3 and 4. In
the case of inverse scattering, the incoming particle 3 is a neutrino, instead
in the case of emission it is an electron or a positron.
The direct Feynman diagram in Fig.~\ref{fig:feynman} is formally identical to
the inverse one in Fig.~\ref{fig:feynman2}, provided that the following
substitutions are made
\begin{itemize}
\item $2\leftrightarrow4$;
\item $1-f_3(E_3)\rightarrow f_3(E_3)$;
\item $q_\mu\rightarrow-q_\mu$.
\end{itemize}
This correspond to the following expression for the inverse reaction (cf.{} Eq.~(2.13) of \citealp{Iwamoto+Pethick.1982}):
\begin{equation}
\label{eq:reaction_inverse}
\frac{\sigma^\mathrm{i}(E_1)}V =\int_{-\infty}^{E_1} \mathrm dq_0f_3(E_3) \tilde W(-q_0,4,2),
\end{equation}
where the superscript ``i'' denotes the cross-section of the inverse process
and $\tilde W$ is defined as in Eq.~\eqref{eq:tildeW}.  In
Eq.~\eqref{eq:reaction_inverse} the transferred momentum is
$q^\mu=p^\mu_3-p^\mu_1$, and therefore the boundaries of the integrals in $q_0$
and $q$ are the same as in Eq.~\eqref{eq:rewrite_eq82}. This would have not been the
case if we had written the upper limit of the integration in $q$ as $2E_1-q_0$
(as in \citealp{Reddy+Prakash+Lattimer.1998}) instead
of as $E_1+E_3$.

Eq.~\eqref{eq:reaction_inverse} has to be solved for the neutrino and
antineutrino emission (i.e., the inverse reactions of the neutrino and
antineutrino absorption), but in the case of the inverse scattering, where particles
$2\equiv4$ and $1\equiv3$, one can use
a simplified expression.
In fact, the principle of detailed balance assures that in a systems
invariant under spatial reflections, the following relation holds for the
transition rate $W_{fi}$ \citep{Iwamoto+Pethick.1982}
\begin{equation}
W_{fi}(\mathbf q,q_0) = \mathrm e^{\frac{q_0}T} W_{fi}(\mathbf q,-q_0),
\end{equation}
from which one has, since $\mu_1\equiv\mu_3$,
\begin{align}
\tilde W(-q_0,4,2)={}&\mathrm e^{\frac{(E_3-\mu_3)-(E_1-\mu_1)}{T}} \tilde W(q_0,2,4),\\
\frac{\sigma^\mathrm{s,i}}V={}&-\mathrm e^{-\frac{E_1-\mu_1}T} \frac{\sigma^\mathrm{s,d}}V,
\end{align}
and, finally,
\begin{equation}
\label{eq:sigma_scattering}
\frac{\sigma^\mathrm{s,tot}(E_1)}V=\frac{\sigma^\mathrm{s,i}(E_1)}V+\frac{\sigma^\mathrm{s,d}(E_1)}V=\frac1{1-f_1(E_1)}
\times\frac{\sigma^\mathrm{s,d}(E_1)}V,
\end{equation}
where the superscript ``s'' implies that we are only considering the scattering
reactions.

\clearpage
\addcontentsline{toc}{chapter}{Bibliography}
\bibliography{bibliografia}
\bibliographystyle{apalike}
\end{document}